\documentclass[11pt,a4paper]{article}
\usepackage{epsfig}
\usepackage[T1]{fontenc}    
\usepackage{graphics}
\usepackage{graphicx}
\usepackage{pstricks,pst-coil,pst-fill,pst-plot}
\usepackage[fleqn]{amsmath}    
\usepackage{amssymb}    
\usepackage{amsfonts}   
\usepackage{verbatim}   
\usepackage{mathrsfs}   
\usepackage{dsfont}
\usepackage{euscript}
\usepackage{yfonts}
\usepackage{enumerate}     
\usepackage{amsthm}         
\usepackage{txfonts}
\usepackage{marvosym}
\usepackage{stmaryrd}
\usepackage{vmargin}        
\usepackage{wasysym}		

\setmarginsrb{1.8cm}{2cm}{1.8cm}{2cm}{1cm}{1cm}{1cm}{1.6cm}
 \makeatletter
 \@addtoreset{equation}{section}
 \makeatother

\providecommand{\bysame}{\leavevmode\hbox to3em{\hrulefill}\thinspace}
\providecommand{\MR}{\relax\ifhmode\unskip\space\fi MR }

\providecommand{\href}[2]{#2}

\let\ua=\uparrow
\let\da=\downarrow
\let\tend=\rightarrow

\long\def\symbolfootnote[#1]#2{\begingroup%
\def\thefootnote{\fnsymbol{footnote}}\footnote[#1]{#2}\endgroup}

\newtheorem{theorem}{Theorem}[section]
\newtheorem{prop}[theorem]{Proposition}
\newtheorem*{theorem*}{Theorem}
\newtheorem{cor}[theorem]{Corollary}

\newtheorem{lemme}[theorem]{Lemma}

\def\Proof{\medskip\noindent {\it Proof --- \ }}

\def\qed{\hfill\rule{2mm}{2mm}}

\newcommand\beq{\begin{equation}}
\newcommand\enq{\end{equation}}
\newcommand\bem{\begin{multline}}
\newcommand\enm{\end{multline}}

\def\beqa{\begin{eqnarray}}
\def\eeqa{\end{eqnarray}}
\def\ba{\begin{array}}
\def\ea{\end{array}}
\def\det{\operatorname{det}}

\newcommand{\f}[2]{{\ensuremath{%
    \mathchoice%
    {\dfrac{#1}{#2}}
    {\dfrac{#1}{#2}}
    {\frac{#1}{#2}}
    {\frac{#1}{#2}}
}}}
\newcommand{\tf}[2]{\ensuremath{#1/#2}}

\def\a{\alpha}

\def\be{\beta}
\def\ga{\gamma}
\def\Ga{\Gamma}

\def\de{\delta}

\def\eps{\epsilon}

\def\la{\lambda}
\def\La{\Lambda}

\def\sg{\sigma}

\def\Sg{\Sigma}

\def\Ups{\Upsilon}
\def\ups{\upsilon}
\def\th{\theta}

\def\Om{\Omega}

\def\vp{\varphi}

\newcommand{\mc}[1]{\ensuremath{\mathcal{#1}}}
\newcommand{\mf}[1]{\ensuremath{\mathfrak{#1}}}
\newcommand{\msc}[1]{\ensuremath{\mathscr{#1}}}

\newcommand{\bs}[1]{\ensuremath{\boldsymbol{#1}}}

\DeclareFontFamily{OT1}{pzc}{}
\DeclareFontShape{OT1}{pzc}{m}{it}{<-> s * [1.10] pzcmi7t}{}
\DeclareMathAlphabet{\mathpzc}{OT1}{pzc}{m}{it}

\def \i{ \mathrm i}

\newcommand{\ov}[1]{\ensuremath{\overline{#1}}}
\newcommand{\wt}[1]{\ensuremath{\widetilde{#1}}}
\newcommand{\wh}[1]{\ensuremath{\widehat{#1}}}

\newcommand{\Int}[2]{\ensuremath{\int\limits_{#1}^{#2}}}

\newcommand{\Fint}[2]{\ensuremath{\fint\limits_{#1}^{#2}}}

\newcommand{\sul}[2]{\ensuremath{\sum\limits_{#1}^{#2}}}
\newcommand{\pl}[2]{\ensuremath{\prod\limits_{#1}^{#2}}}

\newcommand{\R}{\ensuremath{\mathbb{R}}}
\newcommand{\Cx}{\ensuremath{\mathbb{C}}}

\newcommand{\Dp}[1]{\ensuremath{\partial_{#1}}}

\newcommand{\limit}[2]{\ensuremath{\underset{#1 \tend #2}{\longrightarrow} }}

\newcommand{\ex}[1]{\ensuremath{\e{e}^{#1}}}

\newcommand{\op}[1]{ \boldsymbol{ \texttt{#1} } }

\newcommand{\Norm}[1]{\ensuremath{\big| \big| #1 \big|\big| }}
\newcommand{\norm}[1]{\ensuremath{  || #1 || }}

\newcommand{\moy}[1]{\ensuremath{\langle #1 \rangle}}

\newcommand{\dd}{\mathrm{d}}
\newcommand{\e}[1]{\ensuremath{\mathrm{#1}}}

\newcommand{\intff}[2]{\ensuremath{ [  #1 \,; #2 ] }}
\newcommand{\intfo}[2]{\ensuremath{ [  #1 \,; #2 [ }}
\newcommand{\intof}[2]{\ensuremath{ ]  #1 \,; #2 ] }}
\newcommand{\intoo}[2]{\ensuremath{ ]  #1 \,; #2 [ }}

\newcommand{\intn}[2]{\ensuremath{[\![ \, #1 \,;\, #2 \,]\!]}}

\newcommand{\widesim}[2][1.5]{
  \mathrel{\underset{#2}{\scalebox{#1}[1]{$\sim$}}}
}

\begin{document}

\begin{center}
\begin{LARGE}
{\bf On convergence of form factor expansions in the infinite volume quantum Sinh-Gordon model in 1+1 dimensions}
\end{LARGE}

\vspace{1cm}

\vspace{4mm}
{\large Karol K. Kozlowski \footnote{e-mail: karol.kozlowski@ens-lyon.fr}}%
\\[1ex]
Univ Lyon, ENS de Lyon, Univ Claude Bernard Lyon 1, CNRS, Laboratoire de Physique, F-69342 Lyon, France \\[2.5ex]

\par 

\end{center}

\vspace{40pt}

\centerline{\bf Abstract} \vspace{1cm}
\parbox{12cm}{\small}

This paper develops a technique allowing one to prove the convergence of a class of series of multiple integrals which corresponds to the form factor expansion
of space-like separated two-point functions in the 1+1 dimensional massive integrable Sinh-Gordon quantum field theory.

\vspace{20pt}


\vspace{40pt}

\tableofcontents

\section{Introduction}

\subsection{The physical background of the problem}

The $\op{S}$-matrix program, initiated by Heisenberg \cite{HeisenbergSomeAspectsofSMatrixIdeas} and Wheeler \cite{WheelerFirstIntroConceptSmatrix}, was actively investigated in the 60s and 70s.
Its conclusions where rather unsatisfactory in spacial dimensions higher than one:
on the one hand due to the triviality of the $\op{S}$-matrix as soon as a given model exhibits a local conservation law other than the energy-momentum \cite{ColemanMandulaTrivialityOfSMatrixInHigherDWhenManyCharges} and, on the other hand,  due 
to the incapacity of constructing viable, explicit, $\op{S}$-matrices for any model missing such properties. However, the situation turned out to be drastically different for quantum field theories in 1+1 dimensions. 
The pioneering work of Gryanik and Vergeles \cite{GryanikVergelesSMatrixAndOtherStuffForSinhGordon} developed the first aspects of a method allowing one to determine $\op{S}$-matrices for quantum analogues of classical
1+1 dimensional field theories having an infinite set of independent local integrals of motion. It turns out that the existence of analogous conservation laws on the quantum level 
heavily constrains the classes of "in/out" asymptotic states that can be connected by the $\op{S}$-matrix. The reasoning of \cite{GryanikVergelesSMatrixAndOtherStuffForSinhGordon}
applies to the case of models only exhibiting one type of asymptotic particles, the main example being given by the quantum Sinh-Gordon model. In those cases, the $\op{S}$-matrix is diagonal and 
thus fully described by one scalar function of the relative "in" rapidities of the two particles.  

Dealing with the case of models having several types of asymptotic particles, some of equal masses and others realised as bound states of the former, turned out to be more involved. 
An archetype of such models is given by the Sine-Gordon quantum field theory. Building on the factorisability of the $n$-particle $\op{S}$-matrix into two-particle processes
and on the independence of the order in which a three particle scattering process arises from a concatenation of two-particle processes -which is captured by the celebrated Yang-Baxter equation 
\cite{BaxterPartitionfunction8Vertex-FreeEnergy,YangFactorizingDiffusionWithPermutations}-
Zamolodchikov derived the $\op{S}$-matrix of the Sine-Gordon model in the soliton-antisoliton sector \cite{ZalmolodchikovSMatrixSolitonAntiSolitonSineGordon}
and managed to argue the model's asymptotic particles content upon relying on Faddeev-Korepin's \cite{KorepinFaddeevQuantisationOfSolitions} semi-classical quantisation results of the solitons 
in the classical Sine-Gordon model. Later, the $\op{S}$-matrix related to the soliton bound-state sectors, built out from the so-called breathers, was given in \cite{KarowskiThunCompleteSMatrixThirring}. 
This provided the full $\op{S}$-matrix of the model since all asymptotic particles were now taken under consideration. 
Since then, $\op{S}$-matrices of many other models have been found, see \textit{e.g.} \cite{ArinshteinFateyevZamolodchikovSMatrixTodaChain,ZalZalBrosFactorizedSMatricesIn(1+1)QFT}.
In fact, the factorisability of the $n$-particle $\op{S}$-matrix into two-particle processes was 
later established by Iagolnitzer within the $\op{S}$ matrix axiomatics for theories having one  
\cite{IagolnitzerFactorisabilityOfSMatrixIn2D} or several \cite{IagolnitzerFactorisabilityOfSMatrixIn2DMoltiPartModels} asymptotic particles, under the hypotheses
of macrocausality, causal factorisation, absence of particle production, conservation of individual particle momenta throughout the 
scattering\symbolfootnote[2]{the later is implied by the existence of an infinite set of conserved quantities on the quantum level.}
and validity of the Yang-Baxter equation in the case of models having multiple asymptotic particles with internal degrees of freedom.
Later, Parke \cite{ParkeFacotrisabilitySmatrixin1+1theoriesWithConsLaws} argued, on a more loose level of rigour,  the same result solely in the presence of two extra conserved quantities in addition 
to the energy-momentum conservation.

Yet, the main physical interest does not reside in the \textit{per se} calculation of $\op{S}$-matrices -which is just an intermediate tool in reaching the goal-
but rather in being able to obtain a thorough description of correlation functions which are the observables being directly measured in experiments. Doing so may be achieved by obtaining the matrix elements of the local operators in the theory 
taken between the  asymptotic states. Such quantities are called form factors. The full characterisation of the $\op{S}$ matrix of the Sine-Gordon model allowed Weisz \cite{WeisztwoparticleFFForSolitionAntiSolitionInSineGordon}
to argue an expression for a specific 
form factor of the model: the matrix element of the electromagnetic current operator taken between an incoming and outgoing solition. The approach allowing one to calculate systematically
the form factors in massive integrable 1+1 dimensional quantum field theories starting from the $\op{S}$-matrix has been initiated by Karowski, Weisz \cite{KarowskiWeiszFormFactorsFromSymetryAndSMatrices}
who wrote down a set of equation satisfied by a model's n-particle form factors and provided closed expressions for two particle form factors in several models. 
The calculation of form factors was subsequently addressed within the recently developed quantum inverse scattering method \cite{FaddeevSklyaninTakhtajanSineGordonFieldModel}
which allowed to set up a quantum version of the Gelfand-Levitan-Marchenko equations allowing one to describe some of the operators arising in the quantum field theory model by means of solutions 
to certain singular integral equations involving free quantum fields satisfying the Faddeev-Zamolodchikov algebra. Their iterative solution expresses 
formally the quantum fields in terms of series of multiple integrals involving the free fields. The carry out of this program was implemented by Smirnov on the example of 
the Sine-Gordon model in the works \cite{SmirnovUseGLMEqnsTocomputeSineGordonFF,SmirnovGLMEqnsDerivationForSineGordon,SmirnovUseGLMEqnsForSineGordonFF}. 
This allowed him to obtain combinatorial expressions for the form factors of the exponential of the field operator, this for all types of asymptotic particles
of the theory, in particular, reproducing the formulae for certain two-particle form factors derived earlier by Karowski, Weisz \cite{KarowskiWeiszFormFactorsFromSymetryAndSMatrices}. 
Smirnov's approach was improved in the works \cite{SmirnovIntegralRepSolitonFFSineGordonBootstrap,SmirnovProofIntegralRepSolitonFFSineGordonBootstrapFollowUpJPhysAPaper} where 
a set of auxiliary equations satisfied by the form factors was singled out and solved explicitly. 
This provided the first fully explicit representations for \textit{all} the form factors of the Sine-Gordon model\symbolfootnote[3]{The papers only contain soliton form factors
but the soliton/breather and breather/breather form factors can then by directly obtained by simple residue computations.}.
 Around that period Khamitov \cite{KhamitovGLMeqnsForSinhGordonAndProofOfLocalCommutativity} constructed certain local operators for the quantisation of the classical 1+1 dimensional 
Sinh-Gordon field theory by means of formally solving the associated system of Gel'fand-Levitan-Marchenko equations for the exponentials of the field operators.  
He then showed, by using certain identities established by Kirillov \cite{KirillovCombinatorialIdenititesForLocalCommutativityinSinhGordonQFTByKhamitov}, 
that the final combinatorial expressions he got for the form factors do ensure that the associated operators satisfy the CPT invariance and 
the local commutativity, \textit{viz}. that two local operators located at the space-time points $\bs{x}$ and $\bs{y}$, with $\bs{x}-\bs{y}$ being a space-like vector, commute.
I stress that this property is the necessary ingredient for having a causal theory. 
This approach opened up an important change of perspective in the construction of a quantum counterpart of a integrable classical field theory
in that once one is able to propose some expressions, be it combinatoral or fully explicit, for the form factors of local operators taken between asymptotic states and 
then check independently that these satisfy CPT invariance and ensure that the operators satisfy local commutativity, then one may assert that one has built 
a \textit{per se} quantum field theory in that the fundamental requirements thereof are satisfied. 

This point of view was raised to its full glory by Kirillov, Smirnov \cite{KirillovSmirnovFirstCompleteSetBootstrapAxiomsForQIFT,KirillovSmirnovUseOfBootstrapAxiomsForQIFTToGetMassiveThirringFF}
on the example of the massive Thirring model. The authors postulated, as an axiomatic first input of the construction, a set of equations satisfied by the form factors of  
the model and which involve the model's $\op{S}$-matrix: this setting constitutes what is called nowadays the bootstrap program for the form factors.
Those bootstrap equations contain some of the equation already argued in \cite{KarowskiWeiszFormFactorsFromSymetryAndSMatrices}, but also additional ones. 
It was shown by Kirillov, Smirnov \cite{KirillovSmirnovFirstCompleteSetBootstrapAxiomsForQIFT,KirillovSmirnovUseOfBootstrapAxiomsForQIFTToGetMassiveThirringFF} 
that operators whose form factors satisfy the bootstrap equations do satisfy the local commutativity property. 
Hence, in the case of massive quantum integrable field theories, the bootstrap equations can be taken as an axiomatic input of the theory. The 
resolution of the bootstrap equations, if possible, then allows one to define the local operators of the theory as matrix operator.
The resolution of the bootstrap program was systematised over the years and these efforts led to explicit expressions for the form factors of local operators in numerous 1+1 dimensional massive quantum field theories, see \textit{e.g.} 
\cite{SmirnovFormFactors}. The first expressions for the form factors were rather combinatorial in nature. Later, a substantial progress was achieved in simplifying the latter, 
in particular by exhibiting a deeper structure at their root. Notably, one can mention the free field based approach,  also called angular quantisation,
to the calculation of form factors. It was introduced by Lukyanov \cite{LukyanovFirstIntroFreeField} and allowed to obtain convenient representations for certain form factors solving the bootstrap program. 
In particular, the construction led to closed and manageable expressions \cite{BrazhnikovLukyanovFreeFieldRepMassiveFFIntegrable} for the form factors of the exponential of the field operators in the Sinh-Gordon and the Bullough-Dodd models.
While first results were obtained for the form factors 
of operators going to primary operators\symbolfootnote[2]{Those are specific
operators in the model whose ultraviolet behaviour, \textit{viz}. when considering the short distance behaviour of a multi-point correlation function of local operators, 
may be grasped to the leading order by a multi-point correlation functions involving primary operators and computed in the conformal field theory that falls within the ultrafiolet universality class of the considered model.}
the approach was generalised so as to encompass the form factors of the descendant operators \cite{FeiginLashkevichFreeFieldApproachToDescendents,LashkevichPugaiDescendantFFResonanceIdentites}. 
Also, one should mention that the free field construction of the form factors was refined into the $p$-function method, developed for the  Sine-Gordon model in
\cite{BabujianFringKarowskiZapletalExactFFSineGordonBootsstrapI,BabujianKarowskiExactFFSineGordonBootsstrapII,BabujianKarowskiBreatherFFSineGordon}.  
In this last approach, the non-trivial part of the form factors is expressed in terms of the action of some explicit operator on a simple symmetric function $p$ of many variables 
satisfying simple constraints. The choice of different $p$ functions gives rise to different local operators of the model. 

\subsection{The open mathematical problems related to integrable quantum field theories}

Despite its importance, the sole construction of the form factors, even if very explicit, cannot be considered as the end of the story. 
Indeed, one may think of bootstrap program issued explicit expressions for the form factors 
as a way to define the local operators of the theory as matrix operators. Still, from the perspective of physics, 
one would like to go further, on the one hand, by being able to establish that such  operators do enjoy certain properties and, on the other hand, by being able to characterise the model's multi-point vacuum-to-vacuum correlation functions
built out of such operators.  The most essential of such properties is that the form factors obtained as solutions to the bootstrap program do give rise to local matrix operators which enjoy the local commutativity.
As already mentioned,  this was established by Kirillov and Smirnov  
\cite{KirillovSmirnovFirstCompleteSetBootstrapAxiomsForQIFT,KirillovSmirnovUseOfBootstrapAxiomsForQIFTToGetMassiveThirringFF}. More precisely, they considered the matrix elements of the commutator taken between 
asymptotic states and expressed explicitly the products of the two local operators building up the commutator by using their matrix elements, the form factors. This 
thus expressed the commutator as a series of multiple integrals corresponding to a summation over all types of asymptotic states in the theory; namely a summation over all the various types of asymptotic particles, their internal degrees of freedom,
and an integration over all their possible rapidities. Thus, on top of using various algebraic properties issuing from the boostrap equations, the Kirillov and Smirnov  
\cite{KirillovSmirnovFirstCompleteSetBootstrapAxiomsForQIFT,KirillovSmirnovUseOfBootstrapAxiomsForQIFTToGetMassiveThirringFF} approach for establishing local commutativity also demands to rely
on the convergence of the form factor series which are handled in the proof. In fact, this convergence property is what is also needed to ensure that the product of two matrix operators
constructed through the bootstrap program is well-defined. A similar issue with convergence also arises when computing the multi-point vacuum-to-vacuum correlation functions. 
Indeed, by taking explicitly the products of each operator  building up the correlator  and averaging them over the vacuum, as in the case of the commutator, these may be expressed in terms of series of multiple integrals
called form factor expansion.
Hence,  the very possibility to describe vacuum to vacuum multi-point correlation functions through their form factor expansions
also demands to have established the convergence of these series of multiple integrals. While constituting an important mathematical ingredient for the rigorous construction
of the bootstrap program solvable 1+1 dimensional massive quantum field theories, the convergence of form factors based series of multiple integrals is basically a completely open problem. 

Numerical investigations, see \textit{e.g.} \cite{EsslerKonikSpectralFunctionsMAssiveIQFT}, of the magnitude of the higher particle number
form factors contribution to a two-point function seems to indicate that form factor expansions
should converge quickly, this even for moderate separations between the operators. However, this does not constitute by any means a proof thereof. In fact, the sole proof of convergence was achieved by Smirnov, 
in an unpublished note, relatively to the form factor expansion
of a specific two-point function in the Lee-Yang model \cite{CardyMussardoSMatrixForLeeYang,ZalmolodchikovTwoPointFctsLeeYang}. By using the very specific form of the form factors
of certain operators in that model Smirnov was able to bound explicitly the form factors in terms of explicitly summable positive functions. 
The given proof was however lacking a generality that would allow it to be extended to other models where the expressions for the form factors are more intricate.

\subsection{The main result}

The purpose of this work is to develop a technique allowing one to prove the convergence of the form factor expansion associated with of the vacuum-to-vacuum space-like separated two-point functions 
of a large -if not full- class of local operators in the simplest massive 1+1 dimensional integrable quantum field theory: the quantum Sinh-Gordon model. 

Those series takes the form 
\beq
\mc{U}(\mf{r} ) =\sul{N \geq 0}{} \mc{U}_{N}( r ) \qquad  \e{with} \qquad  r  >0 \; ,
\enq
where the $N^{\e{th}}$-summand takes the form
\beq
\mc{U}_N( r  )   \; = \;
\Int{  \R^N }{}   \f{  \dd^N \be  }{ N! ( 2\pi )^{N}  }   
\pl{a\not= b }{N } \Big\{ \ex{ \f{1}{2} \mf{w}(\be_{ab}) } \Big\} \cdot  \pl{a=1}{N}  \Big\{ \ex{-r \,   \cosh(\be_a) }\Big\} \cdot    \mc{K}_{N}^{(\mc{O}_1)}(\bs{\be}_N)
\cdot   \mc{K}_{N}^{(\mc{O}_2)}(\bs{\be}_N)\;. 
\label{introduction forme serie FF}
\enq
%
%
The function $\mc{K}_{N}^{(\mc{O})}\big(  \bs{\be}_{N}\big)$   is given in terms of a combinatorial sum involving a parameter $\mf{b}\in \intoo{0}{\tf{1}{2}}$:
\beq
 \mc{K}_{N}^{(\mc{O})}\big( \bs{\be}_{N} \big) \, = \;  
 \sul{ \bs{\ell}_N \in \{0,1\}^N }{} 
\pl{a<b}{N} \bigg\{ 1 \, - \, \i \f{ (\ell_{a}-\ell_{b}) \cdot \sin[2\pi \mf{b} ] }{ \sinh(\be_{a}-\be_{b})  }  \bigg\} \cdot  p_{N}^{(\mc{O})}\big( \bs{\be}_N\mid \bs{\ell}_N \big) \;, 
\enq
where $\bs{\ell}_N=(\ell_1,\dots, \ell_N)$. The functions $p_{N}^{(\mc{O})}$ depend on the operator considered. Several examples of such functions can be found in 
Subsection \ref{SousSection FF pour plusieurs type de Ops}. 
%
%
%
%
%
%
%
%
%
%

In its turn, the two-body interaction potential $\mf{w}$ is defined, for  $\la \in \R^{*}$, by means of the below Riemann integral of a slowly decaying oscillating integrand\symbolfootnote[2]{Here and in the following,
such integrals understood as $\underset{M_1, M_2 \tend +\infty}{\lim} \Int{-M_1}{M_2} J(x) \ex{\i \la x}$ \;,
in the case of integrands behaving as $J(x) = \f{ C_{\pm} }{ |x|^{\a_{\pm} }  } \Big( 1 \, + \, \e{O}\big( |x|^{-1} \big) \Big) $
when $x\tend \pm \infty$ with $\a_{\pm}>0$ and $C_{\pm}\in \R^*$}:
\beq
\mf{w} (\la)  \; = \;  - 4  \Int{\R }{} \dd x  \, \f{ \sinh(x \mf{b}) \cdot \sinh(x \hat{\mf{b}} ) \cdot \sinh(\tfrac{1}{2}x) \cdot \cosh(x)  }{ x \cdot  \sinh^2(x)  }
\ex{\i \frac{\la x }{ \pi } }   \;,
 \qquad \e{with} \qquad \hat{\mf{b}}=\f{1}{2}-\mf{b} \;. 
\enq
Standard techniques of complex analysis entail that $\mf{w}$ has a logarithmic singularity at the origin $\mf{w} (\la)  \; = \; 2 \ln |\la| + \e{O}(1)$ and decays to zero exponentially fast at $\pm \infty$. 
I refer to Section \ref{Section Form factor series} for more details on why form factor expansions of the vacuum-to-vacuum space-like separated two-point functions 
are represented by the series of multiple integrals given above. There one will also find the discussion of the well-definiteness of the multiple integrals $\mc{U}_N(r)$.

The main result of this paper is gathered in Theorem \ref{Theorem principal de larticle} below.  
\begin{theorem}
\label{Theorem principal de larticle}
Let $p_{N}^{(\mc{O}_1)}, p_{N}^{(\mc{O}_2)}$ be bounded as 
\beq
\big| p_{N}^{(\mc{O}_k)}\big( \bs{\be}_N\mid \bs{\ell}_N \big) \big|  \; \leq \; C_1^N \cdot \pl{a=1}{N} \ex{ C_2 |\be_a|^s }
\label{ecriture borne sup hypothese sur les pn}
\enq
for some $N$-independent constants $C_1, C_2, s>0$.

Then, for any $r_{0}>0$, the series defining the function $\mc{U}( r )$ converges uniformly in
\beq
r\in \intfo{ r_{0} }{ + \infty}
\enq
and one has the upper bound
\beq
 \big| \mc{U}_N( r ) \big|   \; \leq  \; \exp\Bigg[  -   \f{ 3 \pi^4 \mf{b} \, \hat{\mf{b}} \cdot N^2}{ 4 \cdot (\ln N )^3  }  \cdot \bigg\{ 1+ \e{O}\Big(   \f{ 1 }{ \ln N }\Big) \bigg\}    \Bigg] \;.
\label{ecriture upper bound of main theorem}
\enq
\end{theorem}
I stress that the remainder is not uniform in $r \rightarrow {0}^{+} $ in accordance with the expected power-law in $r$ ultraviolet $r \tend 0^+$ behaviour of
$\mc{U}(r)$. Indeed, then one expect the series to loose its convergence properties and $\mc{U}(r)$ should exhibit a power-law behaviour in $r\tend 0^+$.

The technique allowing one to establish this result takes its root in the probabilistic approach to extracting the large-$N$ behaviour of multiple integrals arising in random matrix theory and its refinements to more complex
multiple integrals \cite{AndersonGuionnetZeitouniIntroRandomMatrices,BenArousGuionnetLargeDeviationForWignerLawLEadingAsymptOneMatrixIntegral}, builds on certain features of potential theory \cite{SaffTotikLogarithmicPotential},
singular integral equations of truncated Wiener-Hopf type \cite{NovokshenovSingIntEqnsIntervalGeneral} and the Deift-Zhou non-linear steepest descent \cite{DeiftZhouSteepestDescentForOscillatoryRHPmKdVIntroMethod}
asymptotic analysis of matrix valued Riemann--Hilbert problems. 
To start with, one establishes an upper bound on $ \big| \mc{U}_N(r) \big| $ in terms of certain auxiliary $N$-fold integrals only involving one and two-body interactions between the integration variables.
After some reductions, the large-$N$ behaviour of the auxiliary integral is then estimated in terms of the infimum over the space of probability measures on $\R$ of the quadratic functional 
\beq
\mc{E}_{N}^{(+)}\big[\sg \big]  \, = \, \f{ \varkappa  }{ N }   \Int{}{} \cosh( \ln N s)\,  \dd \sg(s) \, - \, \f{ 1 }{ 2 } \Int{}{} \mf{w}^{(+)}\Big( \ln N (s-u) \Big) \cdot \dd \sg(s) \, \dd \sg(u) \;,  
\qquad \varkappa= \f{r}{2} \;,
\enq
where 
\beq
\mf{w}^{(+)}(u) \; = \; \mf{w}(u) + \f{1}{2} \ln \Bigg( \f{ \sinh(u+2\i\pi \mf{b}) \sinh(u - 2\i\pi \mf{b}) }{   \sinh^2(u) } \Bigg)
\enq
so that $\mf{w}^{(+)}(u)  = \ln |u| +\e{O}(1)$ at the origin. This stage of the analysis ultimately yields the upper bound
\beq
 \big| \mc{U}_N(r) \big|    \; \leq  \; \exp\Bigg\{ - N^2 \cdot \e{inf}\Big\{ \mc{E}_{N}^{(+)}\big[\sg \big]  \; : \; \sg \in \mc{M}^{1}(\R) \Big\} \,  +\, \e{O}\big( (\ln N)^2 N \big) \Bigg\} \;.
\enq
Since $\mc{E}_{N}^{(+)}$ depends itself on $N$, the infimum does depend on $N$ so that to conclude relatively to the convergence of the series one needs to evaluate it explicitly, at least to the leading order in $N$, and then to check 
that the infimum has strictly positive large-$N$ behaviour which   dominates $(\ln N)^2/N$ in the large-$N$ limit.

To start with, it is shown that $\mc{E}_{N}^{(+)}$ 
admits a unique  minimiser $\sg_{\e{eq}}^{(N)}$ on $\mc{M}^{1}(\R)$. $\sg_{\e{eq}}^{(N)}$ is shown to be Lebesgue continuous with a density supported on a 
single interval which satisfies a singular integral equation of truncated Wiener-Hopf type. Such integral operators were extensively studied by Krein's school and 
the solution may be expressed in terms of the solution to a $2 \times 2$ matrix Riemann--Hilbert problem, see \cite{NovokshenovSingIntEqnsIntervalGeneral}. 
The presence of a parameter blowing up with $N$  in this problem allows one to apply the Deift-Zhou non-linear 
steepest descent method so as to produce a solution to this matrix Riemann--Hilbert problem in the large-$N$ regime.
In this way, one is able 
to express $\mc{E}_{N}^{(+)}\big[ \sg_{\e{eq}}^{(N)} \big]$ in terms of the Riemann--Hilbert problem data and then build on the established large-$N$
behaviour of its solution so as to extract the leading large-$N$ behaviour of $\mc{E}_{N}^{(+)}\big[ \sg_{\e{eq}}^{(N)} \big]$. This, finally, yields 
Theorem \ref{Theorem principal de larticle}.

The upper bound on $\mc{U}_N(r)$ obtained in Theorem \ref{Theorem principal de larticle} may be put in parallel with the large-$N$ behaviour arising
in the study of partition functions of $\be$-ensembles or the closely connected random matrix models. These take the form 
\beq
\msc{Z}_{N} \; = \; \Int{\R^N}{} \dd^N \la \pl{a<b}{N} |\la_a-\la_b|^{\be} \cdot \pl{a=1}{N} \ex{-N V(\la_a) } \;. 
\enq
It is a classical result, see \textit{e.g.} \cite{AndersonGuionnetZeitouniIntroRandomMatrices,BorotGuionnetAsymptExpBetaEnsOneCutRegime}, that 
for a large class of potentials $V$ leading to the so-called on-cut regime $\msc{Z}_N$ admits, for any $k \geq 0$, the large-$N$ asymptotic expansion
\beq
\msc{Z}_{N} \; = \; C_N \exp\Bigg\{  -N^2 F_0 \, + \, \sul{p = 1}{ k+ 1 } N^{2-p} F_p  \; + \; \e{O}\big( N^{-k}\big) \Bigg\}
\label{ecriture DA fct part Z_N beta ensembles}
\enq
The first term of the asymptotics is called the free energy and $C_N$ is an explicit, potential $V$ independent sequence related to a Selberg integral.  
In the upper bound \eqref{ecriture upper bound of main theorem} given in Theorem \ref{Theorem principal de larticle}, there appears an additional $\ln N$
scale in the large-$N$ asymptotics. Thus, comparing with the above result for $\be$-ensembles, all looks like one ends up with a free energy that is $N$-dependent,
\textit{viz}. $F_0\hookrightarrow F_0(N)$  such that $F_0(N) \widesim{N\tend +\infty} \tf{f_0}{ (\ln N)^3 }$. One could be tempted to argue that all of such a large-$N$ behaviour is due
to  \eqref{ecriture upper bound of main theorem} being an upped bound so that, if one were to access to the true asymptotic behaviour of $\mc{U}_N(r)$,
then the latter would rather follow the typical random matrix like structure \eqref{ecriture DA fct part Z_N beta ensembles}. I would like to stress that it would notation be so since the
two scales -$N$ and $\ln N$- are natural to this problem. While the leading asymptotic behaviour of $\mc{U}_N(r)$ might surely differ from the
upper bound \eqref{ecriture upper bound of main theorem}, one still expects that
\beq
\ln \big[ \mc{U}_N(r) \big] \; \simeq \; -N^2 (\ln N)^p \Bigg\{ Q_{0,0} \; + \; \sul{ \substack{ (s, \ell) \\  \not= (0,0) } }{}  \f{ Q_{s,\ell} }{ N^{\ell}  (\ln N )^s } \Bigg\}
\enq
in the sense of a two-scaled asymptotic expansion. The presence of the two-scales -$N$ and $\ln N$- stems from the lack of a prefactor of $N$ in the exponentially 
growing at infinity one-body confining potential: $r \,  \cosh(\be) $, which has to compete with $N^2$ terms
issuing from the two-body interaction $\mf{w}(\be-\be^{\prime})$ which is repulsive on short distances. Only after going to a $\ln N$
dilatation scale in between the integration variables do the one and two body interactions become of the same scale in $N$ for typical, bounded, 
values of the new integration variables. The source of the $\tf{1}{\big( \ln N \big)^3}$ prefactor in the upper bound is rather subtle
and I do not have a nice, heuristical interpretation thereof.

\vspace{3mm}

To conclude the presentation of the main results of this work,  I will comment on the generality of the method. 
While developed for the case of a specific two-point function in the Sinh-Gordon model with operators being separated by space-time intervals, 
it is clear that the method is applicable to a wide range of situations. First of all, upon a few additional technical steps, 
the method will allow to also determine convergence in the case of time-like separation between the operators building up a two point function.  
 Moreover, there does not seem to appear any
obstruction so as to apply the method so as to prove the convergence of multi-point correlation functions in the model as well as the one 
of other that the vacuum-to-vacuum expectation values. As such, upon appropriate modifications, this allows to finish the proof, for this model, of the causality property
of the local operators constructed through the bootstrap program. 
Finally, the technique for proving the convergence of form factor expansions developed in this work only relies on very general properties of the form factors
and not their detailed form. Hence, the method looks like a promising path which would allow one to solve
the convergence problem of form factor expansions in more complex integrable quantum field theories such as the Sine-Gordon model.

\subsection*{Outline of the work}

The paper is organised as follows. Section \ref{Section facteurs de forms dans Sinh Gordon} reviews the bootstrap program issued results providing 
explicit expressions for the form factors of local operators in the 1+1 dimensional Sinh-Gordon model.  Section \ref{Section Form factor series} presents the 
general structure of the form factor expansion of two-point functions, in Euclidian space-time, of the model. I establish an upper bound for the $N^{\e{th}}$ summand of this series in the case of space-like separated operators. 
It is shown in Section \ref{Section majorant series FF et pblm minimisation} that the large-$N$ behaviour of this upper bound can be estimated by solving a minimisation problem. The unique solvability 
of this minimisation problem is then established in Section \ref{Section etude pblm minimisation fnelle energie adequate}. 
The characterisation of the minimiser is then reduced to the resolution of a singular integral equation of truncated Wiener-Hopf type in 
Section \ref{Section characterisation minimiseur et eqn integrale singuliere}.  Section \ref{Section auxiliary RHP}
develops the non-linear steepest descent solution of an auxiliary matrix Riemann--Hilbert problem whose solution 
plays a central role in the inversion of the singular integral operator arising in the characterisation of the minimiser. 
 Section \ref{Section Inversion singular integral operator} builds on these results so as to produce a closed form, for $N$ large enough, 
of this operator's inverse. Then, Section \ref{Section caracterisation complete de la mesure eq} utilises these results so as to 
provide an explicit description of the equilibrium measure associated with the minimisation problem. Finally, Section \ref{Section comportement a grand N de la fnelle energi en la mesure eq}
carries out the large-$N$ estimates of the minimisation problem what allows one to  conclude relatively to the convergence of the series.

\section{Form factors in the quantum Sinh-Gordon model}
\label{Section facteurs de forms dans Sinh Gordon}

The classical Sinh-Gordon model describes the evolution of a scalar field $\vp(x,t)$ under the partial differential equation 
\beq
\Big(\Dp{t}^2- \Dp{x}^2\Big)  \vp \, + \, \f{m^2}{g} \sinh(g\vp) \, = \, 0
\enq
which is associated with the extremalisation condition for the action $\mc{S}[\vp]=\Int{}{}\mc{L}(\vp,\Dp{\mu}\vp) \cdot \dd^2x$ 
subordinate to the Lagrangian 
\beq
\mc{L}(\vp,\Dp{\mu}\vp)  \, = \, \f{1}{2} \Dp{\mu} \vp \Dp{}^{\mu} \vp \, - \, \f{ m^2 }{ g^2 } \cosh(g \vp) \;. 
\enq
In infinite volume, the quantum  field theory  underlying to this classical field theory is associated with the Hilbert space
\beq
\mf{h}_{\e{SG}} \, = \, \bigoplus\limits_{n=0}^{+\infty} L^2(\R^n_{>}) \qquad \e{with} \qquad \R^n_{>} \; = \; \Big\{ \bs{\be}_{n}=(\be_1,\dots, \be_n) \in \R^n \; : \; \be_1>\dots>\be_n  \Big\} \;. 
\enq
Here $f\in L^2( \R^n_{>} )$ has the physical interpretation of an "in", \textit{viz}. incoming, asymptotic $n$-particle wave-packet. More precisely, on physical grounds, one interprets elements of the
Hilbert space $\mf{h}_{\e{SG}}$ as parameterised by $n$-particles states, $n\in \mathbb{N}$,  arriving, in the remote past, with well-ordered rapidities $\be_1>\dots>\be_n$. 
Such states are called asymptotic "in" states. Then, within this physical picture, as time goes by, the "in" particles approach each other, interact, scatter and finally travel again as free particles out of the system. Within such a scheme, 
an "out" $n$-particle state is then paramaterised by $n$ well-ordered rapidities  $\be_1<\dots < \be_n$. In fact, one could equivalently associate the model with the Hilbert space 
\beq
 \bigoplus\limits_{n=0}^{+\infty} L^2(\R^n_{<}) \qquad \e{with} \qquad \R^n_{<} \; = \; \Big\{ \bs{\be}_{n}=(\be_1,\dots, \be_n) \in \R^n \; : \; \be_1<\dots<\be_n  \Big\}\; , 
\enq
in which elements of the $L^2(\R^n_{<})$ spaces have the physical interpretation of an "out", \textit{viz}. outgoing in the distant future, asymptotic $n$-particle wave-packet.

\vspace{2mm}

Within the formalism of quantum field theory, the "in" and "out" states are connected by the model's $\op{S}$ matrix. 
In the case of the Sinh-Gordon 1+1 dimensional quantum field theory,  the $\op{S}$-matrix  was first determined in \cite{GryanikVergelesSMatrixAndOtherStuffForSinhGordon}. It corresponds to a diagonal scattering 
between the particles and takes the form 
\beq
\op{S}(\be)\, = \, \f{ \tanh\big[ \tfrac{1}{2}\be - \i \pi  \mf{b}   \big]  }{ \tanh\big[ \tfrac{1}{2}\be + \i \pi  \mf{b}   \big]   }  \qquad \e{with} \qquad  \mf{b}\, = \,   \f{1}{2} \f{ g^2  }{ 8\pi + g^{2}  } \;. 
\enq
This $\op{S}$-matrix satisfies the unitarity $\op{S}(\be)\op{S}(-\be)=1$, and crossing  $\op{S}(\be)=\op{S}(\i \pi-\be)$ symmetries. Moreover, the $\op{S}$ matrix is $2\i\pi$ periodic in $\be$ and has simple poles at 
\beq
\be \, \in \; \i \pi  \, + \,  2 \i\pi \mf{b} \, + \, 2\i \pi \mathbb{Z} \quad \e{and} \quad 
\be \, \in \;  -   2 \i\pi \mf{b} \, + \, 2\i \pi \mathbb{Z} \;. 
\enq
For $g\in \R^*$, one has $0<\mf{b}< \tf{1}{2}$, what ensures that $\op{S}$ has no poles in the physical strip $0<\Im(\be) < \pi$. This property is consistent with the absence of  bound states in the model. 
The $\op{S}$-matrix has simple zeroes at 
\beq
\be \, \in \; \i \pi  \, -  \,  2 \i\pi \mf{b} \, + \, 2\i \pi \mathbb{Z} \quad \e{and} \quad 
\be \, \in \;     2 \i\pi \mf{b} \, + \, 2\i \pi \mathbb{Z}\;. 
\enq
The two zeroes $\i \pi  \, -  \,  2 \i\pi \mf{b}$ and $ 2 \i\pi \mf{b}$ belong to the physical strip and are related by the crossing symmetry 
\beq
\mf{b} \; \hookrightarrow \; \hat{\mf{b}} \, = \, \tf{1}{2} - \mf{b} \;. 
\enq
This is obviously also a symmetry of the model's $\op{S}$-matrix $\op{S}(\be)=\op{S}(\be)_{ \mid \mf{b} \hookrightarrow \hat{\mf{b}} }$. This symmetry reflects the weak/strong duality of the theory, \textit{viz}. the 
invariance of the model's observables under the map $g \hookrightarrow \tf{8\pi}{g}$. 

\vspace{2mm}

In order to realise the quantum field theory of interest on $\mf{h}_{\e{SG}}$ one should provide a thorough and explicit description of the operator content of the
model. In fact, one is interested mainly in so-called local operators which are thought of as objects located at the space-time point $\bs{x}=(x_0,x_1)$. Generically these will be denoted as 
$\mc{O}(\bs{x})$.  The quantum Sinh-Gordon field theory is transitionally  invariant, what means that the model is naturally endowed with a unitary operator $\op{U}_{\op{T}_{\bs{y}}}$ such that 
\beq
\op{U}_{\op{T}_{\bs{y}}} \cdot \mc{O}(\bs{x}) \cdot \op{U}_{\op{T}_{\bs{y}}}^{-1} \, = \, \mc{O}(\bs{x}+\bs{y}) \;. 
\label{ecriture action adjointe operateur de translation}
\enq
The translation operator acts diagonally in the asymptotic states' Hilbert space $\mf{h}_{\e{SG}}$, namely for \newline 
$\bs{f}=(f^{(0)},\dots, f^{(n)},\dots) \in \mf{h}_{\e{SG}}$, it holds 
\beq
\op{U}_{\op{T}_{\bs{y}}} \cdot \bs{f} \; = \; \Big( \op{U}_{\op{T}_{\bs{y}}}^{(0)}\cdot f^{(0)},\dots, \op{U}_{\op{T}_{\bs{y}}}^{(n)}\cdot f^{(n)},\dots \Big)
\quad \e{where} \quad 
\op{U}_{\op{T}_{\bs{y}}}^{(n)} \cdot f^{(n)}(\bs{\be}_{n}) \; = \; \exp\bigg\{\i \sul{a=1}{n}\bs{p}(\be_a)\cdot \bs{y} \bigg\}  f^{(n)}(\bs{\be}_{n})
\enq
in which $\bs{p}(\be) = \big( m\cosh(\be), m \sinh(\be) \big)$ and $\bs{x}\cdot \bs{y}$ stands for the Minkowski $2$-form $\bs{x}\cdot \bs{y} \, = \, x_0y_0-x_1y_1$. 

\vspace{2mm}

In order to describe the action of a local operator $\mc{O}(\bs{x})$ on $\bs{f}=(f^{(0)},f^{(1)},\dots, )$ belonging to an appropriate dense subspace of $\mf{h}_{\e{SG}}$, 
within the bootstrap program, one first introduces the elementary building blocks of the action which arise as
\beq
\Big(\mc{O}(\bs{x})\cdot \bs{f} \Big)^{(0)}\; = \; \sul{m \geq 0}{} \; \Int{ \be_1>\dots > \be_m}{} \hspace{-4mm}  \dd^m \be  \;   \mc{F}_{m;+}^{(\mc{O})}(\bs{\be}_m) \pl{a=1}{m} \Big\{ \ex{- \i \bs{p}(\be_a)\cdot \bs{x} } \Big\}
f^{(m)}\big( \bs{\be}_m \big)   \;. 
\enq
The oscillatory prefactor is simply a consequence of the translation invariance \eqref{ecriture action adjointe operateur de translation}. 
The quantities $\mc{F}_{n}^{(\mc{O})}(\bs{\be}_n)$ are called form factors and are certain meromorphic functions 
in each of the variables $\be_a$ belonging to the so-called physical strip $0\leq \Im (\be_a) \leq \pi$.  $\mc{F}_{m;+}^{(\mc{O})}$
then corresponds to the $+$ boundary value on $\R^m$ of $\mc{F}_{m}^{(\mc{O})}(\bs{\be}_m)$, understood as 
\beq
\mc{F}_{m;+}^{(\mc{O})}(\bs{\be}_m)\; = \; \lim_{ \substack{ \bs{\eps}_m \tend\, \bs{0}  \\ \eps_1>\cdots > \eps_m >0 } } \mc{F}_{m}^{(\mc{O})}(\bs{\be}_m+\i \bs{\eps}_m) \; .
\enq
These quantities, and the axiomatics leading to their characterisation within the bootstrap program, will be discussed below. 

Within the bootstrap approach to quantum integrable field theories, the remaining part of the action of the operator may then be constructed out of the form factors. 
This procedure is, in fact, part of the axioms of the theory. 
 More precisely, the component on higher particle number spaces of the operator's action may be recast, for sufficiently regular functions $\bs{f}$, as 
\beq
\Big(\mc{O}(\bs{x})\cdot \bs{f} \Big)^{(n)}(\bs{\ga}_n) \; = \; \sul{m \geq 0}{} \;   \op{M}_{\mc{O}}^{(m)}\Big( \bs{x} \, \mid     \bs{\ga}_n \Big) \big[f^{(m)}\big] \;,
\enq
in which $\op{M}_{\mc{O}}^{(m)}\big( \bs{x} \, \mid     \bs{\ga}_n \big)$ are certain functions taking values in distributions which act on appropriate spaces of functions in $m$ variables. 
Note that their $x$ dependence follows readily from the translation invariance \eqref{ecriture action adjointe operateur de translation}.  
It is convenient, in order to avoid heavy notations, to represent their action as 
\beq
  \op{M}_{\mc{O}}^{(m)}\Big( \bs{x} \, \mid     \bs{\ga}_n \Big) \big[f^{(m)}\big]  \; = \;
  \pl{a=1}{n} \ex{\i \bs{p}(\ga_a) }\, \cdot \hspace{-3mm}  \Int{ \be_1>\dots > \be_m}{} \hspace{-4mm}  \dd^m \be  \;   \mc{M}_{\mc{O}}\Big(    \bs{\ga}_{n}  ; \bs{\be}_{m} \Big) \cdot \pl{a=1}{m} \ex{-\i \bs{p}(\be_a) \cdot \bs{x} }  \cdot
f^{(m)}\big( \bs{\be}_m \big) \;, 
\enq
and where the integrals should be understood in a distributional sense, \textit{i.e.} the quantities $\mc{M}_{\mc{O}}\big( \bs{x}   \mid   \bs{\ga}_{n}  ; \bs{\be}_{m} \big) $ should be thought of as generalised functions.  
 These generalised functions are postulated to satisfy an 
inductive reduction structure which, again, should be understood in the sense of distributions
\bem
\mc{M}_{\mc{O}}\Big(   \bs{\a}_n ; \bs{\be}_{m} \Big) \; = \;  \mc{M}_{\mc{O}}\Big(    (\a_2,\dots,\a_n) ; (\a_1+\i\pi, \bs{\be}_m) \Big) \\
\; + \; 2\pi \sul{a=1}{m}  \de_{\a_1;\be_a} \pl{k=1}{a-1} \op{S}(\be_k-\a_1) \cdot 
\mc{M}_{\mc{O}}\Big(  (\a_2,\dots,\a_n) ; (\be_1, \dots, \wh{\be}_a , \dots,  \be_m) \Big)\;. 
\label{ecriture des elements de matrices de M entre etars out et in}
\end{multline}
In the above expression, $ \wh{\be}_a$ means that the variable $\be_a$ should be omitted and $\de_{x;y}$ refers to the Dirac mass distribution centred at $x$ and acting on functions of $y$. 
Note in particular that there is no problem with multiplication of distributions in the above formula since the variables are separated. 
Finally, the evaluation at $\a_1+\i\pi$ should be understood in the sense of a meromorphic continuation in the strip $0\leq \Im(z) \leq \pi$ from $\R$ up to $\R+\i\pi$. 
The recursion \eqref{ecriture des elements de matrices de M entre etars out et in} is to be complemented with the initialisation condition $\mc{M}_{\mc{O}}\big( \bs{0}  \mid  \emptyset ; \bs{\be}_{n} \big)\;= \;   \mc{F}_{n}^{(\mc{O})}(\bs{\be}_n)$. 
The recursion may be solved in closed form, see \textit{e.g.} \cite{KirillovSmirnovFirstCompleteSetBootstrapAxiomsForQIFT}, although I will not discuss the form of this solution here
in that it will play no role in the problem to be considered. However, it is clear from the structure of the recursion that the generalised functions 
$\mc{M}_{\mc{O}}\big( \bs{x}  \mid  \bs{\a}_n ; \bs{\be}_{m} \big)$ can be expressed as linear combinations of terms involving
the form factors dressed up by certain products of $\op{S}$-matrices and Dirac masses. This thus justifies the statement that the form factors 
are the elementary building blocks allowing one to define  the action of the operators of the theory.

Finally, one should mention that the generalised functions $\mc{M}_{\mc{O}}\big( \bs{x}  \mid  \bs{\a}_n ; \bs{\be}_{m} \big) $ are invariant under an overall shift of the rapidities
what is a manifestation of the Lorentz invariance of the theory, namely that 
\beq
\mc{M}_{\mc{O}}\Big(    \bs{\a}_n + \theta\,  \ov{\bs{e}}_n ; \bs{\be}_{m} + \theta \, \ov{\bs{e}}_m \Big)  \; = \;
\ex{\theta \op{s}_{\mc{O}} } \cdot \mc{M}_{\mc{O}}\Big(      \bs{\a}_n  ; \bs{\be}_{m}  \Big)  
\label{ecriture invariance FF sous shift global rapidites}
\enq
where $\op{s}_{\mc{O}}$ is called the spin of the operator $\mc{O}$ while  $ \ov{\bs{e}}_{k}=(1,\dots,1)\in \R^k$. 

\vspace{2mm}

One may readily connect this description with the formal picture usually encountered in quantum field theory. In that picture, the 
$\e{in}/\e{out}$ states of the particles with rapidities $\bs{\be}_n$ are formally denoted as $\bs{\mc{A}}_{\e{in}/\e{out}}(\bs{\be}_n)$. Then, 
$\mc{M}_{\mc{O}}\Big( \bs{x} \, \mid     \bs{\a}_n ; \bs{\be}_{m} \Big)$ correspond to the matrix elements of the operator $\mc{O}(\bs{x})$ taken between two asymptotic states
\beq
\mc{M}_{\mc{O}}\Big(     \bs{\a}_n ; \bs{\be}_{m} \Big) \; = \; \Big( \bs{\mc{A}}_{\e{in}}( \bs{\a}_{n}) , \mc{O}(\bs{x})  \bs{\mc{A}}_{\e{in}}(\bs{\be}_{m})\Big)  \; , 
\quad \e{where} \quad \bs{\be}_{k} \, = \, \big( \be_1,\dots, \be_ k \big) \;. 
\enq
In particular, one has that the form factors correspond to the vacuum-to-excited state matrix elements of the operator $\mc{O}(\bs{0})$:
\beq
\mc{F}_{n}^{(\mc{O})}(\bs{\be}_n) \; = \;  \Big( \bs{\mc{A}}_{\e{in}}(\emptyset) , \mc{O}(\bs{0})  \bs{\mc{A}}_{\e{in}}(\be_{1},\dots,\be_n)\Big)  \;. 
\enq
The recursive relation \eqref{ecriture des elements de matrices de M entre etars out et in} may then be interpreted as issuing from the formal LSZ reduction formula \cite{LehmannSymanzikZimmermanLSZReductionFormulaOriginalPaper}. 
See \cite{BabujianFringKarowskiZapletalExactFFSineGordonBootsstrapI,SmirnovFormFactors} for more details.

\subsection{The form factor axioms}

The form factors are postulated to satisfy the below set of bootstrap equations \cite{KirillovSmirnovFirstCompleteSetBootstrapAxiomsForQIFT}:

\begin{itemize}
\item[i)]  $\mc{F}_{n}^{(\mc{O})}(\be_1,\dots, \be_a, \be_{a+1},\dots,  \be_n) \; = \; \op{S}(\be_{a}-\be_{a+1}) \cdot \mc{F}_{n}^{(\mc{O})}(\be_1,\dots, \be_{a+1}, \be_{a},\dots,  \be_n)$;
\item[ii)]  $ \mc{F}_{n}^{(\mc{O})}(\be_1+2\i\pi, \be_2,\dots  ,  \be_n) \; = \; \mc{F}_{n}^{(\mc{O})}(\be_2,\dots,  \be_n,\be_1) 
\, = \, \pl{a=2}{n} \Big\{ \op{S}(\be_{a}-\be_{1}) \Big\}\cdot \mc{F}_{n}^{(\mc{O})}(\be_1,\dots,   \be_n)$; 
\item[iii)] $\mc{F}_{n}^{(\mc{O})}$ is meromorphic in each variable taken singly throughout the strip $0\leq \Im(\be) \leq 2\pi$. Its only poles are simple and located at 
$\i\pi$ shifted rapidities. The residues at these poles enjoy the inductive structure
\beq
-\i \, \e{Res}\bigg(\mc{F}_{n+2}^{(\mc{O})}(\a+\i\pi, \be, \be_1,\dots,   \be_n) \cdot \dd \a \, , \, \a=\be  \bigg) \; = \;
\bigg\{ 1\, - \, \pl{a=1}{n}  \op{S}(\be-\be_{a}) \bigg\} \cdot
\mc{F}_{n}^{(\mc{O})}( \be_1,\dots,   \be_n) \;;
\enq
\item[iv)] $\mc{F}_{n}^{(\mc{O})}$ are boost invariant
\beq
\mc{F}_{n}^{(\mc{O})}( \be_1 + \La,\dots,   \be_n+ \La) \; = \; \ex{\La \op{s}_{\mc{O}} } \cdot \mc{F}_{n}^{(\mc{O})}( \be_1,\dots,   \be_n) \;.
\enq
\end{itemize}

In fact, the above equations can be taken as a set of axioms satisfied by the form factors of the theory.

\subsection{The $2$-particle sector solution}

The form factor axioms in the two-particle sector $n=2$ take the particularly simple form of a scalar Riemann--Hilbert problem in one variable
for a holomorphic function $\op{F}$ in the strip $0\leq \Im(\be) \leq 2\pi $ that has no zeroes in this strip, behaves as $\op{F}(\be) = 1 \, + \, \e{O}\big( \be^{-2} \big)$ as 
$\Re(\be) \tend \pm \infty$ uniformly in $0\leq \Im(\be) \leq 2\pi $ and satisfies 
\beq
\op{F}(\be) \, = \, \op{F}(-\be) \cdot \op{S}(\be)  \qquad \e{and} \qquad 
\op{F}(\i\pi - \be) \, = \, \op{F}(\i\pi + \be)  \;. 
\enq

One can effectively solve these equations by observing that $\op{S}$ admits the integral representation 
\beq
 \op{S}(\be) \; = \; \exp\Bigg\{ 8 \Int{0}{+\infty} \dd x \f{ \sinh(x \mf{b}) \cdot \sinh(x \hat{\mf{b}} ) \cdot \sinh(\tfrac{1}{2}x)   }{ x \sinh(x)  }   \sinh\Big(  \tfrac{ x \be }{ \i \pi } \Big) \Bigg\}
\quad \e{with} \quad \hat{\mf{b}} \, = \, \f{1}{2}\, - \, \mf{b} \; . 
\label{ecriture rep int fct S}
\enq
Following \textit{e.g.} the method of \cite{KarowskiWeiszFormFactorsFromSymetryAndSMatrices}, this yields that 
\beq
\op{F}(\be) \; = \; \exp\Bigg\{ - 4  \Int{0}{+\infty} \dd x \f{ \sinh(x \mf{b}) \cdot \sinh(x \hat{\mf{b}} ) \cdot \sinh(\tfrac{1}{2}x)   }{ x \sinh^2(x)  }   \cos\Big(  \tfrac{ x  }{ \pi }(\i\pi - \be)  \Big) \Bigg\} 
\quad \e{for} \quad 0<\Im(\be) < 2\pi \;. 
\label{ecriture rep int pour F}
\enq

The above integral representation can be obtained by starting from the Cauchy formula valid for  $0 < \Im(\be) < 2\pi$
\beq
\ln \op{F}(\be) \, =   \hspace{-4mm} \Int{ \R \cup \{-\R + 2\i\pi\}  }{} \hspace{-5mm} \f{ \dd s }{ 4\i\pi }  \coth\Big[ \tfrac{1}{2}(s-\be) \Big] \ln \op{F}(s) \; = \; 
\Int{\R}{} \f{ \dd s }{ 4\i\pi }  \coth\Big[ \tfrac{1}{2}(s-\be) \Big] \ln \op{S}(s) 
\enq
and then by taking the $s$ integral by means of the integral representation \eqref{ecriture rep int fct S} for $\ln \op{S}(s)$. 
One may also compute the $s$ integral in a different way by using the Wiener-Hopf factorisation of $\op{S}=\op{S}_{\da}\cdot \op{S}_{\ua}$:
\beq
\op{S}_{\ua}(\be) \; = \; \Ga\left( \ba{c} 1- \mf{z} \\  -\mf{z} \ea \right) \cdot  \Ga\left( \ba{cc} \mf{b} - \mf{z}  &   \hat{\mf{b}} - \mf{z}  \\  1-  \mf{b} - \mf{z}  &  1 - \hat{\mf{b}} - \mf{z}  \ea \right)  
\quad \e{and} \quad 
\op{S}_{\da}(\be) \; = \; \Ga\left( \ba{c} \mf{z} \\ 1+ \mf{z} \ea \right) \cdot  \Ga\left( \ba{cc}  1+\mf{z} - \mf{b}   & 1+\mf{z} -  \hat{\mf{b}}   \\  \mf{z}  + \mf{b}   &  \mf{z} +  \hat{\mf{b}}  \ea \right)  \;, 
\enq
where 
\beq
\mf{z} \, = \, \f{\i \be }{2 \pi } \; \qquad \e{and} \qquad   \Ga\left(\ba{c} a_1,\dots, a_n \\ b_1,\dots, b_{\ell} \ea \right)  \, = \, \f{ \pl{k=1}{n} \Ga(a_k)  }{  \pl{k=1}{n} \Ga(b_k) } \; . 
\label{ecriture convention produit ratios fcts Gamma}
\enq
It is easy to see that $\op{S}_{\ua/\da} \in \mc{O}(\mathbb{H}^{+/-})$ and that $\op{S}_{\ua/\da}(\be) = 1 + \e{O}\big( \mf{z}^{-1} \big)$ when $\mf{z} \tend \infty$\;. 
Then, direct calculations eventually lead to 
\beq
\op{F}(\be) \, = \, \f{1}{  \Ga\Big( 1+\mf{z}, -\mf{z} \Big)}   G\left( \ba{cccc} 1-\mf{b} - \mf{z} \, ,  &   2 - \mf{b} + \mf{z} \,,  &   1- \hat{\mf{b}} - \mf{z}   \, ,  &   2 - \hat{\mf{b}}  + \mf{z}    \\ 
  \mf{b} - \mf{z}   \, ,  &   1+  \mf{b} + \mf{z}   \, ,  &   \hat{\mf{b}} - \mf{z}   \, ,  &   1+  \hat{\mf{b}} + \mf{z}    \ea \right)  \;.
\label{expression twpo body scattering via Barnes}
\enq
Above, $G$ is the Barnes function and I adopted similar product conventions to \eqref{ecriture convention produit ratios fcts Gamma}.

The above formulae easily allow one to check that it holds 
\beq
\op{F}(\i\pi +\be) \op{F}(\be) \,= \, \f{ \sinh(\be)   }{  \sinh(\be) +  \sinh(2\i\pi \mf{b})  } \;. 
\enq

\subsection{The multi-particle sector solution}

The general solution of the form factor axioms  i)-iv) takes the form
\beq
\mc{F}_{n}^{(\mc{O})}(\bs{\be}_n) \, = \, \pl{a<b}{n} \op{F}\big(\be_{ab}\big) \cdot \mc{K}_{n}^{(\mc{O})}\big( \bs{\be}_{n} \big)              \qquad  \e{where} \qquad \be_{ab} \, = \,  \be_{a}-\be_{b} \;, 
\label{ecriture forme generale solution eqns FF}
\enq
in which $\mc{K}_{n}^{(\mc{O})}$ depends on the specific operator whose form factor is being computed. It follows from the form factor axioms that the representation \eqref{ecriture forme generale solution eqns FF}
solves the form factor axioms provided that 
\begin{itemize}
 
 \item $\mc{K}_{n}^{(\mc{O})}$  is a symmetric function of $\bs{\be}_n$

 \item $\mc{K}_{n}^{(\mc{O})}$  is a $2\i\pi $ periodic and meromorphic function of each variable taken singly;

 \item the only poles of $\mc{K}_{n}^{(\mc{O})}$ are simple and located at $\be_a-\be_b\in \i\pi (1+2\mathbb{Z})$. The associated residues are given by 
\beq
 \e{Res}\bigg(\mc{K}_{n}^{(\mc{O})}(\bs{\be}_n) \cdot \dd \be_1 \, , \, \be_1 = \be_2 + \i \pi  \bigg) \; = \;  \f{\i }{ \op{F}(\i\pi) }  \cdot 
 \pl{a=3}{n} \bigg\{  \f{ 1  }{  \op{F}(\be_{2a}+\i\pi) \op{F}( \be_{2a})  }    \bigg\} 
\cdot \bigg\{ 1\, - \, \pl{a=3}{n}  \op{S}(\be_{2a}) \bigg\} \cdot 
\mc{K}_{n-2}^{(\mc{O})}( \bs{\be}_n^{\prime\prime} )
\enq
where $\bs{\be}_n^{(k)} \, = \, \big( \be_{k+1},\dots, \be_n\big)$, \textit{viz}. $\bs{\be}_n^{\prime\prime} \, = \, \big( \be_{3},\dots, \be_n\big)$ . 
\item $\mc{K}_{n}^{(\mc{O})}$ has boost invariance $\mc{K}_{n}^{(\mc{O})}(\bs{\be}_n+\La \ov{\bs{e}}_n) \, = \,
\ex{ \La \op{s}_{\mc{O}} }  \mc{K}_{n}^{(\mc{O})}(\bs{\be}_n)$ with $\ov{\bs{e}}_n=(1,\dots,n)\in \R^n$.
 
\end{itemize}

 There are many ways of solving these equations. In the case of the Sinh-Gordon model, by following the strategy devised for the Lee-Yang model \cite{ZalmolodchikovTwoPointFctsLeeYang},
whose form factors were first argued in  \cite{SmirnovPerturbationCLess1CFTFRomSineGordonAndFFLeeYang,SmirnovReductionsAndClusterPropertyInSineGordonPlusSomeDiscussionsIRLimit},
the works \cite{FringMussardoSimonettiFFSOmeLocalObsSinhGordon,KoubekMussardoFFForMoreOpInSinhGordon} proposed various solutions for $\mc{K}_{n}^{(\mc{O})}$ given in terms of ratios of 
symmetric polynomials satisfying to certain finite difference equations. I will not discuss the form of these 
solutions further in that the resulting expressions do not display an appropriate structure which would  allow one to extract manageable 
upper bounds on the model's form factor.  However, the works \cite{BabujianFringKarowskiZapletalExactFFSineGordonBootsstrapI,BabujianKarowskiExactFFSineGordonBootsstrapII}
developed the so-called kernel method allowing one to systematically construct solutions $\mc{K}_{n}^{(\mc{O})}$ to the above equations in terms of a
weighted symmetrisation operator acting on elementary functions $p_n$. In this approach, it is the choice of the function $p_n$
which determines the operator whose form factors are calculated. The method was originally developed for the Sine-Gordon model.
Still, upon restricting these results to the pure breather excitation sector of the Sine-Gordon model which maps directly onto the Sinh-Gordon sector, 
Babujian and Karowski  \cite{BabujianKarowskiBreatherFFSineGordon} proposed the following general form for the $\mc{K}$-factors
\beq
 \mc{K}_{n}^{(\mc{O})}\big( \bs{\be}_{n} \big) \, = \;  \sul{ \bs{\ell}_n \in \{0,1\}^n }{} (-1)^{\ov{\bs{\ell}}_n}
\pl{a<b}{n} \bigg\{ 1 \, - \, \i \f{ \ell_{ab} \cdot \sin[2\pi \mf{b} ] }{ \sinh(\be_{ab})  }  \bigg\} \cdot p_n^{(\mc{O})}\big(\bs{\be}_n\mid \bs{\ell}_n\big)  \;, 
\label{definition fonction KnO generique}
\enq
where we agree upon the shorthand notations 
\beq
\ov{\bs{\ell}}_n \, = \, \sul{a=1}{n} \ell_k \; , \qquad v_{ab}=v_a-v_b  \;. 
\enq
The function $ \mc{K}_{n}^{(\mc{O})}\big( \bs{\be}_{n} \big)$ so defined will satisfy the equations given above if the functions $p_n^{(\mc{O})}\big(\bs{\be}_n\mid \bs{\ell}_n\big)$
satisfy a the set of constraints
\begin{itemize}

 \item   $\bs{\be}_n\mapsto p_{n}^{(\op{O})}\big(\bs{\be}_n\mid \bs{\ell}_n\big)$ is a collection of $2\i\pi$ periodic holomorphic functions on $\Cx$ that are symmetric in the two sets of variables jointly,  \textit{viz}.
 for any $\sg \in \mf{S}_n$ it holds $ p_{n}^{(\op{O})}\big(\bs{\be}_n^{\sg}\mid \bs{\ell}_n^{\sg} \big)=  p_{n}^{(\op{O})}\big(\bs{\be}_n\mid \bs{\ell}_n\big)$
with $\bs{\be}_n^{\sg}=\big( \be_{\sg(1)}, \dots, \be_{\sg(n)} \big)$;

  \item  $p_{n}^{(\op{O})}\big(\be_2+\i\pi, \bs{\be}_n^{\prime}\mid \bs{\ell}_n\big)\, = \,g(\ell_1,\ell_2) p_{n-2}^{(\op{O})}\big(\bs{\be}_n^{\prime\prime}\mid \bs{\ell}_n^{\prime\prime}\big)
  \, + \, h(\ell_1,\ell_2\mid \bs{\be}_n^{\prime})$
where $h$ does not depend on the remaining set of variables $\bs{\ell}_n^{\prime\prime}$ and
\beq
g(0,1)\, = \, g(1,0) \, = \, \f{ -1 }{ \sin (2\pi \mf{b} ) \,  \op{F}(\i\pi)  } \, ;
\enq

 \item  $ p_{n}^{(\op{O})}\big( \bs{\be}_n + \theta \ov{\bs{e}}_n \mid \bs{\ell}_n \big)  \; = \; \ex{\theta \op{s}_{\op{O}} } \cdot p_{n}^{(\op{O})}\big(\bs{\be}_n\mid \bs{\ell}_n\big)$.

\end{itemize}

 We would like to stress that the $\mc{K}$-transform method allows one to construct functions $p_n$ associated to
\begin{itemize}
 
 \item the conserved current operators $\op{J}_{\ell}^{(\sg)}(\bs{x})$ with $\ell \in 2\mathbb{Z}+1$ and $\sg\in \{\pm\}$ the light-cone index; 
 
 \item the energy-momentum tensor $\op{T}^{\sg,\tau}(\bs{x})$;
 
 \item  the exponential of the field operators $  \bs{:} \ex{ \ga \bs{\vp} }(\bs{x}) \bs{:} $, $\ga \in \Cx$. 

 \end{itemize}
These functions will be discussed below. It is also important to stress that, after an evaluation of the free field vaccuum expectations, the very same combinatorial expressions can be obtained
within the free field approach developed in \cite{LukyanovFirstIntroFreeField} and applied to the case of the Sinh-Gordon model in 
\cite{BrazhnikovLukyanovFreeFieldRepMassiveFFIntegrable} for what concerns the exponential operators and generalised to the case of descendents in 
\cite{FeiginLashkevichFreeFieldApproachToDescendents}.

\subsection{Form factors of various operators of interest}
\label{SousSection FF pour plusieurs type de Ops}

In \cite{BabujianKarowskiBreatherFFSineGordon}, Babujian and Karowski proposed  the following representation for the exponential of the field $\mc{K}$-function part of the form factors 
\beq
 \mc{K}_{n}^{(\ga)}\big( \bs{\be}_{n} \big) \, = \;  \sul{ \bs{\ell}_n \in \{0,1\}^n }{} (-1)^{\ov{\bs{\ell}}_n}
\pl{a<b}{n} \bigg\{ 1 \, - \, \i \f{ \ell_{ab} \cdot \sin[2\pi \mf{b} ] }{ \sinh(\be_{ab})  }  \bigg\} \cdot p_n^{(\ga)}\big(\bs{\be}_n\mid \bs{\ell}_n\big)  \;, 
\enq
where have introduced 
\beq
p_n^{(\ga)}\big(\bs{\be}_n\mid \bs{\ell}_n\big) \; = \;  \Big( \mc{N}^{(\ga)} \Big)^n   \cdot \pl{a=1}{n} \Big\{ \ex{ \f{2\i\pi \mf{b} }{ g } \ga (-1)^{\ell_a} } \Big\} \;. 
\label{definition fonction pn gamma}
\enq
Finally, the normalisation prefactor reads
\beq
\mc{N}^{(\ga)}  \; = \; \f{ - \i  }{ \sqrt{ \op{F}(\i\pi)\sin[2\pi \mf{b}]  }  } \; = \; \f{ - \i  }{ \sqrt{ 2 \sin[\pi \mf{b}]  }  } \exp\bigg\{ \f{1}{2\pi} \Int{0}{2\pi \mf{b} } \f{t \, \dd t }{ \sin(t) }  \bigg\} \;. 
\enq
Note that the expression for $\mc{K}_{n}^{(\ga)}\big( \bs{\be}_{n} \big)$ may be recast in a fully factorised form as
\beq
 \mc{K}_{n}^{(\ga)}\big( \bs{\be}_{n} \big) \, = \;   2^{n\f{n-1}{2}} \hspace{-2mm} \sul{ \bs{\ell}_n \in \{0,1\}^n }{} (-1)^{\ov{\bs{\ell}}_n}
\pl{a<b}{n} \bigg\{  \f{ \sinh\big[ \tfrac{ \be_{ab}}{2}  - \i\pi \mf{b} \ell_{ab} \big] \cdot \cosh\big[ \tfrac{\be_{ab}}{2}  + \i\pi \mf{b} \ell_{ab} \big]   }{ \sinh(\be_{ab})  }  \bigg\} \cdot p_n^{(\ga)}\big(\bs{\be}_n\mid \bs{\ell}_n\big)  \;. 
\label{ecriture K fct FF ex ga phi pour gamma generique}
\enq

Analogously, the $p$ functions associated with the conserved currents $\op{J}_{\ell}^{(\sg)}$, $\sg= \pm$ and $\ell \in 2\mathbb{Z}+1$,  take the form 
\beq
p_n^{(\ell;\sg)}\big(\bs{\be}_n\mid \bs{\ell}_n\big) \; = \; \mc{N}_{n}^{(\ell;\sg)} \cdot  \sg\,  \ex{- \i \f{\pi}{2}\ell} \bigg( \sul{a=1}{n} \ex{\sg \be_a } \bigg) 
\bigg( \sul{a=1}{n} \ex{\ell( \be_a -\i\pi \mf{b} (-1)^{\ell_a}) } \bigg)\cdot \bs{1}_{2\mathbb{N}}(n) \;, 
\enq
where $\bs{1}_{A}$ is the indicator function of the set $A$ while the associated normalisation constant reads 
\beq
\mc{N}_{2p}^{(\ell;\sg)}\, = \, \f{  - m \, \i^{\ell} }{ 4 \sin\big[ \pi \mf{b} \ell \big]  } \cdot \bigg\{ \f{- 1 }{ \op{F}(\i\pi) \sin\big[ 2\pi \mf{b} \big] }\bigg\}^{p} \;, 
\enq
so that the corresponding function $\mc{K}_{n}^{(\ell;\sg)}\big( \bs{\be}_{n} \big)$ takes the form 
\beq
\mc{K}_{2n}^{(\ell;\sg)}\big( \bs{\be}_{2n} \big) \; = \; 
   2^{n(2n-1)} \hspace{-3mm} \sul{ \bs{\ell}_{2n} \in \{0,1\}^{2n} }{} (-1)^{\ov{\bs{\ell}}_{2n} }
\pl{a<b}{2n} \bigg\{  \f{ \sinh\big[ \tfrac{ \be_{ab}}{2}  - \i\pi \mf{b} \ell_{ab} \big] \cdot \cosh\big[ \tfrac{\be_{ab}}{2}  + \i\pi \mf{b} \ell_{ab} \big]   }{ \sinh(\be_{ab})  }  \bigg\} \cdot 
p_{2n}^{(\ell;\sg)}\big(\bs{\be}_{2n}\mid \bs{\ell}_{2n}\big)  \;,  
\enq
while it vanishes for odd $n$.

Finally, the $p$ functions associated with the components of the energy-momentum tensor is expressed as 
\beq
p_n^{(\tau \sg)}\big(\bs{\be}_n\mid \bs{\ell}_n\big) \; = \; \mc{N}_{n}^{(\tau \sg)} \cdot   \tau\,   \bigg( \sul{a=1}{n} \ex{\tau \be_a } \bigg) 
\bigg( \sul{a=1}{n} \ex{\sg( \be_a -\i \f{\pi}{2}(1+2 \mf{b} (-1)^{\ell_a})) } \bigg)\cdot \bs{1}_{2\mathbb{N}}(n) \;, \qquad \e{with} \qquad \tau, \sg \in \{ \pm 1 \} \;. 
\enq
The normalisation constant occurring in this case reads $\mc{N}_{2p}^{(\tau \sg)}=m \mc{N}_{2p}^{(1;\sg)}$. The corresponding function $\mc{K}_{n}^{(\tau \sg)}\big( \bs{\be}_{n} \big)$ vanishes for $n$-odd and, for even $n$s
takes the form 
\beq
\mc{K}_{2n}^{(\tau \sg)}\big( \bs{\be}_{2n} \big) \; = \; 
  2^{n(2n-1)}  \hspace{-3mm}  \sul{ \bs{\ell}_{2n} \in \{0,1\}^{2n} }{} (-1)^{\ov{\bs{\ell}}_{2n} }
\pl{a<b}{2n} \bigg\{  \f{ \sinh\big[ \tfrac{ \be_{ab}}{2}  - \i\pi \mf{b} \ell_{ab} \big] \cdot \cosh\big[ \tfrac{\be_{ab}}{2}  + \i\pi \mf{b} \ell_{ab} \big]   }{ \sinh(\be_{ab})  }  \bigg\} \cdot 
p_{2n}^{( \tau \sg)}\big(\bs{\be}_{2n}\mid \bs{\ell}_{2n}\big)  \;. 
\enq

\section{The form factor series for space-like separated two-point functions}
\label{Section Form factor series}

Observe that owing to \eqref{ecriture des elements de matrices de M entre etars out et in} one has
\beq
\mc{M}_{\mc{O}}\Big( \bs{x}  \mid  \bs{\be}_{N} ;\emptyset  \Big) \, = \,  \mc{M}_{\mc{O}}\Big( \bs{x}  \mid  \emptyset ; \overleftarrow{\bs{\be}}_N+\i\pi \ov{\bs{e}}_N   \Big)
\enq
where $\overleftarrow{\bs{\be}}_N = (\be_N,\dots, \be_1)$.  

Hence, within the form factor bootstrap approach, a vacuum-to-vacuum two-point function of the operators $\mc{O}$ admits the below form factor series expansion
\beq
\moy{ \mc{O}_1(\bs{x}) \mc{O}_2(\bs{0)} } \; = \; \sul{ N \geq 0 }{} \f{ 1 }{ N! } \Int{  \R^N }{} \f{ \dd^N \be }{ (2\pi)^N }  \mc{F}_{N}^{(\mc{O}_1)}(\bs{\be}_N)
\mc{F}_{N}^{(\mc{O}_2)}( \overleftarrow{\bs{\be}}_N+\i\pi \ov{\bs{e}}_N )
\pl{a=1}{N}  \Big\{ \ex{-\i m [ t \cosh(\be_a) - x \sinh(\be_a)] }\Big\}  \; 
\label{ecriture serie FF deux pts op generale}
\enq
where  $\bs{x}=(t,x)$. The series may be recast in form more suited for further handling. From now on, we focus ourselves on the so-called space-like regime, \textit{viz}. $x^2-t^2>0$.

In that case, the Morera theorem allows one to move the contours to $\R+\i\tfrac{\pi}{2}\e{sgn}(x)+\La$ with $\th(\La)=\tf{t}{x}$ what, adjoined to the
overall rapidity shift properties of the form factors \eqref{ecriture invariance FF sous shift global rapidites}, leads to 
\beq
\moy{ \mc{O}_1(\bs{x}) \mc{O}_2(\bs{0)} } \; = \; \ex{ \vp_{x} } \sul{ N \geq 0 }{} \f{ 1 }{ N! } \Int{  \R^N }{} \f{ \dd^N \be }{ (2\pi)^N }
\mc{F}_{N}^{(\mc{O}_1)}(\bs{\be}_N)
\mc{F}_{N}^{(\mc{O}_2)}( \overleftarrow{\bs{\be}}_N  )
\pl{a=1}{N}  \Big\{ \ex{- m \sqrt{x^2-t^2} \cosh(\be_a)  }\Big\}  \;
\enq
in which
\beq
\vp_x \, = \,  \i\pi \op{s}_{\mc{O}_2} \, + \, \big(  \op{s}_{\mc{O}_1} \, + \,   \op{s}_{\mc{O}_2} \big) \cdot  \big[ \i\tfrac{\pi}{2} \e{sgn}(x) + \La ]
\enq
and $\op{s}_{\mc{O}_k}$ is the spin of the operator $\mc{O}_k$.

\vspace{2mm}

The convergence of the above series of multiple integrals is a long-standing open problem whose resolution constitutes the core result of this work. 

\vspace{2mm}

The reason why I focus here only on the space-like regime takes its origin in the desire to discuss the method for proving the convergence of form factor series in the most simple setting.
In the case of the time-like regime, the study of convergence demands to deform the contours in the original series \eqref{ecriture serie FF deux pts op generale} to a non-straight integration curve 
$\be_a\in \R \hookrightarrow \be_a \in \ga(\R)$ where $\ga(u)=u+\i\vartheta(u)$, with $\vartheta$ smooth and such that there exists $M>0$ large enough and $0<\eps < \tf{\pi}{2}$
so that $\vartheta(u)=-\e{sgn}(t) \e{sgn}(u) \eps$ when $|u|\geq M$. The use of such an integration curve then gives rise to additional technical -but not conceptual- complications in the analysis 
outlined in Sections \ref{Section majorant series FF et pblm minimisation}-\ref{Section comportement a grand N de la fnelle energi en la mesure eq} and will not be considered here so as to 
avoid obscuring the main ideas of the method by technicalities.

\begin{figure}[t]

\includegraphics[width=.4\textwidth]{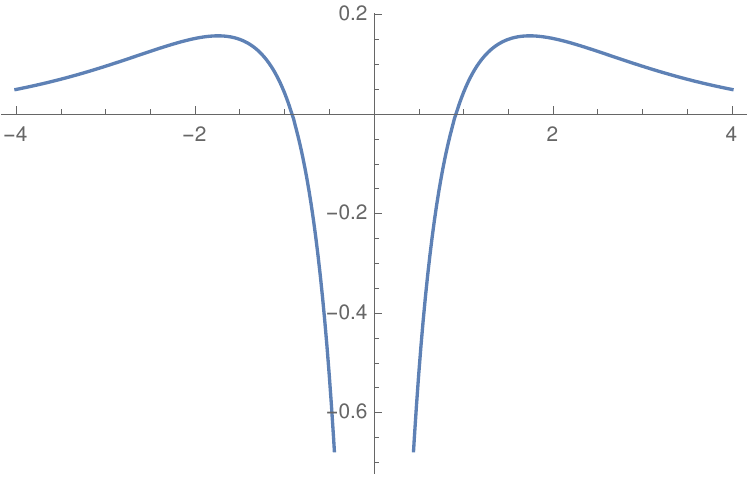}
 
\caption{Typical shape of the potential $\mf{w}$ given in \eqref{definition potentiel w}}
\label{Figure pot deux corps}
 \end{figure}

The form factors of an operator $\mc{O}(\bs{0})$  are given by \eqref{ecriture forme generale solution eqns FF} and \eqref{definition fonction KnO generique}.
By using that, for real $\be$, $\op{F}^{*}(\be)=\op{F}(-\be)$, the series associated with the form factor expansion of space-like separated
two-point function may be recast in a form more suited for further handlings that reads
\beq
\moy{  \mc{O}_1(\bs{x})  \mc{O}_2(\bs{0})   }    \; = \;
\sul{ N \geq 0 }{  } \f{  \mc{Z}_N( m \mf{r} )  }{ N! ( 2\pi )^{N}  }   \quad \mf{r}=\sqrt{x^2-t^2} \;,
\label{ecriture series fct 2 pts exp du champ}
\enq
where, by using $\mc{K}_{n}^{(\mc{O})}\big( \bs{\be}_{n} \big)$ as defined though \eqref{definition fonction KnO generique},  
\beq
\mc{Z}_N(  m \mf{r} )  \; = \; \Int{  \R^N }{}   \dd^N \be   \pl{a\not= b }{N } \Big\{ \ex{ \f{1}{2} \mf{w}(\be_{ab}) } \Big\} \cdot
\pl{a=1}{N}  \Big\{ \ex{-m \mf{r}  \cosh(\be_a)  }\Big\} \cdot
 \mc{K}_{N}^{(\mc{O}_1)}(\bs{\be}_N) \cdot   \mc{K}_{N}^{(\mc{O}_2)}(\bs{\be}_N) \;.
\label{ecriture representation integrale multiple pour ZN de r}
\enq
The expression for $\mc{Z}_N(x,t)$ involves the two-body potential $\mf{w}$ defined as
\beq
 \op{F}(\la)\op{F}(-\la) \; = \; \ex{   \mf{w} (\la) }   \;. 
\enq
It follows readily from \eqref{ecriture rep int pour F} that $\mf{w}$ admits the integral representation valid for $\la \in \R^{*}$
and to be understood in the sense of an oscillatory Riemann integral\symbolfootnote[3]{The well definiteness of the integral may be seen by observing that
$\mf{W}(\la) \, = \,  -\f{1}{|\la|} \Big( 1 + \e{O}\big( \ex{-2\pi |\la| \e{min}(\mf{b},\hat{\mf{b}}) }  \big) \Big) $  so that the integrand decomposes as
$\mf{W}(\la) \ex{\i \frac{\la x }{ \pi } }  \, = \, \mf{W}(\la)  \ex{\i \frac{\la x }{ \pi } } \bs{1}_{\intff{-1}{1}}(\la) \, + \,
\Big(  \mf{W}(\la) \, + \, \tfrac{1}{|\la|}  \Big) \ex{\i \frac{\la x }{ \pi } } \bs{1}_{\intff{-1}{1}^{\e{c}}}(\la)
\, -  \, \tfrac{1}{|\la|}   \ex{\i \frac{\la x }{ \pi } } \bs{1}_{\intff{-1}{1}^{\e{c}}}(\la) $.
The first two functions give rise to absolutely convergent integrals while the convergence of the integral
associated with the last function may be easily inferred by going back to the definition of an oscillatory Riemann-integral
and carrying out an integration by parts.}:
\beq
 \mf{w} (x)  \; = \;   \Int{\R }{} \dd \la  \, \mf{W}(\la)  \ex{\i \frac{\la x }{ \pi } }  \qquad \e{with} \qquad
  \mf{W}(\la)  \; = \; - 4 \f{ \sinh(\la \mf{b}) \cdot \sinh(\la \hat{\mf{b}} ) \cdot \sinh(\tfrac{1}{2}\la) \cdot \cosh(\la)  }{ \la \cdot  \sinh^2(\la)  }   \;.
\label{definition potentiel w}
\enq
The potential $\mf{w}$ takes the typical shape depicted in Figure \ref{Figure pot deux corps} which holds when $\mc{b}$ belongs to compact subsets of $\intoo{0}{\tf{1}{2}}$.
Of course the position of the maximum and its magnitude do depend on $\mf{b}$. 
One can infer from the integral representation \eqref{definition potentiel w} and the representation for $\op{F}$ in terms of  Barnes functions \eqref{expression twpo body scattering via Barnes} that 
\beqa
 \mf{w} (x) & = & 2 \ln |x| \, + \, \e{O}(1) \quad \e{for} \quad \la \tend 0  \label{ecriture comportement de w en 0}\\
 \mf{w} (x) &=&  - 8 \sin[\pi \mf{b}]\sin[\pi \wh{\mf{b}}] \ex{ \mp  x}  \bigg\{ - \f{ \pm x + 1}{ \pi }  \, + \,   \mf{b} \cot[\pi \mf{b}]  \, + \,   \wh{\mf{b}} \cot[\pi \wh{\mf{b}}]  \bigg\}
\; + \; \e{O}\Big(  \ex{\mp (1+\eps)x} \Big) \;,
 \label{ecriture comportement de w en infini}
\eeqa
for $\Re(x) \tend \pm \infty$ and some $\eps>0$. Furthermore, the $ \e{O}(1)$ remainder around $\la=0$ is analytic.

\vspace{2mm}
It is easy to see from the above estimates that each of the multiple integrals $\mc{Z}_N(x,t)$, \textit{c.f.} \eqref{ecriture representation integrale multiple pour ZN de r}, is well-
defined provided that the natural condition on $p_n^{(\mc{O})}$ given in \eqref{ecriture borne sup hypothese sur les pn} holds. Indeed,
it follows from \eqref{ecriture comportement de w en 0}-\eqref{ecriture comportement de w en infini} that, for some $C>0$,  one has the bounds
\beq
\bigg| \pl{a \not= b}{}  \ex{ \f{1}{2}\mf{w}(\be_{ab})  } \bigg| \; \leq \; C^{N^2} \pl{a<b}{N}  \bigg( \f{ | \sinh(\be_{ab})| }{ | \sinh(\be_{ab})| + 1  } \bigg)^2 \;.
\enq
Further, direct bounds on the level of \eqref{definition fonction KnO generique} adjoined to \eqref{ecriture borne sup hypothese sur les pn} ensure that, for some
$C>0$,
\beq
 \Big| \mc{K}_{N}^{(\mc{O}_1)}(\bs{\be}_N)  \Big|\,  \leq \,  C^{N^2}  \pl{a<b}{N}    \f{ | \sinh(\be_{ab})| + 1  }{ | \sinh(\be_{ab})| }   \cdot \pl{a=1}{N} \ex{C_2 |\be_a|^k} \;.
\enq
Hence, all-in-all
\beq
\bigg| \pl{a\not= b }{N } \Big\{ \ex{ \f{1}{2} \mf{w}(\be_{ab}) } \Big\} \cdot
\pl{a=1}{N}  \Big\{ \ex{-m \mf{r}  \cosh(\be_a) } \Big\} \cdot
 \mc{K}_{N}^{(\mc{O}_1)}(\bs{\be}_N) \cdot   \mc{K}_{N}^{(\mc{O}_2)}(\bs{\be}_N)   \bigg|
 \; \leq \;\big[ C^{\prime} \big]^{N^2}\, \pl{a=1}{N}  \Big\{ \ex{-m \mf{r}  \cosh(\be_a)  + C_2 |\be_a|^k }\Big\} \;,
\enq
what does ensure the absolute convergence of the integral \eqref{ecriture representation integrale multiple pour ZN de r}

 \vspace{3mm}

The question of the convergence of the series \eqref{ecriture series fct 2 pts exp du champ} boils down to accessing to the leading in $N$ asymptotics of 
$\ln \big|\mc{Z}_N(x,t)\big|$. 
The multiple integral representation \eqref{ecriture representation integrale multiple pour ZN de r} for $\mc{Z}_N(x,t)$ is not in a form that would allows for an estimation of the leading large-$N$
behaviour of $\ln \big|\mc{Z}_N(x,t)\big|$, at least within the existing techniques. However, building on the representation \eqref{ecriture K fct FF ex ga phi pour gamma generique} one may 
obtain an upper bound for $\big|\mc{Z}_N(x,t)\big|$ in terms of another $N$-fold multiple integral
whose large-$N$ behaviour may be already accessed within the techniques that were pioneered in \cite{BenArousGuionnetLargeDeviationForWignerLawLEadingAsymptOneMatrixIntegral} 
and further developed so as to encompass $N$-dependent interactions in \cite{KozBorotGuionnetLargeNBehMulIntOfToyModelSoVType}. 
 This bound is established below and constitutes the main result of this section.

\begin{prop}
 Let $|x|> |t|$ and $p_n^{(\mc{O}_k)}$ be bounded as in \eqref{ecriture borne sup hypothese sur les pn}. Then, the $N^{\e{th}}$ summand in the form factor expansion admits the  upper bound
\beq
\big| \mc{Z}_N( m \mf{r} ) \big| \; \leq  \;  \big(  C  \ln N \big)^N   \cdot
\e{max}_{p\in \intn{0}{N}} \big| \msc{Z}_{N,p}\big( \tfrac{m \mf{r} }{2}   \big) \big| \:,
\label{ecriture borne sup sur sommant serie FF}
\enq
where $C>0$ is some constant and
\bem
\msc{Z}_{N,p}( \varkappa )  \, = \, \bigg(  \f{1}{\ln N } \bigg)^N \Int{  \R^p }{} \hspace{-1mm} \dd^p \nu  \hspace{-1mm} \Int{\R^{N-p} }{} \hspace{-2mm}  \dd^{N-p} \la  
\pl{a=1}{p}  \Big\{ \ex{- \varkappa \cosh(\nu_a) }\Big\} \cdot  \pl{a=1}{N-p}  \Big\{ \ex{-  \varkappa \cosh(\la_a) }\Big\} \cdot 
\pl{a< b }{p } \Big\{ \ex{  \mf{w}(\nu_{ab}) } \Big\}   \\ 
\times \pl{a< b }{ N - p } \Big\{ \ex{  \mf{w}(\la_{ab}) } \Big\}  \cdot \pl{a=1}{p} \pl{b=1}{N-p}   \bigg\{  \ex{  \mf{w}(\nu_{a}-\la_{b}) }  
\pl{\eps=\pm }{}  \f{ \sinh\big[ \nu_{a}-\la_{b}  - 2\i\pi \eps  \mf{b}  \big]    }{ \sinh(\nu_{a}-\la_{b})  }   \bigg\} 
\;. 
\label{definition fct part bornante}
\end{multline}
\end{prop}

\Proof  
 
 Indeed, by using the obvious bound 
\beq
\Big|\sul{k=1}{p} a_k \Big|^2  \, \leq \, p^2 \sul{k=1}{p} |a_k|^2
\enq
and the upper bound \eqref{ecriture borne sup hypothese sur les pn}, one gets
\beq
\Big| \mc{K}_{N}^{(\mc{O}_1)}(\bs{\be}_N) \cdot \mc{K}_{N}^{(\mc{O}_2)}(\bs{\be}_N) \Big|\; \leq \;
\bigg(  2 \,  C_1 \bigg)^{ 2 N} \pl{a=1}{N} \ex{2 C_2 |\be_a|^k }  \cdot  \sul{ \bs{\ell}_N \in \{0,1\}^N }{}
\pl{a<b}{N}  \bigg\{   \f{ \sinh\big[ \be_{ab}  - 2\i\pi \mf{b} \ell_{ab} \big] \cdot   \sinh\big[ \be_{ab}  + 2\i\pi \mf{b} \ell_{ab} \big]     }{ \sinh^2(\be_{ab})  }   \bigg\} \;. 
\label{ecriture borne sur produits fcts K}
\enq
Here, we stress that the summands ocurring in \eqref{ecriture borne sur produits fcts K} are all real.

Observe that the summand appearing above is clearly symmetric  in $\bs{\be}_N$. Furthermore, for any $ \bs{\ell}_N \in \{0,1\}^N$, there exists $p\in \intn{1}{N}$ and $\sg \in \mf{S}_{N}$
such that\symbolfootnote[1]{Notice that if $p=0$ or $p=N$ then the permutation is trivial and one of the below conditions is trivial}
\beq
\ell_{\sg(a)}=0 \quad \e{for} \quad a=1,\dots, p \qquad \e{and} \qquad \ell_{\sg(a)}=1 \quad \e{for} \quad a=p+1,\dots, N \;. 
\enq
Hence, for this $\bs{\ell}_N$, it holds 
\beq
\pl{a<b}{N}  \pl{\eps=\pm }{} \bigg\{   \f{ \sinh\big[ \be_{ab}  - 2\i\pi  \eps \mf{b} \ell_{ab} \big]      }{ \sinh(\be_{ab})  }   \bigg\}  \; = \;
\pl{a=1}{p} \pl{b=1+p}{N}  \pl{\eps=\pm }{}  \bigg\{   \f{ \sinh\big[ \be_{ \sg(a)\sg(b) }  +  2\i\pi \eps  \mf{b}  \big]    }{ \sinh(\be_{ \sg(a)\sg(b)})  }   \bigg\}  \;. 
\enq
There are $C_{N}^{p}$ different choices of vectors $\bs{\ell}_N\in \{0,1\}^N$ having exactly $p$ entries equal to $0$ and for each such choice one may change
variables under the integral
\beq
\be_{ \sg(a) } \, = \,  \nu_a\, , \;  a=1,\dots,p, \quad \e{and} \quad  \be_{ \sg(a) } \, = \,  \la_{a-p}\, , \;  a=p+1,\dots,N \, ,
\enq
in which $\sg$ is the associated permutation. Thus, upon using that 
\beq
\ex{C_2 |\be|^k} \; \leq  \; C^{\prime} \cdot \ex{  \f{m \mf{r} }{2} \cosh(\be)   }
\enq
one gets the below upper bound valid for some $C>0$
\beq
\big| \mc{Z}_N(m \mf{r} ) \big| \; \leq  \;  \bigg(   C  \f{\ln N}{2}        \bigg)^{   N}   \sul{p=0}{N} C_{N}^{p} \msc{Z}_{N,p}\big( \tfrac{m \mf{r} }{2} \big)
\enq
where $\msc{Z}_{N,p}( \varkappa )$ is as defined in \eqref{definition fct part bornante}. Then, since $ \sul{p=0}{N} C_{N}^{p} =2^N$, one gets the sought upper bound \eqref{ecriture borne sup sur sommant serie FF}. \qed

\vspace{2mm}

Thus, in order to get a bound on the large-$N$ behaviour of $\big| \mc{Z}_N(m \mf{r} ) \big|$, one should access to the leading  one of
$ \ln  \msc{Z}_{N,p}\big( \tfrac{m \mf{r} }{2} \big) $, uniformly in $p\in \intn{0}{N}$.
One may expect, and this will be comforted by the analysis to come, that the leading in $N$ behaviour will be grasped, analogously to the random matrix setting 
\cite{AndersonGuionnetZeitouniIntroRandomMatrices,KozBorotGuionnetLargeNBehMulIntOfToyModelSoVType}, from a
concentration of measure property. To set the latter one should first identify the scale in $N$ at which the integration
variables reach an equilibrium. The latter results from the compensation of the confining effect between the $\cosh$ potential and the repulsive, on
short distances, effect of the potential  $\mf{w}$, all this balanced by the two-body interaction between the $\nu$ and $\la$ integration variables.
One may heuristically convince oneself that the appropriate scale is reached upon dilating all variables by 
\beq
\tau_N=\ln N \, .
\label{definition variable dilatation}
\enq
This observation will be made rigorous and legitimate by the analysis to come. 

Thus, I recast 
\beq
\msc{Z}_{N,p}( \varkappa )  \, = \,    \Int{\R^{N-p} }{} \hspace{-2mm}  \dd^{N-p} \la    \Int{  \R^p }{} \hspace{-1mm} \dd^p \nu  \;  \ov{\varrho}_{N,p}\big(\bs{\la}_{N-p}, \bs{\nu}_p \big)
\label{ecritue rep int rescalee pour ZNp}
\enq
where the integrand takes the form 
\bem
 \ov{\varrho}_{N,p}\big(\bs{\la}_{N-p}, \bs{\nu}_p \big) \; = \; 
\pl{a=1}{p}  \Big\{ \ex{- V_N(\nu_a) }\Big\} \cdot  \pl{a=1}{N-p}  \Big\{ \ex{-  V_N(\la_a) }\Big\}  \\
\times \pl{a< b }{p } \Big\{ \ex{  \mf{w}_N(\nu_{ab}) } \Big\}   
\cdot \pl{a< b }{ N - p } \Big\{ \ex{  \mf{w}_N(\la_{ab}) } \Big\}  \cdot \pl{a=1}{p} \pl{b=1}{N-p}   \bigg\{  \ex{  \mf{w}_{\e{tot};N}(\nu_{a}-\la_{b}) }    \bigg\}  \;. 
\label{definition densite integrande de Z N p majorant}
\end{multline}
The product form of the integrand involves the functions
\beq
V_N(\la) \, = \, \varkappa \cosh(\tau_N \la) \quad , \quad  \mf{w}_{N}(\la) \, = \, \mf{w}(\tau_N \la) \quad, \quad  \mf{w}_{\e{tot};N}(\la) \, = \, \mf{w}_{\e{tot}}(\tau_N \la) \;, 
\enq
upon agreeing that 
\beq
 \mf{w}_{\e{tot}}(\la)  \, = \, \mf{w}(\la) \, + \, \mf{v}_{2\pi \mf{b}, 0^+}(\la) \qquad \e{with} \qquad 
\mf{v}_{ \a , \eta}(\la) \, = \, \ln \Bigg(  \f{  \sinh(\la + \i \a) \cdot \sinh(\la - \i\a ) }{  \sinh(\la + \i \eta) \cdot \sinh(\la - \i \eta ) }   \Bigg) \;. 
\label{definition potentiel varpi tot et correctif v alpha eta}
\enq
In particular, it follows from the asymptotic expansion \eqref{ecriture comportement de w en 0}-\eqref{ecriture comportement de w en infini} 
that $ \mf{w}_{\e{tot}}$ is bounded Lipschitz on $\R$.

Note that 
\beq
\varrho_{N,p}\big(\bs{\la}_{N-p}, \bs{\nu}_p \big) \; = \; \f{ \ov{\varrho}_{N,p}\big(\bs{\la}_{N-p}, \bs{\nu}_p \big) }{ \msc{Z}_{N,p}( \varkappa )  } \;, 
\enq
gives rise to the density of a probability measure on $\R^{N-p}\times \R^p$.

\section{Leading large-$N$ behaviour of $\msc{Z}_{N,p}( \varkappa )$ in terms of a minimisation problem}
\label{Section majorant series FF et pblm minimisation}

In this section, I obtain an upper bound on the leading order of the exponential large-$N$ asymptotics of the bounding partition function $\msc{Z}_{N,p}( \varkappa )$. 
Prior to stating the theorem, I need to introduce an auxiliary functional on $\mc{M}^{1}(\R)\times \mc{M}^{1}(\R)$, with $\mc{M}^{1}(\R)$ referring to the space of probability measures on $\R$:
\bem
\mc{E}_{N,t}[\mu,\nu]\; = \; \f{1}{N} \bigg\{ t \Int{}{} V_N(s) \dd \nu(s) \, + \,  (1-t) \Int{}{} V_N(s) \dd \mu(s) \bigg\} \; - \; \f{t^2}{2} \Int{}{} \mf{w}_N(s-u) \cdot \dd \nu(s) \dd \nu(u)  \\ 
 \; - \; \f{ (1-t)^2}{2} \Int{}{} \mf{w}_N(s-u) \cdot \dd \mu(s) \dd \mu(u)  \, - \, t(1-t)  \Int{}{} \mf{w}_{\e{tot};N}(s-u) \cdot \dd \mu(s) \dd \nu(u)   \;. 
\label{definition fonctionelle E N t}
\end{multline}
The parameters  $N\in \mathbb{N}^*$ and $t \in \intff{0}{1}$ appearing above should be considered as fixed in the minimisation problem to come.    Note that for $t=0$, resp. $t=1$, $\mc{E}_{N,t}$
only depends on one of its two variables and hence effectively induces a functional only on $\mc{M}^{1}(\R)$. 
Below, I will refer to $\mc{E}_{N,t}$, $0<t<1$, or to its effective restrictions to $\mc{M}^{1}(\R)$, as "energy functional".

\begin{theorem}
 
\label{Theorem estimation comportement gd N Z N p majorant}

The multiple integral \eqref{ecritue rep int rescalee pour ZNp} admits the below estimate:
\beq
\msc{Z}_{N,p}( \varkappa )  \,  \leq \,  \exp\bigg\{ -N^2   \e{inf} \Big\{ \,  \mc{E}_{N,\frac{p}{N}}[\mu,\nu]  \; : \;  (\mu,\nu)\in  \mc{M}^{1}(\R)\times \mc{M}^{1}(\R)   \Big\} \; + \; \e{O}\Big( N \tau_N^2 \Big)  \bigg\} \,,  
\label{ecriture borne sup sur fct part majorante}
\enq
in which the control is uniform in $p\in \intn{0}{N}$. 

\end{theorem}

Note that since $V_N$, $-\mf{w}$ and $ -\mf{w}_{\e{tot}}$ are all bounded from below, $\mc{E}_{N,t}[\mu,\nu]$ is bounded from below on $\mc{M}^{1}(\R)\times \mc{M}^{1}(\R)$
and so the infimum appearing in \eqref{ecriture borne sup sur fct part majorante} is also bounded from below.  The proof of the characterisation of the large-$N$ behaviour of $\msc{Z}_{N,p}( \varkappa )$
through a minimisation problem goes in two steps. First, one introduces $\ov{\mu}_{N,p}$
the measure on $\R^{N-p}\times \R^p$ induced by $\ov{\varrho}_{N,p}\big(\bs{\la}_{N-p}, \bs{\nu}_p \big)$ given in \eqref{definition densite integrande de Z N p majorant}
and shows that it concentrates with super-Gaussian precision on the interval $\intff{-2+\eps}{2+\eps}^{N}$. Then, one builds on this result to get an upper bound on
$\msc{Z}_{N,p}( \varkappa )$. 
The techniques allowing one to establish such results are rather standard nowadays, see \textit{e.g.} \cite{AndersonGuionnetZeitouniIntroRandomMatrices,KozBorotGuionnetLargeNBehMulIntOfToyModelSoVType}. 

In fact, I have little doubt that the upper bound in \eqref{ecriture borne sup sur fct part majorante} is optimal, \textit{viz}. that the inequality $\leq$ can be replaced by an $=$. However, 
following \cite{AndersonGuionnetZeitouniIntroRandomMatrices,KozBorotGuionnetLargeNBehMulIntOfToyModelSoVType}, proving this equality would demand to obtain a lower bound on
$\msc{Z}_{N,p}( \varkappa )$ by estimating the contribution to the integral \eqref{ecritue rep int rescalee pour ZNp} issuing from integration variables located in the vicinity of the
configuration whose empirical measure is sufficiently close to the minimiser in \eqref{ecriture borne sup sur fct part majorante}. For this, one needs to have some quantitative 
information on the minimier's density. While posing no conceptual problem to be obtained, doing so would demand much more work than what will be developed in 
Sections  \ref{Section characterisation minimiseur et eqn integrale singuliere}-\ref{Section caracterisation complete de la mesure eq} relatively to solving a simpler minimisation problem
which is already enough in what concerns the main goal of this work, \textit{viz}. establishing the convergence of the form factor series. Hence, I omit establishing the lower-bound here.

\subsection{Concentration on compact subsets}

\begin{lemme}
\label{Lemme localisation a compact pour integration fction partition}
Let $ \ov{\mu}_{N,p}$ be the measure on $\R^{N-p}times \R^p$ with density $\ov{\varrho}_{N,p}$ given by \eqref{definition densite integrande de Z N p majorant}
and let $\eps>0$. Then, the partition function $\msc{Z}_{N,p}( \varkappa )$ enjoys the estimates
\beq
\msc{Z}_{N,p}( \varkappa )  \, = \,  \ov{\mu}_{N,p}\big[  \Om^{(\eps)}  \big] \; + \; \e{O}\Big( \ex{ - \f{\varkappa}{4}N^{2+\eps} } \Big) \;, 
\enq
where 
\beq
\Om^{(\eps)} \; = \; \Big\{  (\bs{\la}_{N-p}, \bs{\nu}_p) \in \R^N \; : \;   |\la_a| < 2 + \eps \;, \; \;
a=1,\dots, N-p \quad and \quad  |\nu_a| < 2 + \eps \;, \; \; a=1,\dots, p \Big\}
\label{definition domaine integration Omega epsilon}
\enq
and $\ov{\mu}_{N,p}$ refers to the measure on $\R^{N-p}\times \R^p$ with density $\ov{\varrho}_{N,p}\big(\bs{\la}_{N-p}, \bs{\nu}_p \big)$ introduced in \eqref{definition densite integrande de Z N p majorant}. 
 
\end{lemme}

\Proof 
Given $a \in \intn{1}{N-p}$ and $b \in \intn{1}{p}$, let
\beq
O_{1;a}^{(\eps)} \; = \; \Big\{  (\bs{\la}_{N-p}, \bs{\nu}_p) \in \R^N   \; : \;  | \la_a| \,  \geq \,  2 + \eps \Big\} \quad \e{and} \quad 
O_{2;b}^{(\eps)} \; = \; \Big\{  (\bs{\la}_{N-p}, \bs{\nu}_p) \in \R^N    \; : \;  | \nu_b| \,  \geq \,  2 + \eps \Big\}  \;.
\enq
Then, since 
\beq
 \Big[ \Om^{(\eps)} \Big]^{\e{c}}  \subseteq  \Big\{ \bigcup\limits_{a=1}^{N-p} O_{1;a}^{(\eps)} \Big\} \, \bigcup\,   \Big\{ \bigcup\limits_{a=1}^{p} O_{2;a}^{(\eps)}  \Big\} \;,
\enq
 it  follows that 
\beq
\ov{\mu}_{N,p}\Big[ \,   \big[ \Om^{(\eps)} \big]^{\e{c}}  \,   \Big]    \;  \leq  \;  \sul{a=1}{N-p}  \ov{\mu}_{N,p}\Big[  O_{1;a}^{(\eps)}  \Big] \, + \,  \sul{a=1}{p}  \ov{\mu}_{N,p}\Big[  O_{2;a}^{(\eps)}  \Big] \;. 
\enq
By using that $\mf{w}$ is bounded from above, one gets 
\bem
\ov{\mu}_{N,p}\Big[ \,    O_{k;a}^{(\eps)}  \,   \Big]    \;  \leq  \; \exp\Bigg\{ \frac{1}{2} \Big[ p(p-1) \, + \,(N-p)(N-p-1) \Big] \sup_{\R}\{ \mf{w}(\la) \}  \; + \; p(N-p) \norm{ \mf{w}_{\e{tot}} }_{ L^{\infty}(\R) }  \Bigg\} \\
\times \bigg(\Int{\R}{} \hspace{-1mm} \dd s \,  \ex{-V_{N}(s) } \bigg)^{N-1} \cdot 
\Int{ |s|> 2+\eps  }{} \hspace{-1mm} \dd s \,  \ex{-V_{N}(s) }   \\
\; \leq \; \f{ \exp\Big\{ c N^2 \Big\} }{  \tau_N^{N-1}  } \cdot  \bigg(\Int{\R}{} \dd s \, \ex{-V(s) } \bigg)^{N-1} \cdot  2 \,\ex{-V_N(2+\eps) }\cdot  \Int{2+\eps}{+\infty} \hspace{-1mm} \dd s \, \ex{ -(s-2-\eps)V_N^{\prime}(2+\eps)} \\
\; \leq \; \exp\Big\{ c^{\prime} N^2 \Big\} \cdot  \exp\Big\{- \varkappa \cosh[\tau_N(2+\eps)] \Big\} \; \leq \; \f{C^{\prime\prime} }{N}\ex{ - \f{\varkappa}{4}N^{2+\eps} } \;,  
\end{multline}
for some $N$-independent constants $c, c^{\prime}, C^{\prime\prime}>0$. Also, in the intermediate steps I used that 
\beq
\cosh (x) \, = \, \sul{n\geq 0}{} \f{ \cosh^{(n)}(y) }{ n! } \cdot (x-y)^n \; \geq \; \cosh(y) \,  +\, (x-y) \sinh(y)
\enq
which holds provided that $x\geq y \geq 0$. \qed

\subsection{Upper bound}

\begin{lemme}

One has the uniform in $p\in \intn{0}{N}$ upper bound 
\beq
\msc{Z}_{N,p}( \varkappa ) \; \leq \; \exp\bigg\{  - N^2 \e{inf} \Big\{ \,  \mc{E}_{N,\frac{p}{N}}[\mu,\nu]  \; : \;  (\mu,\nu)\in  \mc{M}^{1}(\R)\times \mc{M}^{1}(\R)   \Big\}  \; + \; \e{O}\Big(N \tau_N^{2}\Big) \bigg\} \;. 
\enq

\end{lemme}

\Proof

To start with, one introduces a regularised vector $\wt{\bs{\be}}_{\ell}\in \R^{\ell}$ associated with any vector $\bs{\be}_{\ell} \in \R^{\ell}$. 
Namely, given $\be_1 \leq \dots \leq \be_{\ell}$, define $\wt{\be}_1< \dots < \wt{\be}_{\ell}$ as 
\beq
\wt{\be}_1 \, = \, \be_1 \quad \e{and} \quad  \wt{\be}_{k+1}\, = \,  \wt{\be}_{k} \, + \, \max \big( \, \be_{k+1}- \be_k  , \ex{- \tau_N^2 } \big) \,.
\enq
For any $\bs{\be}_{\ell}\in \R^{\ell}$, one picks $\sg \in \mf{S}_{\ell}$, such that $\be_{\sg(1)} \leq \dots \leq \be_{\sg(\ell)}$ and then obtains 
$\wt{\be}_1^{\,(\sg)}< \dots < \wt{\be}_{\ell}^{\,(\sg)}$ by the above procedure. Then, the vector $\wt{\bs{\be}}_{\ell} \in \R^{\ell}$ has coordinates
$\big( \wt{\bs{\be}}_{\ell} \big)_k \, = \, \wt{\be}_{ \sg^{-1}(k) }^{\, (\sg)}$. 
The new configuration has been constructed such that, for $\ell \not=k$, 
\beq
\big| \wt{\be}_{k} - \wt{\be}_{\ell} \big| \, \geq \, \ex{-\tau_N^2 } \qquad , \quad
\big| \be_{k} - \be_{\ell} \big| \, \leq \, \big| \wt{\be}_{k} - \wt{\be}_{\ell} \big| \quad \e{and} \quad
\big| \be_{k} - \wt{\be}_{k} \big| \leq (\sg^{-1}(k)-1) \cdot \ex{-\tau_N^2 } \, \leq \, (N-1)\ex{-\tau_N^2} \;, 
\label{ecriture bornes sur les vars be k tilde}
\enq
with $\sg$ as introduced above. Here, the main point is that, as such, the variables $\wt{\be}_k$ exhibit much better spacing properties than the original ones,
while still remaining close to the original variables $\be_k$. Note that
the scale of regularisation $\ex{-\tau_N^2 }$ is somewhat arbitrary, but in any case should be taken negligible compared to an algebraic decay. 

To proceed further, one needs to introduce the empirical measure associated with an $\ell$-dimensional vector $\bs{\nu}_{\ell}\in \R^{\ell}$ : 
$$
L_{\ell}^{(\bs{\nu})} = \f{1}{\ell} \sum_{a = 1}^{\ell} \delta_{\nu_{a}}
$$
where $\de_{x}$ refers to the Dirac mass at $x$. Further,  denote by $L_{\ell;u}^{ ( \bs{\nu} ) } $ the convolution of $L_{\ell}^{ ( \bs{\nu} ) }$ with the uniform 
probability measure on $\intff{0}{ \tfrac{1}{N} \ex{-\tau_N^2 } }$.  The main advantage of the convoluted empirical measure $L_{\ell;u}^{(\bs{\nu})}$ is that it is Lebesgue continuous,
this for any $\bs{\nu}_{\ell}\in \R^{\ell}$; as such it can appear in the argument of $\mc{E}_{N,\frac{p}{N}}$ and yield finite results.

By using the empirical measures associated with $\bs{\la}_{N-p}$ and $\bs{\nu}_p$, one may recast the unnormalised integrand 
$\ov{\varrho}_{N,p}$ introduced in \eqref{definition densite integrande de Z N p majorant},  
\bem
\ov{\varrho}_{N,p}\big(\bs{\la}_{N-p}, \bs{\nu}_p \big) \, = \,
\exp\Bigg\{ -p \Int{}{} V_N(s) \dd L_{p}^{(\bs{\nu})}(s) \, - \,  (N-p) \Int{}{} V_N(s) \dd L_{N-p}^{(\bs{\la})}(s)  \, + \, \f{ p^2 }{ 2 } \Int{ x \not= y }{} \mf{w}_{N}(x-y)  \dd L_{p}^{(\bs{\nu})}(x)\dd L_{p}^{(\bs{\nu})}(y)  \\
\, + \, \f{ (N-p)^2 }{ 2 } \Int{ x \not= y }{} \mf{w}_{N}(x-y)  \dd L_{N-p}^{(\bs{\la})}(x)\dd L_{N-p}^{(\bs{\la})}(y)    
\, + \, p(N-p) \Int{  }{} \mf{w}_{\e{tot};N}(x-y)  \dd L_{p}^{(\bs{\nu})}(x)\dd L_{N-p}^{(\bs{\la})}(y)     \Bigg\} 
\end{multline}
To get an upper bound on $\msc{Z}_{N,p}(\varkappa)$ one needs
to relate this expression, up to some controllable errors, to an evaluation of the energy functional $\mc{E}_{N,\f{p}{N}}$ on some well-built convoluted empirical measure.  
Following Lemma \ref{Lemme localisation a compact pour integration fction partition}, one may limit the reasoning to integration variables belonging to $\intff{-2-\eps}{2+\eps}$
for some $\eps>0$.

Pick $\bs{\nu}_{p}\in \intff{-2-\eps}{2+\eps}^{p}$ with $p \in \intn{1}{N}$. Further, set $w_{p;a}^{(\sg)}= \Big[ (\sg^{-1}(a)-1) + \tfrac{1}{N} \Big]  \ex{-\tau_N^2}$ with 
$\sg \in \mf{S}_{p}$ such that $\nu_{\sg(1)}\leq \cdots \leq \nu_{\sg(p)}$ and observe that
\beq
 \Big| \nu_{a} \, - \,  \wt{\nu}_a \, - \, \tfrac{ \ex{-\tau_N^2} }{ N } s \Big| \; \leq \; w_{p;a}^{(\sg)} \quad \e{when} \quad s \in \intff{0}{1} \;.
\enq
Then, by the mean value theorem, one gets the bound
\bem
\bigg| \Int{}{} V_N(s) \dd L_{p}^{(\bs{\nu})}(s) \, - \, \Int{}{} V_N(s) \dd L_{p;u}^{(\, \wt{\bs{\nu}}\, )}(s)  \bigg| \; \leq \; 
\f{1}{p} \sul{a=1}{p} \Int{0}{1} \! \dd s \, \Big| V_N(\nu_{a}) \, -  \,  V_N\Big( \wt{\nu}_a \, + \, \tfrac{ \ex{-\tau_N^2} }{ N } s \Big) \Big| \\
\leq \; \f{1}{p} \sul{a=1}{p} w_{N;a}^{(\sg)}  \cdot 
\underset{ t\in \intff{0}{1} }{ \e{sup} }\bigg\{ \varkappa \tau_N \Big| \sinh\Big( \tau_N  [\nu_a + tw_{p;a}^{(\sg)} ]  \Big)  \Big| \bigg\} 
\; \leq \;  p \tau_N C^{\prime} N^{2+\eps} \ex{-\tau_N^2} \;, 
\end{multline}
for some $C^{\prime}>0$ and where we used that $w_{p;a}^{(\sg)}\, \leq p \ex{-\tau_N^2}$ and that, for $|\nu|\leq 2+\eps$, it holds 
\beq
\big| \sinh\big[ \tau_N(\nu  + N \ex{-\tau_N^2} ) \big] \big| \; \sim \; \f{ N^{2+\eps} }{ 2} \; . 
\enq
Likewise, using similar bounds   and the fact that $\mf{w}_{\e{tot}}$ is bounded Lipschitz, one gets 
\bem
\bigg| \Int{}{} \mf{w}_{\e{tot};N}(x-y) \, \dd L_{p}^{(\bs{\nu})}(x)  \dd L_{N-p}^{(\bs{\la})}(y)  \, - \,
\Int{}{}\mf{w}_{\e{tot};N}(x-y)\,  \dd L_{p;u}^{(\, \wt{\bs{\nu}} \, )}(x)  \dd L_{N-p;u}^{(\, \wt{\bs{\la}}\, )}(y)  \bigg| \\
\; \leq \; 
\f{1}{p(N-p)} \sul{a=1}{p} \sul{b=1}{N-p} \Int{0}{1} \! \dd s \dd u \bigg| \mf{w}_{\e{tot};N}(\nu_a-\la_b) \, - \, \mf{w}_{\e{tot};N}\Big( \wt{\nu}_a-\wt{\la}_b \, + \,  \tfrac{ \ex{-\tau_N^2} }{ N } (s-u) \Big) \Big| \\ 
\leq \; \f{1}{p(N-p)} \sul{a=1}{p} \sul{b=1}{N-p}  \tau_N \norm{ \mf{w}_{\e{tot}}^{\prime} }_{L^{\infty}(\R) }\Big\{ p-1 + N-p-1+\tfrac{1}{N} \Big\}  \ex{-\tau_N^2}
\; \leq \; C N \tau_{N} \ex{-\tau_N^2} \;, 
\end{multline}
for some $C>0$.

Finally, the estimate on the integrals involving the $\mf{w}_N$ interaction require more care due to the presence of a logarithmic behaviour at the origin: $\mf{w}(\la)\, =\, 2\ln | \la | \, + \, \mf{w}_{\e{reg}}(\la) $, 
with $\mf{w}_{\e{reg}}$ analytic in some open neighbourhood of $0$, \textit{c.f.} \eqref{ecriture comportement de w en 0}. This upper bound is obtained in two steps, depending one whether $|\nu_a-\nu_b|$ is small enough or not.

\noindent (i) If $|\nu_a-\nu_b| \, \leq  \, \tf{\eps}{\tau_N}$ then, by using that $\la \mapsto \ln \la$ is increasing on $\R^+$ and the shorthand notation $\nu_{ab}=\nu_a-\nu_b$, one has
for $N$-large enough
\bem
\mf{w}_N(\nu_{ab}) \; \leq \; 2 \ln \big( \tau_N |\, \wt{\nu}_{ab} |\big) \, + \, \mf{w}_{\e{reg};N}\big( \, \wt{\nu}_{ab} \big)
\; + \; \Big( |\nu_a-\wt{\nu}_a | \, + \, |\nu_b-\wt{\nu}_b | \Big)\cdot  \tau_N  \cdot
\underset{ s\in \intff{0}{1} }{ \e{sup} }\bigg|\mf{w}_{\e{reg}}^{\prime}\Big( \tau_N( \, \wt{\nu}_{ab} + s [ \nu_{ab} \, - \, \wt{\nu}_{ab} ]) \Big) \bigg| \\
\; \leq \; \mf{w}_N( \, \wt{\nu}_{ab} ) \, + \, 2\tau_N \, (p-1) \, \norm{ \mf{w}_{\e{reg}}^{\prime} }_{L^{\infty}(\intff{-2\eps}{2\eps}) } \cdot \ex{-\tau_N^2} \;. 
\end{multline}
(ii) If $|\nu_a-\nu_b| \, \geq  \, \tf{\eps}{\tau_N}$ then it holds 
\bem
\mf{w}_N(\nu_{ab}) \; \leq  \;  \mf{w}_{N}\big( \, \wt{\nu}_{ab} \big)
\; + \; \Big( |\nu_a-\wt{\nu}_a | \, + \, |\nu_b-\wt{\nu}_b | \Big)\cdot  \tau_N  \cdot \underset{ s\in \intff{0}{1} }{ \e{sup} }\bigg\{\mf{w}^{\prime}\Big( \tau_N( \, \wt{\nu}_{ab} + s [ \nu_{ab} \, - \, \wt{\nu}_{ab} ])  \Big) \bigg\} \\
\; \leq \; \mf{w}_N( \, \wt{\nu}_{ab} ) \, + \, 2 \tau_N \, (p-1) \, \norm{ \mf{w}^{\prime} }_{L^{\infty}( \R \setminus \intff{ - \frac{\eps}{2} }{ \frac{\eps}{2} } ) } \cdot \ex{-\tau_N^2} \;. 
\end{multline}
This leads to the upper bound 
\beq
\Int{ x \not= y }{} \mf{w}_{N}(x-y) \, \dd L_{p}^{(\bs{\nu})}(x)  \dd L_{p}^{(\bs{\nu})}(y)  \, - \,
\Int{ x \not= y}{} \mf{w}_{N}(x-y) \, \dd L_{p}^{(\, \wt{\bs{\nu}} \,  )}(x)  \dd L_{p}^{(\, \wt{\bs{\nu}} \, )}(y)   
\; \leq \; C \, p \,  \tau_N  \, \ex{ -\tau_N^2 } \;,
\enq
in which $C>0$ is an $\eps$-dependent constant. Further, one has that
\bem
 \de \Sg \; = \; \Int{ x \not= y }{} \mf{w}_{N}(x-y) \, \dd L_{p}^{(\, \wt{\bs{\nu}} \,  )}(x)  \dd L_{p}^{(\, \wt{\bs{\nu}} \, )}(y)    \, - \,
\Int{ x \not= y}{} \mf{w}_{N}(x-y) \, \dd L_{p;u}^{(\, \wt{\bs{\nu}} \,  )}(x)  \dd L_{p;u}^{(\, \wt{\bs{\nu}} \, )}(y)  \\
\; =  \;  \Int{ x \not= y }{} \dd L_{p}^{(\, \wt{\bs{\nu}} \,  )}(x)  \dd L_{p}^{(\, \wt{\bs{\nu}} \, )}(y)  \Int{0}{1} \dd s \dd u 
\bigg\{ \mf{w}_{N}(x-y)  \, - \,  \mf{w}_{N}\Big( x - y    \, + \,  \tfrac{ \ex{-\tau_N^2} }{ N } (s-u)  \Big) \bigg\} \; - \; \f{1}{p}\Int{0}{1} \dd s \dd u \mf{w}_{N}\Big(    \tfrac{ \ex{-\tau_N^2} }{ N } (s-u)  \Big) \;. 
\label{ecriture delta SG}
\end{multline}
Then, using the expressions 
\beq
\mf{w}_N(x) \, = \, \left\{ \ba{ccc}  2\ln (\tau_N|x|) \, + \, \mf{w}_{\e{reg};N}(x) &  \e{for} &   |x| \leq \tf{ \eps }{ \tau_N }    \vspace{2mm} \\
					  \mf{w}(\tau_N x)  			&  \e{for}   & |x| \geq \tf{ \eps }{ \tau_N } \ea \right. 
\enq 
one gets the upper bound
\beq
\Big| \mf{w}_N(x) \, - \, \mf{w}_N(y)  \Big| \leq C \bigg( 1 +  \underset{t\in \intff{0}{1} }{ \e{max} } \f{1}{|x+t(y-x)|} \bigg) \cdot  |x-y| \;,
\enq
which leads, for any $|x-y| \geq \ex{-\tau_N^2}$ and $s,u \in \intff{0}{1}$, to 
\beq
\bigg| \mf{w}_{N}(x-y)  \, - \,  \mf{w}_{N}\Big( x - y    \, + \,  \tfrac{ \ex{-\tau_N^2} }{ N } (s-u)  \Big) \bigg| \; \leq \; \f{C}{N} |s-u| \;. 
\enq
Hence, all-in-all, taking into account \eqref{ecriture bornes sur les vars be k tilde},   $\big| \de \Sg  \big| \, = \, \e{O}\Big( \tfrac{\tau_N^2}{p} \Big)$,
where the control issues from the last term in \eqref{ecriture delta SG}.

Thus, upon invoking Lemma \ref{Lemme localisation a compact pour integration fction partition}, one gets that, for some constant $C>0$ and with $\Om^{(\eps)}$ as introduced in \eqref{definition domaine integration Omega epsilon},   
\bem
\msc{Z}_{N,p}( \varkappa )   \; \leq \; \e{O}\Big( \ex{ - \f{\varkappa}{4}N^{2+\eps} } \Big) \; + \;   \Int{   \Om^{(\eps)}   }{} \dd^{p}\nu \dd^{N-p}\la \\ 
\exp\bigg\{  - \Int{}{}   V_N(s)\Big[  p \dd L_{p;u}^{(\, \wt{\bs{\nu}} \,  )}(s) \, + \, (N-p) \dd L_{N-p;u}^{(\, \wt{\bs{\la}} \,  )}(s) \Big]  
+ C \tau_N \ex{-\tau_N^2} N^{2+\eps} \Big(p^2+(N-p)^2\Big) \bigg\} \\
\times \exp\bigg\{ p(N-p)\Int{}{}\mf{w}_{\e{tot};N}(x-y) \, \dd L_{p;u}^{(\, \wt{\bs{\nu}} \, )}(x)  \dd L_{N-p;u}^{(\, \wt{\bs{\la}}\, )}(y) \; + \; C p(N-p)N \tau_N \ex{-\tau_N^2}  \bigg\} \\
\times \exp\bigg\{ \tfrac{1}{2}\Int{}{}\mf{w}_{N}(x-y) \Big[ p^2  \dd L_{p;u}^{(\, \wt{\bs{\nu}} \, )}(x) \dd L_{p;u}^{(\, \wt{\bs{\nu}} \, )}(y) \, + \, 
(N-p)^2  \dd L_{N-p;u}^{(\, \wt{\bs{\la}} \, )}(x) \dd L_{N-p;u}^{(\, \wt{\bs{\la}} \, )}(y) \Big]    \\
\; + \; C \big[ p^3+(N-p)^3\big] \tau_N \ex{-\tau_N^2}  \; + \; C \big[ p+(N-p)\big] \tau_N^2 \bigg\} \;. 
\end{multline}
Thus, one gets 
\bem
\msc{Z}_{N,p}( \varkappa )   \; = \; \e{O}\Big( \ex{ - \f{\varkappa}{4}N^{2+\eps} } \Big) \; + \;   \Int{   \Om^{(\eps)}   }{} \dd^{p}\nu \dd^{N-p}\la 
\exp\bigg\{  - N^2 \mc{E}_{N,\frac{p}{N}}\Big[L_{N-p;u}^{(\, \wt{\bs{\la}} \, )} \, , \, L_{p;u}^{(\, \wt{\bs{\nu}} \, )}\Big] \; + \; \e{O}\Big(N \tau_N^{2}\Big) \bigg\} \\ 
\; \leq \; \Big(4+2\eps\Big)^N \cdot  \exp\bigg\{  - N^2 \e{inf} \Big\{ \,  \mc{E}_{N,\frac{p}{N}}[\mu,\nu]  \; : \;  (\mu,\nu)\in  \mc{M}^{1}(\R)\times \mc{M}^{1}(\R)   \Big\}  \; + \; \e{O}\Big(N \tau_N^{2}\Big) \bigg\}
 \; + \;  \e{O}\Big( \ex{ - \f{\varkappa}{4}N^{2+\eps} } \Big)   \\
\; \leq \;   \exp\bigg\{  - N^2 \e{inf} \Big\{ \,  \mc{E}_{N,\frac{p}{N}}[\mu,\nu]  \; : \;  (\mu,\nu)\in  \mc{M}^{1}(\R)\times \mc{M}^{1}(\R)   \Big\}  \; + \; \e{O}\Big(N \tau_N^{2}\Big) \bigg\}
 \; + \;  \e{O}\Big( \ex{ - \f{\varkappa}{4}N^{2+\eps} } \Big)  \;.
\end{multline}
To get the claim, it is enough to get a Gaussian lower bound on the partition $\msc{Z}_{N,p}( \varkappa )$ what can be done by using Jensen's inequality
applied to the probability measure on $\R^N$:
\beq
\dd \mf{p}_N \, = \, \pl{a=1}{N-p} \Big\{ \dd \la_a \; \ex{-V_N(\la_a)}  \Big\} \cdot \pl{a=1}{p} \Big\{ \dd \nu_a  \; \ex{-V_N(\nu_a)} \Big\} \cdot \bigg\{ \Int{\R}{} \dd s \ex{-V_N(s)} \bigg\}^{-N}
\enq
what yields 
\bem
\msc{Z}_{N,p}( \varkappa )   \; \geq \; \exp\Bigg\{ \Int{\R^N}{} \dd \mf{p}_N  \bigg(\sul{a<b}{N-p} \mf{w}_N(\la_{ab}) \, + \, \sul{a<b}{p} \mf{w}_N(\nu_{ab})
\, + \, \sul{a=1}{p}\sul{b=1}{N-p} \mf{w}_{N;\e{tot}}(\nu_{a}-\la_{b})
    \bigg)  \Bigg\} \cdot  \bigg\{ \Int{\R}{} \dd s \ex{-V_N(s)} \bigg\}^{N}  \\
\; \geq \; \exp\Bigg\{ - N^2  \Int{\R^2}{} \f{ \dd s \dd u  \, \ex{-V(s)-V(u) }  }{ \Big( \Int{\R}{} \dd x \, \ex{-V(x) } \Big)^2}  \cdot \Big[ \,  |\mf{w}(s-u) | \, + \, |\mf{w}_{\e{tot}}(s-u) |  \Big]  \Bigg\}
\cdot  \bigg\{ \Int{\R}{} \f{\dd s}{\tau_N} \ex{-V(s)} \bigg\}^{N}  \; \geq \; \ex{- c N^2} \;. 
\end{multline}
Hence, it holds that 
\beq
\msc{Z}_{N,p}( \varkappa ) \cdot \bigg( 1  \, - \,  \e{O}\Big(  \f{\ex{ - \f{\varkappa}{4}N^{2+\eps} } }{ \msc{Z}_{N,p}( \varkappa )  }  \Big)    \bigg)
\; \leq \; \exp\bigg\{  - N^2 \e{inf} \Big\{ \,  \mc{E}_{N,\frac{p}{N}}[\mu,\nu]  \; : \;  (\mu,\nu)\in  \mc{M}^{1}(\R)\times \mc{M}^{1}(\R)   \Big\}  \; + \; \e{O}\Big(N \tau_N^{2}\Big) \bigg\}
\enq
what leads to the claim since 
\beq
\ln \bigg( 1  \, - \, \e{O}\Big(     \msc{Z}_{N,p}^{-1}( \varkappa ) \cdot  \ex{ - \f{\varkappa}{4}N^{2+\eps} }    \Big) \bigg) \, = \,  \e{O}\Big(   \ex{ - \f{\varkappa}{8}N^{2+\eps} } \Big) 
\enq
\qed

\section{The energy functional associated with the bounding partition function}
\label{Section etude pblm minimisation fnelle energie adequate}

As shown in Theorem \ref{Theorem estimation comportement gd N Z N p majorant},
the large-$N$ behaviour of $\msc{Z}_{N,p}(\varkappa)$ can be controlled, up to $\e{O}(N \tau_N^2)$ corrections, by solving the minimisation problem associated with the energy functional $\mc{E}_{N,t}$ on $\mc{M}^{1}(\R)\times \mc{M}^{1}(\R) $
introduced in \eqref{definition fonctionelle E N t}. In this section, I will establish the unique solvability of this problem as well as a lower bound for the value of the minimum.

 \begin{prop}
 \label{Proposition LSC et cpct level sets for E N t} 
  For any $0<t<1$, the functional $\mc{E}_{N,t}$ is lower semi-continuous on $\mc{M}^{1}(\R)\times \mc{M}^{1}(\R)$, has compact level sets, is not identically $+\infty$ and is bounded from below. 
 The same properties hold for $\mc{E}_{N,0}$ and $\mc{E}_{N,1}$ seen as functionals on $\mc{M}^{1}(\R)$.

 \end{prop}

\Proof 
Let $0<t<1$. The functional $\mc{E}_{N,t}$ is lower semi-continuous as supremum of a family of continuous functionals in respect to the bounded-Lipschitz topology on $\mc{M}^{1}(\R)\times \mc{M}^{1}(\R)$:
$\mc{E}_{N,t}[\mu,\nu] = \e{sup}_{M \nearrow +\infty} \mc{E}_{N,t}^{(M)}[\mu,\nu] $ with 
\bem
\mc{E}_{N,t}^{(M)}[\mu,\nu]\; = \; \f{1}{N}  \Int{ \R }{} \e{min}\big( V_N(s), M ) \cdot \Big\{ t \dd \nu(s) \, + \,  (1-t) \dd \mu(s) \Big\} \; + \;
\f{t^2}{2} \Int{ \R^2 }{} \e{min}\big( -\mf{w}_N(s-u),M \big) \cdot \dd \nu(s) \dd \nu(u)  \\
 \; + \; \f{ (1-t)^2}{2} \Int{ \R^2 }{}  \e{min}\big( -\mf{w}_N(s-u),M\big) \cdot \dd \mu(s) \dd \mu(u)  \, - \, t(1-t)  \Int{\R^2 }{} \mf{w}_{\e{tot};N}(s-u) \cdot \dd \mu(s) \dd \nu(u)   \;,
\end{multline}
where we used that  $ \la \mapsto \mf{w}_{\e{tot};N}(\la) $ is already bounded Lipschitz on $\R$. 
Further, $\mc{E}_{N,t}$ is not identically $+\infty$ as follows from evaluating it on the probability measures 
\beq
( \dd \mu(s) , \dd \nu(s) ) \, = \,  ( \psi(s) \cdot \dd s, \psi(s) \cdot \dd s) \quad \e{with} \quad \psi \in \mc{C}^{\infty}_{\e{c}}(\R) \quad \e{such}\, \e{that} \quad \Int{\R }{} \psi(s) \dd s =1 \;. 
\enq

Let 
\beq
f_t(s,u)= \f{t}{2N} \Big( V_N(s) \, +\, V_N(u) \Big) \, - \, \f{t^2}{2} \mf{w}_N(s-u) \;. 
\enq
Clearly, $f_t$ is bounded from below on $\R^2$. Denote by $c_t$ a lower bound on it, \textit{viz}. $f_t\geq c_{t}$. Owing to $\norm{ \mf{w}_{\e{tot}} }_{L^{\infty}(\R)} $ being finite, 
the existence of a lower bound on $f$ entails that $\mc{E}_{N,t}$ is bounded from below as
\beq
\mc{E}_{N,t}[\mu,\nu] \geq c_t + c_{1-t} \, - \, t(1-t) \norm{ \mf{w}_{\e{tot}} }_{L^{\infty}(\R)} \;.
\enq
It remains to establish that $\mc{E}_{N,t}$ has compact level sets. For that purpose, first observe that $f_t$ blows up 
\begin{itemize}
\item[i)] at $\infty$ due to the exponential blow up of $V_N$, 
\item[ii)] along the diagonal due to the logarithmic singularity of $\mf{w}_N$. 
\end{itemize}
Hence, there exists $K_L\tend +\infty$ such that 
\beq
\Big\{(s,u) \, : \, f_{t}(s,u) \geq K_L \;\; \e{and} \;\;  f_{1-t}(s,u) \geq K_L  \Big\} \supset 
B_L = \Big\{ (s,u) \, : \, |s|>L \;\; \e{or} \;\; |u|>L \Big\} \cup \Big\{ (s,u) \, : \, |s-u|<L^{-1}  \Big\} \;. 
\enq
 Let $M>0$ and $(\mu,\nu)$ be such that $\mc{E}_{N,t}[\mu,\nu]\leq M$. Then, for $L$ large enough, it holds
\beq
\big( \mu\big[ \R \setminus \intff{-L}{L}\big] \big)^2 \, + \, \big( \nu\big[ \R \setminus \intff{-L}{L} \big] \big)^2 \; \leq \;
\mu\otimes\mu\Big[ \big\{(s,u) \, : \,   f_{1-t}(s,u) \geq K_L  \big\} \Big] \, + \, \nu\otimes\nu\Big[ \big\{(s,u) \, : \,   f_{t}(s,u) \geq K_L  \big\} \Big]  \;.
\enq
Observe that owing to $K_{L}-c_{1-t}>0$  for $L$ large enough and $c_{1-t}$ being a lower bound for $f_{1-t}(s,u)$ one gets the lower bounds
\beq
\f{ f_{1-t}(s,u) - c_{1-t}  }{ K_L - c_{1-t} } \; \geq \; 1 \quad \e{on} \quad \Big\{ (s,u) \, : \, f_{1-t}(s,u) \geq K_L \Big\}
\quad \e{and} \quad
\f{ f_{1-t}(s,u) - c_{1-t}  }{ K_L - c_{1-t} } \; \geq \; 0 \quad \e{on} \quad \R^2
\enq
 Hence, it holds that
\beq
\big( \mu\big[ \R \setminus \intff{-L}{L}\big] \big)^2  \; \leq \; \hspace{-1cm}
\Int{  \big\{ (s,u) \, : \, f_{1-t}(s,u) \geq K_L \big\}  }{} \hspace{-1cm} \dd \mu(s) \dd \mu(u)
\; \leq \; \hspace{-1cm}
\Int{  \big\{ (s,u) \, : \, f_{1-t}(s,u) \geq K_L \big\}  }{} \hspace{-1cm} \dd \mu(s) \dd \mu(u) \f{ f_{1-t}(s,u) - c_{1-t}  }{ K_L - c_{1-t} }
\; \leq \; \Int{  \R^2  }{}  \dd \mu(s) \dd \mu(u) \f{ f_{1-t}(s,u) - c_{1-t}  }{ K_L - c_{1-t} } \;.
\enq
An analogous bound holds for $\nu$. From there it follows that
\bem
\big( \mu\big[ \R \setminus \intff{-L}{L}\big] \big)^2 \, + \, \big( \nu\big[ \R \setminus \intff{-L}{L} \big] \big)^2
\; \leq \; \Int{\R^2 }{} \hspace{-2mm} \dd \mu(s) \dd \mu(u)  \, \f{ f_{1-t}(s,u) - c_{1-t}  }{ K_L - c_{1-t} } \; +
\Int{ \R^2 }{} \hspace{-2mm} \dd \nu(s) \dd \nu(u) \, \f{ f_{t}(s,u) - c_{t}  }{ K_L - c_{t} } \\
\; \leq \;  \f{ 1  }{ K_L - \e{max}\big\{c_{t},  c_{1-t} \big\}  }  \cdot \Big\{ \mc{E}_{N,t}[\mu,\nu]\, - \, c_{t} \, - \, c_{1-t} + t(1-t) \norm{ \mf{w}_{\e{tot}} }_{ L^{\infty}(\R) }  \Big\}
\; \leq \;  \f{ C_M }{ K_L } \;, 
\end{multline}
for some $C_M>0$ and $L>L_0$ with $L_0$ being $M$-independent and where $c_t$ is the lower bound on $f_t$. Hence, 
\beq
(\mu,\nu) \in \mc{K}_{M}\times \mc{K}_{M} \qquad \e{with}  \qquad \mc{K}_M\; = \; \bigcap\limits_{ \substack{ L\in \mathbb{N} \\  >L_0} } \bigg\{  \mu \in \mc{M}^{1}(\R) \, : \,
\mu \big[ \R \setminus \intff{-L}{L}\big] \, \leq \, \sqrt{\tfrac{ C_M }{K_L} }  \bigg\} \;.
\label{ecriture inclusion niveaux de fnell energie dans compact}
\enq
Since $\mc{K}_M$ is uniformly tight by construction and closed as an intersection of level sets of the lower semi-continuous functions
$\mu \mapsto \hspace{-3mm} \Int{ \R \setminus \intff{-L}{L}}{} \hspace{-3mm} \dd\mu(s)$, by virtue of Prokhorov's theorem,
$\mc{K}_M$ is compact. Since the level sets of $\mc{E}_{N,t}$ are closed, \eqref{ecriture inclusion niveaux de fnell energie dans compact} entails that they are compact as well.  \qed

\vspace{2mm}

For further purpose, I introduce two functions

\beq
\mf{w}_N^{(\pm)}(u) \; = \; \mf{w}^{(\pm)}(\tau_N u ) \qquad \e{with} \qquad 
\left\{ \ba{ccc}   \mf{w}^{(+)}(u  )& = &   \mf{w}(u) + \f{1}{2} \mf{v}_{2\pi \mf{b},0^+}(u)   \\
  \mf{w}^{(-)}(u ) &  =  & - \f{1}{2}\mf{v}_{2\pi \mf{b},0^+}(u)   \ea \right. \;, 
\enq
and where $\mf{v}_{\a,\eta}$ has been introduced in \eqref{definition potentiel varpi tot et correctif v alpha eta}. 

Both $\mf{w}^{(\pm)}$ have strictly negative Fourier transforms, namely
\beq
\mf{w}^{(\ups)}(x) \, = \, - \Int{\R}{} \hspace{-1mm} \dd \la   \f{ R^{(\ups)}(\la)  }{ \la }  \, \ex{- \i \la x } \; ,  \quad x \not= 0\,, 
\label{ecriture rep int pour varpi plus et moins}
\enq
where the integration is to be understood in the sense of an oscillatory Riemann--integral, and where 
\beq
R^{(+)}(\la) \; = \;  \f{ \sinh( \pi \mf{b} \la ) \cdot \sinh(\pi \hat{\mf{b}} \la ) \cdot \sinh(\tfrac{\pi}{2}\la)   }{  \cosh^2(\tfrac{\pi}{2} \la)   }  \qquad \e{and} \qquad 
R^{(-)}(\la) \; = \;  \f{ \sinh( \pi \mf{b} \la ) \cdot \sinh(\pi \hat{\mf{b}} \la )   }{  \sinh(\tfrac{\pi}{2}\la)    }  \;. 
\label{definition R plus et moins}
\enq
The formulae for $R^{(\ups)}$ follow from the integral representation for $\mf{w}$ given in \eqref{definition potentiel w} and the explicit form of the Fourier transform
of the function $\mf{v}_{ \a , \eta}$ which was defined in \eqref{definition potentiel varpi tot et correctif v alpha eta}:
\beq
\mc{F}\big[ \mf{v}_{ \a , \eta} \big](\la) \, = \, \Int{ \R }{} \dd x  \,  \ex{\i x \la } \mf{v}_{ \a , \eta}(x) \, = \, - 4\pi
\f{   \sinh\Big( \la \tfrac{\eta-\a}{2}\Big)  \sinh\Big( \la \tfrac{\pi - \eta-\a}{2}\Big)   }{  \la \sinh\Big( \tfrac{\pi \la }{ 2 }  \Big) }  \;. 
\label{ecriture TF de v alpha eta}
\enq
I refer to Lemma \ref{Lemme calcul integrales elementaires} given in Appendix \ref{Appendix Section resultats auxiliaires} for the details
on how to get \eqref{ecriture TF de v alpha eta}.

The functions $\mf{w}^{(\pm)}$ give rise to the below two functionals 
\beqa
\mc{E}_{N}^{(+)}\big[\sg \big]  & = & \f{ 1 }{ N }   \Int{}{} V_N(s)\,  \dd \sg(s) \, - \, \f{ 1 }{ 2 } \Int{}{} \mf{w}_N^{(+)}(s-u) \cdot \dd \sg(s) \, \dd \sg(u) \, ,  \label{definition fonctionnelle energie +}\\
\mc{E}_{N}^{(-)}\big[ \sg \big]  & = &  - \f{ 1 }{ 2 } \Int{}{} \mf{w}_N^{(-)}(s-u) \cdot \dd \sg(s) \, \dd \sg(u) \;  .
\eeqa
$\mc{E}_{N}^{(+)}$ is a functional on $\mc{M}^{1}(\R)$ while $\mc{E}_{N}^{(-)}$  is a functional on $\mc{M}^{(2t-1)}_{\mf{s}}(\R)$, the space of signed, bounded, measures on $\R$ of total mass $2t-1$.

\begin{lemme}
\label{Lemme stricte convexite fnelle E N et t} 

Let $0<t<1$.  $\mc{E}_{N,t}$ is strictly convex and may be recast as
\beq
\mc{E}_{N,t}\big[\mu,\nu\big] \; = \; \sul{\ups = \pm }{}   \mc{E}_{N}^{(\ups)}\big[ \sg^{(\ups)}_t \big] \qquad  with  \qquad 
\sg^{(\pm)}_{t} \, = \, t \nu \pm (1-t) \mu \;. 
\label{ecriture decomposition fnell energie}
\enq
 Moreover, it holds 
\beq
\mc{E}_{N,t}\big[\a \mu + (1-\a) \sg , \a \nu + (1 - \a) \varrho \big] \, - \, \a \mc{E}_{N,t}\big[\mu , \nu \big] \, - \, (1 - \a) \mc{E}_{N,t}\big[ \sg , \varrho \big]\; = \; - \a (1-\a) \mf{D}_{N,t}\big[ \mu  - \sg, \nu - \varrho \big] 
\label{ecriture convexite fnelle energie}
\enq
where 
\beq
\mf{D}_{N,t}\big[ \mu ,\nu \big] \; = \; -t(1-t)  \Int{ \R^2 }{} \mf{w}_{\e{tot};N}(s-u) \cdot \dd \mu(s) \dd \nu(u)
\; - \; \f{1}{2} \Int{ \R^2 }{} \mf{w}_N(s-u) \cdot \Big( t^2 \dd \nu(s) \dd \nu(u) \; + \; (1-t)^2 \dd \mu(s) \dd \mu(u) \Big) \, \;.
\label{definition distance induite par mesure proba}
\enq
For compactly supported signed measures of zero total mass,  $\mf{D}_{N,t}$ satisfies 
\beq
\mf{D}_{N,t}\big[ \mu,\nu \big] \; \geq \; 0  \quad   and \quad  \mf{D}_{N,t}\big[ \mu , \nu \big] \, = \, 0 \quad if\, and\, only\, if \quad (\mu,\nu)=(0,0) \;. 
\enq

 \noindent Analogous statements hold, upon restricting to $\mc{M}^{1}(\R)$, when $t=0$ or $t=1$. 
\end{lemme}
 
 \Proof 
Equations \eqref{ecriture decomposition fnell energie}-\eqref{ecriture convexite fnelle energie} follows from straightforward calculations upon implementing in $\mc{E}_{N,t}$
the change of unknown measure given by \eqref{ecriture decomposition fnell energie}.  
To establish the properties of $\mf{D}_{N,t}$ and hence the strict convexity of $\mc{E}_{N,t}$, one implements the change of variables given in \eqref{ecriture decomposition fnell energie} so as to get, 
for any signed measure on $\R$ of zero mass, that 
\beq
\mf{D}_{N,t}\big[ \mu,\nu \big] \; = \;  - \, \f{ 1 }{ 2 } \Int{ \R^2 }{} \mf{w}_N^{(+)}(s-u) \cdot \dd \sg^{(+)}_{t}(s) \, \dd \sg^{(+)}_{t}(u)
 \, - \, \f{ 1 }{ 2 } \Int{ \R^2 }{} \mf{w}_N^{(-)}(s-u) \cdot \dd \sg^{(-)}_{t}(s) \, \dd \sg^{(-)}_{t}(u) \;.
\enq
 Thus, one gets that 
\beq
\mf{D}_{N,t}\big[ \mu,\nu \big] \; = \;   \f{1}{2} \Int{ \R }{} \hspace{-1mm} \dd \la \, \bigg\{ \;  \big| \mc{F}\big[ \sg^{(+)}_{t}\big](\la)\big|^2 \f{ R^{(+)}(\la) }{ \la }
\, + \, \big| \mc{F}\big[ \sg^{(-)}_{t}\big](\la)\big|^2 \f{ R^{(-)}(\la) }{ \la }  \bigg\}\;, 
\enq
which is a number in $\intff{0}{+\infty}$ and where $\sg_{t}^{(\pm)}$ are related to $\mu, \nu$ through  \eqref{ecriture decomposition fnell energie}. 
Here, I remind that the Fourier transform of a signed measure $\mu$ such that $|\mu|$ has finite total mass is expressed as  
\beq
\mc{F}[\mu](\la) \; = \; \Int{ \R }{} \dd\mu(s) \ex{\i s \la} \;.
\enq

Observe that if $\mu, \nu$ are signed, compactly supported, measures on $\R$ with zero total mass then for any $t\in \intff{0}{1}$,  $\la \mapsto  \mc{F}\big[ \sg^{(\ups)}_{t}\big](\la)$
is entire. 

Now, given two such measures, since 
\begin{itemize}
 
 \item $ \tf{R^{(\pm)}(\la)}{\la}$  are both positive on $\R$,
 
 \item the only zero of $ \tf{R^{(+)}(\la)}{\la}$ on $\R$ is at $\la=0$ and is double,

 \item  $\tf{R^{(-)}(\la)}{\la}$ has no zeroes on $\R$,

\end{itemize}

 \noindent it follows that if $ \mf{D}_{N,t}\big[ \mu , \nu \big] \, = \, 0$, then it holds
\beq
\mc{F}\big[ \sg^{(\ups)}_{t}\big](\la) \; = \; 0 \quad \e{on} \; \R^* \;. 
\enq
However, since these are entire functions, it holds $\mc{F}\big[ \sg^{(\ups)}_{t}\big]=0$. This entails that $\sg^{(\ups)}_{t}=0$  and thus, for $0<t<1$, that $\mu=\nu=0$.

The reasoning when $t=0$ or $t=1$ are quite analogous. \qed

 \begin{theorem}
  
  For $0<t<1$ $\mc{E}_{N,t}$ admits a unique minimiser of $\mc{M}^{1}(\R)\times \mc{M}^{1}(\R)$. Similarly, $\mc{E}_{N,0}$  and $\mc{E}_{N,1}$ admit unique minimisers of $\mc{M}^{1}(\R)$.
 
 \end{theorem}

\Proof
 $\mc{E}_{N,t}$ is lower-continuous with compact level-sets, bounded from below and not identically $+\infty$ by virtue of Proposition \ref{Proposition LSC et cpct level sets for E N t}. 
Hence,  it attains its minimum. Since it is strictly convex by virtue of Lemma \ref{Lemme stricte convexite fnelle E N et t}, this minimum is unique.  
\qed

 \vspace{2mm}
 
 From now on, we will denote the unique minimiser for $0<t<1$ as $\big( \mu_{\e{eq}}^{(N,t)}, \nu_{\e{eq}}^{(N,t)} \big)$, \textit{viz}. 
\beq
 \e{inf} \bigg\{  \mc{E}_{N,t}[\mu,\nu] \; : \;  (\mu,\nu)   \in \mc{M}^{1}(\R) \times \mc{M}^{1}(\R)    \bigg\} \; = \;  \mc{E}_{N,t}\Big[ \mu_{\e{eq}}^{(N,t)}, \nu_{\e{eq}}^{(N,t)} \Big] \;. 
\enq
By virtue of \eqref{ecriture decomposition fnell energie}, the minimisation problem may thus be recast as  
\beq
  \mc{E}_{N,t}\Big[ \mu_{\e{eq}}^{(N,t)}, \nu_{\e{eq}}^{(N,t)} \Big] \; = \; \e{inf} \left\{   \sul{\ups = \pm }{}   \mc{E}_{N}^{(\ups)}\big[ \sg^{(\ups)} \big]  \; : \; 
\ba{c}  \big( \sg^{(+)}, \sg^{(-)} \big)   \in \mc{M}^{1}(\R) \times \mc{M}^{(2t-1)}_{\mf{s}} ( \R )  \vspace{2mm} \\
  \Big( \tfrac{ \sg^{(+)} +  \sg^{(-)} }{2t} , \tfrac{ \sg^{(+)} -  \sg^{(-)} }{2(1-t)} \Big) \in    \mc{M}^{1}(\R) \times \mc{M}^{1}(\R)  
\ea \right\}  \;. 
\label{ecriture auxiliaire du minimiseur a deux mesures}
\enq
 Above $ \mc{M}^{(\a)}_{\mf{s}}(\R)$ stands for the space of real signed measures on $\R$ of total mass $\a \in \R$. The constraint appearing in the second line
does not allow one,  \textit{a priori},  to reduce the two-dimensional vector equilibrium problem associated with $\mc{E}_{N,t}$ into two one-dimensional 
problems. However, \eqref{ecriture auxiliaire du minimiseur a deux mesures} allows one to obtain a lower bound for $  \mc{E}_{N,t}\big[ \mu_{\e{eq}}^{(N,t)}, \nu_{\e{eq}}^{(N,t)} \big]$
in terms of the minimum of  $\mc{E}_{N}^{(+)}$ which can be characterised by solving a one-dimensional equilibrium problem. The estimates obtained by computing this minimum turn out to be 
enough for the purpose of this analysis. Indeed, 
it is readily seen that $\mc{E}_{N}^{(+)}$ is lower-continuous, has compact level sets, is strictly convex on $\mc{M}^{1}(\R)$, bounded from below and not identically $+\infty$. Thus, there exists a unique
probability measure $\sg_{\e{eq}}^{(N)}$ such that 
\beq
 \mc{E}_{N}^{(+)}\big[ \sg_{\e{eq}}^{(N)} \big] \; = \; \e{inf} \Big\{   \mc{E}_{N}^{(+)}\big[ \sg \big] \; : \;   \sg    \in \mc{M}^{1}(\R) \Big\} \;. 
\enq
Moreover, it follows from the positivity of $\tf{ R^{(-)}(\la) }{ \la }$ that, for any signed measure $\sg$ on $\R$,  
\beq
\mc{E}_{N}^{(-)}\big[ \sg \big] \; = \;  \f{1}{2} \Int{ \R }{} \hspace{-1mm} \dd \la \, \big| \mc{F}\big[ \sg \big](\la)\big|^2 \f{ R^{(-)}(\la) }{ \la }  \; \geq \; 0 \; .
\enq
As a consequence, 
\beq
  \mc{E}_{N,t}\Big[ \mu_{\e{eq}}^{(N,t)}, \nu_{\e{eq}}^{(N,t)} \Big] \; \geq  \;   \mc{E}_{N}^{(+)}\big[ \sg_{\e{eq}}^{(N)} \big] \;. 
\enq

\begin{cor}
 
 One has the upper bound 
\beq
\msc{Z}_{N,p}(\varkappa) \, \leq \, \exp\bigg\{ - N^2  \mc{E}_{N}^{(+)}\big[ \sg_{\e{eq}}^{(N)} \big] \; + \; \e{O}(N \tau_N^2 \big) \bigg\}
\enq
 with a remainder that is uniform in $p$. 
\end{cor}

The analysis carried so far brings the estimation of the bounding partition function $\msc{Z}_{N,p}(\varkappa)$ to the characterisation of the 
minimiser $ \sg_{\e{eq}}^{(N)}$ of $\mc{E}_{N}^{(+)}$ and the evaluation of the large-$N$ behaviour of $\mc{E}_{N}^{(+)}\big[ \sg_{\e{eq}}^{(N)} \big]$.

\section{Characterisation of $\sg_{\e{eq}}^{(N)}$'s density by a singular integral equation}
\label{Section characterisation minimiseur et eqn integrale singuliere}

We now characterise the minimiser $\sg_{\e{eq}}^{(N)}$ of $\mc{E}_{N}^{(+)}$ introduced in \eqref{definition fonctionnelle energie +} by solving the singular integral equation satisfied by its density. 
It is standard to see, \textit{c.f.} \textit{e.g.} \cite{DeiftOrthPlyAndRandomMatrixRHP} for the details, that this minimiser is the unique solution 
on $\mc{M}^{1}(\R)$ to the problem:
\begin{itemize}
 
 \item there exist a constant $C_{\e{eq}}^{(N)}$ such that,  $\forall \nu \in \mc{M}^{1}(\R)$ with compact support and such that $\mc{E}_N^{(+)}[\nu]<+\infty$, it holds 
\beq
\Int{ \R }{} \bigg\{ \f{1}{N} V_N(x)  \; - \;  \Int{ \R }{} \mf{w}^{(+)}\Big(\tau_N(x-t)\Big)   \dd \sg_{\e{eq}}^{(N)}(t)  \bigg\} \dd \nu(x) \; \geq \; C_{\e{eq}}^{(N)}
\label{ecriture minimisation en dehors du support}
\enq

\item $ \dd \sg_{\e{eq}}^{(N)}$ almost everywhere 
\beq
\f{1}{N}V_N(x)  \; - \;  \Int{ \R }{} \mf{w}^{(+)}\Big(\tau_N(x-t)\Big)   \dd \sg_{\e{eq}}^{(N)}(t) \; = \; C_{\e{eq}}^{(N)}   \;.
\label{eqn minimiseur sur le support}
\enq

\end{itemize}

Starting from there, one may already establish several properties of the minimiser $ \sg_{\e{eq}}^{(N)}$.
We shall establish a set of conditions which, if satisfied, ensure that $ \sg_{\e{eq}}^{(N)}$ is supported on a
single segment $\intff{a_N}{b_N}$. For that purpose, it is useful to introduce the
singular integral operator on $\mc{H}_s(\intff{a_N}{b_N})$:
\beq
\mc{S}_N\big[ \phi \big](\xi) \; = \;  \Fint{a_N}{b_N} \big( \mf{w}^{(+)}\big)^{\prime}\big[ \tau_N (\xi-\eta) \big]  \cdot   \phi(\eta) \dd \eta \; . 
\label{ecriture op int sing origine}
\enq
Also, it is of use to introduce the effective potential subordinate to a function $\phi\in \mc{H}_s(\intff{a_N}{b_N})$
\beq
V_{N;\e{eff}}[\phi](\xi) \; = \; \f{1}{N} V_N(\xi) \, - \, \Fint{a_N}{b_N} \mf{w}^{(+)}\big[ \tau_N (\xi-\eta) \big]  \cdot   \phi(\eta) \dd \eta 
\label{ecriture potentiel effectif associee a une solution phi}
\enq

\begin{prop}
\label{Proposition characterisation qqes ptes mesure eq}
Let $a_N<b_N$ and $\varrho_{\e{eq}}^{(N)}\in \mc{H}_s(\intff{a_N}{b_N})$, $\tf{1}{2}<s<1$ be a solution to the equation

\beq
\f{ 1 }{ N \tau_N } V_N^{\prime}(x)  \; = \; \mc{S}_N\Big[ \varrho_{\e{eq}}^{(N)} \Big](x)  \qquad  \; on \quad \intoo{a_N}{b_N}
\label{ecriture eqn int sing lin pour densite mesure eq}
\enq
subject to the conditions 
\beq
\varrho_{\e{eq}}^{(N)}(\xi) \geq  0 \quad for \quad \xi \in \intff{a_N}{b_N} \qquad, \qquad \Int{a_N}{b_N}\varrho_{\e{eq}}^{(N)}(\xi) \dd \xi \; = \; 1 
\label{ecriture normalisation a unite}
\enq
and 
\beq
V_{N;\e{eff}}[ \varrho_{\e{eq}}^{(N)} ](\xi) \; > \; \e{inf}\Big\{  V_{N;\e{eff}}[ \varrho_{\e{eq}}^{(N)} ](\eta) \, : \, \eta \in \R \Big\} \qquad for \; any \quad \xi \in \R \setminus \intff{a_N}{b_N} \;. 
\label{ecriture positivite stricte du potentiel effectif}
\enq
Then, the equilibrium measure $\sg_{\e{eq}}^{(N)}$ is supported on the segment $\intff{a_N}{b_N}$, is continuous in respect to Lebesgue's measure with density $\dd \sg_{\e{eq}}^{(N)}= \varrho_{\e{eq}}^{(N)}(\xi) \dd \xi $. 
Moreover, it holds that $a_N=-b_N$ and the density takes the form
\beq
 \varrho_{\e{eq}}^{(N)}(\xi) \, = \, \sqrt{(b_N-\xi)(\xi-a_N)} \cdot h_N(\xi) 
\label{ecriture forme densite mesure equilibre}
\enq
for a smooth function  $h_N(\xi)$ on $\intff{a_N}{b_N}$. 

\end{prop}

We would like to point out that the condition $\varrho_{\e{eq}}^{(N)}\in \mc{H}_s(\intff{a_N}{b_N})$, $\tf{1}{2}<s<1$, enforces
a certain regularity on the function and, in particular, tames it behaviour at the endpoints $a_N, b_N$. In fact, as well be discussed
later on, the very fact of solving \eqref{ecriture eqn int sing lin pour densite mesure eq} in this class of functions imposes a constraint on $a_N$ and $b_N$.
The normalisation to unit condition \eqref{ecriture normalisation a unite} imposes a second constraint on $a_N$ and $b_N$.

\Proof

Given a solution  $\varrho_{\e{eq}}^{(N)}\geq 0$ in $\mc{H}_s(\intff{a_N}{b_N})$, $\tf{1}{2}< s < 1$ to
\beq
\f{ 1 }{ N \tau_N } V_N^{\prime}(x)  \; = \; \mc{S}_N\big[ \varrho \big](x) \quad 
\e{and}\, \e{such}\, \e{that} \quad \Int{a_N}{b_N} \varrho_{\e{eq}}^{(N)}(\xi)\dd \xi \, = \, 1 \; ,
\enq
  one
has that $\dd \sg(\xi)= \varrho_{\e{eq}}^{(N)}(\xi) \dd \xi$ is a probability measure. Moreover, the associated effective potential is constant on $\intff{a_N}{b_N}$
as follows from taking the antiderivative of the singular integral equation satisfied by $\varrho_{\e{eq}}^{(N)}$, so that \eqref{eqn minimiseur sur le support} holds.
Moreover, it follows from \eqref{ecriture positivite stricte du potentiel effectif} that $C_{\e{eq}}^{(N)}=\e{inf}\Big\{  V_{N;\e{eff}}[ \varrho_{\e{eq}}^{(N)} ](\eta) \, : \, \eta \in \R \Big\} $. 
Finally, the strict positivity in \eqref{ecriture positivite stricte du potentiel effectif} ensures that 
\eqref{ecriture minimisation en dehors du support} holds as well. Since the measure $\sg$ satisfies \eqref{ecriture minimisation en dehors du support}-\eqref{eqn minimiseur sur le support},
it must coincide with the equilibrium measure $\sg_{\e{eq}}^{(N)}$.

Finally, the fact that $\sg_{\e{eq}}^{(N)}$ is Lebesgue continuous and that its density is of the form \eqref{ecriture forme densite mesure equilibre} follows from  Lemma 2.5 in \cite{KozBorotGuionnetLargeNBehMulIntMeanFieldTh} whose statement, transcripted to the notations of this work, is relagated to Appendix \ref{Appendix Section resultats auxiliaires}
Lemma \ref{Lemme rappel article BGK15}. In order to identify the present setting with the one of the lemma, one sets
\beq
T(x,y) \; = \; \wt{\mf{w}}^{(+)}\big[\tau_N(x-y)\big] \, - \, \f{1}{N} \Big[ V_N(x)+V_N(y) \Big] \quad \e{with} \quad \wt{\mf{w}}^{(+)}(x) = \mf{w}^{(+)}(x) - 2 \ln \Big( \f{|x|}{\tau_N} \Big) \;.
\enq
Observe that the logarithmic singularity of $\mf{w}^{(+)}(x)$ at $x=0$ exactly cancels with the logarithmic counterterm, so that $\wt{\mf{w}}^{(+)}$ is continuous on $\R$. Moreover, since
$\mf{w}^{(+)}(x)$ is bounded at infinity, one gets that for $K>0$ large enough,
\beq
\big| \wt{\mf{w}}^{(+)}(x) \big| \; \leq \; C \quad \e{on} \; \; \intff{-K}{K} \quad \e{and} \quad
 \wt{\mf{w}}^{(+)}(x)  \; \leq \; C-\ln \Big( \f{ |x| }{ \tau_N} \Big) \quad \e{on} \; \; \intff{-K}{K}^{\e{c}}
\enq
Hence,
\bem
T(x,y) \; = \;\, - \, \f{1}{N} \Big[ V_N(x)+V_N(y) \Big]  \; + \; C \bs{1}_{ \intff{-K}{K}  }\big[\tau_N (x-y)\big] \; + \;
  \bs{1}_{ \intff{-K}{K}  }\big[\tau_N (x-y)\big] \cdot \Big(  C \, - \, \ln |x-y| \Big) \\
  \; \leq \; - \big[ f(x)+f(y)\big] \qquad \e{with} \quad f(x) \,  =  \, \f{ V_N(x) }{ N } \, - \,   C_1 \;,
\end{multline}
for some $C_1 > 0$.  This function $f$ does enjoy that $\liminf_{x \tend \pm \infty} \Big\{ \tf{ f(x) }{ \ln |x|}  \Big\} \, =\,  +\infty$. Moreover,
a direct calculation shows that the $\mc{E}_{T}$ defined in Lemma \ref{Lemme rappel article BGK15} satisfies
$\mc{E}_{T}[\mu]=\mc{E}_N^{(+)}[\mu]$ and hence admits a unique minimiser. Thus \eqref{ecriture forme densite mesure equilibre} follows.

It remains to establish the symmetry property of the support. Given $\sg_{\e{eq}}^{(N)}$ the unique solution to the variational problem
\eqref{ecriture minimisation en dehors du support}-\eqref{eqn minimiseur sur le support}, the image measure $\mf{s}\#\sg_{\e{eq}}^{(N)}$
with $\mf{s}(x)=-x$ satisfies, for any $\nu \in \mc{M}^1(\R)$ compactly supported and such that $\mc{E}_N^{(+)}[\nu]<+\infty$,
\bem
\Int{ \R }{} \bigg\{ \f{1}{N} V_N(x)  \; - \;  \Int{ \R }{} \mf{w}^{(+)}\Big(\tau_N(x-t)\Big)   \dd \Big(\mf{s}\# \sg_{\e{eq}}^{(N)}\Big)(t)  \bigg\} \dd \nu(x) \\
\; = \; \Int{ \R }{} \bigg\{ \f{1}{N} V_N(x)  \; - \;  \Int{ \R }{} \mf{w}^{(+)}\Big(\tau_N(x-t)\Big)   \dd  \sg_{\e{eq}}^{(N)}(t)  \bigg\} \dd \Big(\mf{s}\# \nu \Big)(x)
\;  \geq \;
 C_{\e{eq}}^{(N)}
\end{multline}
since $\mf{s}\# \nu \in \mc{M}^1(\R)$, is compactly supported and such that $\mc{E}_N^{(+)}[\mf{s}\# \nu]\, = \, \mc{E}_N^{(+)}[\nu]<+\infty$.
Likewise, $\dd \dd \Big(\mf{s}\# \sg_{\e{eq}}^{(N)}\Big)$ almost everywhere,
\beq
\f{1}{N}V_N(  x)  \; - \;  \Int{ \R }{} \mf{w}^{(+)}\Big(\tau_N(x-t)\Big)   \dd \Big(\mf{s}\# \sg_{\e{eq}}^{(N)}\Big) (t) \; = \;
\f{1}{N}V_N\big( \mf{s}^{-1}(x) \big)  \; - \;  \Int{ \R }{} \mf{w}^{(+)}\Big(\tau_N(\mf{s}^{-1}(x) -t)\Big)   \dd  \sg_{\e{eq}}^{(N)}  (t)
\;= \;  C_{\e{eq}}^{(N)} \;.
\enq
Thus, $\mf{s}\# \sg_{\e{eq}}^{(N)}$ also solves the same variational problem. By uniqueness of its solutions, $\mf{s}\# \sg_{\e{eq}}^{(N)} \, =\,  \sg_{\e{eq}}^{(N)}$.
In particular, for a measure supported on the single interval, this entails that $\intff{a_N}{b_N} \, = \, \intff{-b_N}{-a_N}$, \textit{viz}. $a_N=-b_N$. \qed

\subsection{An intermediate regularisation of the operator $\mc{S}_N$}

For technical purposes, it is convenient to introduce the singular integral operator $\mc{S}_{N;\ga}$ which provides one with a convenient regularisation of the 
singular integral operator $\mc{S}_N$ arising in \eqref{ecriture op int sing origine}:
\beq
\mc{S}_{N;\ga}[\phi](\xi) \; = \; \f{1}{2}\Fint{a_N}{b_N} S_{\ga}\big(\tau_N(\xi-\eta)  \big)  \phi(\eta) \cdot  \dd \eta
\label{definition op int sing S N gamma}
\enq
where, recalling that $\tau_N=\ln N$,
\beq
S_{\ga}(\xi) \, = \, 2 \big( \mf{w}^{(+)}\big)^{\prime}(\xi) \cdot \bs{1}_{ \intff{-\ga \ov{x}_N }{ \ga \ov{x}_N } }(\xi) \qquad \e{and} \qquad \ov{x}_N \; = \; \tau_N x_N \qquad \e{with} \qquad x_N \, = \, (b_N-a_N)  \;. 
\label{definition constante xN et XN bar}
\enq

There, $\ga>1$ but is otherwise  arbitrary. For technical reasons, it is easier to invert  $\mc{S}_{N;\ga}$ at finite $\ga>1$. Then, one obtains the \textit{per se} inverse of 
$\mc{S}_N$ by sending $\ga \tend +\infty$ at the level of the obtained answer. This inverse will be constructed in Section \ref{Section Inversion singular integral operator}.
Below, I characterise explicitly the distributional Fourier transform $S_{\ga}$ 
\beq
\mc{F}\big[ S_{\ga} \big](\la) \; = \; 2 \Fint{-\ga \ov{x}_N}{ \ga \ov{x}_N} \big( \mf{w}^{(+)}\big)^{\prime}(\xi)\,\ex{ \i \la \xi}\dd \xi  \;, 
\label{ecriture TF distributionnelle}
\enq
a result that will be of use later on.

\begin{figure}[h]
\begin{center}

\includegraphics[width=.7\textwidth]{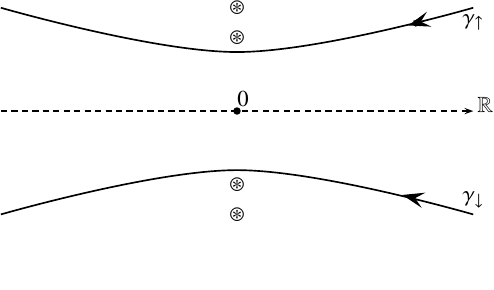}

\caption{Contours $\mc{C}_{\ua}\cup \mc{C}_{\da}$.
$\mc{C}_{\ua/\da}$ separates all the poles of $ R(\la)$ in $\mathbb{H}^{+/-}$ from $\R$. These poles are indicated by $\circledast$
and the closest poles are at $\pm \i$. The contours $\mc{C}_{\ua/\da}$ are chosen such that $\e{dist}(\mc{C}_{\ua/\da}, \R)  =  (1 -\a )$ with $\a$ small enough, the minimum of the distance being attained on the imaginary axis.
\label{contour pour les ctrs gamma up et down dans def frak r N}}
\end{center}
\end{figure}

\begin{lemme}
\label{Lemme calcul TF gN}
The distributional Fourier transform $\mc{F}[S_{\ga}](\la)$ defined by \eqref{ecriture TF distributionnelle} admits the representation
\beq
\frac{\mc{F}[S_{\ga}](\la)}{2 \i \pi } \; = \; R(\la) \;  +  \; \sul{\sg = \pm }{} \mf{r}_{N}^{(\sg)}(\la) \ex{ \i\la \sg \ga \ov{x}_N} \;, \quad \la \in \R \;,
\label{ecriture forme precise TF gN}
\enq
where
\beq
R(\la) \; = \; 2 R^{(+)}(\la) \; = \;  2 \f{ \sinh( \pi \mf{b} \la ) \cdot \sinh(\pi \hat{\mf{b}} \la ) \cdot \sinh(\tfrac{\pi}{2}\la)   }{  \cosh^2(\tfrac{\pi}{2} \la)   }  \;,
\label{definition TF de w prime principale}
\enq
and, upon introducing the convention $\mc{C}_+\equiv \mc{C}_{\ua}$ and $\mc{C}_- \equiv \mc{C}_{\da}$,
\beq
\mf{r}_N^{(\sg)}(\la) \; = \;   \sg  \Int{ -\mc{C}_{-\sg} }{ } \hspace{-2mm}   R(y) \f{ \ex{-\i\sg y\ga \ov{x}_N }  }{ \la - y } \cdot \f{ \dd y }{ 2 \i \pi }\; , \qquad \e{for} \quad \sg \in \{ \pm \} \;.
\label{definition fonction rN}
\enq
The contours $\mc{C}_{\ua/\da}$ appearing above are as depicted in Figure \ref{contour pour les ctrs gamma up et down dans def frak r N} and $-\mc{C}_{\ua/\da}$ refers to the contour
endowed with the opposite orientation that $\mc{C}_{\ua/\da}$. The integral representation for $\mf{r}_N^{(\sg)}$ holds for $\la \in \Cx \setminus \mc{C}_{-\sg}$.

Besides, there exists $C_{\epsilon} > 0$ independent of $N$ such that, uniformly in $\la \in \mathbb{H}^{\sg}$, it holds:
\beq
|\mf{r}_N^{(\sg)}(\la)| \; \leq \; \f{ C_{\eps} }{ 1+ |\la| } \cdot
\exp\big\{- \ga \ov{x}_N(1  -\a   ) \big\} \;,  
\enq
for any $\a>0$ and uniformly in $N$. Finally, one has that $\mf{r}_N^{(\sg)} \in \mc{O}\big( \, \ov{\mathbb{H}^{\sg}} \, \big)$, with $\sg \in \{ \pm \}$. 
\end{lemme}

\Proof We first start by providing a more convenient representation for $\mf{w}^{(+)}(x)$
\beq
\mf{w}^{(+)}(x) \; = \; - \bs{1}_{\R^+}(x) \Int{ -\mc{C}_{\da} }{} \dd \la \, \f{ R^{(+)}(\la) }{ \la } \ex{-\i \la x} \; - \;
\bs{1}_{\R^-}(x) \Int{ -\mc{C}_{\ua} }{} \dd \la \, \f{ R^{(+)}(\la) }{ \la } \ex{-\i \la x}
\label{ecriture representation reguliere pour w+}
\enq
Here, we focus on the $x>0$ case, the $x<0$ case can be treated quite analogously.
Consider the contour
\beq
\Ga_{\e{tot};\da}^{(M)} \; = \; \big\{- \mc{C}_{\da; M} \big\} \cup \Ga^{(M)}_{-;\da} \cup \intff{M}{-M} \cup \Ga^{(M)}_{+;\da} \quad \e{with} \quad
\mc{C}_{\da; M} = \mc{C}_{\da} \cap \{ -M \leq \Re(\la) \leq M  \} \;.
\enq
Here, $\intff{M}{-M}$ is the segment $\intff{-M}{M}$ oriented from $M$ to $-M$ and
$ \Ga^{(M)}_{\pm;\da}$ is as depicted in Figure \ref{contour pour les ctrs gamma up et down dans def frak r N}.

By Morera's theorem, since $\la \mapsto \ex{-\i x \la } \tf{ R^{(+)}(\la) }{ \la }$ is meromorphic on an open neighbourhood of
the bounded domain delimited by $\Ga_{\e{tot};\da}^{(M)}$, it holds that
\beq
\Int{ \Ga_{\e{tot};\da}^{(M)} }{} \dd \la  \f{ R^{(+)}(\la) }{ \la } \cdot \ex{-\i x \la } \; = \;0
\label{ecriture Morera pour rep int de w+}
\enq
It remains to estimate the contribution of each subcontour to the above integral as $M \tend +\infty$.
By definition, the integral along $\intff{-M}{M}$ will simply converge to $\mf{w}^{(+)}(x)$. Further, since
\beq
2 R^{(+)}(\la) \, =\, \e{sgn}\big[ \Re(\la) \big]\Big( 1 \, + \, \e{O}\big( \ex{-2\pi \mf{c} |\Re(\la)| }  \big) \Big)
\qquad \e{with} \qquad \mf{c} \; = \; \e{min}\Big\{ \tfrac{1}{2}, \mf{b}, \hat{\mf{b}} \Big\}
\enq
one readily has that, for $M$ large enough,
\beq
\bigg| \Int{ \Ga^{(M)}_{ \pm;\da} }{} \hspace{-1mm} \dd \la \,  \f{ R^{(+)}(\la) }{ \la } \cdot \ex{-\i x \la } \bigg| \, \leq \,
\Int{ 0 }{  \varkappa_{\pm}^{(M)}  } \f{\dd s }{ |\pm M + \i s | } \ex{-xs} \, \leq \; \f{C}{M} \; \limit{M}{+\infty} \; 0 \;.
\enq
Finally, one has $\la \mapsto \big| \ex{-\i x \la } \tf{ R^{(+)}(\la) }{ \la } \big| \in L^1\big( \mc{C}_{\da} \big)$ and that
$ \ex{-\i x \la } \cdot \bs{1}_{\intff{-M}{M}}\big( \Re(\la) \big) \cdot  \tf{ R^{(+)}(\la) }{ \la } $  converges pointwise to $ \ex{-\i x \la } \tf{ R^{(+)}(\la) }{ \la }$
on $ \mc{C}_{\da}$. Since it is dominated on $ \mc{C}_{\da}$ by the modulus of its limit which is in $ L^1\big( \mc{C}_{\da} \big)$, by dominated convergence, one has that
\beq
 \Int{ -\mc{C}_{\da;M} }{} \hspace{-1mm} \dd \la \,  \f{ R^{(+)}(\la) }{ \la } \cdot \ex{-\i x \la }  \;  \limit{M}{+\infty} \;
 \Int{ -\mc{C}_{\da} }{} \hspace{-1mm} \dd \la \,  \f{ R^{(+)}(\la) }{ \la } \cdot \ex{-\i x \la } \;.
\enq
The claim follows by taking the $M\tend + \infty$ limit of \eqref{ecriture Morera pour rep int de w+}.

The representation \eqref{ecriture representation reguliere pour w+} is the starting point for calculating the Fourier transform of $S_{\ga}$.
One has that
\bem
\frac{\mc{F}[S_{\ga}](\la)}{2 \i \pi } \; = \; \lim_{\eps\tend 0^+} \Bigg\{ \Int{-\ga \ov{x}_N}{-\eps}+\Int{\eps}{ \ga \ov{x}_N}  \Bigg\} \f{ \dd \xi }{2\i\pi}
2 \big( \mf{w}^{(+)}\big)^{\prime}(\xi)\,\ex{ \i \la \xi}   \\
\; = \; \lim_{\eps\tend 0^+} \Bigg\{ \i \Int{-\ga \ov{x}_N}{-\eps} \f{ \dd \xi }{2\i\pi}   \ex{\i\la \xi} \Int{ -\mc{C}_{\ua} }{} \dd \mu R(\mu) \ex{-\i\xi \mu}
\; + \; \i \Int{\eps}{ \ga \ov{x}_N} \f{ \dd \xi }{2\i\pi}   \ex{\i\la \xi} \Int{ -\mc{C}_{\da} }{} \dd \mu R(\mu) \ex{-\i\xi \mu}  \Bigg\} \\
 \; = \; \lim_{\eps\tend 0^+} \Bigg\{  \Int{ -\mc{C}_{\ua} }{} \f{ \dd \mu }{2\i\pi}  R(\mu) \f{ \ex{-\i \eps (\la- \mu) } \, - \,  \ex{-\i \ga \ov{x}_N (\la- \mu) }   }{ \la-\mu }
\; + \;   \Int{ -\mc{C}_{\ua} }{} \dd \mu R(\mu) \f{  \ex{\i \ga \ov{x}_N (\la- \mu) } \, - \,  \ex{ \i \eps (\la- \mu) }    }{ \la-\mu } \Bigg\} \\
 \; = \;  \Int{\R}{} \dd \mu  R(\mu) \f{  \ex{ \i (\la-\mu) \ga \ov{x}_N}  \, - \,   \ex{ -\i (\la-\mu) \ga \ov{x}_N}   }{2 \i \pi(\la-\mu) }
\; + \; \lim_{\eps\tend 0^+}\Bigg\{ \Int{\R}{} \dd \mu  R(\mu) \f{  \ex{ -\i (\la - \mu ) \eps }  \, - \,   \ex{ \i (\la-\mu) \eps}   }{2 \i \pi(\la-\mu) }    \Bigg\} \;.
\label{ecriture calcul regularise de TF de S gamma}
\end{multline}
There, dealing with absolutely convergence integrals, we have applied Fubini in the second line, what allowed us to take the $\xi$ integrals in the third line.
To get the fourth line, we have deformed the contours to the real axis then gathered all under the same integral and, finally split again in the two integrals
appearing there. These two integrals can be further simplified.

It holds that
\bem
 \Int{\R}{} \dd \mu  R(\mu) \f{  \ex{ \i (\la-\mu) \ga \ov{x}_N}  \, - \,   \ex{ -\i (\la-\mu) \ga \ov{x}_N}   }{2 \i \pi(\la-\mu) } \; = \; \lim_{M\tend + \infty} \lim_{\eps \tend 0^+}
\Int{-M}{M} \f{\dd \mu }{2 \i \pi } R(\mu)  \f{  \ex{ \i (\la-\mu+\i\eps ) \ga \ov{x}_N}  \, - \,   \ex{ -\i (\la-\mu+\i \eps ) \ga \ov{x}_N}   }{ \la-\mu + \i \eps   } \\
\; = \;  \lim_{M\tend + \infty} \lim_{\eps \tend 0^+} \Bigg\{
 \hspace{-4mm} \Int{ \Ga_{+;\da}^{(M)} \cup \{-\mc{C}_{\da}^{(M)}\} \cup  \Ga_{-;\da}^{(M)}   }{} \hspace{-8mm}  \f{\dd \mu }{2 \i \pi }  R(\mu)
\f{  \ex{ \i (\la-\mu+\i\eps ) \ga \ov{x}_N}   }{ \la-\mu + \i \eps   }
\;\; + \hspace{-4mm} \Int{ \Ga_{+;\ua}^{(M)} \cup \{ \mc{C}_{\ua}^{(M)} \} \cup  \Ga_{-;\ua}^{(M)}   }{} \hspace{-8mm}  \f{\dd \mu }{2 \i \pi }  R(\mu)
\f{  \ex{ - \i (\la-\mu+\i\eps ) \ga \ov{x}_N}   }{ \la-\mu + \i \eps   }  \; + \;  R(\la+\i\eps) \Bigg\} \;.
\end{multline}
There, we agree that the curve $\Ga_{+;\ua}^{(M)}$, resp. $ \Ga_{-;\ua}^{(M)}$, is obtained by reflecting $\Ga_{+;\da}^{(M)}$, resp. $ \Ga_{-;\da}^{(M)}$,
on $\R$.
At this stage, one may take the $\eps \tend 0^+$ limit by either invoking continuity or applying the dominated convergence theorem.
The $M\tend +\infty$ limit of the resulting integrals may be computed exactly as it was discussed earlier on, leading to
\beq
\Int{\R}{} \dd \mu  R(\mu) \f{  \ex{ \i (\la-\mu) \ga \ov{x}_N}  \, - \,   \ex{ -\i (\la-\mu) \ga \ov{x}_N}   }{2 \i \pi(\la-\mu) }  \; = \;
R(\la) \; - \; \sul{\sg \in \{\pm\}}{} \sg  \Int{ - \mc{C}_{ \sg} }{} \hspace{-1mm} \dd \mu \, R(\mu)  \f{  \ex{ -\i \sg (\la - \mu) \ga \ov{x}_N }     }{2 \i \pi (\la-\mu) } \;,
\enq
where $\mc{C}_{\ua/\da}$ are as given in Fig.~\ref{contour pour les ctrs gamma up et down dans def frak r N}.

In order to take the $\eps \tend 0^+$ limit of the remaining integral in \eqref{ecriture calcul regularise de TF de S gamma}, one needs to regularise
the integrand prior to applying dominated convergence. First of all, upon invoking Morera's theorem and the definition of a Riemann oscillatory integral
one gets
\bem
\lim_{\eps\tend 0^+}\Bigg\{ \Int{\R}{} \hspace{-2mm} \dd \mu \,   R(\mu) \f{  \ex{ -\i (\la - \mu ) \eps }  \, - \,   \ex{ \i (\la-\mu) \eps}   }{2 \i \pi(\la-\mu) }    \Bigg\}  \\
\; = \;
\lim_{\eps\tend 0^+} \lim_{M\tend + \infty} \Bigg\{  \hspace{-3mm}  \Int{-M}{ -M+\i(\eta + \Im(\la) ) }  \hspace{-1mm} +  \hspace{-1mm}
\Int{ -M+\i(\eta + \Im(\la) )  }{  M+\i(\eta + \Im(\la) )  }
\hspace{-2mm} +   \Int{M+\i(\eta + \Im(\la) )}{M}  \Bigg\} \dd \mu R(\mu) \sul{\sg = \pm }{} \sg  \f{ \ex{-\i\sg \eps (\la-\mu) } }{ 2\i\pi (\la-\mu)  } \;.
\end{multline}
There, since the lengths of the segments $\intff{\pm M}{  \pm M+\i(\eta + \Im(\la) ) } $
is uniformly bounded in $M$ and $R$ is also bounded there,
direct bounds yield
\beq
\bigg| \Int{ \pm M }{ \pm M+\i(\eta + \Im(\la) ) } \hspace{-2mm} \dd \mu \, R(\mu) \sul{\sg = \pm }{} \sg  \f{ \ex{-\i\sg \eps (\la-\mu) } }{ 2\i\pi (\la-\mu)  }   \bigg| \; \leq \;
\f{ C }{ |\la \mp M | }  \limit{M}{+\infty} 0
\enq
In its turn, the integral along the segment $ \intff{  - M + \i(\eta + \Im(\la) )  }{  M + \i(\eta + \Im(\la) )} $ converges, by definition,
to the Riemann oscillatory integral along $\R+\i (\eta + \Im(\la) )$, leading eventually to
\beq
\lim_{\eps\tend 0^+}\Bigg\{ \Int{\R}{} \dd \mu  R(\mu) \f{  \ex{ -\i (\la-\mu) \eps }  \, - \,   \ex{ \i (\la-\mu) \eps}   }{2 \i \pi(\la-\mu) }    \Bigg\} \; = \;
\lim_{\eps\tend 0^+}\Bigg\{ \Int{\R + \i \eta + \i \Im(\la) }{} \hspace{-4mm} \dd \mu
\;   R(\mu)  \cdot  \f{  \ex{ -\i (\la-\mu) \eps }  \, - \,   \ex{ \i (\la-\mu) \eps}   }{2 \i \pi(\la-\mu) }    \Bigg\}
\enq
Observe that the large-argument asymptotics of $R$ may be presented in the form
\beq
R(\mu) = \e{sgn}\Big( \Re (\mu-\la)\Big) \; + \; \e{O}\Big(  \ex{ - 2\pi\mf{c} |\Re(\mu)| }  \Big) \qquad \e{with} \qquad \mf{c} \; = \; \e{min}\Big\{ \tfrac{1}{2}, \mf{b}, \hat{\mf{b}} \Big\} \;,
\enq
and whenever $\la \in \Cx$ is fixed while $|\Re(\mu)| \tend +\infty$.  By dominated convergence, this entails that
\beq
\lim_{\eps\tend 0^+}\Bigg\{ \Int{\R + \i \eta + \i \Im(\la) }{} \hspace{-4mm}  \dd \mu  \Big[ R(\mu) \, - \, \e{sgn}\Big( \Re (\mu-\la)\Big) \Big] \cdot
\f{  \ex{ -\i (\la - \mu) \eps }  \, - \,   \ex{ \i (\la - \mu) \eps}   }{2 \i \pi(\la-\mu) }    \Bigg\} \; = \; 0 \;.
\enq
Putting the pieces together and then re-centering the integral at $\la$, gives
\bem
 \lim_{\eps\tend 0^+}\Bigg\{ \Int{\R}{} \! \dd \mu  R(\mu) \f{  \ex{ -\i (\la-\mu) \eps }  \, - \,   \ex{ \i (\la-\mu) \eps}   }{2 \i \pi(\la-\mu) }    \Bigg\} \; = \;
\lim_{\eps\tend 0^+}\Bigg\{ \Int{\R + \i \eta }{} \hspace{-2mm} \dd \mu   \;  \e{sgn}\big[ \Re (\mu)\big]  \cdot  \f{  \ex{ \i \mu \eps }  \, - \,   \ex{ -\i \mu \eps}   }
{- 2 \i \pi \mu }    \Bigg\}  \\
\; = \; \lim_{\eps\tend 0^+} \lim_{M\tend + \infty} \Bigg\{ - \hspace{-2mm} \Int{-M + \i \eta }{ M + \i \eta }  \f{ \dd \mu }{2\i\pi }  \e{sgn}\big[ \Re (\mu)\big]  \f{ \ex{\i\eps \mu} }{ \mu }
\; + \hspace{-1mm}  \Int{-M + \i \eta }{ M + \i \eta }  \f{ \dd \mu }{2\i\pi }  \e{sgn}\big[ \Re (\mu)\big]  \f{ \ex{-\i\eps \mu} }{ \mu }  \Bigg\}
\end{multline}
The remaining integrals may now be transformed again by appling the residue theorem.
\beq
  \hspace{-2mm} \Int{-M + \i \eta }{ M + \i \eta }  \f{ \dd \mu }{2\i\pi }  \e{sgn}\big[ \Re (\mu)\big]  \f{ \ex{-\i\eps \mu} }{ \mu } \;  = \;
\Bigg\{ - \Int{ \op{C}^{(\da)}_{M;R}  }{  } - \Int{ \mc{I}_R }{}  + \Int{\mc{I}_L}{} + \Int{ \op{C}^{(\da)}_{M;L}  }{  } \Bigg\} \f{ \dd \mu }{2\i\pi }
   \f{ \ex{-\i\eps \mu} }{ \mu }
\enq
There, the integrals run over the oriented quarter discs
\beq
\op{C}^{(\da)}_{M;R}\; = \; \Big\{ \i\eta + M \ex{\i\th} \; , \; \th \in \intff{ 0 }{ -\tf{\pi}{2} }  \Big\} \quad \e{and} \quad
\op{C}^{(\da)}_{M;L}\; = \; \Big\{ \i\eta - M \ex{\i\th} \; , \; \th \in \intff{ \tf{\pi}{2} }{0}  \Big\}
\enq
and the deformed segments
\beq
\ba{ccc}  \mc{I}_R & = & \intff{ \i(\eta-M) }{ \i\eta } \setminus \intff{-\i r }{ \i r } \cup \Big\{ r \ex{\i\th} \; , \; \th \in \intff{-\tf{\pi}{2} }{ \tf{\pi}{2} } \Big\} \vspace{2mm} \\
     \mc{I}_L & = & \intff{ \i\eta }{ \i(\eta-M) } \setminus \intff{\i r }{ -\i r } \cup \Big\{ r \ex{\i\th} \; , \; \th \in \intff{\tf{\pi}{2} }{ 3\tf{\pi}{2} } \Big\}
\ea  \;.
\enq
The integral along $\op{C}^{(\da)}_{M;R}$ may be estimated as follows
\bem
\bigg|  \Int{ \op{C}^{(\da)}_{M;R}  }{  }  \f{ \dd \mu }{2\i\pi }    \f{ \ex{-\i\eps \mu} }{ \mu }  \bigg| \; \leq \;
\bigg|  \Int{ 0  }{ -\tfrac{\pi}{2} }  \dd \th  \f{ M \i \ex{\i\th} }{2\i\pi \big( \i\eta + M \ex{\i\th} \big) }   \ex{\eps \eta -\i\eps M \ex{\i\th}}   \bigg|  \\
\; \leq \;  \Int{ -\tfrac{\pi}{2} }{ 0  }   \dd \th  \f{  M  \ex{\eps \eta +M \eps \sin(\th) }   }{2\pi \big[ (\eta+ M \sin(\th))^2+M^2 \cos^2(\th) \big]^{\f{1}{2}} }
\;  \leq  \; C \ex{\eps \eta} \Int{0}{\tf{\pi}{2}} \ex{- \tfrac{2}{\pi}M\eps \th} \; \limit{M}{+\infty} \; 0
\end{multline}
The integral along $\op{C}^{(\da)}_{M;L}$ may be estimated analogously and vanishes as well in the $M\tend + \infty$ limit. Finally, one has
\bem
 \Int{\mc{I}_L}{} - \Int{ \mc{I}_R }{}   \f{ \dd \mu }{2\i\pi } \f{ \ex{-\i\eps \mu} }{ \mu } \; = \; \f{1}{\i \pi}\Int{\eta }{\eta-M} \f{ \dd s }{ s } \ex{\eps s } \bs{1}_{\intff{-r}{r}^{\e{c}}}(s)
\; - \; \Int{ - \tfrac{\pi}{2} }{ \tfrac{\pi}{2} } \f{ \dd \th }{ 2\pi } \ex{-\i \eps r \ex{\i\th} } \; + \;
\Int{  \tfrac{\pi}{2} }{ 3 \tfrac{\pi}{2} } \f{ \dd \th }{ 2\pi } \ex{-\i \eps r \ex{\i\th} } \\
\limit{r}{0^+} \f{1}{\i\pi} \Fint{ \eta }{\eta -M } \f{\dd s }{ s } \ex{\eps s } \; \limit{ M }{ + \infty } \; \f{1}{\i\pi} \Fint{ \eta }{ - \infty } \f{\dd s }{ s } \ex{\eps s }
\; = \; \f{1}{\i\pi} \Fint{ -\eta }{ + \infty } \f{\dd s }{ s } \ex{ - \eps s }
\end{multline}
Thus, all-in-all,
\beq
  \lim_{M\tend + \infty} \Int{-M + \i \eta }{ M + \i \eta }   \hspace{-2mm}  \f{ \dd \mu }{2\i\pi }  \e{sgn}\big[ \Re (\mu)\big]  \f{ \ex{-\i\eps \mu} }{ \mu } \;  = \;
 \f{1}{\i\pi} \Fint{ -\eta }{ + \infty } \f{\dd s }{ s } \ex{ - \eps s }
\enq
The very same reasoning shows that
\beq
  \lim_{M\tend + \infty} \Int{-M + \i \eta }{ M + \i \eta } \hspace{-2mm} \f{ \dd \mu }{2\i\pi }  \e{sgn}\big[ \Re (\mu)\big]  \f{ \ex{\i\eps \mu} }{ \mu } \;  = \;
 \f{- 1}{\i\pi} \Fint{ \eta }{ + \infty } \f{\dd s }{ s } \ex{ - \eps s }
\enq
Thus, by putting these results together, one gets that
\beq
 \lim_{\eps\tend 0^+}\Bigg\{ \Int{\R}{} \dd \mu  R(\mu) \f{  \ex{ -\i (\la-\mu) \eps }  \, - \,   \ex{ \i (\la-\mu) \eps}   }{2 \i \pi(\la-\mu) }    \Bigg\} \; = \;
\f{-1}{\i\pi} \lim_{\eps\tend 0^+}\Bigg\{ \Fint{\eta}{+\infty}  \hspace{-1mm} \dd \mu \, \f{ \ex{-\eps \mu} }{ \mu } \, - \, \Fint{-\eta}{+\infty}  \hspace{-1mm} \dd \mu \,   \f{ \ex{ - \eps \mu} }{ \mu } \Bigg\}
\; = \; \f{1}{\i\pi} \lim_{\eps\tend 0^+} \Fint{\eps \eta}{- \eps \eta} \dd \mu \f{\ex{\mu}}{\mu} \, = \, 0 \;.
\enq

\qed

\section{An auxiliary Riemann--Hilbert problem}
\label{Section auxiliary RHP}

This section carries out the Deift-Zhou non-linear steepest descent analysis of a Riemann--Hilbert problem for a $2\times2$ piecewise analytic matrix $\chi$. This matrix 
allows one to construct the inverse, on an appropriate functional space, of the operator $\mc{S}_{N;\ga}$. See
\cite{KozBorotGuionnetLargeNBehMulIntOfToyModelSoVType,NovokshenovSingIntEqnsIntervalGeneral}
for more details on the connection between these problems. At this stage, I would like to provide a summary of the strategy that will be followed.
The implementation of the Deift-Zhou non-linear steepest descent method on the level of the Riemann--Hilbert for $\chi$ first demands to
solve an auxiliary scalar Riemann--Hilbert problem which will be discussed in Sub-section \ref{SousSection Opening Scalar RHP}.
The latter utilises the properly normalised at $\infty$ Winer-hopf factors $R_{\ua/\da}$ of the distributional Fourier transform $R$
of $2 \big(\mf{w}^{(+)}\big)^{\prime}$ introduced in Lemma \ref{Lemme calcul TF gN}. Since one works with an integral operator
whose integral kernel is a truncation of a the compact interval, \textit{c.f.} \eqref{ecriture TF distributionnelle}, there are "finite-size" corrections
to the distributional Fourier transform given in \eqref{ecriture forme precise TF gN}. These have to be treated separately in the course of the
large-$N$ analysis of the the Riemann--Hilbert problem for $\chi$ what requires the introduction of two auxiliary matrices $\mc{P}_{L;\ua/da}$,
\textit{c.f.} \eqref{definition facteurs matriciels extra P L ua et da}  below. A standard ingredient of the Deift-Zhou non-linear steepest descent consists
in finding an appropriate factorisation of the jump matrix $G_{\chi}$ for the sought solution $\chi$. In the present case, this is done
with the help of the auxiliary matrices $\mc{R}_{\ua/\da}$ and $M_{\ua/\da}$ introduced below in \eqref{definition matrices R ua da} and \eqref{definition matrices M ua da},
along with their approxmiants $\mc{R}_{\ua/\da}^{(\infty)}$ valid uniformly away from the real axis, see  \eqref{definition matrices R ua da infty}.
The use of these auxiliary matrices provides one with an auxiliary Riemann--Hilbert problem which deals with solutions
having prescribed poles at $0$. To satisfy the pole condition one needs to introduce a certain auxilary matrix, $\mc{P}_{R}$, that will take care of it, see \eqref{definition matrice PR}.
Once all this is provided, the original Riemann--Hilbert problem for $\chi$ is recast in the form of a Riemann--Hilbert problem for an
unknown matrix $\Pi$ whose jump matrix satisfies $G_{\pi}-I_2=\e{O}\big( \ex{-2 \ov{x}_N(1-\a)}\big)$, for some $\a>0$. This problem
is then solved in terms of a Neumann series expansion of the associated singular linear integral equation.

\subsection{An opening scalar Riemann--Hilbert problem}
\label{SousSection Opening Scalar RHP}

From now on, we fix $\eps>0$ and small enough and  introduce the solution $\ups$ to the following scalar Riemann--Hilbert problem:
\begin{itemize}
\item $\ups \in \mc{O}\big( \Cx \setminus \{\R + \i \eps \} \big) $ and has continuous $\pm$-boundary values on 
$\R+  \i \eps$ ;
\item $\ups(\la) = \left\{ 
\ba{cc}    \i \,  \big(- \i \la \big)^{-\f{3}{2}} \cdot   \big(1 \; + \; \e{O}\big(\la^{-1}\big) \big)   & 
			 \,\,\mathrm{if}\,\,\Im(\la) > \eps  \\
  \big( \i \la \big)^{-\f{3}{2}} \cdot  \big(1 \; + \; \e{O}\big(\la^{-1}\big) \big) &
				\,\,\mathrm{if}\,\, \Im(\la) < \eps  \ea\right. $

  when $\la \tend \infty$, uniformly up to the boundary;
\item $ \ups_+(\la) \cdot R(\la) \; = \;   \ups_-(\la) $ \quad for \quad  $\la \in \R + \i \eps$ \;. 
\end{itemize}
This problem admits a unique solution given by 
\beq
\ups(\la) \; = \; \left\{  \ba{cc}   R_{\ua}^{-1}(\la)   &  \,\,\mathrm{if} \quad  \Im(\la)\, > \, \eps  \\
							  R_{\da}(\la)   &   \,\,\mathrm{if} \quad \Im(\la) \,  <  \, \eps   \ea \right.
\label{definition fonction alpha}
\enq
where $R_{\ua/\da}$ are the Wiener-Hopf factors of $R$ given in \eqref{definition TF de w prime principale}: $R=R_{\ua}\, R_{\da}$, where 
\beq
R_{\ua}(\la) =   \sqrt{ \pi \mf{b}  \hat{\mf{b}} } \cdot \la^3  \cdot \mf{b}^{-\i \mf{b} \la} \cdot \hat{\mf{b}}^{-\i \hat{\mf{b}} \la} \cdot  2^{ -\i \frac{\la}{2} } \cdot 
\Ga\left( \ba{c}  \tfrac{1}{2} - \i \tfrac{\la}{2} \, ,  \tfrac{1}{2}  - \i \tfrac{\la}{2} \vspace{1mm} \\
1-\i \mf{b} \la \, , 1-\i \hat{\mf{b}} \la \, ,  1-\i \tfrac{\la}{2} \ea \right) 
  \label{ecriture explicite R up} 
\enq
and
\beq
R_{\da}(\la) =  \f{  \i  }{ \la^3  }  \sqrt{ \f{ 4 \pi }{ \mf{b}  \hat{\mf{b}}  }   }   \cdot \mf{b}^{  \i \mf{b} \la} \cdot \hat{\mf{b}}^{ \i \hat{\mf{b}} \la} \cdot  2^{ \i \frac{\la}{2} } \cdot 
\Ga\left( \ba{c}  \tfrac{1}{2} + \i \tfrac{\la}{2}  \, ,  \tfrac{1}{2} + \i \tfrac{\la}{2}  \vspace{1mm} \\
\i \mf{b} \la \, , \i \hat{\mf{b}} \la \, ,  \i \tfrac{\la}{2}  \ea \right)  \;.
\label{ecriture explicite R moins}
\enq
Above, I employed hypergeometric like notations for ratios of products of $\Ga$-functions, \textit{c.f.} \eqref{ecriture convention produit ratios fcts Gamma}. 
Note that $R_{\ua}$ and $R_{\da}$ satisfy to the relations
\beq
R_{\ua}(-\la) \; = \; - \la^3 \cdot R_{\da}(\la)  \qquad i.e. \qquad R_{\da}(-\la) \; = \; \f{  R_{\ua}(\la) }{ \la^3 }  \;. 
\label{ecriture identite conjugaison R plus et R moins}
\enq
Moreover, it holds 
\beq
R_{\da}(0) \; = \;  \Big( \f{ \pi^3 \mf{b} \hat{\mf{b}}\la^3 }{ R_{\ua}(\la)} \Big)_{\mid \la=0} \; = \;   \pi^{\f{3}{2}}\cdot  \Big( \mf{b} \hat{\mf{b}} \Big)^{\f{1}{2}}  \, = \, 
\Big( \f{ R_{\ua}(\la) }{  \la^3 } \Big)_{\mid \la=0}   \;. 
\enq

Finally, $R_{\ua/\da}$ exhibit the asymptotic behaviour 
\beqa
R_{\ua}(\la) &  =  & -\i \big( - \i  \la \big)^{ \f{3}{2}} \cdot  \Big( 1\; + \; \e{O}\big( \la^{-1} \big) \Big) \qquad \e{for} \quad 
\la \underset{  \la \in  \mathbb{H}^+   }{ \longrightarrow }  \infty   \\
\label{427bis} R_{\da}(\la) & = &       \big( \i \la \big)^{-\f{3}{2}}  \cdot \Big( 1\; + \; \e{O}\big( \la^{-1} \big) \Big) \hspace{1.4cm} \e{for} \quad 
\la \underset{  \la \in  \mathbb{H}^-   }{ \longrightarrow }  \infty 
\eeqa
as it should be. The subscripts $\ua$ and $\da$ indicate the direction, in respect to $\R+\i\eps$, in the complex plane where  $R_{\ua/\da}$ has no poles or zeros.

\subsection{Preliminary definitions} 
I need to define few other objects before describing the Riemann--Hilbert problem of interest. Let:
\beq
\mc{R}_{\ua}(\la) \; = \; \left( \ba{cc}  0   &   -1  \\
									1    &    -R(\la) \ex{  \i \la \ov{x}_N }     \ea \right) \qquad \e{and} \qquad 
\mc{R}_{\da}(\la) \; = \; \left( \ba{cc}  -1   &  R(\la) \ex{- \i \la \ov{x}_N }  \\ 
									0    &   1    \ea \right)\;,
\label{definition matrices R ua da}
\enq
as well as their "asymptotic" variants:
\beq
\mc{R}_{\ua}^{(\infty)} \; = \; \left( \ba{cc}  0   &   -1  \\
									1    &   0    \ea \right) \qquad \e{and} \qquad 
\mc{R}_{\da}^{(\infty)} \; = \; \left( \ba{cc}  -1   &  0  \\ 
									0    &   1    \ea \right)  \;. 
\label{definition matrices R ua da infty}
\enq
Further, let 
\beq
M_{\ua}(\la)  \; = \; \left( \ba{cc}  1   &   0  \\
									 -\f{ 1- R^2(\la) }{ \ups^2(\la) \cdot R(\la) } \ex{ \i \la \ov{x}_N }  & 1    \ea \right) \qquad \e{and} \qquad 
M_{\da}(\la) \; = \; \left( \ba{cc}  1   &  \ups^2(\la) \cdot \f{ 1- R^2(\la) }{  R(\la) }  \ex{- \i \la \ov{x}_N }  \\ 
									0    &   1    \ea \right)\;,
\label{definition matrices M ua da}
\enq
where $\ups$ is given by \eqref{definition fonction alpha} and $\ov{x}_N$ is as introduced in \eqref{definition constante xN et XN bar}. It will also be useful to introduce the two matrices
\beq
\left\{ \ba{c} \mc{P}_{L;\ua} (\la) \; = \; I_2 \; + \; \mf{r}^{(+)}_N(\la) \,\ex{ \i (\ga-1)\la \ov{x}_N }\cdot  \sg^-  \vspace{2mm} \\
			\mc{P}_{L;\da} (\la) \; = \; I_2 \; + \; \mf{r}^{(-)}_N(\la)  \,\ex{ - \i (\ga-1)\la \ov{x}_N }\cdot  \sg^-  \ea \right. \;\;,
\label{definition facteurs matriciels extra P L ua et da}
\enq
with $\mf{r}^{(\pm)}_N$ as defined through \eqref{definition fonction rN}.
It is readily seen that $\mc{P}_{L;\ua} (\la)$ is analytic   above $\mc{C}_{\da}$ while $\mc{P}_{L;\da} (\la)$ is analytic below $\mc{C}_{\ua}$ where
the curves $\mc{C}_{\ua/\da}$ have been depicted in Fig.~\ref{contour pour les ctrs gamma up et down dans def frak r N}.
In a later part of this section, we will construct a piecewise analytic matrix $\Pi$, \textit{c.f.}  \eqref{ecriture eqn int sing pour matrice Pi moins} and \eqref{ecriture rep int matrice Pi},
which satisfies $\Pi(\la)=I_2  \, + \, \e{O}\Big(       \ex{-2 \ov{x}_N (1-\a) } \Big) $ uniformly on $\Cx$. Here, we already use it in the construction of an auxiliary matrix $\mc{P}_R$
defined by
\beq
\mc{P}_R(\la) \; = \; I_2 \; + \;  \sul{ \ell = 0 }{ 2 } \f{ 1 }{ \la^{3-\ell} } \,  \Pi^{-1}(0) \cdot  \mc{Q}_{\ell}   \cdot \Pi(0)  \;, 
\label{definition matrice PR}
\enq
in which $  \mc{Q}_{\ell} \in \mc{M}_2(\Cx)$ are constant  matrices defined later on through \eqref{reparametrisation matices Qk} and \eqref{ecriture systeme pour matrices Mk}.  
They admit the large-$N$ behaviour 
\beq
\mc{Q}_{\ell} \, = \, c_{\ell} \sg^{+} \, + \, \e{O}\Big(  ( \, \ov{x}_N)^2 \cdot \ex{-2 \ov{x}_N (1-\a) } \Big)  
\label{ecriture DA matrice Q ell}
\enq
for any $0<\a<1$ and where the coefficients $c_{\ell}$ are defined by the below Laurent series expansion 
\beq
\f{ R_{\da}^2(\la) }{ R(\la) } \ex{ - \i \la \ov{x}_N } \; = \;  \sul{\ell=0}{3} \f{ c_{\ell} }{ \la^{3-\ell} } \; + \; \e{O}\big( \la \big) \quad \e{as} \quad \la \tend 0 \;. 
\label{definition coefficients c ell}
\enq
One should note that
\beq
c_k \; = \; (-\i)^k w_k \;\; \e{where} \;\; w_k \in \R \;\; \e{and}\;\;\e{that} \qquad w_k \, = \,  \f{ 1 }{ k!}  (\,\ov{x}_N )^k \cdot \Big\{ 1\, + \, \e{O}\Big(\tfrac{1}{\ov{x}_N}  \Big)  \Big\}\;. 
\label{definition des constantes wk et de leur asymptotiques}
\enq
Hence, $w_{\ell}>0$ provided that $\ov{x}_{N}$ is large enough. Finally, a direct computation of the coefficients $c_1, c_2$ yields that 
\beq
c_0 = 1 \quad \e{and} \quad  c_1^2=2c_2 \quad  i.e. \quad  w_1^2 = 2w_2 \,. 
\label{ecriture relations entre les coeffs ck}
\enq

\subsection{  $2 \times 2$ matrix Riemann--Hilbert problem for $\chi$}
\label{Sous Section RHP pour chi}

Below, I   adopt the shorthand notation
\beq
\mf{s}_{\la}=\e{sgn}\big( {\rm Re}\,\la \big)  \;. 
\label{definition signe s lambda}
\enq
Consider the below auxiliary Riemann--Hilbert problem for a $2\times 2$ matrix function $\chi\in \mc{M}_2\Big( {\cal O}(\Cx\setminus\R) \Big)$:
\begin{itemize}

\item $\chi$ has continuous $\pm$-boundary values on  $\R$;
\item there exist constant matrices $\chi^{(a)}$ with  $\chi^{(1)}_{12}\not= 0$ so that $\chi$ has the $|\la| \tend \infty$ asymptotic expansion
\beq
\hspace{-1.5cm}
\chi(\la) = \left\{ 
\ba{cc} \mc{P}_{L;\ua}(\la) \cdot  \left( \ba{cc} -\mf{s}_{\la} \cdot \ex{ \i\la \ov{x}_N }  & 1 \\
										-1 &  0     \ea \right)  
		      \cdot \ex{ \i\f{3 \pi}{2}\sg_3} \big(-\i \la \big)^{  \f{3}{2} \sg_3 } \cdot 
  \Big(I_{2}    +   \f{\chi^{(1)}}{\la}   +   \f{\chi^{(2)}}{\la^2}   +   \e{O}\big(\la^{-3}\big) \Big) \cdot \op{Q}(\la)  & 
			 \la \in \mathbb{H}^+ \vspace{3mm} \\
\mc{P}_{L;\da}(\la) \cdot  \left( \ba{cc} -1  & \mf{s}_{\la} \cdot \ex{- \i\la \ov{x}_N }   \\
										0  & 1    \ea \right)
			  \cdot \big( \i \la \big)^{\f{3 }{2} \sg_3 }  \cdot 
		     \Big(I_{2}    +   \f{\chi^{(1)}}{\la}   +   \f{\chi^{(2)}}{\la^2}   +   \e{O}\big(\la^{-3}\big) \Big) \cdot \op{Q}(\la) &
				\la \in \mathbb{H}^-  \ea\right. 
\label{ecriture DA chi dans RHP}
\enq

in which the matrix $\op{Q}$ takes the form
\beq
 \op{Q}(\la) \; = \; \left( \ba{cc}  0   & - \chi^{(1)}_{12}  \\
   \Big\{  \chi^{(1)}_{12} \Big\}^{-1}   &   \mf{q}_1 \, + \, \la   \ea \right)  \qquad \e{where} \qquad \mf{q}_1 \, = \, \Big( \chi^{(1)}_{11} \chi^{(1)}_{12} \, - \, \chi^{(2)}_{12}  \Big)\cdot  \Big\{  \chi^{(1)}_{12} \Big\}^{-1}    \;;
\label{definition matrice Q}
\enq
\item $\chi_+(\la) \; = \;   G_{\chi}(\la) \cdot \chi_-(\la) $ \quad for \quad  $\la \in \R$  where
\beq
G_{\chi}(\la) \; = \; \left( \ba{cc}   \ex{ \i \la \ov{x}_N }  & 0  \\  
							\frac{ 1 }{  2\i\pi  }	\cdot \mc{F}\big[ S_{\ga} \big](\la)    &  -\ex{- \i \la \ov{x}_N }  \ea \right)   \;. 							 
\label{definition matrice de saut G chi}
\enq
\end{itemize}

Here and in the following, an $\e{O}$ remainder appearing in a matrix equality should be understood to hold entrywise. 
Also, one should remark that the matrix $\op{Q}$ appearing in the asymptotic expansion \eqref{ecriture DA chi dans RHP} is chosen such that $\chi$
has the large-$\la$ behaviour 
\beq
\chi(\la) \, = \, \chi_{\ua/\da}^{(\infty)}(\la) \cdot  \big(\mp\i \la \big)^{  \f{1}{2} \sg_3 }\qquad \la \in \mathbb{H}^{\pm} \;, 
\label{ecriture forme asymptotiques chi}
\enq
with $\chi_{\ua/\da}^{(\infty)}(\la)$ bounded at $\infty$.

\begin{prop}
\label{Proposition ecriture forme asymptotique matrice chi}

Let $a_N$, $b_N$ be such that $x_N=b_N-a_N\geq \de$, uniformly in $N$  for some $\de>0$. Then, there exists $N_0$ such that, for any $N \geq N_0$, the  Riemann--Hilbert problem for $\chi$ has a unique solution
which takes the form given in Figure \ref{Figure definition sectionnelle de la matrice chi}. 
Furthermore, the unique solution to the above Riemann--Hilbert problem satisfies $\det \chi(\la) = \e{sgn}\big( \e{Im}(\la) \big)$ for any $\la \in \Cx\setminus\R$.
\end{prop}

\Proof

Uniqueness follows from standard reasonings in matrix valued Riemann--Hilbert problems since $G_{\chi}$ is invertible. One obtains the relation  $\det \chi(\la) = \e{sgn}\big( \e{Im}(\la) \big)$
by solving the scalar Riemann-Hilbert problem satisfied by the determinant. The existence of solutions at large-$N$ will be discussed in subsection \ref{SousSousSection resolution RHP pour chi}
and will build on the chain of transformations  discussed in subsections
\ref{SousSousSection chi vers Psi}, \ref{SousSousSection RHP pour Pi},  \ref{SousSousSection RHP pour Psi} to come.

\subsubsection{Transformation $\chi \rightsquigarrow \Psi$}
\label{SousSousSection chi vers Psi}

\begin{figure}[h]
\begin{center}

\includegraphics[width=.8\textwidth]{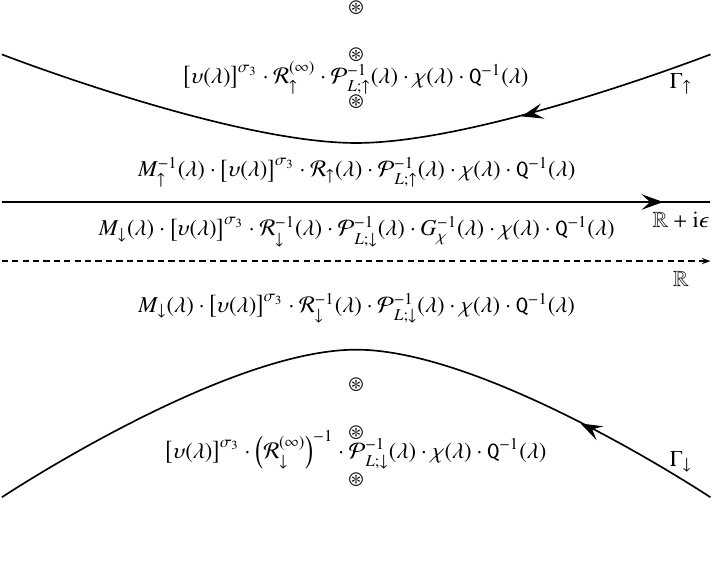}

\caption{Contour $\Sg_{\Psi}=\Ga_{\ua}\cup \Ga_{\da}$ in the Riemann--Hilbert problem for $\Psi$.
$\Ga_{\ua/\da}$ separates all the poles of $ R^{-1}$ in $\mathbb{H}^{\pm}$ from $\R$. These poles are indicated by $\circledast$
and the closest poles are at $\pm 2 \i$. The contours $\Ga_{\ua/\da}$ are chosen such that $\e{dist}(\Ga_{\ua/\da}, \R) = 2(1 -\a )$ with $\a$ small enough and where the minimum is attained on the imaginary axis.   
\label{contour pour le RHP de Phi}}
\end{center}
\end{figure}

Define a piecewise analytic matrix $\Psi$ out of the matrix $\chi$
according to Figure \ref{contour pour le RHP de Phi}. It is readily checked that the Riemann--Hilbert problem for $\chi$ is equivalent to the following Riemann--Hilbert problem for $ \Psi $:

\begin{itemize}

\item $\Psi \in \mc{O}(\Cx^{*}\setminus \Sg_{ \Psi } )$ and has continuous boundary values on 
$\Sg_{\Psi} = \Ga_{\ua} \cup \Ga_{\da} $ ;
\item The matrix $  \left( \ba{cc} -1   &  R^{-1}(\la) \ex{-\i \la \ov{x}_N  } \\ 
				0    & 1 \ea \right) \cdot \big[\ups(\la) \big]^{-\sg_3} \cdot \Psi(\la)$ has a limit when $\la \rightarrow 0$ ;
\item $\Psi(\la) = I_{2} \; + \; \e{O}\big(\la^{-1}\big) $ when $\la \tend \infty$ non-tangentially to $\Sg_{\Psi}$ ;
\item $\Psi_+(\la) \; = \;   G_{\Psi}(\la) \cdot \Psi_-(\la) $ \quad for \quad  $\la \in \Sg_{\Psi}$.
\end{itemize}
The jump matrix $G_{\Psi}$ takes the form:
\begin{eqnarray}
\e{for}\,\,\la \in \Ga_{\ua} \qquad G_{\Psi}(\la) & = & I_2 \; + \;   \f{ \ex{ \i\la \ov{x}_N } }{  \ups^2(\la) R(\la) } \cdot \sg^-\;, \label{ecriture saut Psi sur gamma up} \\
\label{ecriture saut Psi sur gamma down}  \e{for}\,\,\la \in \Ga_{\da} \qquad G_{\Psi}(\la) & = & I_2  \;  - \;    \f{  \ups^2(\la)\,\ex{-\i\la \ov{x}_N }   }{  R(\la) } \cdot \sg^+ \;. 
\end{eqnarray}

Note that there is a freedom of choice of the curves $\Ga_{\ua/\da}$, provided that these
avoid (respectively from below/above) all the poles of $R^{-1}(\la)$ in $\mathbb{H}^{+/-}$.  
Furthermore, we stress that within our choice of conventions, $-$ corresponds to the "upper" boundary value on $\Ga_{\ua/\da}$.

\subsubsection{The auxiliary Riemann--Hilbert problem for $\Pi$}
\label{SousSousSection RHP pour Pi}
 
\noindent Given contours $\Ga_{\ua/\da}$ at distance at least $2 (1-\a) $ from $\R$, for some $\a>0$ small and fixed, the jump matrix $G_{\Psi}$ satisfies, uniformly in $N$,
\beq
\label{eoim}\norm{ G_{\Psi}-I_2 }_{\mc{M}_2(L^2(\, \Sg_{\Psi} )) } \, + \,  \norm{ G_{\Psi}-I_2 }_{ \mc{M}_2(L^{\infty}(\, \Sg_{\Psi} )) }   \; = \;
 \e{O}\Big( \ex{- 2 \ov{x}_N  (1-\a) } \Big) \;,
\enq
where $\ov{x}_N$ has been introduced in \eqref{definition constante xN et XN bar}. Indeed, taken that
\beq
\bigg|   \f{ \ex{ \i\la \ov{x}_N } }{  \ups^2(\la) R(\la) }  \bigg| \; \leq \; C (1+|\la|^3)\ex{-\Im(\la) \ov{x}_N } \quad \e{along} \quad \Ga_{\ua}
\quad \e{and} \quad
\bigg|  \f{  \ups^2(\la)\,\ex{-\i\la \ov{x}_N }   }{  R(\la) } \bigg|  \; \leq \; \f{C \ex{-\Im(\la) \ov{x}_N } }{ 1+|\la|^3 } \quad \e{along} \quad \Ga_{\da}
\enq
the estimates \eqref{eoim} follow from the exponential decay of $G_{\Psi}-I_2$ along $\Ga_{\ua/\da}$ along with the minimal/maximal value
of the imaginary part of $\la$ along these curves.

These bounds are enough so as to solve the below auxiliary Riemann--Hilbert problem for $\Pi$:
\begin{itemize}
\item $\Pi \in \mc{O}(\Cx\setminus \Sg_{ \Psi } )$ and has continuous $\pm$ boundary values on 
$\Sg_{\Psi}$ ;
\item $\Pi(\la) = I_{2} \; + \; \e{O}\big(\la^{-1}\big) $ when $\la \tend \infty$ non-tangentially to $\Sg_{\Psi}$ ;
\item $\Pi_+(\la) \; = \;   G_{\Psi}(\la) \cdot \Pi_-(\la)$ for $\la \in \Sg_{\Psi}$ \;. 
\end{itemize}
Above, $\Sg_{\Psi}$ is as given in Fig. \ref{contour pour le RHP de Phi} and $G_{\Psi}$ is defined in \eqref{ecriture saut Psi sur gamma up}-\eqref{ecriture saut Psi sur gamma down}
Since the jump matrix $G_{\Psi}$ has unit determinant and is exponentially in $\ov{x}_N$ close to the identity on $\Sg_{\Psi}$, the setting developed in \cite{BealsCoifmanScatteringInFirstOrderSystemsEquivalenceRHPSingIntEqnMention}
ensures that the Riemann--Hilbert problem for $\Pi$ is uniquely solvable for $N$ large enough.  To construct its solution, one first, introduces the singular integral operator on the space $\mc{M}_2\big( L^2( \Sg_{\Psi} ) \big)$ 
of $2\times 2$ matrix-valued $L^2\big( \Sg_{\Psi} \big)$ functions by 
\beq
\mc{C}^{(-)}_{ \Sg_{\Psi} }\big[ \Pi \big] (\la) \; = \; \lim_{ \substack{ z \tend \la \\ z \in - \e{side}  \, \e{of} \,\Sg_{\Psi} } } 
\int_{ \Sg_{\Psi} }{} \f{ (G_{\Psi}-I_2)(t) \cdot \Pi(t) }{t-z} \cdot \f{ \dd t}{ 2 \i \pi } \;. 
\enq
Since $ G_{\Psi}-I_2 \in \mc{M}_2\Big( \big(L^{\infty} \cap L^2 \big) \big( \Sg_{\Psi} \big) \Big)$ and $\Sg_{\Psi}$ is a Lipschitz curve, 
 ${\cal C}^{(-)}_{\Sigma_{\Psi}}$ is continuous on  $\mc{M}_2(L^2\big( \Sg_{\Psi}))$ 
and  satisfies:
\beq
\big| \big| \big|  \mc{C}^{(-)}_{\Sg_{\Psi} }   \big| \big| \big|_{ \mc{M}_2(L^2( \Sg_{\Psi} )) }  \; \leq  \; 
C  \ex{- 2 \ov{x}_N(1-\a) } \;, 
\enq
for some constant $C>0$ and where the control in $N$ originates from \eqref{eoim}. Hence, since 
\beq
G_{\Psi}-I_2 \in \mc{M}_2\Big( L^2\big( \Sg_{\Psi} \big) \Big)  \quad \e{and} \quad 
\mc{C}^{(-)}_{\Sg_{\Psi} }[I_2] \in  \mc{M}_2\Big( L^2\big( \Sg_{\Psi} \big) \Big) 
\enq
 provided that $N$ is large enough, it follows that the singular integral equation
\beq
\Big(I_2\otimes \e{id}\,  - \, \mc{C}^{(-)}_{\Sg_{\Psi} }  \Big)\big[ \Pi_- \big] \; = \; I_2
\label{ecriture eqn int sing pour matrice Pi moins}
\enq
admits a unique solution $\Pi_{-}$ such that $\Pi_{-} - I_2 \in \mc{M}_2\big( L^2( \Sg_{\Psi} ) \big)$.  It is then a standard fact \cite{BealsCoifmanScatteringInFirstOrderSystemsEquivalenceRHPSingIntEqnMention} in the theory of Riemann--Hilbert problems 
that the matrix  
\beq
\Pi(\la) \; = \; I_2 \; + \; \Int{ \Sg_{\Psi} }{} \f{ (G_{\Psi}-I_2)(t) \cdot \Pi_-(t) }{ t-\la } \cdot \f{ \dd t }{ 2\i\pi }
\label{ecriture rep int matrice Pi}
\enq
 is the unique solution to the Riemann--Hilbert problem for $\Pi$. 
We close this section by establishing that, for any open neighbourhood $U$ of $\Sigma_{\Psi}$ such that $\e{dist}(\Sg_{\Psi}, \Dp{}U)> \de >0$, there exists a constant $C > 0$ such that:
\beq
\label{ecriture bornes en N pour Pi moins Id}
\forall \la \in \Cx\setminus  U,\qquad \max_{a,b \in \{1,2\}} \Big| \big[\Pi(\la) - I_{2}\big]_{ab} \Big|  \leq \frac{C\,  \ex{- 2\ov{x}_N (1-\a) } }{1 + |\la|}\;.
\enq
Indeed, direct bounds and the use of the Cauchy-Schwartz identity yield that
\bem
 \Big| \big[\Pi(\la) - I_{2}\big]_{ab} \Big|  \; \leq \; \Int{ \Sg_{\Psi} }{} \!  \f{ |\dd t|  }{ 2\pi |\la-t|}\big| \, \big[ G_{\Psi}(t)-I_2 \big]_{ab} \big|
\; + \;  \Int{ \Sg_{\Psi} }{} \! \f{ |\dd t|  }{ 2\pi |\la-t|} \,  \Big| \Big[ \big( G_{\Psi}(t)-I_2 \big) \big( \Pi(t)-I_2 \big) \Big]_{ab} \big| \\
\; \leq \; \Int{ \Sg_{\Psi} }{} \!  \f{ |\dd t|  }{ 2\pi |\la-t|}\big| \, \big[ G_{\Psi}(t)-I_2 \big]_{ab} \big|
\; + \; C \cdot  \max_{a,b \in \{1,2\}} \Bigg\{\Int{ \Sg_{\Psi} }{} \!  \f{ |\dd t|  }{   |\la-t|^2 }\big| \, \big[ G_{\Psi}(t)-I_2 \big]_{ab} \big|^2  \Bigg\}^{\f{1}{2} }
\cdot \norm{ \Pi - I_2 }_{ \mc{M}_2\big( L^2(\Sg_{\Psi}) \big)  }
\label{ecritrue borne sup sur element matrice de Pi moins Id}
\end{multline}
It follows from a previous discussion that one has the bounds
\beq
 \Big|  \big[ G_{\Psi}(t)-I_2 \big]_{ab}   \Big|^k  \; \leq \; C \mc{B}_{k}(t) \; = \; C \bigg\{ \big( 1+|t|^3 \big)^k \ex{-k \Im(t) \ov{x}_N } \bs{1}_{\Ga_{\ua}}(t) \; + \;
 \f{ \ex{k \Im(t) \ov{x}_N }  }{ \big( 1+|t|^3 \big)^k }  \bs{1}_{\Ga_{\da}}(t) \bigg\} \;.
\enq
It is easy to see that, pointwise in $t\in \Sg_{\Psi}$,
\beq
\Phi_{k}(\la,t) \; = \; \big( 1+|\la| \big)^k \big( 1+|t|^3 \big)^k  \f{ \ex{-k \Im(t) \ov{x}_N } }{ |t-\la|^k } \bs{1}_{\Ga_{\ua}}(t) \; + \;
  \big( 1+|\la| \big)^k \f{ \ex{k \Im(t) \ov{x}_N }  }{ \big( 1+|t|^3 \big)^k \cdot  |t-\la|^k   }  \bs{1}_{\Ga_{\da}}(t) \limit{ |\la|}{+ \infty} \mc{B}_k(t) \in L^1(\Sg_{\Psi})
\enq
Moreover, by distinsguishing the regions $|t-\la| \leq  \tf{|\la| }{ 2}$ and $|t-\la| >   \tf{ | \la|  }{ 2} $, one even obtains
the uniform in $\la\in \Cx\setminus U$ bound
\beq
\big| \Phi_{k}(\la,t)  \big| \; \leq \; C \, \mc{B}_{\tf{k}{2}}(t) \in L^1(\Sg_{\Psi}) \;.
\enq
Thus, by dominated convergence applied to the second integral,
\beq
 \big( 1+|\la| \big) \cdot \Bigg\{\Int{ \Sg_{\Psi} }{} \!  \f{ |\dd t|  }{   |\la-t|^k }\big| \, \big[ G_{\Psi}(t)-I_2 \big]_{ab} \big|^k  \Bigg\}^{\f{1}{k} }
\; \leq \;   \Bigg\{\Int{ \Sg_{\Psi} }{} \!   |\dd t|  \, \Psi_{k}(\la,t)  \Bigg\}^{\f{1}{k} } \limit{|\la|}{+\infty}
 \Bigg\{\Int{ \Sg_{\Psi} }{} \!   |\dd t|  \, \mc{B}_{k}(t)  \Bigg\}^{\f{1}{k} } \;  =  \; \e{O}\Big( \ex{-2 \ov{x}_N(1-\a) } \Big) \;.
\enq
The above, adjoined to \eqref{ecritrue borne sup sur element matrice de Pi moins Id}, ensures that \eqref{ecriture bornes en N pour Pi moins Id} holds for $\la\in \Cx \setminus U$, with $|\la|$
large enough. When $|\la| \leq K$ for some $K$ large enough and  $\la\in \Cx \setminus U$, then one has the direct bound
\beq
 \Bigg\{\Int{ \Sg_{\Psi} }{} \!  \f{ |\dd t|  }{   |\la-t|^k }\big| \, \big[ G_{\Psi}(t)-I_2 \big]_{ab} \big|^k  \Bigg\}^{\f{1}{k} }
 \; \leq  \;
\f{ C }{\de}\,  \Bigg\{\Int{ \Sg_{\Psi} }{} \!   |\dd t| \, \mc{B}_{k}(t)  \Bigg\}^{\f{1}{k} }\;  =  \; \e{O}\Big( \ex{-2 \ov{x}_N(1-\a) } \Big) \;.
\enq
The latter entails \eqref{ecriture bornes en N pour Pi moins Id}.

\subsubsection{Solution of the Riemann--Hilbert problem for $\Psi$}
\label{SousSousSection RHP pour Psi}

The Riemann--Hilbert problem for $\Psi$ and $\Pi$ have the same jump matrix $G_{\Psi}$, but $\Psi$ must have at least a third order pole at $\la=0$ in order to fulfil the regularity condition
at $\la=0$. Thus, one looks for $\Psi$ in the form
\beq
\label{PsiPi}\Psi(\la) \; = \; \Pi(\la) \cdot \mc{P}_R(\la)
\enq
with $\mc{P}_R$ as defined in \eqref{definition matrice PR} with matrices $\mc{Q}_{\ell}$ yet to be fixed.
Clearly the matrix $\Pi(\la) \cdot \mc{P}_R(\la)$ has the desired asymptotic behaviour at $\infty$ and it satisfies the
jump conditions on $\Sg_{\Psi}$ -this for any choice of the $\mc{Q}_{\ell}$s- with jump  matrix $G_{\Psi}$.
Thus, in order to satisfy the regularity condition at $\la=0$ it is enough that the matrices $\mc{Q}_{\ell}$ be chosen so that the matrix
\beq
\mc{T}(\la) \; = \; \bigg(I_2\, - \, \sul{\ell=0}{2} \f{ c_{\ell} }{ \la^{3-\ell} } \sg^+ \bigg) \cdot  \Ups(\la) \cdot \bigg( I_2\, + \, \sul{\ell=0}{2} \f{ \mc{Q}_{\ell} }{ \la^{3-\ell} }  \bigg) \;, \qquad
\Ups(\la)\, = \, \Pi(\la)\Pi^{-1}(0) \;,
\label{definition matrice T pour construction Psi}
\enq
is regular at $\la=0$. Below, we establish the unique solvability for such matrices $\mc{Q}_{\ell}$ in the large-$N$ limit.

\begin{lemme}
\label{Lemme solvabilite pour matrices Q ell}

Let $N\geq N_0$ with $N_0$ large enough. There exists a unique choice of $2\times 2$ matrices $\mc{Q}_{\ell}$, $\ell=0,1,2$ such that
the matrix $\mc{T}$ given by \eqref{definition matrice T pour construction Psi} is regular at $\la=0$. Moreover, one has that
\beq
\mc{Q}_{\ell} \; = \; c_k \sg^+ \, + \, \e{O}\Big(  (\ov{x}_N)^4 \ex{-2 \ov{x}_N(1-\a) } \Big)
\enq
where $c_k$s have been introduced in \eqref{definition coefficients c ell}.

\end{lemme}

It is clear that $\mc{P}_R$ given by \eqref{definition matrice PR}
with $\mc{Q}_{\ell}$ defined through Lemma \ref{Lemme solvabilite pour matrices Q ell}, in particular
\eqref{reparametrisation matices Qk} and \eqref{ecriture systeme pour matrices Mk} to come, does give rise to  a
piecewise analytic matrix $\Psi$ through \eqref{PsiPi} which satisfies the regularity condition.

\Proof

First of all, note that it follows from the integral representation \eqref{ecriture rep int matrice Pi} for $\Pi$, that the latter has a finite value $\Pi(0)$ at $\la=0$ and that
$\Pi(0) = I_2 \, + \, \e{O}\big(  \ex{- 2\ov{x}_N (1-\a) } \big)$. The matrix  $\Ups(\la) \, = \, \Pi(\la)\Pi^{-1}(0)$ is thus regular around $0$
and admits the expansion  around $\la=0$:
$\Ups(\la)=I_2\, + \, \sul{\ell=1}{5} \la^{\ell} \Ups_{\ell}\, + \,\e{O}(\la^6)$ for some matrices $\Ups_{\ell}\, = \, \e{O}\big(  \ex{- 2\ov{x}_N (1-\a) } \big)$.
Starting from the expansion 
\beq
\mc{T}(\la) \, = \, \Ups(\la)\, + \, \sul{\ell=0}{2} \f{ 1 }{ \la^{3-\ell} } \bigg\{ \Ups(\la) \mc{Q}_{\ell} \, - \,  c_{\ell} \sg^{+} \Ups(\la)  \bigg\} \, - \,
\sul{s=0}{4}  \f{ 1 }{ \la^{6-s} } \sul{ \substack{ r,\ell =0 \\ r+\ell=s } }{2} c_{\ell} \sg^{+} \Ups(\la) \mc{Q}_{r} \,,  
\enq
one gets $\mc{T}(\la) \, = \, \sul{s=0}{5} \la^{s-6} \mc{T}_s\, + \, \e{O}(1)$, where
\beq
\mc{T}_{0}\, = \, - c_0 \sg^{+} \mc{Q}_0 \; , \quad \mc{T}_{1}\, = \, - \Big\{ c_0 \sg^{+} \Ups_1 \mc{Q}_0 + c_1 \sg^+\mc{Q}_0\, + \, c_0 \sg^{+} \mc{Q}_1 \Big\} \; ,
\enq
\beq
 \mc{T}_{2}\, = \, -\Big\{ c_0 \sg^{+} \Ups_2 \mc{Q}_0 + c_1 \sg^+\Ups_{1}\mc{Q}_0\, + \, c_0 \sg^{+} \Ups_{1}\mc{Q}_1 \, + \,  c_2 \sg^+\mc{Q}_0\, + \, c_1 \sg^{+} \mc{Q}_1  \, + \, c_0 \sg^{+} \mc{Q}_2 \Big\} \;.
\enq
The remaining three expressions are more bulky 
\bem
\mc{T}_{3}\,= \, \mc{Q}_0 \, - \, c_0 \sg^+ \, - \, c_0 \sg^+\Big(\Ups_3 \mc{Q}_0 \, + \,
\Ups_2 \mc{Q}_1  \, + \, \Ups_1 \mc{Q}_2 \Big)  \, - \, c_1 \sg^+\Big(\Ups_2 \mc{Q}_0 \, + \, \Ups_1 \mc{Q}_1  \, + \,  \mc{Q}_2 \Big) \\
\, - \, c_2 \sg^+ \Big(\Ups_2 \mc{Q}_0 \, + \, \mc{Q}_1   \Big) \;, 
\end{multline}
\bem
\mc{T}_{4}\,= \, \mc{Q}_1 \, - \,  c_1  \sg^+  \, + \, \Ups_1 \mc{Q}_0  \, - \, c_0 \sg^+\Big(\Ups_1 \, + \, \Ups_4 \mc{Q}_0 \, + \, \Ups_3 \mc{Q}_1  \, + \, \Ups_2 \mc{Q}_2 \Big)   \\
\, - \, c_1 \sg^+\Big(\Ups_3 \mc{Q}_0 \, + \, \Ups_2 \mc{Q}_1  \, + \, \Ups_1 \mc{Q}_2 \Big) 
\, - \, c_2 \sg^+ \Big(\Ups_2 \mc{Q}_0 \, + \, \Ups_1 \mc{Q}_1  \, + \, \mc{Q}_2  \Big) \;, 
\end{multline}
\bem
\mc{T}_{5}\,= \, \mc{Q}_2 \, - \,  c_2  \sg^+  \, + \, \Ups_1 \mc{Q}_1 \, + \, \Ups_2 \mc{Q}_0    \, - \, c_0 \sg^+\Big(\Ups_2 \, +\, \Ups_5 \mc{Q}_0 \, + \, \Ups_4 \mc{Q}_1  \, + \, \Ups_3 \mc{Q}_2 \Big)  \\
\, - \, c_1 \sg^+\Big(\Ups_1 \, + \, \Ups_4 \mc{Q}_0 \, + \, \Ups_3 \mc{Q}_1  \, + \, \Ups_2 \mc{Q}_2 \Big) 
\, - \, c_2 \sg^+ \Big(\Ups_3 \mc{Q}_0 \, + \, \Ups_2 \mc{Q}_1  \, + \, \Ups_1 \mc{Q}_2  \Big) \;. 
\end{multline}

The equations $\mc{T}_0=0$, $\mc{T}_1=0$, $\mc{T}_{2}=0$, which ensure the vanishing of the poles of order $6, 5$ and $4$ of $\mc{T}$ at $\la=0$ may be solved as
\beq
\mc{Q}_0 \, =\, \mc{M}_0 \;, \quad \mc{Q}_1 \, =\, -\Ups_{1}\mc{M}_0 \, +\,  \mc{M}_1 \qquad \e{and} \qquad \mc{Q}_2 \, =\, -\Ups_{1}\mc{M}_1 \, +\, \Big( \Ups_1^2 - \Ups_2 \Big)\cdot \mc{M}_0 \, +\,  \mc{M}_2  \; , 
\label{reparametrisation matices Qk}
\enq
where 
\beq
\mc{M}_k\; = \; \left(\ba{cc} a_k & b_k \\ 0  &  0   \ea  \right) 
\enq
are yet to be fixed.  The vanishing of the third, second and first order poles of $\mc{T}$ at $\la=0$ leads to the below system on the matrices $\mc{M}_k$:
\beq
\Big( I_3 \otimes I_2 \, - \, \op{W} \Big) \cdot \bs{M}\; = \; \bs{S} \qquad \e{with} \qquad \bs{M}\;= \; \left(\ba{c} \mc{M}_0 \\ \mc{M}_1 \\ \mc{M}_2 \ea \right)  \qquad \e{and} \qquad 
\bs{S}\;= \; \left(\ba{c} c_0 \sg^+  \\ c_1 \sg^+  \,  +\, c_0 \sg^+ \Ups_1    \\ \sg^+ \Big(c_2  \,  +\, c_1 \Ups_1 \,  +\, c_0 \Ups_2  \Big)  \ea \right) 
\label{ecriture systeme pour matrices Mk}
\enq
while 
\beq
\Big( \op{W}_{00} \,, \, \op{W}_{01} \,, \, \op{W}_{02}  \Big) \; = \; c_0 \, \sg^{+} \, \Big( \Ups_3 \,  - \,  2 \Ups_2 \Ups_1 \, + \, \Ups_1^3 \, , \,  \Ups_2 \,  - \,  \Ups_1^2 \, , \, \Ups_1 \Big)
\enq
in which the matrix product by $\sg^+$ should be distributed coordinate-wise and
\bem
\Big( \op{W}_{10} \,, \, \op{W}_{11} \,, \, \op{W}_{12}  \Big) \; = \;  \sg^{+} \, \Big( c_0 \big[ \Ups_4 \,  + \,   \Ups_2(\Ups_1^2- \Ups_2 ) \, - \, \Ups_3 \Ups_1 \big]  
\, +\, c_1\big[ \Ups_3 \,  - \,  2 \Ups_2 \Ups_1 \, + \, \Ups_1^3\big] \, , \, \\
c_0\big[ \Ups_3 \,  - \,  \Ups_2 \Ups_1 \big] \, + \, c_1\big[ \Ups_2 \,  - \,  \Ups_1^2 \big]  \, , \, c_0 \Ups_2 \, + \, c_1 \Ups_1  \Big) \;. 
\end{multline}
Finally, 
\beqa
 \op{W}_{20} & = & \sg^{+} \, \Big\{ c_0 \big[ \Ups_5    - \Ups_4 \Ups_1  +   \Ups_3(\Ups_1^2 - \Ups_2 ) \big]  
  +  c_1\big[ \Ups_4   -    \Ups_3 \Ups_1  +   \Ups_2 ( \Ups_1^2 - \Ups_2 )  \big]  
  +   c_2 \big[ \Ups_3  -    \Ups_2 \Ups_1  +    \Ups_1 ( \Ups_1^2 - \Ups_2 )  \big]   \Big\} \;, \nonumber \\
 \op{W}_{21} & = & \sg^{+} \, \Big\{ c_0 \big[ \Ups_4  \, - \, \Ups_3 \Ups_1   \big]  
\, +\, c_1\big[ \Ups_3 \,  - \,   \Ups_2 \Ups_1  \big]  
\, + \, c_2 \big[   \Ups_2  - \Ups_1^2  \big]   \Big\} \;,  \nonumber \\
 \op{W}_{22} & = & \sg^{+} \, \Big\{ c_0 \Ups_3  \, + \, c_1 \Ups_2  \, + \, c_2 \Ups_1 \Big\} \;. 
\nonumber
\eeqa
The system is uniquely solvable, at least for $N$-large enough, since $\Ups_k=\e{O}\Big( \ex{-2\ov{x}_N(1-\a) } \Big)$ pointwise in $k$
while $c_k=\e{O}\Big( (\ov{x}_N)^{k} \Big)$. Moreover, it is clear from these estimates that, when $N\tend +\infty$, it holds
\beq
 \mc{M}_k \; = \; c_k \sg^+  \, + \, \e{O}\Big( (\, \ov{x}_N)^4 \ex{-2\ov{x}_N(1-\a) } \Big) \;, 
\enq
 where the loss of the control on the remainder is due to the polynomial blow-up in $\ov{x}_N$ of the coefficients $c_k$, \textit{c.f.} \eqref{definition des constantes wk et de leur asymptotiques}.

\subsubsection{Solution of the Riemann--Hilbert problem for $\chi$}
\label{SousSousSection resolution RHP pour chi}

\begin{figure}[h]
\begin{center}

\includegraphics[width=.8\textwidth]{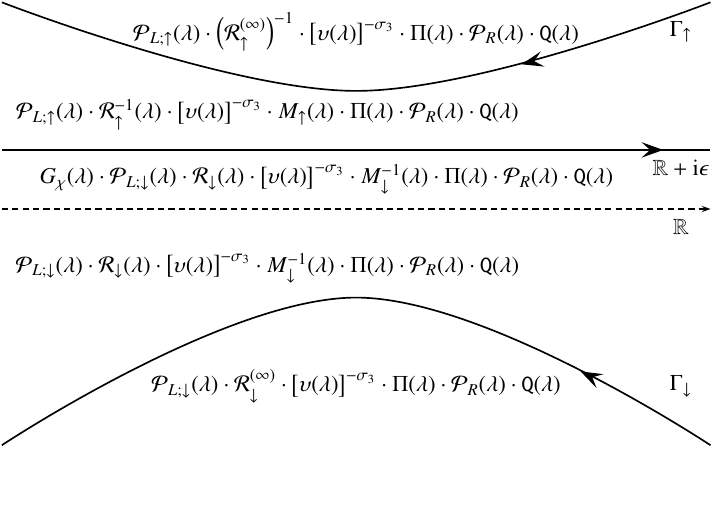}

\caption{Piecewise definition of the matrix $\chi$.
The curves $\Ga_{\ua/\da}$ separate all poles of $\la \mapsto R^{-1}(\la)$ from $\R$ and are such that $\e{dist}(\Ga_{\ua/\da}, \R) =2(1-\a) > \epsilon > 0$
for a sufficiently small $\a$. The matrix $\Pi$ appearing here is defined through \eqref{ecriture rep int matrice Pi}.  
\label{Figure definition sectionnelle de la matrice chi}}
\end{center}
\end{figure}

Tracking back the transformations $\Pi \rightsquigarrow   \Psi \rightsquigarrow \chi$, gives the construction of the solution $\chi$ of 
the Riemann--Hilbert problem of Proposition~\ref{Proposition ecriture forme asymptotique matrice chi}, summarised in Figure~\ref{Figure definition sectionnelle de la matrice chi}. 
It is clear that the solution so-constructed is holomorphic on $\Cx \setminus \R$, has continuous $\pm$ boundary values on $\R$ and the desired system of jumps. What remains to check, however, is the form 
taken by the asymptotic expansion, \textit{viz}. that the \textit{same} matrices $\chi^{(a)}$ occur in the $\la \tend \infty$ asymptotics of $\chi$ on $\mathbb{H}^{+}$ and $\mathbb{H}^{-}$. 
It follows readily from the asymptotic expansion
\beq
R(\la) \, = \, \mf{s}_{\la} \, + \, \e{O}\big(\la^{-\infty}\big),  
\enq
with $\mf{s}_{\la}$ as in \eqref{definition signe s lambda}, 
that $\chi$ will have the asymptotic expansion given in Subsection \ref{Sous Section RHP pour chi} with \textit{a priori} two sets of matrices $\chi^{(a)}_{\ua/\da}$
grasping the expansion in $\mathbb{H}^{+/-}$. These are obtained from the below large-$\la$ asymptotic expansions
\beqa
 \ex{ - \i \f{3 \pi }{ 2 } \sg_3 } \cdot \big(- \i \la \big)^{ - \f{ 3 }{ 2 } \sg_3 } \cdot \big[ \ups(\la) \big]^{-\sg_3 } \cdot \Pi(\la) \cdot \mc{P}_R(\la)  
& = &    I_{2}  \; + \; \f{\chi^{(1)}_{\ua}}{\la} \; + \; \cdots  \qquad \la \in \mathbb{H}^{+} \\
  \big( \i \la \big)^{ - \f{ 3 }{ 2 } \sg_3 } \cdot \big[ \ups(\la) \big]^{-\sg_3 } \cdot \Pi(\la) \cdot \mc{P}_R(\la)  
& = &    I_{2}  \; + \; \f{\chi^{(1)}_{\da}}{\la} \; + \; \cdots \qquad \la \in \mathbb{H}^{-} \;. 
\eeqa
It follows from the explicit expression for $\ups(\la)$ for $\Im(\la)\, >\, \eps$ and $\Im(\la)\, <\, \eps$ that there exists constants $\ups_{\ua/\da}^{(a)}$ such that
\beq
\ups(\la) \; = \; \left\{ \ba{cc}  \i \big( -\i \la \big)^{-\tfrac{3}{2}} \cdot \bigg( 1  \, + \, \f{1 }{ \la}  \ups_{\ua}^{(1)} \, + \,  \f{ 1 }{ \la^2 } \ups_{\ua}^{(2)}  \, + \, \cdots \bigg)  &  \Im(\la)> \eps \vspace{2mm} \\
\big( \i \la \big)^{-\tfrac{3}{2}} \cdot \bigg(1 \, + \,\f{1 }{ \la}  \ups_{\da}^{(1)} \, + \,  \f{ 1 }{ \la^2 } \ups_{\da}^{(2)}   \, + \, \cdots \bigg)  &  \Im(\la) <\eps  \ea \right. \;. 
\enq
The jump condition for $\ups$: $\ups_+(\la)\, = \, R(\la) \ups_{-}(\la)$ may be meromorphically  extended to $\Cx$. Then, since 
\beq
 \i \big( -\i \la \big)^{-\tfrac{3}{2}} = \mf{s}_{\la} \big( \i \la \big)^{-\tfrac{3}{2}} \, ,
\enq
the large $\la$ behaviour of $R$ entails that $ \ups_{\ua}^{(a)} =   \ups_{\da}^{(a)} $ for any $a\in \mathbb{N}^{*}$.

Recall  the integral representation \eqref{ecriture rep int matrice Pi} for $\Pi$ and observe that owing to the analyticity in the neighbourhood of $\Sg_{\Psi}$
 of the jump matrix $G_{\Psi}$ and the enponential decay at infinity of $G_{\Psi}$, one may deform the slope of the curves $\Ga_{\ua}\cup \Ga_{\da}$ in that integral representation -without changing its value-.
 That entails that, uniformly in all directions
$\Pi$ admits the large-$\la$ asymptotic expansion 
\beq
\Pi(\la) \; = \; I_2 \, + \, \f{1}{\la} \Pi^{(1)} \, + \, \f{1}{\la} \Pi^{(2)} \, + \, \cdots 
\enq
for some constant matrices $\Pi^{(a)}$.

\vspace{2mm}
Putting all of these expansions together concludes the proof of Proposition~\ref{Proposition ecriture forme asymptotique matrice chi}. \qed

\subsection{Properties of the solution $\chi$}
\label{SousSection Proprietes generales de la solution chi}

\begin{lemme}
\label{Lemme Ecriture diverses proprietes solution RHP chi}
The solution $\chi$ to the Riemann--Hilbert problem satisfies
\beq
\chi(-\la) \; = \; 
\left(  \ba{cc} 1 & 0 \\ 
							0 & -1 \ea \right) \cdot \chi(\la) \cdot 
										\left( \ba{cc} 1 & 0 \\
												 \la  & 1 \ea \right) 
\quad and \qquad 
\Big(\chi(\la^*)\Big)^* \; = \; 
\left(  \ba{cc} -1 & 0 \\ 
							0 & 1 \ea \right) \cdot \chi(-\la) \cdot 
										\left( \ba{cc} 1 & 0 \\
												 0  & -1 \ea \right) \;. 									
\label{ecriture propriete conjugaison et reflection solution RHP chi}
\enq

\end{lemme}

\Proof Since $G_{\chi}(-\la) \; = \; \ex{ \f{\i\pi\sg_3}{2} } G^{-1}_{\chi} (\la) \ex{- \f{\i\pi\sg_3}{2} }$ as can be inferred from \eqref{definition constante xN et XN bar} and $\mf{w}^{(+)}$ being even, the matrix:
\beq
\Xi(\la) \; = \; \chi^{-1}(\la) \cdot \ex{- \f{\i\pi\sg_3}{2} } \cdot \chi(-\la)
\enq
is continuous across  $\R$ and is thus an entire function. Further, denoting for short the asymptotic series appearing in \eqref{ecriture DA chi dans RHP} as
\beq
\chi_{\infty}(\la) \; = \; \bigg( I_2 \, + \,  \f{1}{\la} \chi^{(1)} \,  + \, \f{1}{\la^2} \chi^{(2)}\, + \, \cdots \bigg) \cdot \op{Q}(\la) \; = \;
\left( \ba{cc}  \frac{ 1 }{ \la } \,   L_{11}(\la)   &  \frac{ 1 }{ \la^2 } \,   L_{12}(\la)  \vspace{2mm} \\
  L_{21}(\la)   &  \la \,  L_{22}(\la)   \ea \right) \;,
\enq
one readily gets that, for $a\in \{1,2\}$,  
\beq
L_{aa}(\la)\, = \, 1 \, +\,  \e{O}\big( \la^{-1} \big)  \quad \e{while} \quad
L_{12}(\la) \, = \, \mf{l}_{12} \, +\,  \e{O}\big( \la^{-1} \big)  \quad \e{and} \quad
L_{21}(\la) \, = \, \mf{l}_{21} \, +\,  \e{O}\big( \la^{-1} \big)  \;,
\label{ecriture forme matrice M ab}
\enq
for some coefficients $\mf{l}_{12}$ and $\mf{l}_{21}$ that can be explicitly expressed in terms of the entries of $\chi^{(1)}, \chi^{(2)}$ and $\chi^{(3)}$.
These bounds are due to the very specific form taken by the coefficient $\mf{q}_1$ appearing in $\op{Q}(\la)$ introduced in \eqref{definition matrice Q}.
Observe that evaluating the relation $\det\big[ \chi(\la) \big]=\e{sgn}\Big( \Im(\la) \Big)$, \textit{c.f.} Proposition \ref{Proposition ecriture forme asymptotique matrice chi},
on $\chi$'s asymptotic expansion
implies that $\det\chi_{\infty}(\la)=1$ in the sense of a formal power series in $\la^{-1}$ at infinity. Moreover, it holds
\beq
\big( \pm \i \la \big)^{ \frac{3}{2} \sg_3} \cdot  \chi_{\infty}(\la)  \; = \; 
\left( \ba{cc} -\i    L_{11}(\la)   &  -  L_{12}(\la) \\
  -\frac{1}{\la^2} \cdot L_{21}(\la)   &  \i  L_{22}(\la) \ea \right)  \cdot   \big(  \pm \i \la \big)^{ \frac{1}{2} \sg_3}  \;.
\label{ecriture DA chi infty amenage}
\enq
All of this entails that $\Xi$ admits the large-$\la$ expansion 
\beq
\Xi(\la) \; = \;  \left( \ba{cc}  \i  \mf{l}_{12}     &  0  \\
  - \i \la    &  \i  \mf{l}_{12}   \ea \right) \; + \; \e{O}\Big( \la^{-1} \Big) \;.
\enq
In principle, as follows from the formulation of the Riemann--Hilbert problem for $\chi$, this asymptotic expansion
is valid only in $\mathbb{H}^+\cup\mathbb{H}^-$, non-tangentially to $\R$. However, it is easy to see from the form taken by the solution, \textit{c.f.}
Section \ref{SousSousSection resolution RHP pour chi}, that the stated asymptotic expansion is actually uniform up to $\R$.

Since $\Xi(\la)$ is entire, by Liouville theorem this asymptotic expression is exact, namely
\beq
\label{ecriture expression explicite Xi} \Xi(\la) \; = \;  \i   \left( \ba{cc}   \mf{l}_{12}     &  0  \\
  - \la    &   \mf{l}_{12}   \ea \right) \;.
\enq
By taking the $\la\tend 0^+$ limit of this expression and upon computing independently $\Xi(0)$ by using the jump condition $ \chi_-(0) \; = \;  \sg_3 \cdot \chi_+(0)$,
as ensured by $\mc{F}[S_{\ga}]$ being odd, one infers that
$\mf{l}_{12}=-1$. This yields the first identity

The second identity is established analogously by considering 
\beq
\wt{\Xi}(\la) \, = \, \chi^{-1}(-\la) \cdot \ex{ \i \f{\pi}{2} \sg_3 } \cdot \Big( \chi(\la^*) \Big)^{*} \;. 
\enq

\qed





\section{Inversion of the singular integral operator}
\label{Section Inversion singular integral operator}

In this section, I establish the solvability criterion on $ \mc{H}_s(\intff{ a_N }{ b_N })$, with $0<s<1$, of the equation
$\mc{S}_{N;\ga}[\varphi] = H$ as well as the explicit form of the solution. The inverse of $\mc{S}_{N;\ga}$ is constructed with the help of the solution $\chi$
to the Riemann--Hilbert problem discussed in the previous section.  Here, the inversion of the operator is established through direct calculations. 
One is referred to \cite{KozBorotGuionnetLargeNBehMulIntOfToyModelSoVType} for a more detailed construction of the inverse. 
 
 Furthermore, in the following, I agree upon the notations:
\beq
\ov{a}_N=\tau_N a_N \;, \quad \ov{b}_N=\tau_N b_N \;, \quad \ov{x}_N=\tau_N x_N \;. 
\label{definition de b bar N a bar N x bar N}
\enq

\subsection{Solving the regularised equation $\mc{S}_{N;\ga}[\varphi] = H$ on $ \mc{H}_s(\intff{ a_N }{ b_N })$, $0<s<1$}

With the $2 \times 2$ matrix $\chi$ now constructed, I can come back to the inversion of the integral operator $\mc{S}_{N;\gamma}$
defined in \eqref{definition op int sing S N gamma}.
For any  $\varphi \in  \mc{H}_s(\intff{ a_N }{  b_N  })$ one has the Fourier transform identity
\beq
\mc{F}\Big[ \mc{S}_{N;\ga}[\varphi]  \Big](\tau_N \la) \; = \;  \f{1}{2\tau_N } \mc{F}\big[ S_{\ga}\big](\la) \cdot \mc{F}\big[ \varphi \big](\tau_N \la) \;. 
\label{ecriture TF SN gamma de varphi}
\enq
Hence, the boundedness of  $\mc{F}\big[ S_{\ga}\big]$ on $\R$, \textit{c.f.} \eqref{ecriture forme precise TF gN}, ensures that  $\mc{S}_{N;\ga}:  \mc{H}_s(\intff{ a_N }{  b_N  }) \mapsto  \mc{H}_s(\intff{ a_N -\ga x_N }{  b_N + \ga x_N })$
is a bounded operator.

To characterise the image of $\mc{S}_{N;\ga}$, I first need to introduce the functional 
\beq
\mc{J}_{12}[H] \; = \; \Int{\R  }{} \chi_{12;+}(\mu) \, \mc{F}[H](\tau_N\mu)\, \ex{- \i  \mu \ov{b}_N }\cdot\f{\dd\mu}{ (2 \i \pi)^2 }
\enq
on $\mc{H}_s(  \intff{ a_N - \ga x_N  }{ b_N + \ga x_N  }  )$.

\begin{lemme}
 
 Let $0<s<1$. Then the subspace
\beq
\mf{X}_{s}\Big(  \intff{ a_N - \ga x_N  }{ b_N + \ga x_N  }   \Big) \; = \; \bigg\{   
H \in \mc{H}_s\Big(  \intff{ a_N - \ga x_N  }{ b_N + \ga x_N  }   \Big) \; : \;\mc{J}_{12}[H] \, = \, 0
\bigg\} \, . 
\enq
is a closed subspace of  $\mc{H}_s\Big(  \intff{ a_N - \ga x_N  }{  b_N +  \ga   x_N  }  \Big)$ and it holds
\beq
\mc{S}_{N;\ga}\Big( \mc{H}_s\intff{ a_N }{  b_N  }  \Big) \; \subseteq \; \mf{X}_{s}\Big(  \intff{ a_N - \ga x_N  }{ b_N + \ga x_N  }   \Big) .
\enq

\end{lemme}

\Proof

It follows from the asymptotic behaviour of the solution $\chi$ \eqref{ecriture DA chi dans RHP}
and from its explicit construction in Subsection \ref{SousSousSection resolution RHP pour chi}
 along with the identities \eqref{ecriture forme matrice M ab}-\eqref{ecriture DA chi infty amenage} that 
$\chi_{12;+}(\mu)=\e{O}\big(|\mu|^{-\f{1}{2}}\big)$ uniformly in $N$ and for $\mu \in \R$. Hence, one has the bounds
\bem
\Big| \mc{J}_{12}[H] \Big| \; \leq  \;  C \Bigg\{ \Int{\R}{} \dd \mu  \f{ \big| \chi_{12;+}(\mu) \big|^{2} }{ \big[ 1 + |\mu \tau_N|\big]^{2s}  } \Bigg\}^{\f{1}{2}} \;  \cdot  \;
 \Bigg\{ \Int{\R}{} \dd \mu  \big|  \mc{F}[H](\tau_N\mu) \big|^{2} \cdot \big[ 1 + |\mu \tau_N|\big]^{2s}   \Bigg\}^{\f{1}{2}}  \\
\, \leq \, C^{\prime} \cdot \norm{ H }_{ \mc{H}_s( \intff{ a_N - \ga x_N  }{ b_N + \ga x_N  }   ) } \;.
\end{multline}
The above entails that $\mc{J}_{12}$ is continuous. As a consequence,  $\mf{X}_{s}\Big(  \intff{ a_N - \ga x_N  }{ b_N + \ga x_N  }   \Big)$ is closed.

Moreover, for any $\varphi \in  \mc{C}^1(\intff{ a_N }{  b_N  })$, one has 
\beq
\mc{J}_{12}\Big[ \mc{S}_{N;\ga}[\varphi]  \Big] \; = \;     \Int{\R  }{} \bigg\{ \chi_{22;+}(\mu) \,  \mc{F}\big[ \varphi \big](\tau_N \mu)\, \ex{- \i  \mu \ov{a}_N}\bigg\} \cdot \f{\dd\mu}{ 4 \i \pi \tau_N  }
\;  +\;  
 \Int{\R  }{} \bigg\{ \chi_{22;-}(\mu) \,  \mc{F}\big[ \varphi \big](\tau_N \mu)\, \ex{- \i \mu \ov{b}_N} \bigg\} \cdot \f{\dd\mu}{ 4 \i \pi \tau_N  }
\enq
where, starting from \eqref{ecriture TF SN gamma de varphi}, I made use of the jump relation:
\beq
\f{1}{2 \i \pi  } \mc{F}\big[ S_{\ga}\big](\la) \chi_{1a;+}(\la) \, = \, \ex{\i\la \ov{x}_N} \, \chi_{2a;+}(\la) \, + \, \chi_{2a;-}(\la)  \;. 
\label{jump relation for chi1a avec Sgamma}
\enq
Since $\vp$ has support in $\intff{a_N}{b_N}$,
the function $ \mu \mapsto \mc{F}\big[ \varphi \big](\tau_N \mu)\, \ex{- \i \mu \ov{a}_N}$, resp.  $\mu \mapsto  \mc{F}\big[ \varphi \big](\tau_N \mu)\, \ex{- \i \mu \ov{b}_N}$,
is bounded on $\mathbb{H}^+$,  resp. $\mathbb{H}^-$. Moreover, since $\vp$ is $\mc{C}^1$ these functions are a $\e{O}(\mu^{-1})$ at infinity uniformly in the half-planes of interest.
Hence, given that  $\chi_{22}(\mu)=\e{O}\big(|\mu|^{-\f{1}{2}}\big)$ uniformly on $\Cx$, one may take the first integral
by the residues in the upper-half plane and the second one by the residues in the lower half-plane. The integrands being analytic in the respective domains, one finds $\mc{J}_{12}\big[ \mc{S}_{N;\ga}[\varphi]  \big] \; = \; 0$. 
By density of $\mc{C}^{1}(\intff{a_N}{b_N})$ in $\mc{H}_s(\intff{a_N}{b_N})$ and continuity on  $\mc{H}_s(\intff{a_N}{b_N})$ of $\mc{S}_{N;\ga}$ and $\mc{J}_{12}$, it follows that
$\mc{J}_{12}\big[ \mc{S}_{N;\ga}[\varphi]  \big]=0$ for any $\vp \in \mc{H}_s(\intff{a_N}{b_N})$. \qed

\vspace{3mm}

I now introduce the operator $\mc{W}_{N;\ga} : \mf{X}_s\Big(  \intff{ a_N - \ga x_N  }{  b_N +  \ga  x_N  }  \Big) \tend \mc{H}_s\Big(  \intff{ a_N   }{  b_N   }  \Big)  $ through the  Fourier transform of its action
\beq
\mc{F}\Big[ \mc{W}_{N;\ga}[H] \Big] (\tau_N \la ) \; = \; \f{ \tau_N }{ \i \pi } 
\,\Int{ \R  }{} \f{ \dd \mu }{ 2 \i \pi }\, 
\f{  \ex{\i \la \ov{a}_N }  \ex{- \i \mu \ov{b}_N}  }{ \mu- \la }
\bigg\{    \f{ \mu }{ \la }  \cdot \chi_{11}(\la) \chi_{12;+}(\mu) \, -  \, \chi_{11;+}(\mu) \chi_{12}(\la) \bigg\}
\cdot \mc{F}\big[H\big](\tau_N \mu ) 
\label{definition Fourier de operateur W N gamma}
\enq
where $\la \in \mathbb{H}^+$. Note that the formula extends to $\la\in \R$ by taking the $+$ boundary value of $\chi_{1a}(\la)$, $a=1,2$, which are, in fact,
holomorphic in a neighbourhood of $\R$ owing to the form taken by the jump matrix $G_{\chi}$ given in \eqref{definition matrice de saut G chi}. Note that since
one works in Sobolev spaces, only the Fourier transform matters for defining the operators. The fact that \eqref{definition Fourier de operateur W N gamma}
does give rise to a continuous map between
appropriate Sobolev spaces is established later on, in Theorem \ref{Theoreme invertibilite operateur S N gamma} and
\eqref{ecriture finitude norme sobolev de WN gamma sur fct sobolev} in particular.

This action may be recast in a slightly different form by decomposing $\mu=\mu-\la+\la$ appearing in the numerator and then
observing that the factorised term not involving the denominator $\mu-\la$ vanishes by using that $H  \in \mf{X}_{s}\big(  \intff{ a_N - \ga x_N  }{ b_N + \ga x_N  }   \big)$, \textit{viz}. that
$\mc{J}_{12}[H]=0$:
\beq
\mc{F}\Big[ \mc{W}_{N;\ga}[H] \Big] (\tau_N \la ) \; = \; \f{ \tau_N }{ \i \pi } 
\,\Int{ \R  }{} \f{ \dd \mu }{ 2 \i \pi }\, 
\f{  \ex{\i \la \ov{a}_N }  \ex{- \i \mu \ov{b}_N}  }{ \mu- \la }
\bigg\{      \chi_{11}(\la) \chi_{12;+}(\mu) \, -  \, \chi_{11;+}(\mu) \chi_{12}(\la) \bigg\}
\cdot \mc{F}\big[H\big](\tau_N \mu )  \;. 
\label{ecriture noyau WNgamma forme reduite}
\enq

I now establish key properties of this integral operator.

\begin{theorem}
 
\label{Theoreme invertibilite operateur S N gamma}

Let $0<s<1$.  The operator $\mc{W}_{N;\ga}$ is a continuous left inverse of $\mc{S}_{N;\ga}$ on $\mc{H}_s(\intff{a_N}{b_N})$, \textit{viz}.
\beq
\mc{W}_{N;\ga}\circ\mc{S}_{N;\ga} = \e{id} \quad  on  \quad \mc{H}_s(\intff{a_N}{b_N}) \;  .
\label{WNgamma comme inerse a gauche}
\enq
Moreover, it holds that 
\beq
\mc{S}_{N;\ga} \circ \mc{W}_{N;\ga}\big[ H \big](\xi) \, = \, H(\xi) \qquad almost \; everywhere \; on \; \intff{a_N}{b_N} \;,
\label{WNgamma comme inerse a droite apres restriction}
\enq
for any $H \in \mf{X}_s( \intff{ a_N-\ga x_N }{ b_N + \ga x_N} )$. 
 
\end{theorem}

\Proof 

The proof goes in three steps. I first establish that indeed $\mc{W}_{N;\ga} \big[ \,  \mf{X}_s\big(  \intff{ a_N - \ga x_N  }{  b_N +  \ga  x_N  }  \big) \,  \big] \subset \mc{H}_s\Big(  \intff{ a_N   }{  b_N   }  \Big)$,
and that $\mc{W}_{N;\ga}$ is a continuous operator. Then, I establish the left inverse property \eqref{WNgamma comme inerse a gauche} and, finally, 
the right inversion property restricted to the interval $\intff{a_N}{b_N}$ expressed in \eqref{WNgamma comme inerse a droite apres restriction}. 

\vspace{2mm}

{$\bullet$ \bf Support of $\mc{W}_{N;\ga}\big[ H \big]$ } 

\vspace{2mm}

Since $H$ has compact support, $\mc{F}[H]$ is entire so that the jump relation  
\beq
\ex{-\i \mu \ov{b}_N} \chi_{1a;+}(\mu) \, = \, \ex{-\i \mu \ov{a}_N} \chi_{1a;-}(\mu)
\label{relation saut chi 1a avec exponentielle oscillante}
\enq
adjoined to the decay properties of the integrand at infinity  allows one to  deform the $\mu$ integral in \eqref{ecriture noyau WNgamma forme reduite}
up to $\R-\i\eps^{\prime}$ with $\eps^{\prime}>0$ and small enough. Thus, for $\la \in \R$, one has that
\beq
\mc{F}\big[ \mc{W}_{N;\ga}[H] \big] (\tau_N \la ) \; = \; \f{ \tau_N  \ex{\i \la \ov{a}_N } }{ \i \pi } \Big\{  G^{(1)}_{+}(\la) \, - \,   G^{(2)}_{+}(\la)  \Big\}
\enq
where 
\beqa 
G^{(1)}(\la)  & = & \chi_{11}(\la) \op{C}_{\R-\i\eps^{\prime}}\Big[  \ex{-\i \ov{a}_N *} \chi_{12}(*)  \mc{F}\big[H\big](\tau_N * )   \Big](\la) \; , \\
G^{(2)}(\la) & = & \chi_{12}(\la) \op{C}_{\R-\i\eps^{\prime}}\Big[  \ex{-\i \ov{a}_N *} \chi_{11}(*)  \mc{F}\big[H\big](\tau_N * )   \Big](\la) \;,  
\eeqa
in which I made use of the Cauchy transform subordinate to a curve $\msc{C}$
\beq
\op{C}_{ \msc{C} } [f](\la) \; = \; \Int{ \msc{C} }{} \hspace{-1mm} \f{\dd \mu }{ 2\i\pi } \f{ f(\mu) }{ \mu - \la }  \;.
\enq
Also, above, $*$ stands for the running variable on which the Cauchy transform acts. Since the functions $G^{(k)}$ are analytic in the upper half-plane, it follows that 
\beq
\e{supp}\Big\{ \mc{F}^{-1}\big[ \ex{\i a_N * }  G^{(k)}(\tau_N^{-1} * ) \big] \Big\} \;\subset  \;  \intfo{a_N}{+\infty}  \;. 
\enq
Analogously, since the integrand has no pole at $\la=\mu$, by moving the $\mu$-integration contour to $\R+\i\eps^{\prime}$ and using the jump relations for $\chi$, 
\eqref{ecriture noyau WNgamma forme reduite} may be recast as 
\beq
\mc{F}\big[ \mc{W}_{N;\ga}[H] \big] (\tau_N \la ) \; = \; \f{ \tau_N  \ex{\i \la \ov{b}_N } }{ \i \pi } \Big\{  \wt{G}^{(1)}_{-}(\la) \, - \,   \wt{G}^{(2)}_{-}(\la)  \Big\}
\enq
where 
\beqa
\wt{G}^{(1)}(\la) & = & \chi_{11}(\la) \op{C}_{\R + \i\eps^{\prime}}\Big[  \ex{-\i \ov{b}_N *} \chi_{12}(*)  \mc{F}\big[H\big](\tau_N * )   \Big](\la) \;,  \\ 
\wt{G}^{(2)}(\la) & = & \chi_{12}(\la)  \op{C}_{\R + \i\eps^{\prime}}\Big[  \ex{-\i \ov{b}_N *} \chi_{11}(*)  \mc{F}\big[H\big](\tau_N * )   \Big](\la) \;. 
\eeqa
 Since the functions $\wt{G}^{(k)}$ are analytic in the lower half-plane, it follows that 
\beq
\e{supp}\Big\{ \mc{F}^{-1}\Big[ \ex{ \i b_N * }  \wt{G}^{(a)}(\tau_N^{-1} * ) \Big] \Big\} \;\subset  \;  \intof{-\infty}{b_N}  \;. 
\enq
Hence, $\e{supp}\Big\{\mc{W}_{N;\ga}[H] \Big\} \subset \intff{a_N}{b_N}$. 

I now establish that $\mc{W}_{N;\ga}$ is continuous. One may recast 
\bem
\mc{F}\big[ \mc{W}_{N;\ga}[H] \big] (\tau_N \la )   =  \f{ \tau_N }{ \i \pi }   \ex{\i \la \ov{a}_N } \Bigg\{  \f{\chi_{11}(\la) }{\la}  \op{C}_{\R}\Big[  \wh{\chi}_{12}(*) \mc{F}\big[H_{\eps^{\prime}} \big](\tau_N * )  \Big](\la-\i \eps^{\prime}) \\
 -  \chi_{12}(\la)  \op{C}_{\R}\Big[  \wh{\chi}_{11}(*) \mc{F}\big[H_{\eps^{\prime}} \big](\tau_N * )  \Big](\la-\i \eps^{\prime}) \Bigg\}
\nonumber
\end{multline}
in which 
\beq
\wh{\chi}_{12}(\mu) \, = \,   ( \mu + \i \eps^{\prime}) \cdot \chi_{12}( \mu + \i \eps^{\prime}) \, \ex{-\i ( \mu + \i \eps^{\prime}) \ov{b}_N } \quad , \qquad 
\wh{\chi}_{11}(\mu) \, = \,   \chi_{11}( \mu + \i \eps^{\prime}) \, \ex{-\i ( \mu + \i \eps^{\prime}) \ov{b}_N } 
\enq
and $H_{\eps^{\prime}}(\xi) \, = \, \ex{-\tau_N \eps^{\prime} \xi } H(\xi)$. 
This entails that 
\bem
\mc{F}\Big[  \ex{-2 \tau_N \eps^{\prime} * }  \mc{W}_{N;\ga}[H] \Big] (\tau_N \la ) \; = \; \f{ \tau_N }{ \i \pi }   \ex{\i ( \la  + 2 \i \eps^{\prime}) \ov{a}_N } 
\Bigg\{  \f{ \chi_{11}( \la  + 2 \i \eps^{\prime}) }{ \la  + 2 \i \eps^{\prime}}  \op{C}_{\R}\Big[  \wh{\chi}_{12}(*) \mc{F}\big[H_{\eps^{\prime}} \big](\tau_N * )  \Big](\la+\i \eps^{\prime})  \\
 \, - \, \chi_{12}( \la  + 2 \i \eps^{\prime})  \op{C}_{\R}\Big[  \wh{\chi}_{11}(*) \mc{F}\big[H_{\eps^{\prime}} \big](\tau_N * )  \Big](\la+\i \eps^{\prime}) \Bigg\} \;. 
\end{multline}
Since it holds that $\la^{-1} \chi_{11}(\la)=\e{O}\big( |\la|^{-\f{1}{2}} \big)$ and $\chi_{12}(\la)=\e{O}\big( |\la|^{-\f{1}{2}} \big)$, one has that 
\bem
\norm{  \mc{W}_{N;\ga}[H]  }_{ \mc{H}_s( \intff{a_N}{b_N} ) } \, = \,
\Norm{  \ex{ 2 \tau_N \eps^{\prime} * } \cdot  \ex{-2 \tau_N \eps^{\prime} * }  \mc{W}_{N;\ga}[H]  }_{ \mc{H}_s( \intff{a_N}{b_N} ) } \; \leq \; \wt{C} \Norm{  \ex{-2 \tau_N \eps^{\prime} * }  \mc{W}_{N;\ga}[H]  }_{ \mc{H}_s(\R) }   \\
\; \leq \; C \cdot \bigg\{ 
\Norm{  \op{C}_{\R}\big[  \wh{\chi}_{12}(*) \mc{F}\big[H_{\eps^{\prime}} \big](\tau_N * )  \big](* + \i \eps^{\prime}) }_{  \mc{F}[\mc{H}_{ s - \frac{1}{2} }(\R)] }
+ 
\Norm{  \op{C}_{\R}\big[  \wh{\chi}_{11}(*) \mc{F}\big[H_{\eps^{\prime}} \big](\tau_N * )  \big](* + \i \eps^{\prime}) }_{  \mc{F}[\mc{H}_{ s - \f{1}{2} }(\R)] }  \bigg\} \\
\, \leq \,  C^{\prime} \cdot \bigg\{ 
\Norm{    \wh{\chi}_{12}(*) \mc{F}\big[H_{\eps^{\prime}} \big](\tau_N * )   }_{  \mc{F}[\mc{H}_{ s - \f{1}{2} }(\R)] }
\; + \; 
\Norm{     \wh{\chi}_{11}(*) \mc{F}\big[H_{\eps^{\prime}} \big](\tau_N * )    }_{  \mc{F}[\mc{H}_{ s - \f{1}{2} }(\R)] }  \bigg\} \\
\, \leq \,  C^{\prime \prime } \cdot  \Norm{   \mc{F}\big[H_{\eps^{\prime}} \big](\tau_N * )   }_{  \mc{F}[\mc{H}_{ s }(\R)] }  \, \leq \,
C^{(3)} \cdot  \norm{   H_{\eps^{\prime}}    }_{  \mc{H}_{ s }( \intff{ a_N - \ga x_N }{ b_N + \ga x_N  } ) }
\, \leq \, 
C^{(4)}  \cdot  \norm{  H   }_{  \mc{H}_{ s }(\R) }   \;.
\label{ecriture finitude norme sobolev de WN gamma sur fct sobolev}
\end{multline}
In the first and last bound I made use of Lemma \ref{Lemme bornage produit dans norme Hs}. Also, in the intermediate bound,
I used that $\op{C}_{\R-\i \eps^{\prime} } $ is a continuous operator on $\mc{H}_{s}(\R)$ for $|s|<\tf{1}{2}$.

\vspace{2mm}

{$\bullet$ \bf  The left inversion}

\vspace{2mm}

Pick $\vp \in \mc{C}^1\cap \mc{H}_s(\intff{a_N}{b_N})$. Since  $\mc{S}_{N;\ga}[\vp]\in \mf{X}_s\Big(\intff{a_N-\ga x_N}{b_N+\ga x_N} \Big)$, the form of
the Fourier transform \eqref{ecriture TF SN gamma de varphi} yields that, for $\la \in \mathbb{H}^+$,  
\beq
\mc{F}\Big[ \mc{W}_{N;\ga}\circ\mc{S}_{N;\ga}[\vp] \Big] (\tau_N \la )  =     
 \Int{ \R  }{} \f{ \dd \mu }{ 2 \i \pi }\, 
\f{  \ex{\i \la \ov{a}_N  - \i  \mu \ov{b}_N}  }{ \mu- \la }
\bigg\{      \chi_{11}(\la) \chi_{12;+}(\mu) \, -  \, \chi_{11;+}(\mu) \chi_{12}(\la) \bigg\}
\cdot \f{ \mc{F}\big[S_{\ga}\big](\mu ) }{ 2\i\pi } \mc{F}\big[\vp\big](\tau_N \mu ) \;.  
\enq
Then, by using the jump relation \eqref{jump relation for chi1a avec Sgamma} one gets 
\bem
\mc{F}\Big[ \mc{W}_{N;\ga}\circ\mc{S}_{N;\ga}[\vp]  \Big] (\tau_N \la ) \; = \;  
\,\Int{ \R  }{} \f{ \dd \mu }{ 2 \i \pi }\, 
\f{  \ex{\i  \la \ov{a}_N  }     }{ \mu- \la }
\bigg\{       \chi_{11}(\la) \chi_{22;+}(\mu) \, -  \, \chi_{21;+}(\mu) \chi_{12}(\la) \bigg\}
 \ex{- \i   \mu \ov{a}_N }  \mc{F}\big[\vp\big](\tau_N \mu )  \\
\, + \,\Int{ \R  }{} \f{ \dd \mu }{ 2 \i \pi }\, 
\f{  \ex{\i  \la \ov{a}_N  }     }{ \mu- \la }
\bigg\{       \chi_{11}(\la) \chi_{22;-}(\mu) \, -  \, \chi_{21;-}(\mu) \chi_{12}(\la) \bigg\}
 \ex{- \i   \mu \ov{b}_N }  \mc{F}\big[\vp\big](\tau_N \mu ) \;.  
\end{multline}
Indeed each of these integrals is well defined since $\mc{F}[\vp]\in L^{\infty}(\R)$ for continuous $\vp$ while 
$\chi_{22}(\mu)=\e{O}\big( |\mu|^{-\f{1}{2} } \big)$ and  $\chi_{21}(\mu)=\e{O}\big( |\mu|^{-\f{3}{2} } \big)$, \textit{c.f.} \eqref{ecriture DA chi dans RHP} and \eqref{ecriture forme matrice M ab}-\eqref{ecriture DA chi infty amenage},
so that the integrand is a $\e{O}\big( |\mu|^{-\f{3}{2} } \big)$ pointwise in $\la$. Since $\mu \mapsto  \ex{- \i   \mu \ov{a}_N}  \mc{F}\big[\vp\big](\tau_N \mu ) $, resp.  
$\mu \mapsto  \ex{- \i   \mu \ov{b}_N }  \mc{F}\big[\vp\big](\tau_N \mu ) $, is bounded and analytic on $\mathbb{H}^+$, resp. $\mathbb{H}^-$, one may take the 
two $\mu$ integrals by the residues in $\mathbb{H}^{+}$, resp $\mathbb{H}^{-}$. This yields, for $\la \in \mathbb{H}^+$, 
\beq
\mc{F}\Big[ \mc{W}_{N;\ga}\circ\mc{S}_{N;\ga}[\vp] \Big] (\tau_N \la ) \; = \; \Big[    \chi_{11}(\la) \chi_{22}(\la) \, -  \, \chi_{21}(\la) \chi_{12}(\la)  \Big]\cdot  \mc{F}\big[\vp\big](\tau_N \la ) \;. 
\enq
Since $\det\big[ \chi(\la) \big]=1$ for any $\la \in \mathbb{H}^+$, one gets that $ \mc{W}_{N;\ga}\circ\mc{S}_{N;\ga}= \e{id}$ on $(\mc{C}^1\cap \mc{H}_{s})(\intff{a_N}{b_N})$.
Then, by continuity of the operators and density of $\mc{C}^1(\intff{a_N}{b_N})$ in $\mc{H}_s(\intff{a_N}{b_N})$, the left inversion identity holds.

\vspace{2mm}

{$\bullet$ \bf  The restricted right inversion}

\vspace{2mm}

By using \eqref{ecriture TF SN gamma de varphi}, simplifying the kernel owing to $H \in \mf{X}_s\Big(\intff{a_N-\ga x_N}{b_N+\ga x_N} \Big)$ and moving the $\mu$-integration 
to $\R-\i\eps^{\prime}$ with the help of the jump relation \eqref{relation saut chi 1a avec exponentielle oscillante}, one gets that, for $\la \in \R$, 
\bem
\mc{F}\Big[ \mc{S}_{N;\ga}\circ\mc{W}_{N;\ga}[H]  \Big] (\tau_N \la ) \; = \;  
\,\Int{ \R - \i \eps^{\prime}  }{} \f{ \dd \mu }{ 2 \i \pi }\, 
\f{   \ex{- \i \mu \ov{a}_N}  }{ \mu- \la }
\bigg\{     \Big[  \ex{\i   \la \ov{b}_N }  \chi_{21;+}(\la) \, + \, \ex{\i \la \ov{a}_N }  \chi_{21;-}(\la) \Big]\,  \chi_{12}(\mu) \\ 
\, -  \, \Big[  \ex{\i \la \ov{b}_N }  \chi_{22;+}(\la) \, + \, \ex{\i \la \ov{a}_N }  \chi_{22;-}(\la) \Big]  \, \chi_{11}(\mu)  \bigg\} \cdot  \mc{F}\big[ H \big](\tau_N \mu ) \\
\; = \;  \ex{\i \la \ov{b}_N } G_+^{(\ua)}(\la) \; + \;  \ex{\i \la \ov{a}_N } G_-^{(\da)}(\la) \; + \; \Big(  \chi_{21;-}(\la)  \chi_{12;-}(\la) \, - \, \chi_{22;-}(\la)  \chi_{11;-}(\la)  \Big) \cdot   \mc{F}\big[ H \big](\tau_N \la ) \;. 
\end{multline}
where the last term evaluates to $ \mc{F}\big[ H \big](\tau_N \la )$ by virtue of $\det\big[ \chi_-(\la) \big]=-1$ and stems from the contribution of the pole at $\mu=\la$
when deforming the contours from $\R+\i\eps^{\prime}$ to $\R-\i\eps^{\prime}$ in the integral which, eventually, gives rise to $ G_-^{(\da)}$. Also, above, we have introduced
\beqa
G^{(\ua)}(\la) & = & \Int{ \R - \i \eps^{\prime}  }{} \f{ \dd \mu }{ 2 \i \pi }\, 
\f{   \ex{- \i \mu \ov{a}_N}  }{ \mu- \la }
\Big\{     \chi_{21}(\la) \chi_{12}(\mu) \, -\,    \chi_{22}(\la) \chi_{11}(\mu) \Big\} \cdot  \mc{F}\big[ H \big](\tau_N \mu )  \; ,  \\ 
G^{(\da)}(\la) & = & \Int{ \R + \i \eps^{\prime}  }{} \f{ \dd \mu }{ 2 \i \pi }\, 
\f{   \ex{- \i \mu \ov{b}_N}  }{ \mu- \la }
\Big\{     \chi_{21}(\la) \chi_{12}(\mu) \, -\,    \chi_{22}(\la) \chi_{11}(\mu) \Big\} \cdot  \mc{F}\big[ H \big](\tau_N \mu ) \;. 
\eeqa
Since $G^{(\ua )}\in \mc{O}\big( \mathbb{H}^{+ } \big)$ and $G^{(\da )}\in \mc{O}\big( \mathbb{H}^{- } \big)$, one has that
\beq
\e{supp}\bigg\{  \mc{F}^{-1}\big[ \ex{\i * b_N } G_+^{(\ua)}( * \tau_N^{-1} ) \big] \bigg\} \, \subseteq \, \intfo{b_N}{+\infty}  \qquad \e{and} \qquad
\e{supp}\bigg\{  \mc{F}^{-1}\big[ \ex{\i * a_N } G_+^{(\da)}( * \tau_N^{-1} ) \big] \bigg\} \, \subseteq \, \intof{-\infty}{a_N} \;,
\enq
and thus $ \mc{S}_{N;\ga}\circ\mc{W}_{N;\ga}[H] = H$ on $\intoo{a_N}{b_N}$.  This entails the claim. \qed

\subsection{The \textit{per se} inverse of $\mc{S}_N$}

\begin{figure}[h]
\begin{center}

\includegraphics[width=.8\textwidth]{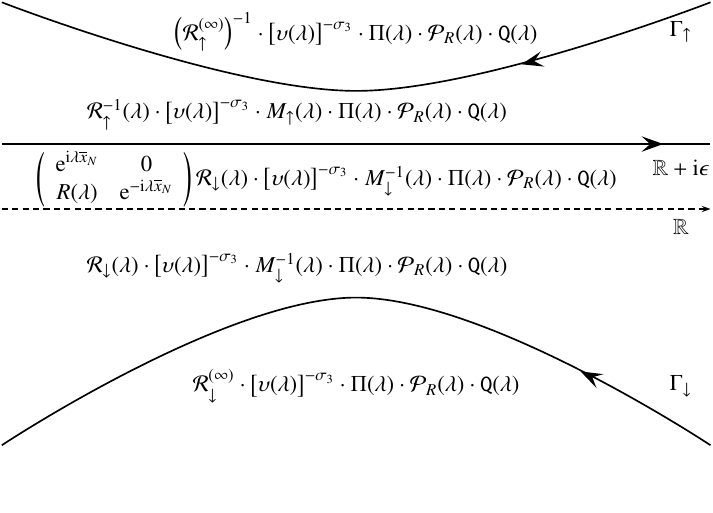}

\caption{Piecewise definition of the matrix $\chi$ in the $\gamma \rightarrow +\infty$ limit.
The curves $\Ga_{\ua/\da}$ separate all poles of $\la \mapsto \la R(\la)$ from $\R$ and are such that $\e{dist}(\Ga_{\ua/\da}, \R) =2(1-\a)$
for some $\a >0$ and sufficiently small. 
\label{Figure definition sectionnelle de la matrice chi a gamma infty} }
\end{center}
\end{figure}

In order to construct the inverse to $\mc{S}_N$, we should take the limit $\ga \tend +\infty$ in the previous formulae. It so happens
that this limit is already well-defined on the level of the  
solution to the Riemann--Hilbert problem for $\chi$ as defined through Figure~\ref{Figure definition sectionnelle de la matrice chi}. 
More precisely, from now on, let $\chi$ be as defined in Figure~\ref{Figure definition sectionnelle de la matrice chi a gamma infty}
where the matrix $\Pi$ is as defined through \eqref{ecriture eqn int sing pour matrice Pi moins}-\eqref{ecriture rep int matrice Pi}. 
It solves the Riemann--Hilbert problem introduced in Section \ref{Sous Section RHP pour chi} with the sole difference that one should replace
$\tf{\mc{F}[S_{\ga}]}{2\i\pi}$ appearing in the jump matrix $G_{\chi}$ by $R$.

\begin{prop}
\label{Proposition invertibilite operateur S}

Let $0 < s < 1$. The operator 
$\mc{S}_N \; : \; \mc{H}_{s}\big( \intff{a_N}{b_N} \big) \longrightarrow \mc{H}_{s}\big( \R \big)$
is continuous and invertible on its image:
\beq
\label{x}
\mf{X}_{s}\big( \R \big) = \Bigg\{H \in \mc{H}_{s}(\R)\;\; :\;\;   \Int{\R+\i \eps^{\prime} }{} \chi_{12}(\mu) \, \mc{F}[H](\tau_N\mu)\, \ex{- \i \mu \ov{b}_N}\cdot\f{\dd\mu}{ (2 \i \pi)^2 }  \; = \; 0\Bigg\}\;.
\enq 
More specifically, one has the left and right inverse relations 
\beq
\mc{W}_N\circ \mc{S}_N \, = \, \e{id}  \quad on \quad \mc{H}_s(\intff{a_N}{b_N}) \quad and \quad
\mc{S}_N\circ \mc{W}_N[H](\xi) \, = \, H(\xi)   \quad a.e. \; on\;  \intff{a_N}{b_N}
\enq
for any $H \in \mf{X}_s(\R)$. The operator $\mc{W}_N\,:\,\mf{X}_{s}(\R) \longrightarrow \mc{H}_s(\intff{a_N}{b_N})$ is given, whenever it makes sense, as an encased oscillatorily convergent Riemann integral transform
\beq
\label{forumule explicite pour WN}
\mc{W}_N[H](\xi) \; = \; \f{ \tau_N^{2} }{ \pi }  
\Int{ \R + 2 \i \eps^{\prime} }{} \f{ \dd \la }{ 2 \i \pi }\,\Int{ \R + \i \eps^{\prime} }{} \f{ \dd \mu }{ 2 \i \pi }\, 
 \ex{- \i \la \tau_N(\xi-a_N) } W(\la,\mu)
\,\ex{- \i  \mu \ov{b}_N} \mc{F}\big[H\big](\tau_N \mu ) \;, 
\enq
 where $\eps^{\prime}> 0 $ is small enough. 
The integral kernel 
\beq
W(\la,\mu) \, = \, \f{ 1 }{ \mu- \la }
\bigg\{    \f{ \mu }{ \la }  \cdot \chi_{11}(\la) \chi_{12}(\mu) \, -  \, \chi_{11}(\mu) \chi_{12}(\la) \bigg\} \;, 
\label{definition noyau W}
\enq
is expressed in terms of the entries of the matrix  $\chi$ which is understood to be defined as in Figure~\ref{Figure definition sectionnelle de la matrice chi a gamma infty}. 
\end{prop}

 \Proof
The proof of the statement is very analogous to the one given in Theorem \ref{Theoreme invertibilite operateur S N gamma}, upon using the
definition of $\chi$ in the limit $\ga\tend +\infty$ provided in Figure~\ref{Figure definition sectionnelle de la matrice chi a gamma infty}
which has only jumps on $\R$ given by the jump matrix
\beq
\left( \ba{cc} \ex{\i\la \ov{x}_N}   &   0  \\
                    R(\la)        &   \ex{-\i\la \ov{x}_N } \ea \right) \;.
\enq
To obtain the  representation \eqref{forumule explicite pour WN}, one starts from the representation
\beq
\mc{W}_N[H](\xi) \; = \; \f{ \tau_N^{2} }{ \pi }
\Int{ \R   }{} \f{ \dd \la }{ 2 \i \pi }\,\Int{ \R   }{} \f{ \dd \mu }{ 2 \i \pi }\,
 \ex{- \i \la \tau_N(\xi-a_N) } W_{+,+}(\la,\mu)
\,\ex{- \i  \mu \ov{b}_N} \mc{F}\big[H\big](\tau_N \mu ) \;,
\enq
with $W_{+,+}(\la,\mu)= \lim_{\eps, \eps^{\prime} \tend 0^+} W\big( \la + \i\eps, \mu + \i\eps^{\prime} \big)$ and then deforms the contours to
$\big\{ \R + 2 \i \eps^{\prime} \big\} \times \big\{ \R +   \i \eps^{\prime} \big\}$. The well definiteness of the former, for $H$ regular enough, as well as the one of the
contour deformation procedure can be seen by implementing the techniques for dealing with oscilatorily convergent Riemann integrals which were outlined
earlier in the paper. The details are left to the reader. \qed

\section{Complete characterisation of the equilibrium measure} 
 \label{Section caracterisation complete de la mesure eq}

The main result of this section is Theorem \ref{Theorem caracterisation minimiseur fnelle EN +} which provides a closed expression for
the density of the equilibrium measure $\sg_{\e{eq}}^{(N)}$
along with an explicit characterisation of its support. It turns out that some of the expressions obtained below take a simpler form when the point $\i$, resp. $-\i$,
is located above, resp. below, the curve $\Ga_{\ua}$, resp. $\Ga_{\da}$. Hence, from now on, I shall
choose the constant $\a$ characterising the distance of the contours $\Ga_{\ua/\da}$ to $\R$ to be such that
\beq
\tfrac{1}{2} < \a < 1\; \qquad \e{so}\; \e{that} \quad \a=\tfrac{1}{2}(1+\tilde{\a}) \quad \e{with} \quad \tilde{\a} \in \intoo{0}{1} \;.
\label{definition alpha prime}
\enq
This range of $\tilde{\a}$ will be tacitly assumed in all the results stated below. Moreober, such a choice of $\a$ ensures that
the points $+/- \i$ are indeed located in the part of the complex plane lying above/below the curve $\Ga_{\ua/\da}$.
Also, one gets, when $N\tend+\infty,$ the new parameterisation of  the control on the remainders arising in the matrix $\Pi$ \eqref{ecriture bornes en N pour Pi moins Id} and the matrices $\mc{Q}_{\ell}$ \eqref{ecriture DA matrice Q ell}:
\beq
\Pi(\la) \; = \; I_2 \,  + \, \e{O}\bigg( \f{ \ex{ -\ov{x}_N(1-\tilde{\a}) } }{  1 + |\la|} \bigg)   \qquad \e{and} \qquad
\mc{Q}_{\ell}\; = \; c_{\ell} \sg^{+ }  \,  + \, \e{O}\bigg(  (\, \ov{x}_N )^4  \ex{ -\ov{x}_N(1-\tilde{\a}) }\bigg)  \;,
\label{ecriture estimee modifiee sur Pi et Q ell}
\enq
with a control on $\Pi(\la)$ which has the same uniformness properties as stated in \eqref{ecriture bornes en N pour Pi moins Id}.

\begin{theorem}
\label{Theorem caracterisation minimiseur fnelle EN +}
 Let $N\geq N_0$ with $N_0$ large enough. Then, the unique minimiser $\sg_{\e{eq}}^{(N)}$
 of the functional $\mc{E}_N^{(+)}$ introduced in \eqref{definition fonctionnelle energie +} is absolutely continuous in respect to the Lebesgue measure
 and is supported on the segment  $\intff{a_N}{b_N}$. The endpoints are the unique solutions to the equations 
\beq
a_N+b_N=0 \qquad and \qquad    \vartheta  \cdot \f{    ( \ov{b}_N)^2   \, \ex{ \ov{b}_N }  }{    N  } \cdot \mf{t}( 2 \ov{b}_N) \cdot
\bigg\{ 1   \, + \, \e{O}\bigg(  (\ov{b}_N)^5 \ex{ -2\ov{b}_N(1-\tilde{\a})  }   \bigg) \bigg\} \, = \, 1 \;, 
\label{ecriture solution pour bornes mesure equilibre}
\enq
where    the remainder is smooth and differentiable in $\ov{b}_N$. Above, one has
\beq
\vartheta \, = \, \f{ 2 \, \varkappa }{ 3  (2\pi)^{ \f{5}{2} } } \cdot \f{ \Ga\big( \mf{b}, \hat{\mf{b}} \big) }{ \mf{b}^{\mf{b}} \, \hat{\mf{b}}^{\hat{\mf{b}}} }\; , 
\label{definition vartheta}
\enq
while, upon using the constants $w_k$ introduced in \eqref{definition des constantes wk et de leur asymptotiques},  
\beq
\mf{t}( \ov{x}_N) \, = \,  \f{ 6 }{ (\ov{x}_N)^2 } \bigg\{  2  \, + \, w_2  \, - \, w_1 \, -\, \f{w_1 w_3}{ w_2 }  \bigg\} \; \widesim{  \ov{x}_N \tend + \infty } \; 1 \, + \, \e{O}\Big( \tfrac{ 1 }{ \ov{x}_N } \Big) \;. 
\label{definition polynome frak t}
\enq
In particular,  $\ov{b}_N$ is uniformly away from zero and admits the large-$N$ expansion 
\beq
\ov{b}_N\, = \, \ln N  \,-  \, 2  \ln \ln N  \, - \,  \ln \vartheta   \; + \; \e{O}\Big( \f{ \ln \ln N }{ \ln N }  \Big) \;. 
\label{ecritue DA b_N}
\enq

 Finally, the density $\varrho_{\e{eq}}^{(N)}$ of the equilibrium measure $\dd \sg_{\e{eq}}^{(N)}(\xi) \,  =  \, \varrho_{\e{eq}}^{(N)}(\xi) \bs{1}_{\intff{a_N}{b_N}}(\xi) \dd \xi $ 
 is expressed in terms of the integral transform of the potential as
\beq
\varrho_{\e{eq}}^{(N)}(\xi)\, = \,  \mc{W}_N[H_N](\xi) 
\enq
where the operator $\mc{W}_N$ is as introduced in \eqref{forumule explicite pour WN} while 
\beq
H_N(\eta)\;  = \; \f{1}{N \tau_N} V^{\prime}_N(\eta) \; = \;  \f{ \varkappa }{ N } \sinh(\tau_N \eta). 
\label{definition fct HN}
\enq

\end{theorem}

\Proof 


Let $a_N<b_N$ be such that $x_N=b_N-a_N>\de$ for some $\de>0$. The lower bound  $x_N=b_N-a_N>\de$ entails, by virtue of Prop.\ref{Proposition ecriture forme asymptotique matrice chi}, that the Riemann--Hilbert problem for $\chi$
is uniquely solvable provided that $N$ is large enough. By virtue of Proposition \ref{Proposition invertibilite operateur S}, 
the function $H_N$ as defined in \eqref{definition fct HN}  belongs to the image of $\mc{H}_s(\intff{a_N}{b_N})$ by $\mc{S}_N$, \textit{c.f.} $\mf{X}_{s}(\R)$  as defined in \eqref{x},
provided that 
\beq
\Int{\R+\i \eps^{\prime} }{} \chi_{12}(\mu) \, \mc{F}[H_N](\tau_N\mu)\, \ex{- \i \mu \ov{b}_N}\cdot\f{\dd\mu}{ (2 \i \pi)^2 }  \, = \, 0 \;. 
\label{ecriture integrale de contrainte sur HN}
\enq
The above integral is evaluated in Proposition \ref{Proposition evaluation integrale de contrainte} and the result obtained there entails that  \eqref{ecriture integrale de contrainte sur HN}
holds if and only if $a_N+b_N=0$.

 By virtue of Proposition \ref{Proposition contrainte de normalisation}, eq. \eqref{ecriture DA integrale de normalisation}, 
\beq
 \Int{a_N}{b_N}  \mc{W}_N[H_N](\eta) \dd \eta  \; = \;  \f{ \vartheta }{ N }  \cdot (\ov{b}_N)^2 \cdot \ex{ \ov{b}_N } \cdot \mf{t}(2 \ov{b}_N)\, \bigg( 1\, + \,  \e{O}\Big( (\,\ov{b}_N)^5 \ex{ - 2\ov{b}_N(1-\tilde{\a}) } \Big) \bigg)  \;, 
\enq
in which $\mc{W}_N$ is as defined in \eqref{forumule explicite pour WN} while the remainder is smooth in $b_N$ and differentiable provided that $b_N>\de^{\prime}$ form some $\de^{\prime}>0$. 
There, $\vartheta$ is as in \eqref{definition vartheta} while $\mf{t}(\ov{x}_N)$ has been introduced in \eqref{definition polynome frak t}. The map 
\beq
x \mapsto x^2 \ex{x}
\enq
is a smooth diffeomorphism from $\R^+$ onto $\R^+$. As a consequence, taking the remainder to be as above, it follows that for $N$ large enough 
\beq
 \vp \; : \; b \mapsto \vp(b)= \vartheta   \cdot b^2 \cdot \ex{ b } \cdot \mf{t}(2 b)\, \bigg( 1\, + \,  \e{O}\Big( b^5 \ex{ - 2b(1-\tilde{\a}) } \Big) \bigg) 
\enq
is a smooth diffeomorphism from $\intfo{C}{+\infty}$ onto $\intfo{\vp(C)}{+\infty}$ with $C$ large enough. In particular, there exists a unique $\ov{b}_N=\ln N \cdot b_N$ such that $\vp(\ov{b}_N)=N$. 
It is direct to see that $\ov{b}_N$ admits the large-$N$ behaviour \eqref{ecritue DA b_N}. 

Now, let $b_N$ be as defined above, $a_N=-b_N$ and define $\varrho_{\e{eq}}^{(N)}=\mc{W}_N[H_N] \in \mc{H}_s(\intff{a_N}{b_N})$, for any $0<s<1$.
It follows from the $\ga \tend + \infty$ limit of the result stated in Theorem \ref{Theoreme invertibilite operateur S N gamma} that 
$\e{supp}\big[ \varrho_{\e{eq}}^{(N)} \big]= \intff{a_N}{b_N}$.

By construction  the density $\varrho_{\e{eq}}^{(N)}$ satisfies the singular integral equation $\mc{S}_N[\varrho_{\e{eq}}^{(N)}]=H_N$ with $H_N$ as defined above.
It follows from Proposition \ref{Proposition propriete densite mesure eq dans son support} that $\varrho_{\e{eq}}^{(N)}(x)>0$ on $\intoo{a_N}{b_N}$. 
Moreover, it follows from Proposition \ref{Proposition propriete pot effectif} that the effective potential \eqref{ecriture potentiel effectif associee a une solution phi} 
subordinate to $\varrho_{\e{eq}}^{(N)}$ does satisfy \eqref{ecriture positivite stricte du potentiel effectif}. As a consequence, by virtue of Proposition 
\ref{Proposition characterisation qqes ptes mesure eq},  $\varrho_{\e{eq}}^{(N)}$ is the density of the equilibrium measure $\sg_{\e{eq}}^{(N)}$ and the latter is supported on 
the single interval $\intff{a_N}{b_N}$. \qed



\subsection{Explicit expressions for the asymptotic expansion of the solution $\chi$}
\label{SousSection DA de chi explicite dans le plan complexe}

To obtain the large-$N$ expansion of $\chi$, one should first establish the one of the coefficients $\mf{q}_1$ and $\chi_{12}^{(1)}$ appearing in the matrix elements of $\op{Q}$. 
The matrices $\chi^{(a)}$ arise in the large $\la$ expansion of 
\beq
(\i \la)^{\f{3}{2} \sg_3} \cdot \big[\ups(\la)\big]^{-\sg_3} \cdot \Pi(\la) \cdot \mc{P}_{R}(\la)
\enq
Upon setting $\wt{\mc{Q}}_{\ell}=\Pi^{-1}\!(0) \, \mc{Q}_{\ell} \, \Pi(0)$ and using the large-$\la$ expansions
\beq
\Pi(\la)\; = \; I_2 + \f{1}{\la} \Pi^{(1)} + \f{1}{\la^2} \Pi^{(2)} \, + \, \e{O}\Big(\la^{-3}\Big) \quad \e{and} \quad
\ups(\la) \; = \; (\i \la)^{-\f{3}{2}}\bigg(  1 + \f{\ups^{(1)}}{\la}  + \f{ \ups^{(2)} }{\la^2}  \, + \, \e{O}\Big(\la^{-3}\Big)  \bigg)
\enq
a direct computation yields 
\beq
\left\{ \ba{l} 
\chi^{(1)}\; = \; \Pi^{(1)} + \wt{\mc{Q}}_{2}- \ups^{(1)} \sg_3   \vspace{2mm} \\ 
\chi^{(2)}\; = \; \Pi^{(2)} + \wt{\mc{Q}}_{1}+ \Pi^{(1)}\wt{\mc{Q}}_{2} + \f{I_2+\sg_3 }{2} \big[ \ups^{(1)}  \big]^2- \ups^{(2)} \sg_3  
\,- \, \ups^{(1)} \sg_3 \big( \Pi^{(1)} + \wt{\mc{Q}}_{2} \big)  \ea  \right. \;. 
\enq
Hence, owing to \eqref{ecriture estimee modifiee sur Pi et Q ell}
\beq
\chi_{12}^{(1)}\; = \; c_2 \; + \; \e{O}\Big(  (\, \ov{x}_N)^4 \ex{- \ov{x}_N (1- \tilde{\a})}  \Big) \qquad \e{and} \qquad 
\mf{q}_1 \; = \; - \f{ c_1}{c_2 }\; + \; \e{O}\Big(  (\, \ov{x}_N)^4 \ex{- \ov{x}_N (1-\tilde{\a})}  \Big) \;. 
\enq
The coefficients $c_k$ appearing above have been introduced in \eqref{definition coefficients c ell}.  

Thus, starting from the expression for $\chi$ given in Fig.~\ref{Figure definition sectionnelle de la matrice chi a gamma infty}, one gets that for $\la$ located above the curve $\Ga_{\ua}$
it holds 
\beq
\chi(\la) \; = \; \left( \ba{cc} \mf{q}_2^{-1} R_{\ua}^{-1}(\la) \, L_{22}(\la)   &  R_{\ua}^{-1}(\la)\,  \Big[ (\la+\mf{q}_1) \, L_{22}(\la) - \mf{q}_2 L_{21}(\la) \Big]  \vspace{1mm}  \\  
			       -\mf{q}_2^{-1} R_{\ua}(\la)\,  L_{12}(\la)   & - R_{\ua}(\la)\,  \Big[ (\la+\mf{q}_1) L_{12}(\la) - \mf{q}_2 L_{11}(\la) \Big]    \ea \right) 
\enq
where I have set $\mf{q}_2=\chi^{(1)}_{12}$ and  
\bem
L(\la) \; = \; \left( \ba{cc} \Pi_{11}(\la) (\mc{P}_R(\la))_{11} + \Pi_{12}(\la) (\mc{P}_R(\la))_{21}   &    \Pi_{11}(\la) (\mc{P}_R(\la))_{12} + \Pi_{12}(\la) (\mc{P}_R(\la))_{22}  \vspace{1mm} \\
		\Pi_{21}(\la) (\mc{P}_R(\la))_{11} + \Pi_{22}(\la) (\mc{P}_R(\la))_{21}   &    \Pi_{21}(\la) (\mc{P}_R(\la))_{12} + \Pi_{22}(\la) (\mc{P}_R(\la))_{22}       \ea \right) \\ 
\, = \, I_2 \, + \; \sul{\ell=0}{2}c_{\ell}\la^{\ell-3} \sg^+ \; + \; \e{O} \Bigg(  \f{  (\, \ov{x}_N)^4 \ex{- \ov{x}_N (1-\tilde{\a})}  }{  1+|\la | } \Bigg)
\end{multline}
with a control that is uniform throughout the domain located above of $\Ga_{\ua}$. Therefore, by taking the matrix products explicitly, one gets that in this domain 
\beq
\chi_{11}(\la) \; = \; \f{1}{c_{2} R_{\ua}(\la)} \Big\{ 1+ \e{O}(\de_N) \Big\} \quad , \quad 
 \chi_{12}(\la) \; = \;  \f{1}{  R_{\ua}(\la) } \Big\{  \big(\la-\tfrac{ c_{1} }{ c_{2} } \big) + \e{O}(\de_N) \Big\} \;,  
\enq
where I agree upon $\de_{N}=(\, \ov{x}_N)^4 \ex{- \ov{x}_N (1-\tilde{\a})}$. Likewise, 
\beq
 \chi_{21}(\la) \; = \;  - \f{ R_{\ua}(\la) }{ c_{2} } \bigg\{  \sul{\ell=0}{2}  \f{ c_{\ell} \la^{\ell} }{ \la^{3} }  + \e{O}\bigg( \f{ (\, \ov{x}_N)^2 \de_N }{ 1+|\la| } \bigg) \bigg\}   \; , \; \; 
  \chi_{22}(\la) \; = \;    -  R_{\ua}(\la) \cdot \bigg\{ -c_2 \, + \,  \f{ \la-\tfrac{ c_{1} }{ c_{2} } }{ \la^{3} } \sul{\ell=0}{2} c_{\ell} \la^{\ell}  + \e{O}\Big( (\, \ov{x}_N)^2 \de_N \Big)  \bigg\} \; .
\nonumber 
\enq
The control on the remainder in the above formulae is not precise enough. From the general setting \eqref{ecriture DA chi dans RHP} and \eqref{ecriture forme matrice M ab}-\eqref{ecriture DA chi infty amenage}, 
 one knows that $\chi_{2a}(\la) = C |\la|^{-\f{1}{2}}(1+\e{o}(1))$
for some constant $C$ and when $\la \tend \infty$ this means that the remainder must decay as $\la^{-2}$ at $\infty$. This decay has its amplitude still controlled by $(\, \ov{x}_N)^2 \de_{N}$, 
as follows from the very construction of the terms that contribute to the remainder. Thus, one may improve the control on both remainders appearing in the last equation by $\e{O}\bigg( \f{ (\, \ov{x}_N)^2 \de_N }{ (1+|\la|)^2 } \bigg)$. 
This structure may of course also be checked directly by considering the explicit expression for the remainders and extracting from there the precise bounds. 
Then, upon using that 
\beq
-c_2 \, + \, \big(\la-\tfrac{ c_{1} }{ c_{2} } \big) \sul{\ell=0}{2} c_{\ell} \la^{\ell-3}  \; = \; - \la^{-3} \cdot \big(\la+\tfrac{ c_{1} }{ c_{2} } \big) 
\enq
one gets  
\beq
 \chi_{21}(\la) \; = \;  - \f{ R_{\ua}(\la) }{ c_{2} } \cdot \bigg\{  \sul{\ell=0}{2} \f{ c_{\ell} \la^{\ell} }{ \la^{3} }  + \e{O}\bigg( \f{ (\, \ov{x}_N)^2 \de_N }{ (1+|\la|)^2  } \bigg) \bigg\}   \quad, \quad 
  \chi_{22}(\la) \; = \;       R_{\ua}(\la) \cdot \bigg\{ \f{ \la+\tfrac{ c_{1} }{ c_{2} } }{ \la^{3} }  +  \e{O}\bigg( \f{ (\, \ov{x}_N)^2 \de_N }{ (1+|\la|)^2  } \bigg) \bigg\} \;. 
\enq
Thus, in each of the matrix entries, one may factor out the leading term since it does not vanish in the considered domain, what yields 
\beq
\chi(\la) \; = \; \left(\ba{cc}  \big[ c_{2} R_{\ua}(\la) \big]^{-1} \cdot \Big( 1+ \e{O}(\de_N) \Big)  &   \big(\la-\tfrac{ c_{1} }{ c_{2} } \big) \cdot  \big[ R_{\ua}(\la) \big]^{-1} \cdot \Big( 1+ \e{O}(\de_N) \Big)   \\ 
 - c_{2}^{-1} R_{\ua}(\la) \cdot \Big\{ \sul{\ell=0}{2} c_{\ell} \la^{\ell-3} \Big\} \cdot \Big( 1+ \e{O}(\de^{\prime}_N) \Big) &      \f{ R_{\ua}(\la) }{\la^{3} } \cdot \big(\la+\tfrac{ c_{1} }{ c_{2} } \big) \cdot \Big( 1+ \e{O}(\de^{\prime}_N) \Big)   \ea \right)\;.  
\label{forme DA de chi au dessus de gamma up} 
\enq
Here, the control on the remainder is uniform on throughout the domain. Also, I agree upon $\de^{\prime}_N = (\, \ov{x}_N)^2 \de_{N}$.  

By carrying out a similar analysis, one gets that for $\la$ located in between $\R$ and the curve $\Ga_{\da}$
\beq
\chi(\la) \; = \; \left(\ba{cc}    \big[  c_{2} R_{\da}(\la) \big]^{-1} \cdot \Big\{ u_{\e{reg}}(\la)+ \e{O}(\de^{\prime}_N) \Big\}  & 
\big[R_{\da}(\la) \big]^{-1}   \Big[ c_2 \, + \, \big(\la-\tfrac{ c_{1} }{ c_{2} } \big) u_{\e{reg}}(\la)  + \e{O}(\de^{\prime}_N)   \Big]    \\ 
   c_{2}^{-1} R_{\da}(\la) \Big( 1+ \e{O}(\de_N) \Big)  &     R_{\da}(\la)\big(\la-\tfrac{ c_{1} }{ c_{2} } \big)   \Big( 1+ \e{O}(\de_N) \Big)  \ea \right)\;.  
\label{forme DA de chi entre R et gamma down} 
\enq
The function $u_{\e{reg}}$ appearing above is defined as 
\beq
u_{\e{reg}}(\la)\, = \; \f{ \ups^2(\la) }{ R(\la)  } \ex{-\i\la \ov{x}_N } \, - \, \sul{\ell=0}{2} c_{\ell} \la^{\ell-3} \; = \; 
\f{ R_{\da}(\la) }{ R_{\ua}(\la)  } \ex{-\i\la \ov{x}_N } \, - \, \sul{\ell=0}{2} c_{\ell} \la^{\ell-3}  \;. 
\enq

I stress that in each of the expressions appearing above, the remainders are differentiable, with the caveat that each derivative produces a loss of $\ov{x}_N$ in the  precision
on the control.



\subsection{The constraint integral} 
\label{Soussection Integrale de contraintes}

\begin{prop}
\label{Proposition evaluation integrale de contrainte}
Given  $H_N$ as in \eqref{definition fct HN}, it holds 
\beq
\msc{J}_{12}[H_N] \, = \, 
\Int{\R+\i \eps^{\prime} }{} \chi_{12}(\mu) \, \mc{F}[H_N](\tau_N\mu)\, \ex{- \i  \mu \ov{b}_N}\cdot\f{\dd\mu}{ (2 \i \pi)^2 } 
\, = \,  -\f{    \varkappa  \chi_{12}(\i) }{ 4 \pi N  \tau_N } \cdot  \Big( \ex{\ov{b}_N } \, - \, \ex{-\ov{a}_N } \Big) \;. 
\enq

Moreover, one has that $ \chi_{12}(\i)\not=0$ for $N$ large enough, provided that $x_N$ is uniformly away from $0$.  
\end{prop}

The above result allows one to deduce that the endpoints of the support of the equilibrium measure satisfy $b_N =-a_N$ as expected from the evenness of the confining potential and the one of the two-body interaction.

\Proof  

A direct calculation yields 
\beq
\Int{a_N}{b_N} \ex{ \i\tau_N \mu (\eta-b_N) } H_N(\eta) \dd \eta \; = \; \f{ -  \i  \varkappa }{ 2 N  \tau_N } \Bigg\{ \bigg[ \f{ \ex{\ov{b}_N }}{\mu-\i } \, - \, \f{ \ex{-\ov{b}_N }}{\mu+\i }   \bigg]
\; - \; \ex{- \i \mu \ov{x}_N} \bigg[ \f{ \ex{\ov{a}_N }}{\mu-\i } \, - \, \f{ \ex{-\ov{a}_N }}{\mu+\i }   \bigg] \Bigg\} \;. 
\label{ecriture TF de H}
\enq
Thus, 
\beq
\msc{J}_{12}[H_N] \, = \,  -\f{    \varkappa    }{ 4 \pi N  \tau_N }
\Int{\R+\i \eps^{\prime} }{} \hspace{-2mm} \f{\dd\mu}{  2 \i \pi   }   \Bigg\{  \chi_{12}(\mu)  \sul{\sg= \pm }{} \f{ \sg \ex{ \sg \ov{b}_N }}{\mu- \sg \i } 
\; - \; \chi_{12}(\mu) \ex{- \i \mu \ov{x}_N} \sul{\sg= \pm }{} \f{ \sg \ex{ \sg \ov{a}_N }}{\mu- \sg \i }  \Bigg\} \;, 
\label{reecriture integral definition J12}
\enq
It follows from the asymptotics given in \eqref{ecriture forme asymptotiques chi} (see the conjunction of \eqref{ecriture DA chi dans RHP} and \eqref{ecriture forme matrice M ab}-\eqref{ecriture DA chi infty amenage}) that
\beq
\chi_{12}(\mu) \; = \; \f{ \chi^{(\infty)}_{\ua ;12}(\mu) }{ (- \i \mu)^{\frac{1}{2}} } \; , \quad \e{resp}. \quad \chi_{12}(\mu) \; = \; \f{ \chi^{(\infty)}_{\da ;12}(\mu) }{ (  \i \mu)^{\frac{1}{2}} } \; ,   
\enq
for $|\mu|$ large enough in $\mathbb{H}^{+}$, resp. $\mathbb{H}^{-}$,  and where $\chi^{(\infty)}_{\ua/\da;12}(\mu)$ is bounded at $\infty$. Furthermore, the jump conditions satisfied by $\chi_{12}$:
$\ex{- \i \mu \ov{x}_N} \chi_{12;+}(\mu) \; = \; \chi_{12;-}(\mu)$ allow one to recast the integral \eqref{reecriture integral definition J12} as 
\beq
\msc{J}_{12}[H_N] \, = \,  -\f{    \varkappa    }{ 4 \pi N  \tau_N }
\Int{\R+\i \eps^{\prime} }{} \hspace{-2mm} \f{\dd\mu}{  2 \i \pi   }  \,   \chi_{12}(\mu) \sul{\sg= \pm }{} \f{ \sg \ex{ \sg \ov{b}_N }}{\mu- \sg \i }
\; + \; \f{    \varkappa    }{ 4 \pi N  \tau_N }
\Int{\R-\i \eps^{\prime} }{} \hspace{-2mm}  \f{\dd\mu}{  2 \i \pi   }  \, \chi_{12}(\mu) \sul{\sg= \pm }{} \f{ \sg \ex{ \sg \ov{a}_N }}{\mu- \sg \i }   \;.
\enq
The resulting integrals can then be taken by the residues at $\mu=\i$, relatively to the first one, and $\mu=-\i$, relatively to the second one. One gets 
\beq
\msc{J}_{12}[H_N] \, = \,  -\f{    \varkappa    }{ 4 \pi N  \tau_N } \Big\{\chi_{12}(\i)   \ex{\ov{b}_N } \, - \,  \chi_{12}(-\i)   \ex{-\ov{a}_N }  \Big\}  \;. 
\enq
Finally, Lemma \ref{Lemme Ecriture diverses proprietes solution RHP chi} entails that $\chi_{12}(-\la)=\chi_{12}(\la)$.   
Moreover, it follows from the asymptotic expansion for $\chi$ in the region above $\Ga_{\ua}$, \textit{c.f.} \eqref{forme DA de chi au dessus de gamma up}, that 
\beq
\chi_{12}(\la) \; = \; \f{ \la-\tfrac{c_1}{c_2} }{  R_{\ua}(\la) }\bigg( 1  \;  + \; \e{O}\Big( (\, \ov{x}_N )^4  \ex{ -\ov{x}_N(1-\tilde{\a}) }   \Big) \bigg) \;. 
\enq
Since $ R_{\ua}(\i)\not=0$ and it holds that $\tf{c_1}{c_2}\, = \, 2\i \cdot  \big( \ov{x}_N\big)^{-1} \cdot \big(1+\e{O}(\ov{x}_N^{-1}) \big)$ as $N\tend +\infty$,
one has that $\chi_{12}(\i)$ is uniformly away from $0$, provided that $x_N$ is also uniformly away
from zero, \textit{viz}. that $\ov{x}_N \tend + \infty$. \qed

\subsection{The normalisation constraint}
\label{Soussection Integrale de normalisation}

\begin{prop}
\label{Proposition contrainte de normalisation} 
 
Let, for short, $ \ov{c}_N^{  (+)}= \ov{b}_N$ and $ \ov{c}_N^{ (-)}=- \ov{a}_N$. Then, it holds
\beq
\mc{W}_N[H_N](\xi) \; = \; \f{\varkappa \tau_N }{ 2 \i  \pi N  }  
\Int{ \R + 2 \i \eps^{\prime} }{} \f{ \dd \la }{ 2 \i \pi } 
 \sul{\sg= \pm }{}  \f{  \sg \, \ex{ \ov{c}_N^{(\sg)} } }{  \sg \i- \la }
\bigg\{    \f{ \sg \i }{ \la }  \cdot \chi_{11}(\la) \chi_{12}( \sg \i ) \, -  \, \chi_{11}(\sg \i ) \chi_{12}(\la) \bigg\} \cdot \ex{-\i\la \tau_N (\xi-a_N)}\;. 
\label{expression explicite pour densite mesure equilibre}
\enq
Moreover, one has 
\bem
 \Int{a_N}{b_N}  \mc{W}_N[H_N](\eta) \dd \eta  \; = \; \f{\varkappa \i  }{ 2 \pi  N} \bigg\{  \i  \chi_{12}(\i) \,  \chi_{11;-}^{\prime}(0) \cdot \big[ \ex{ \ov{b}_N } \, - \, \ex{ - \ov{a}_N }  \big]
 \, + \, \ex{ \ov{b}_N }  \Big[ \chi_{12}(\i) \chi_{11;-}(0) \, - \,  \chi_{12;-}(0) \chi_{11}(\i)  \Big]  \\ 
\, + \,   \ex{ - \ov{a}_N }  \Big[ \chi_{12}(-\i) \chi_{11;-}(0) \, - \,  \chi_{12;-}(0) \chi_{11}(-\i)  \Big] \bigg\} \;. 
\end{multline}
Furthermore, in the case when $a_N=-b_N$, this integral admits the large-$N$ expansion 
\beq
 \Int{a_N}{b_N}  \mc{W}_N[H_N](\eta) \dd \eta  \; = \;  \f{ \vartheta }{ N }  \cdot (\ov{b}_N)^2 \cdot \ex{ \ov{b}_N } \cdot \mf{t}(2 \ov{b}_N)\, \bigg( 1\, + \,  \e{O}\Big( (\,\ov{b}_N)^4 \ex{ - 2\ov{b}_N(1-\tilde{\a}) } \Big) \bigg)  \;,
\label{ecriture DA integrale de normalisation}
\enq
in which the constant $\vartheta$ is expressed as 
\beq
\vartheta \, = \, \f{ 2 \, \varkappa }{ 3  (2\pi)^{ \f{5}{2} } } \cdot \f{ \Ga\big( \mf{b}, \hat{\mf{b}} \big) }{ \mf{b}^{\mf{b}} \, \hat{\mf{b}}^{\hat{\mf{b}}} }\; , 
\label{definition constante vartheta}
\enq
while, upon using the constants $w_k$ introduced in \eqref{definition des constantes wk et de leur asymptotiques},  
\beq
\mf{t}( \ov{x}_N) \, = \,  \f{ 6 }{ (\ov{x}_N)^2 } \bigg\{  2  \, + \, w_2  \, - \, w_1 \, -\, \f{w_1 w_3}{ w_2 }  \bigg\} \; \widesim{  \ov{x}_N \tend + \infty } \; 1 \, + \, \e{O}\Big( \tfrac{ 1 }{ \ov{x}_N } \Big) \;. 
\label{definition fonction frak t}
\enq
Finally, the remainder in \eqref{ecriture DA integrale de normalisation} is smooth and differentiable in respect to the endpoint of the support $b_N$. 

\end{prop}

We stress that the integral \eqref{expression explicite pour densite mesure equilibre} which gives $\mc{W}_N[H_N](\xi)$ cannot be taken by residues and
further simplified for $\xi \in \intoo{a_N}{b_N}$ due to the "wrong" exponential growth of the integrand in the $\Im(\la) \tend \pm \infty$ directions.

\Proof 

By inserting the expression \eqref{ecriture TF de H} for the Fourier transform of $H_N$ into the one for $\mc{W}_N[H_N]$ \eqref{forumule explicite pour WN}, one gets that 
\beq
\mc{W}_N[H_N](\xi) \; = \; \f{ -\i \varkappa \tau_N }{ 2 \pi N  }  
\! \Int{ \R + 2 \i \eps^{\prime} }{}  \hspace{-2mm}\f{ \dd \la }{ 2 \i \pi } \, \ex{-\i\la \tau_N (\xi-a_N) }
\Int{ \R + \i \eps^{\prime} }{} \hspace{-2mm} \f{ \dd \mu }{ 2 \i \pi }\,
 W(\la,\mu)
 \sul{\sg= \pm }{}  \sg   \bigg\{  \f{\ex{ \sg \ov{b}_N }}{\mu- \sg \i } \, - \,  \ex{-\i\mu \ov{x}_N }  \f{\ex{ \sg \ov{a}_N }}{\mu- \sg \i }  \bigg\}
\enq
in which $W$ has been introduced in \eqref{definition noyau W}. The $\mu$ integrals can now be taken by the residues. For that one splits the  integrand in two pieces, depending 
on whether the integrand contains the explicit term $ \ex{-\i\mu \ov{x}_N } $ or not. In the integral not involving $ \ex{-\i\mu \ov{x}_N } $, one has that
$\mu \mapsto W(\la,\mu)$ is analytic in $\mathbb{H}^{+}$ and decays, for fixed $\la$, as $\e{O}(|\mu|^{-\f{1}{2}})$.
As a consequence, the associated integrand decays, as a whole, as   $\e{O}(|\mu|^{-\f{3}{2}})$.
It is meromorphic on  $\mathbb{H}^{+}$ and has a single pole there, located at $\mu=\i$. This pole is simple. 

Now focusing on the integral involving $ \ex{-\i\mu \ov{x}_N }$, one observes that owing to the jump condition $ \ex{-\i\mu \ov{x}_N }  \chi_{1a;+}(\mu) \, = \, \chi_{1a;-}(\mu)$,
and the point-wise in $\la$ decay of the integrand in the lower-half plane controlled as  $\e{O}(|\mu|^{-\f{3}{2}})$, 
one may evaluate the corresponding integral by taking the residues in the lower half, the sole pole being located at $\mu=-\i$. This computation then yields \eqref{expression explicite pour densite mesure equilibre}.

Taking explicitly the $\la$ integral by using the representation \eqref{expression explicite pour densite mesure equilibre} leads to 
\beq
 \Int{a_N}{b_N}  \mc{W}_N[H_N](\xi) \dd \xi  \; = \; \f{\varkappa \tau_N }{ 2 \i  \pi N  }   \sul{\sg= \pm }{}  
\Int{ \R + 2 \i \eps^{\prime} }{} \f{ \dd \la }{ 2 \i \pi } \f{ 1 - \ex{-\i \la \ov{x}_N }  }{ \i \tau_N \la }
\f{  \sg \, \ex{ \ov{c}_N^{(\sg)} } }{  \sg \i- \la }
\bigg\{    \f{ \sg \i }{ \la }  \cdot \chi_{11}(\la) \chi_{12}( \sg \i ) \, -  \, \chi_{11}(\sg \i ) \chi_{12}(\la) \bigg\}  \;. 
\enq
The $\la$ integral may be taken, analogously to the previous case, by splitting the integral in $2$ and taking the residues in $\mathbb{H}^{+} + 2\i\eps^{\prime}$ or  $\mathbb{H}^{-} + 2 \i \eps^{\prime}$, depending whether $\ex{-\i \la \ov{x}_N }$
is present or not in the integrand. A direct inspection shows that the only pole present is  at $\la=0$. One gets 
\bem
 \Int{a_N}{b_N}  \mc{W}_N[H_N](\xi) \dd \xi  \; = \; \f{ - \varkappa }{ 2  \pi N  } 
\Bigg\{ \ex{\ov{b}_N} \Big[ \chi_{12}(\i) \chi_{11;-}^{\prime}(0) \, - \, \i  \chi_{12}(\i) \chi_{11;-}(0) \, + \, \i  \chi_{12;-}(0) \chi_{11}(\i) \Big] \\
\, - \,  \ex{-\ov{a}_N} \Big[ \chi_{12}(-\i) \chi_{11;-}^{\prime}(0) \, + \, \i  \chi_{12}(-\i) \chi_{11;-}(0) \, - \, \i  \chi_{12;-}(0) \chi_{11}(-\i) \Big] \Bigg\} \;. 
\end{multline}
At this stage it remains to invoke the relations $\chi_{12}(\i)=\chi_{12}(-\i)$ and $\chi_{11}(-\i)=\chi_{11}(\i)+\i \chi_{12}(\i)$ following from Lemma \ref{Lemme Ecriture diverses proprietes solution RHP chi} so as to 
conclude that, for $a_N=-b_N$,  it holds 
\beq
 \Int{a_N}{b_N}  \mc{W}_N[H_N](\eta) \dd \eta  \; = \; \f{\i \varkappa  \ex{ \ov{b}_N } }{ 2 \pi  N} \bigg\{  
2 \Big[ \chi_{12}(\i) \chi_{11;-}(0) \, - \,  \chi_{12;-}(0) \chi_{11}(\i)  \Big]  \, - \, \i  \chi_{12;-}(0) \chi_{12}(\i)  \Big] \bigg\} \;. 
\enq
Then to get the leading asymptotics, it remains to use the explicit form of the asymptotic expansion of $\chi(\la)$
for $\la$ between $\R$ and $\Ga_{\da}$ given in \eqref{forme DA de chi entre R et gamma down}:
\beq
\chi_{11}(\la) \; = \;  \big[  c_{2} R_{\da}(\la) \big]^{-1} u_{\e{reg}}(\la) \; + \; \e{O}\Big(  (\, \ov{x}_N )^4  \ex{ -\ov{x}_N(1-\tilde{\a}) } \Big) 
\enq
and
\beq
\chi_{12}(\la) \; = \;  \big[R_{\da}(\la) \big]^{-1}   \Big[ c_2 \, + \, \big(\la-\tfrac{ c_{1} }{ c_{2} } \big) u_{\e{reg}}(\la)   \Big]  \; + \; \e{O}\Big(  (\, \ov{x}_N )^6  \ex{ -\ov{x}_N(1-\tilde{\a}) } \Big)\;, 
\enq
and the one for $\la$ above of $\Ga_{\ua}$ given in \eqref{forme DA de chi au dessus de gamma up}:
\beq
\chi_{11}(\i) \; = \;  \big[  c_{2} R_{\ua}(\i) \big]^{-1}  \; + \; \e{O}\Big(  (\, \ov{x}_N )^2  \ex{ -\ov{x}_N(1-\tilde{\a}) } \Big) \qquad \e{and} \qquad
\chi_{12}(\i) \; = \;  \big[R_{\ua}(\i) \big]^{-1}  \big(\i-\tfrac{ c_{1} }{ c_{2} } \big)   \; + \; \e{O}\Big(  (\, \ov{x}_N )^4  \ex{ -\ov{x}_N(1-\tilde{\a}) } \Big)\;. 
\enq
One gets
\bem
 \Int{a_N}{b_N}  \mc{W}_N[H_N](\eta) \dd \eta  \; = \; \f{\i \varkappa  \ex{ \ov{b}_N } }{ 2 \pi  N R_{\da}(0) R_{\ua}(\i) } \bigg\{ 
2 \bigg[ \f{c_3}{c_2}\Big(\i -\f{c_1}{c_2} \Big) \, - \, \f{1}{c_2} \Big( c_2 -\f{c_1 c_3}{c_2} \Big) \Big] \, - \i \, \Big( c_2 -\f{c_1 c_3}{c_2} \Big)\Big(\i -\f{c_1}{c_2} \Big) \\
\, + \, \e{O}\Big( (\, \ov{x}_N )^6 \ex{-2\ov{b}_N(1-\tilde{\a} )}\Big) \bigg\}
\end{multline}
where I used that $u_{\e{reg}}(0)=c_3$ as follows from the very definition of the constant $c_3$, \textit{c.f.} \eqref{definition coefficients c ell}.  
The expansion \eqref{ecriture DA integrale de normalisation} then follows upon direct algebra and after using the rewriting of the coefficients $c_k$ in terms of the $w_k$
as introduced in \eqref{definition des constantes wk et de leur asymptotiques}. \qed

\subsection{Positivity constraints}
\label{Soussection contraintes de positivite}

For the purpose of this section, it is convenient to introduce the auxiliary functions:
\beq
W_{a}(\la) \; = \;  \f{2\i }{ 1 + \la^2 } \bigg\{ \chi_{11}( \i ) \chi_{a2}(\la) \, - \,  \chi_{12}(  \i )  \chi_{a1}(\la)  \bigg\} 
\; - \; \f{   \i  }{ \i +  \la }\chi_{12}(  \i ) \chi_{a2}(\la) \;, 
\label{definition des fcts Wa}
\enq
and agree upon 
\beq
u_N \, = \, \f{ \varkappa \tau_N \ex{ \ov{b}_N }  }{  2\i \pi N } \;.
\enq
Also, it will be of use to introduce the function 
\beq
\varrho_{\e{bd}}\big(x \big) \, = \,   -\pi \Int{\R+\i\eps^{\prime} }{}  \hspace{-2mm} \f{ \dd \la }{2\i\pi} \f{\mf{r}(-\i\la)}{R(\la) } \ex{\i\la x}
\label{definition varrho bd}
\enq
where $\mf{r}$ is defined by the formula
\beq
\qquad \mf{r}(\a) \, = \,   \f{ 3 \pi \mf{b}  \hat{\mf{b}} \,  \a  }{ 2 \, (\a-1) }
\cdot \f{  \i R_{\ua}( \i\a)  }{ \a^3 \sqrt{\pi \mf{b} \hat{\mf{b}} }   }  
\; = \; \f{ 3 \pi \mf{b}  \hat{\mf{b}} \,  \a  }{ 2 \, (\a-1) } \,  \mf{b}^{\a \mf{b} }  \,  \hat{\mf{b}}^{\a \hat{\mf{b}} } \, 2^{\frac{1}{2}\a} \cdot 
\Ga \left( \ba{c} \tfrac{1+\a}{2} \, , \, \tfrac{1+\a}{2} \\ 1 + \mf{b} \a \, , \,  1 + \hat{\mf{b}} \a    \, , \, 1 + \tfrac{\a}{2}  \ea \right) \;.  
\label{definition frak r}
\enq
Note that the integral defining $\varrho_{\e{bd}}(x)$ converges uniformly on $\R^+$ since, uniformly in $ x \in \R^+$, the integrand decays as a $\e{O}\big( |\la|^{-\frac{3}{2}} \big)$.

\begin{prop}
\label{Proposition propriete densite mesure eq dans son support}
 
 Let $b_N+a_N=0$. Then, for $\xi \in \intoo{a_N}{b_N}$ it holds 
\beq
\mc{W}_N[H_N](\xi) \; = \; \varrho_{\e{bd}}^{(N)}\big(\tau_N (b_N-\xi) \big) \, - \, \varrho_{\e{bd}}^{(N)}\big( 0  \big)  \, + \, \varrho_{\e{bd}}^{(N)}\big(\tau_N (\xi-a_N) \big)
\, - \, \varrho_{\e{bd}}^{(N)}\big(  \ov{x}_N  \big) \, + \, \varrho_{\e{bk}}^{(N)}(\xi)
\label{ecriture density local comme somme densite bulk et bord}
\enq
 where, for $x>0$,  
\beq
\varrho_{\e{bd}}^{(N)}\big(x \big)   =   u_N \Int{ \R + \i \eps^{\prime} }{} \f{ \dd \la }{2\i \pi } \f{ \ex{\i \la x} }{ R(\la) } W_{2}(\la)  \label{ecriture rep int densite bord}
\enq
and is real valued on $\R$ while
\beq
\varrho_{\e{bk}}^{(N)}(\xi) \; = \;  \f{3}{4} \cdot \mc{V}_N \cdot \big( \xi - a_N \big)\big( b_N \, - \, \xi \big) \qquad with \qquad 
\mc{V}_N\, = \, - \f{ 2 u_N \tau_N^2 }{ 3 \pi^3 \mf{b} \hat{\mf{b}} } W_{2;-}(0) \in \R \;. 
\label{expression densite bulk}
\enq
Moreover, one has the large-$N$ asymptotics
\beq
 \mc{V}_N \, = \, \vartheta \cdot \f{ \ov{b}_N^2 \ex{\ov{b}_N }  \mf{t}(2\ov{b}_N) }{ N } \cdot 
\f{ \wt{w}_1 }{ \wt{w}_2 b_N^3 \mf{t}(2\ov{b}_N) } \cdot \Big( 1+ \e{O}\big( \de_N \big)  \Big) 
\label{ecriture DA facteur VN}
\enq
in which $\de_N= (\, \ov{x}_N)^4 \ex{-\ov{x}_N (1-\tilde{\a}) }$, $\vartheta$ and $\mf{t}(2\ov{b}_N)$ are as introduced in \eqref{definition vartheta}, \eqref{definition polynome frak t}.

Finally, when $b_N$ satisfies the constrains given in \eqref{ecriture solution pour bornes mesure equilibre}, one has that for $N$ large enough 
\beq
\mc{W}[H_N](\xi) >0 \qquad for \qquad \xi \in \intoo{a_N}{b_N}\;. 
\label{ecriture equation positivite transfo WN de HN sur aN bN}
\enq

\end{prop}

\Proof 

In order to obtain the representation \eqref{ecriture density local comme somme densite bulk et bord}, one first observes that
\beq
\la \mapsto \sul{\sg= \pm }{}  \f{  \sg \, \ex{ \ov{c}_N^{(\sg)} } }{  \sg \i- \la }
\bigg\{    \f{ \sg \i }{ \la }  \cdot \chi_{11}(\la) \chi_{12}( \sg \i ) \, -  \, \chi_{11}(\sg \i ) \chi_{12}(\la) \bigg\} \cdot \ex{-\i\la \tau_N (\xi-a_N)}
\enq
admits a holomorphic extension to an tubular neighbourhood of $\R$, has no singularity at $\la=0$ and is a $\e{O}( |\la|^{-\f{3}{2} }$ as $\Re(\la) \tend \pm \infty$,
so that the integral defining $\mc{W}_N[H_N]$ in  \eqref{expression explicite pour densite mesure equilibre} is absolutely convergent, uniformly in $\xi \in \intff{a_N}{b_N}$.
One may thus readily squeeze the integration contour in \eqref{expression explicite pour densite mesure equilibre} to $\R-\i\eps^{\prime}$, for some $\eps^{\prime}>0$ as it is clear that the pieces of the
deformation contour at $\infty$ do not contribute in the process. One gets that
\beq
\mc{W}_N[H_N](\xi) \; = \; \f{\varkappa \tau_N }{ 2 \i  \pi N  }
\Int{ \R -\i\eps^{\prime}  }{} \f{ \dd \la }{ 2 \i \pi }
 \sul{\sg= \pm }{}  \f{  \sg \, \ex{ \ov{c}_N^{(\sg)} } }{  \sg \i- \la }
\bigg\{    \f{ \sg \i }{ \la }  \cdot \chi_{11,+}(\la) \chi_{12}( \sg \i ) \, -  \, \chi_{11}(\sg \i ) \chi_{12,+}(\la) \bigg\} \cdot \ex{-\i\la \tau_N (\xi-a_N)} \;.
\label{ecriture premiere expression pour WN de HN}
\enq
There, one should understand $ \chi_{1a,+}$ as the analytic continuation to $0\geq \Im(\la)\geq 2 \eps^{\prime}$ of the $+$ boundary values.
These analytically continued boundary values satisfy the relation
\beq
\chi_{1a;+}(\la) \, = \, \f{1}{R(\la)} \Big\{  \ex{\i\la \ov{x}_N }\chi_{2a;+}(\la) \, + \,  \chi_{2a;-}(\la) \Big\}
\label{ecriture reexpression chi 1a + via chi 2a} 
\enq
in which $\chi_{2a;-}(\la)\equiv \chi_{2a}(\la)$ for $\Im(\la)<0$. One then proceeds
to compute the sums over $\sg=\pm$ in \eqref{ecriture premiere expression pour WN de HN} by explicitly implementing the relation $a_N+b_N=0$
and using Lemma \ref{Lemme Ecriture diverses proprietes solution RHP chi} which ensures that
$\chi_{12}(-\la) \, = \, \chi_{12}(\la)$ and $\chi_{11}(-\la) \, = \, \chi_{11}(\la) \, + \, \la \chi_{12}(\la)$:
\beq
 \sul{\sg= \pm }{}  \f{  \i  }{  \la( \sg \i- \la ) } \chi_{12}( \sg \i )  \; = \;  -2 \i \,  \f{ \chi_{12}( \i ) }{\la^2 +1}
\quad \e{and} \quad
\sul{\sg= \pm }{}  \f{  \sg }{  \la( \sg \i- \la ) } \chi_{11}( \sg \i )  \; = \;  -2 \i \,  \f{ \chi_{11}( \i ) }{\la^2 +1} \; + \; \i \f{ \chi_{12}(\i) }{ \i + \la }  \;.
\enq
All of this recasts $\mc{W}_N[H_N]$ in the form
\beq
\mc{W}_N[H_N](\xi) \; = \; u_N
\Int{ \R -\i\eps^{\prime}  }{} \f{ \dd \la }{ 2 \i \pi }  \f{1}{R(\la) } \bigg\{ \ex{\i\la \tau_N (b_N-\xi)} W_{2,+}(\la)  \, + \,  \ex{-\i\la \tau_N (\xi-a_N)} W_2(\la) \bigg\} \;,
\enq
in which $W_2$ is as defined through \eqref{definition des fcts Wa} and $W_{2,+}(\la)$ is to be understood as the analytic continuation of the $+$ boundary value of $W_2$ on $\R$ up to
$\R-\i\eps^{\prime}$. Now, one splits the integral in two pieces (one depending on $(b_N-\xi)$ and the other on  $(\xi-a_N)$) and deforms the contours to
$\R+\i \eps^{\prime}$ in the piece involving the analytic continuation of $W_{2;+}$, what leads to
\beq
\mc{W}_N[H_N](\xi) \; = \; \varrho_{\e{bd}}^{(N)}\big(\tau_N (b_N-\xi) \big) 
\, + \, u_N \hspace{-1mm} \Int{ \R - \i \eps^{\prime} }{}  \hspace{-1mm} \f{ \dd \la }{2\i \pi } \f{ \ex{-\i \la \tau_N(\xi-a_N)} }{ R(\la) } W_{2}(\la)  \, + \,
u_N \cdot \e{Res}\bigg( \f{ \ex{\i \la \tau_N(b_N-\xi)} }{ R(\la) } W_{2;+}(\la) \, , \,  \la=0  \bigg) \;. 
\label{expression partielle pour densite WN de HN}
\enq
The second term produces $\varrho_{\e{bd}}^{(N)}\big(\tau_N (\xi-a_N) \big)$ upon implementing the change of variables $\la \hookrightarrow - \la$ 
and using that $W_{2}(-\la)=-W_{2}(\la)$ as can be seen from Lemma \ref{Lemme Ecriture diverses proprietes solution RHP chi}. 
To evaluate the last term, one may use the small $\la$-expansion 
\beq
\f{1}{R(\la) } \, = \,  \f{ 1 }{ \mf{b} \hat{\mf{b}} \pi^3 \la^3} \; + \; \f{1}{2\pi \mf{b} \hat{\mf{b}}  \la} \cdot \Big( \tfrac{5}{12} - \tfrac{ \mf{b}^2 +\hat{\mf{b}}^2 }{3}  \Big) \; + \; \e{O}(\la) \;, 
\enq
the relation \eqref{ecriture reexpression chi 1a + via chi 2a} and the fact that $\chi_{1a;+}$ does not have a pole at $\la=0$ so as to evaluate the residue in \eqref{expression partielle pour densite WN de HN} as 
\bem
u_N \e{Res}\bigg( \f{ \ex{\i \la \tau_N(b_N-\xi)} }{ R(\la) } W_{2;+}(\la) \, , \,  \la=0  \bigg)  \; = \; - u_N \e{Res}\bigg( \f{ \ex{-\i \la \tau_N(\xi-a_N)} }{ R(\la) } W_{2;-}(\la) \, , \,  \la=0  \bigg)  \\
\; = \; d_N^{(2)} (\xi-a_N)^2 \, + \, d_N^{(1)} (\xi-a_N) \, + \, d_N^{(0)}\;, 
\end{multline}
in which 
\beq
d_N^{(2)} \, = \, \f{  u_N  \tau_N^2  W_{2;-}(0) }{ 2\pi^3  \mf{b} \hat{\mf{b}} } 
\enq
and $d_{N}^{(1)}$, $d_N^{(0)}$ admit explicit expressions involving at most second order derivatives of  $W_{2;-}$ at $\la=0$. 
 This yields
\beq
\mc{W}_N[H_N](\xi) \; = \; \varrho_{\e{bd}}^{(N)}\big(\tau_N (b_N-\xi) \big) \, + \, \varrho_{\e{bd}}^{(N)}\big(\tau_N (\xi-a_N) \big) 
\, + \,  d_N^{(2)} (\xi-a_N)^2 \, + \, d_N^{(1)} (\xi-a_N) \, + \, d_N^{(0)} \;. 
\label{ecriture quasi explicite de WN avec termes bord et polynome bulk}
\enq
Observe that $ \varrho_{\e{bd}}^{(N)}$ is continuous on $\R$ owing to the $\e{O}(|\la|^{-\tf{3}{2}})$ decay of the integrand, uniformly in $x$.
Since all functions in \eqref{expression partielle pour densite WN de HN} are continuous on $\intff{a_N}{b_N}$ and since $\e{supp}\Big[ \mc{W}_N[H_N] \Big] \subset \intff{a_N}{b_N}$
by virtue of Theorem \ref{Theoreme invertibilite operateur S N gamma}, one has that $\mc{W}_N[H_N](a_N)=\mc{W}_N[H_N](b_N)=0$ by continuity. This yields that 
\beq
 \varrho_{\e{bd}}^{(N)}\big( 0 \big) \, + \, \varrho_{\e{bd}}^{(N)}\big(\tau_N (b_N-a_N) \big)  \, + \, d_N^{(0)} \; = \; 0 \quad \e{and} \quad 
  d_N^{(1)} \, = \, - d_N^{(2)} (b_N-a_N) \;. 
\enq
Inserting the latter in \eqref{ecriture quasi explicite de WN avec termes bord et polynome bulk} yields \eqref{ecriture density local comme somme densite bulk et bord}. 

\vspace{2mm}

To establish that $\mc{V}_N \in \R$ and $\varrho_{\e{bd}}^{(N)}(x) \in \R $ for $x\geq 0$, one invokes the second relation in \eqref{ecriture propriete conjugaison et reflection solution RHP chi}
given in Lemma \ref{Lemme Ecriture diverses proprietes solution RHP chi}
what ensures that $\big( W_2(\la) \big)^{*}\, = \, -  W_2(-\la^*)$. From there, one infers that
\bem
\Big( \varrho_{\e{bd}}^{(N)}\big(x \big) \Big)^* \, =\,    u_N^* \Int{ \R  }{} \f{ \dd \la }{ - 2\i \pi } \f{ \ex{-\i (\la - \i \eps^{\prime} ) x} }{ R(\la - \i \eps^{\prime} ) }
\times -W_{2}\big(-(\la- \i \eps^{\prime})\big) \\
\, = \,  u_N  \Int{ \R  }{} \f{ \dd \la }{   2\i \pi } \f{ \ex{\i (\la + \i \eps^{\prime} ) x} }{ - R(\la + \i \eps^{\prime} ) }
\times -W_{2}\big( \la+ \i \eps^{\prime} \big) \; = \; \varrho_{\e{bd}}^{(N)}\big(x \big)
\end{multline}
Similarly,
\beq
\mc{V}_N^* \; =  \; - \bigg( \f{ 2 u_N \tau_N^2 }{ 3 \pi^3 \mf{b} \hat{\mf{b}} }  W_2(-\i\eps )  \bigg)_{\mid \eps=0^+}^{*}
\; = \; \f{ 2 u_N \tau_N^2 }{ 3 \pi^3 \mf{b} \hat{\mf{b}} } \times  -W_2(-\i\eps )_{\mid \eps=0^+} \; = \; \mc{V}_N \; .
\enq

\vspace{2mm}

Now to get the large-$N$ behaviour stated in \eqref{ecriture DA facteur VN} one uses the large-$N$ expansion of $\chi$ between $\R$ and $\Ga_{\da}$
given in \eqref{forme DA de chi entre R et gamma down}
so as to conclude that 
\beq
W_{2}(\la) \, = \, \f{ \i R_{\da}(\la) }{ c_2 R_{\ua}(\i) (\la+\i)  } \cdot \Big(  \i c_1 \, - \, \la c_2 \Big[ \i  - \tfrac{ c_1 }{ c_2 } \Big]  \Big) \cdot \Big( 1 \, + \, \e{O}( \de_N )  \Big)  \;.
\enq
This expansion of $W_2$ is uniform throughout the mentioned domain.
A direct calculation then yields the claimed form of the large-$N$  behaviour of $\mc{V}_N$. Further, if $\la$ is located between $\R$ and $\Ga_{\ua}$, then $-\la$ is located between $\R$
and $\Ga_{\da}$, so that upon using $W_{2}(-\la)=-W_2(\la)$, one infers from the above that for $\la$ located between $\R$ and $\Ga_{\ua}$, it holds
\beq
W_{2}(\la) \, = \, \f{ \i R_{\ua}(\la) }{ c_2 \la^3 R_{\ua}(\i) (\la - \i)  }
\cdot \Big(  \i c_1 \, + \, \la c_2 \Big[ \i  - \tfrac{ c_1 }{ c_2 } \Big]  \Big) \cdot \Big( 1 \, + \, \e{O}( \de_N )  \Big) \;,
\enq
uniformly throughout the mentioned domain. From there, one infers the decomposition
\beq
W_{2}(\la) \, = \, W_{2}^{(\infty)}(\la) \, + \, \de W_{2}(\la) \, ,
\enq
where
\beq
W_{2}^{(\infty)}(\la) \; = \; - \f{   R_{\ua}(\la) }{ \la^2 R_{\ua}(\i) (\la - \i)  }   \qquad \e{and} \qquad
\de W_{2}(\la) \; = \;  \f{  - R_{\ua}(\la) }{ \la^2 R_{\ua}(\i) (\la - \i)  } \e{O}( \de_N ) \, -  \,
\f{ \i c_1 R_{\ua}(\la)  }{ c_2 \la^3 R_{\ua}(\i)   } \cdot \Big( 1 \, + \, \e{O}( \de_N )  \Big) \;.
\enq
Observe that
\beq
\f{ R_{\ua}(\la) }{ (\la-\i) \la^2 } \; = \; \f{ 2 \mf{r}(-\i\la)  }{ 3 \sqrt{\pi \mf{b} \hat{\mf{b}} } } \;.
\enq
Further, by using the constraint \eqref{ecriture solution pour bornes mesure equilibre}, it follows that\symbolfootnote[2]{The loss of control on the remainder stems from the form of the asymptotic expansion for $b_N$.}
\beq
\f{ -2 u_N }{ 3 \sqrt{\pi \mf{b} \hat{\mf{b}} } R_{\ua}(\i)  }
\; = \; - \f{ \pi }{ \tau_N } \big( 1+\de \mf{c} \big) \qquad \e{with} \qquad \de \mf{c} \; = \; \e{O}\Big( \f{ \ln \tau_N }{ \tau_N } \Big) \;.
\enq
Hence, one ends up with the decomposition
\beq
u_N \f{ W_2(\la) }{ R(\la) } \; = \; -  \f{\pi}{\tau_N}  \cdot \f{ \mf{r}(-\i\la) }{ R(\la) } \; + \; \f{ \de \msc{W}_{2}(\la) }{ \tau_N }
\enq
where the remainder term takes the form
\beq
\de \msc{W}_{2}(\la) \; = \; - \pi  \cdot \f{ \mf{r}(-\i\la) }{ R(\la) } \de \mf{c} \, - \, \pi \f{ 1 + \de \mf{c} }{ R(\la) }\cdot  \mf{r}(-\i\la)
\cdot \Bigg\{     \e{O}( \de_N ) \, +  \,
\f{ \i c_1 (\la-\i)   }{ c_2 \la  R_{\ua}(\i)   } \cdot \Big( 1 \, + \, \e{O}( \de_N )  \Big) \Bigg\}
\enq
and satisfies, uniformly away from $\la=0$ to the bounds
\beq
\big| \de \msc{W}_2(\la) \big| \; \leq \; C \f{ \ln \tau_N }{ \tau_N |\la|^{\tf{3}{2}}}  \;.
\label{ecriture borne sup sur delta W2}
\enq
The above  leads to the decomposition
\beq
\varrho_{\e{bd}}^{(N)}\big(x \big)   \; = \; \f{1}{\tau_N} \Big[ \varrho_{\e{bd}}\big(x \big)  \, + \,  \de\varrho_{\e{bd}}^{(N)}\big(x \big)   \Big]
\qquad \e{with} \qquad \de\varrho_{\e{bd}}^{(N)}\big(x \big)  \; = \;  \Int{ \R + \i \eps^{\prime} }{} \f{ \dd \la }{2\i \pi }   \ex{\i \la x} \de \msc{W}_{2}(\la)
\label{ecriture decomposition densite bord N dpdt parte dom et correction}
\enq
and where $\varrho_{\e{bd}}$ has been introduced in \eqref{definition varrho bd}. Note that $ \de\varrho_{\e{bd}}^{(N)}$ is well defined, continuous on $\R$
and satisfies $\de\varrho_{\e{bd}}^{(N)}(x) \; = \; \e{O}\Big( \tf{(\ln \tau_N)}{\tau_N} \Big)$ uniformly on $\R$.

\vspace{2mm}
We are now in position to establish the positivity property given in \eqref{ecriture equation positivite transfo WN de HN sur aN bN}.
To start with observe that $\varrho_{\e{bd}}^{(N)}$ is continuous
on $\R ^+$ due to the uniform in $x$ $\e{O}(|\la|^{-\tf{3}{2}})$ estimates on the decay of the integrand. Moreover, one has that
$ \varrho_{\e{bd}}^{(N)}\big( x  \big) \; = \; \e{O}\Big( \ex{-2 x} \Big)$ as $x\tend +\infty$ with a control that is uniformly in $N$.
This can be inferred by deforming the integration contour in \eqref{ecriture rep int densite bord} to $\R+2\i(1+\eps^{\prime})$, $\eps^{\prime}>0$ and small enough,
and, in doing so, picking the residue at $\la=2\i$, which is the pole of the integrand that is closest to the real axis in $\mathbb{H}^+$.

Next, it is easy to deduce from the uniform estimates given in \eqref{ecriture borne sup sur delta W2}, and from the decomposition
\eqref{ecriture decomposition densite bord N dpdt parte dom et correction} along with the explicit expression for $\varrho_{\e{bd}}$  \eqref{definition varrho bd},
that  one has the uniform in $N$ bound  $\varrho_{\e{bd}}^{(N)}< \tf{B}{\tau_N}$ for some $B>0$. 
Further, since it holds  
\beq
\varrho_{\e{bk}}^{(N)}(\xi) \, = \, \f{3}{4} (\xi-a_N)(b_N-\xi) \Big( 1 + \e{O}\big( \tau_N^{-1} \big) \Big)
\enq
there exists $C>0$ such that for any $\xi \in \intff{ a_N+C\tau_N^{-1} }{ b_N-C\tau_N^{-1} }$ one has $\varrho_{\e{bk}}^{(N)}(\xi) > 5 \tf{B}{\tau_N}$. 
The above bounds applied to \eqref{ecriture density local comme somme densite bulk et bord} thus ensures that
\beq
\mc{W}_N[H_N](\xi) \; >  \; \tf{B}{\tau_N} \qquad \e{throughout} \qquad \xi \in \intff{ a_N+C\tau_N^{-1} }{ b_N-C\tau_N^{-1} }\;. 
\enq
It remains to establish positivity on a sufficiently large on the $\tau_N^{-1}$ scale neighbourhood of $a_N$ and $b_N$. Furthermore, by symmetry, it is enough
to focus on the neighbourhood of $b_N$. In order to deal with this point, one first needs to discuss the local behaviour at $0$ of $\varrho_{\e{bd}}$,
$\varrho_{\e{bd}}^{(N)}$ and $\de\varrho_{\e{bd}}^{(N)}$.

The integrand arising in the expression \eqref{definition varrho bd} for $\varrho_{\e{bd}}$ behaves as
\beq
 -\pi      \f{\mf{r}(-\i\la)}{R(\la) } \;  \widesim{\Re(\la) \tend \pm \infty} \;  \f{ C_{\pm} }{ (\pm \la)^{\f{3}{2}} } \Big( 1+ \e{O}\big( \tfrac{1}{\la  } \big) \Big)\;.
\enq
A classical Fourier analysis then ensures that
\beq
\varrho_{\e{bd}}(x)-\varrho_{\e{bd}}(0) \, = \, C \sqrt{x} \,  \Big( 1 \, + \, \e{o}(1) \Big)
\enq
as $x\tend 0^+$ for some $C\not= 0$. A similar analysis based on the estimates \eqref{ecriture borne sup sur delta W2} allows one to infer that
\beq
\de \varrho_{\e{bd}}^{(N)}(x)-\de \varrho_{\e{bd}}^{(N)}(0) \; = \; \sqrt{x} \f{\ln \tau_N }{ \tau_N } \Big( C_N \, + \, \e{o}(1) \Big)
\enq
for some $C_N \in \R$. Further, observe that
owing to Lemma \ref{Lemme positivite densite asymptotiques bord et pot effectif asymptotique bord}
$\varrho_{\e{bd}}(x)-\varrho_{\e{bd}}(0) >0$ and vanishes only at $x=0$ on $\R^+$, \textit{c.f.} \eqref{ecriture positivite difference densite bord asymptotique}.
Since $\varrho_{\e{bd}}(x)\, = \, \e{O}\big( \ex{-2x} \big)$ as $x\tend +\infty$, by continuity of $\varrho_{\e{bd}}(x)-\varrho_{\e{bd}}(0)$ and the square root behaviour
at $x=0$, one gets the uniform in $x \geq 0$ estimate
\beq
  \f{  \de \varrho_{\e{bd}}^{(N)}(x) - \de \varrho_{\e{bd}}^{(N)}(0)  }{ \varrho_{\e{bd}}(x) - \varrho_{\e{bd}}(0) } \, = \, \e{O}\Big( \f{ \ln \tau_N }{ \tau_N }  \Big) \;. 
\enq
Finally, it is easy to infer from the asymptotic behaviour of  $\varrho_{\e{bd}}^{(N)}$ and the mean value theorem that
\beq
\varrho_{\e{bd}}^{(N)}\big(\tau_N (\xi-a_N) \big) \, - \, \varrho_{\e{bd}}^{(N)}\big(   \ov{x}_N  \big) \; = \; (\xi-b_N) \cdot \e{O}\Big( \tau_N^{-1}  \Big)\;.
\enq
 Hence, it holds 

\beq
\mc{W}_N[H_N](\xi) \; = \;  \f{1}{\tau_N} \Big\{ \varrho_{\e{bd}}\big(\tau_N (b_N-\xi) \big) \, - \, \varrho_{\e{bd}}\big( 0  \big) \Big\}
\cdot \bigg( 1 + \e{O}\Big( \f{ \ln \tau_N }{ \tau_N }  \Big)  \bigg) \; + \; 
\varrho_{\e{bk}}^{(N)}(\xi)\cdot \bigg( 1 + \e{O}\Big( \f{ 1 }{ \tau_N }  \Big)  \bigg) 
\enq
The second term is readily seen to be strictly positive on  $\intfo{ b_N-C\tau_N^{-1} }{b_N}$, while the positivity of the first one follows from  
eq. \eqref{ecriture positivite difference densite bord asymptotique} established in Lemma 
\ref{Lemme positivite densite asymptotiques bord et pot effectif asymptotique bord} leading to 
\beq
\mc{W}_N[H_N](\xi) \; >  \; 0 \qquad \e{throughout} \qquad \xi \in \intfo{ b_N-C\tau_N^{-1} }{b_N}\;. 
\enq
This entails the claim. \qed

\vspace{2mm}

One may obtain a similar characterisation of the effective potential subordinate to $\mc{W}_N\big[H_N\big]$
\beq
V_{N;\e{eff}} \Big[ \mc{W}_N\big[H_N\big] \Big] (\xi) \, = \, \f{1}{N} V_N(\xi) \, - \, \Int{a_N}{b_N} \hspace{-1mm} \dd \eta  \, \mf{w}^{(+)}_N(\xi-\eta) \cdot \mc{W}_N\big[H_N\big](\eta) \;,
\enq
on $\R \setminus \intff{a_N}{b_N}$.
In that characterisation, it is convenient to introduce the auxiliary function 
\beq
 \msc{J}_{\e{ext}}(x) \, = \,  - \f{\pi^2}{4} \Int{ \R + \i \eps^{\prime} }{} \hspace{-2mm} \f{\dd \la }{ 2\i\pi } \, \mf{l}(-\i \la) R(\la) \ex{\i\la x} \;.
\label{definition J ext}
\enq
Its definition involves
\beq
\mf{l}(\a) \; = \;  \f{ 6    \a  }{\a+ 1  }  \f{   \sqrt{\pi \mf{b} \hat{\mf{b}} }   } {  \i R_{\ua}( \i\a)  }  \; = \; 
\f{ 6  }{ \a^2 (\a + 1)  } \,  \mf{b}^{-\a \mf{b} }  \,  \hat{\mf{b}}^{-\a \hat{\mf{b}} } \, 2^{-\frac{1}{2}\a} \cdot 
\Ga \left( \ba{c} 1 + \mf{b} \a \, , \,  1 + \hat{\mf{b}} \a    \, , \, 1 + \tfrac{\a}{2} \\   \tfrac{1+\a}{2} \, , \, \tfrac{1+\a}{2} \ea \right) \; .
\label{definition fonction mf l}
\enq

 More precisely, one has the  

\begin{prop}
\label{Proposition propriete pot effectif}
Let $a_N+b_N=0$. 
One has, for $\xi < a_N$, 
\beq
V_{N;\e{eff}}^{\prime}\Big[ \mc{W}_N\big[H_N\big] \Big](\xi) \, = \, \tau_N \bigg\{  H_N(\xi) \, - \,  \msc{J}_{ext}^{(N)}\Big(\tau_N (a_N-\xi) \Big)  \bigg\} 
\enq
where, for $\eps^{\prime}>0$ and small enough,
\beq
\msc{J}_{ext}^{(N)}\big(x \big) =  \f{ \varkappa \ex{\ov{b}_N} }{2 N } \Int{ \R +\i\eps^{\prime} }{} \f{ \dd \la }{2\i \pi }   \ex{\i \la x}   R(\la)   W_{1;+}(\la)    \\
\label{ecriture DA en série pot effectif prime}
\enq
and is real valued on $\R^+$. 

\vspace{2mm}

Finally,   the potential satisfies to the symmetry
\beq
V_{N;\e{eff}}^{\prime}\Big[ \mc{W}_N\big[H_N\big] \Big](\xi) \, = \,- V_{N;\e{eff}}^{\prime}\Big[ \mc{W}_N\big[H_N\big] \Big](a_N+b_N-\xi )  \quad for \quad  \xi> b_N \; , 
\enq
and as soon as \eqref{ecriture solution pour bornes mesure equilibre} is fulfilled, one has 
\beq
V_{N;\e{eff}}\Big[ \mc{W}_N\big[H_N\big] \Big](\xi) \; > \; V_{N;\e{eff}}\Big[ \mc{W}_N\big[H_N\big] \Big](a_N) \qquad for \qquad \xi \in \R \setminus \intff{a_N}{b_N} \;. 
\label{ecriture positivite potentiel effectif}
\enq

\end{prop}

 \Proof 
It is direct to obtain that 
\beq
V_{N;\e{eff}}^{\prime}\Big[ \mc{W}_N\big[H_N\big] \Big](\xi) \, = \, \tau_N \bigg\{  H_N(\xi) \, - \,  \mc{S}_N\Big[ \mc{W}_N[H_N]\Big](\xi)  \bigg\}  \;. 
\enq
Moreover, upon using the $\ga \tend + \infty$ limit of \eqref{ecriture TF SN gamma de varphi}, 
one gets 
\beq
\mc{S}_N\Big[ \mc{W}_N[H_N]\Big](\xi) \, = \,  \f{ \varkappa \ex{\ov{b}_N} }{2 N } \Int{ \R  }{} \f{ \dd \la }{2\i \pi }   \ex{\i \la \tau_N (a_N-\xi) }   R(\la)   W_{1;+}(\la) \; = \; 
 \f{ \varkappa \ex{\ov{b}_N} }{2 N } \Int{ \R  }{} \f{ \dd \la }{2\i \pi }   \ex{-\i \la \tau_N (\xi-b_N) }    R(\la)   W_{1;-}(\la)
\enq
The integrand arising in the defintion of $\msc{J}_{ext}^{(N)}(x)$ behaves as $\e{O}\big( |\la|^{-\tf{3}{2}} \big)$ as $\Re(\la) \tend \pm \infty$, and uniformly in $x$.
This ensures that $\msc{J}_{ext}^{(N)}$ is continuous on $\R^+$.
Moreover, upon defoming the integration contour to $\R+\i (1+\eps^{\prime})$, $\eps^{\prime}>0$ and small enough,
and picking the residue of the second order pole at $\la=\i$,  one gets that  $\msc{J}_{ext}^{(N)}(x) \, = \, \e{O}( x \ex{-x})$, uniformly in $N$.
The fact that $\msc{J}_{ext}^{(N)}(x)\in \R$ follows from the second identity in
\eqref{ecriture propriete conjugaison et reflection solution RHP chi} which ensures that $\Big( W_1(\la) \Big)^{*}=  W_1(-\la^*)$. It then remains
to compute the complex conjugate of the integral representation \eqref{ecriture DA en série pot effectif prime}.

\vspace{2mm}

Further, by using that $W_{1}(\la)=W_1(-\la)$ as can be inferred from Lemma \ref{Lemme Ecriture diverses proprietes solution RHP chi}, a direct calculation yields
$\mc{S}_N\Big[ \mc{W}_N[H_N]\Big](a_N+b_N-\xi)=- \mc{S}_N\Big[ \mc{W}_N[H_N]\Big](\xi)$ for $\xi>b_N$, what then allows one to infer the sought reflection
property of the effective potential.

\vspace{2mm}

Since the effective potential is constant on $\intff{a_N}{b_N}$  by virtue of the linear integral equation satisfied by $\mc{W}_N[H_N]$,
in order to establish the positivity of the effective potential \eqref{ecriture positivite potentiel effectif},
it is enough to establish that $V_{N;\e{eff}}^{\prime}\Big[ \mc{W}_N\big[H_N\big] \Big](\xi) <0$ on $\intof{-\infty}{a_N}$
what will entail the claim upon invoking the potential's symmetry, its continuity at $a_N$ and $b_N$ which follows from the square root vanishing of $\mc{W}_N\big[H_N\big] $
at $a_N$ and $b_N$ and an integration of the potential's derivative.

It order to control the sign of $V_{N;\e{eff}}^{\prime}\Big[ \mc{W}_N\big[H_N\big] \Big](\xi) <0$ on $\intof{-\infty}{a_N}$,
it appears convenient to obtain a representation for $\msc{J}_{ext}^{(N)}$ that would be more suited for studying its large $N$ behaviour.
First observe that the large-$N$ asymptotic behaviour of the integrand given in \eqref{ecriture DA en série pot effectif prime}
can be inferred by using the form of the asymptotic expansion of $\chi$ above $\Ga_{\ua}$ given in \eqref{forme DA de chi entre R et gamma down}, 
what yields 
\beq
W_{1}(\la) \, = \, \f{ \i   }{ c_2  R_{\ua}(\i) R_{\ua}(\la) (\la + \i)   } \cdot \Big(  \i c_1 \, - \, \la c_2 \Big[ \i  - \tfrac{ c_1 }{ c_2 } \Big]  \Big)
\cdot \Big( 1 \, + \, \e{O}( \ov{x}_N^2 \de_N)  \Big)
\enq
with a remainder that is uniform above the curve $\Ga_{\ua}$ and differentiable. Thus, provided that \eqref{ecriture solution pour bornes mesure equilibre} holds, one gets that
\beq
\f{ \varkappa \ex{\ov{b}_N} }{ 2N } R(\la) W_1(\la) \; = \; \f{1}{\tau_N^2} \bigg\{ - \f{\pi^2 }{ 4 } \mf{l}(-\i\la) R(\la) \; + \; \de \msc{W}_1(\la) \bigg\} \;,
\enq
with
\beq
\big| \de \msc{W}_1(\la) \big| \; \leq \; C \f{ \ln \tau_N  }{  \tau_N \cdot |\la|^{\f{3}{2}} } \;,
\enq
provided that one is located above of $\Ga_{\ua}$ and uniformly away from the poles of $R(\la)$. Thus, upon deforming the contour in the integral representation
for $\msc{J}_{ext}^{(N)}$ from $\R+\i\eps^{\prime}$ up to a contour $\wh{\Ga}_{\ua}$ that is located slightly above of $\Ga_{\ua}$ but which circumvents
the poles of $R$ at $(2p+1)\i$, $p\in \mathbb{N}$, from below, one gets
\beq
\msc{J}_{ext}^{(N)}(x) \; = \; \f{-\pi^2 }{ 4 \tau_N^2 } \Int{ \wh{\Ga}_{\ua} }{} \f{ \dd \la }{2\i\pi} \mf{l}(-\i\la) R(\la) \ex{\i \la  x} \; +  \;  \f{1}{\tau_N^2 } \de\msc{J}_{ext}^{(N)}(x)
\label{ecriture intermediaire pour J ext N}
\enq
with
\beq
\de\msc{J}_{ext}^{(N)}(x) \; = \; \Int{\wh{\Ga}_{\ua} }{} \f{ \dd \la }{2\i\pi}  \de \msc{W}_1(\la) \ex{\i \la  x}  \;.
\enq
Then, one may redeform back the contour $ \wh{\Ga}_{\ua}$ up to $\R+\i\eps^{\prime}$ in the first integral appearing in \eqref{ecriture intermediaire pour J ext N}, what gives
\beq
\msc{J}_{\e{ext}}^{(N)}(x) \, = \, \f{ 1 }{ \tau_N^2 } \cdot \Big\{ \msc{J}_{\e{ext}}(x) \, + \, \de \msc{J}_{\e{ext}}^{(N)}(x) \Big\} \, ,
\enq
where $\msc{J}_{\e{ext}}$ has been introduced in \eqref{definition J ext}.

Moreover, it holds for $\xi<a_N$ tat 
\beq
H_N(\xi) \, = \, - \f{\varkappa \ex{\ov{b}_N} }{2N} \ex{ \tau_N(a_N-\xi)  } \cdot \Big\{ 1 - \ex{- 2 \ov{b}_N} \cdot \ex{ -2 \tau_N(a_N-\xi)  }   \Big\} 
 \, = \, \f{1}{\tau_N^2} \msc{J}_{\e{ext}}(0) \ex{ \tau_N(a_N-\xi)  }  j_N \cdot \Big\{  1 + \de H_N(\xi)  \Big\} \;. 
\enq
where, by using $\mf{t}$ as introduced in  \eqref{definition polynome frak t}, 
\beq
\msc{J}_{\e{ext}}(0) \; = \; - \f{3}{4}  \f{ (2\pi)^{\f{5}{2}} \, \mf{b}^{\mf{b} } \, \hat{\mf{b}}^{ \hat{\mf{b}} }  }{ \Ga(\mf{b}, \hat{\mf{b}}) } 
\quad , \quad j_N \, = \, \f{ 1 }{ b_N^2 \, \mf{t}(2\ov{b}_N) } \quad \e{and} \quad
 \de H_N(\xi)  \, = \, \e{O}\Big(\tau_N^5 \ex{-2\ov{b}_N (1-\tilde{\a}) } \Big)\;. 
\enq
Note that the value of $\msc{J}_{\e{ext}}$ at $0$ follows from the results gathered in Lemma \ref{Lemme positivite densite asymptotiques bord et pot effectif asymptotique bord}  to come. 

Those rewriting allow one to recast the effective potential's derivative as 
\beq
V_{N;\e{eff}}^{\prime}\Big[ \mc{W}_N\big[H_N\big] \Big](\xi) \, = \, \f{ j_N }{ \tau_N }   \Big(  1 + \de H_N(\xi)  \Big) \cdot 
\bigg\{  \msc{J}_{\e{ext}}(0) \ex{ \tau_N(a_N-\xi)  }   \, - \,  \msc{J}_{\e{ext}}\big[ \tau_N(a_N-\xi) \big]  \, + \, \de V_{N;\e{eff}}^{\prime}(\xi) \bigg\} \;. 
\label{ecriture forme asymptotique pour V N eff prime}
\enq
where 
\beq
 \de V_{N;\e{eff}}^{\prime}(\xi)  \, = \,  - \,  \msc{J}_{\e{ext}}\big[ \tau_N(a_N-\xi) \big]  \bigg\{  \f{1 }{  j_N   \big[ 1 + \de H_N(\xi)  \big] } \, -\, 1  \bigg\}
\, - \,  \f{  \de \msc{J}_{\e{ext}}^{(N)}\big[ \tau_N(a_N-\xi) \big] }{  j_N   \big[ 1 + \de H_N(\xi)  \big] } \;. 
\enq
It is established in Lemma \ref{Lemme positivite densite asymptotiques bord et pot effectif asymptotique bord}
that $\msc{J}_{\e{tot}}(x) \,  = \,  \msc{J}_{\e{ext}}(0) \ex{x}   \, - \,  \msc{J}_{\e{ext}}(x) < 0$ for $x>0$. Moreover,
since
\beq
-\f{ \pi^2 }{ 4 } \mf{l}(-\i\la) R(\la) \; = \; \wt{C} \f{ \e{sgn}\big[ \Re(\la) \big] }{ (-\i\la) ^{\f{3}{2} } } \cdot \Big(  1 \, + \, \e{O}(\la^{-1} ) \Big)
\enq
for some constant $ \wt{C} \not=0$, one infers from standard estimates of Fourier integrals that, for a constant $C\not=0$
\beq
\msc{J}_{\e{ext}}(x) \; = \; \msc{J}_{\e{ext}}(0) \, + \, C \sqrt{x} \, + \, \e{O}(x)
\enq
as $x \tend 0^+$, so that $\msc{J}_{\e{tot}}(x) \,  = \, - C \sqrt{x}\, \big( 1+\e{O}(\sqrt{x}) \big)$. Since, $\msc{J}_{\e{tot}}(x)<0$, one has that $C>0$.
In a similar manner, one argues that
\beq
\de \msc{J}_{\e{ext}}^{(N)}(x) \; = \; C^{\prime\prime} \f{\ln \tau_N}{ \tau_N} \, - \, \f{ \ln \tau_N  }{\tau_N}\Big( C^{\prime}  \sqrt{x} \, + \, \e{O}(x) \Big)
\enq
as $x \tend 0^+$ and uniformly in $N$. This implies that
\beq
 \de V_{N;\e{eff}}^{\prime}(\xi) \; = \; \check{C}\f{\ln \tau_N}{ \tau_N} \, + \, \f{ \ln \tau_N  }{\tau_N}\Big( \check{C}^{\prime}  \sqrt{ \tau_N(a_N-\xi) } \, + \, \e{O}\big( \tau_N(a_N-\xi) \big) \Big)
\enq
with remainders that are uniform in $N$ and for some constants $\check{C}$, $\check{C}^{\prime}$ bounded in $N$.
The square root singularity of $\mc{W}_N[H_N]$ at $a_N$  and the continuity of $V_{N;\e{eff}}^{\prime}\Big[ \mc{W}_N\big[H_N\big] \Big]$
on $\R$ and its vanishing on $\intoo{a_N}{b_N}$ entail that  $V_{N;\e{eff}}^{\prime}\Big[ \mc{W}_N\big[H_N\big] \Big]=\e{O}\Big( \sqrt{a_N-\xi} \Big)$ when $\xi \tend a_N$ from below.
Upon substituing $V_{N;\e{eff}}^{\prime}\Big[ \mc{W}_N\big[H_N\big] \Big](a_N)$ in \eqref{ecriture forme asymptotique pour V N eff prime},
one deduces that $\check{C}=0$. Thus, all in all, one gets that
\beq
\f{ \de V_{N;\e{eff}}^{\prime}(\xi)  }{ \msc{J}_{\e{tot}} \big(\tau_N(a_n-\xi))  } \, = \, \e{O}\Big(  \f{ \ln \tau_N  }{ \tau_N }   \Big) \;.
\enq
 Thus, it holds uniformly in $N$ that
\beq
V_{N;\e{eff}}^{\prime}\Big[ \mc{W}_N\big[H_N\big] \Big](\xi) \, = \, \f{ j_N }{ \tau_N }   \Big(  1 + \de H_N(\xi)  \Big) \cdot \msc{J}_{\e{tot}}\big[ \tau_N(a_N-\xi) \big] 
\cdot \Big\{  1 \, + \, \e{O}\Big(  \tfrac{ \ln \ln N  }{\ln N } \Big)  \Big\} \;. 
\enq
Thus for $N$ large enough $V_{N;\e{eff}}^{\prime}\Big[ \mc{W}_N\big[H_N\big] \Big]$ is strictly negative on $\intoo{-\infty}{a_N}$ and hence that 
$V_{N;\e{eff}}\Big[ \mc{W}_N\big[H_N\big] \Big](\xi) > V_{N;\e{eff}}\Big[ \mc{W}_N\big[H_N\big] \Big](a_N)$ on this interval.  \qed

I now establish some of the auxiliary properties that were used in the proofs of
Propositions \ref{Proposition propriete densite mesure eq dans son support} and \ref{Proposition propriete pot effectif}.

\begin{lemme}
\label{Lemme positivite densite asymptotiques bord et pot effectif asymptotique bord} 
 
Let  $\varrho_{\e{bd}}$ and $\msc{J}_{ext}$ be respectively defined as in \eqref{definition varrho bd} and \eqref{definition J ext}. 
Then, it holds 
\beq
\msc{J}_{\e{ext}}(0) \; = \; - \f{3}{4}  \f{ (2\pi)^{\f{5}{2}} \, \mf{b}^{\mf{b} } \, \hat{\mf{b}}^{ \hat{\mf{b}} }  }{ \Ga(\mf{b},\hat{\mf{b}}) } 
\enq
and, for $x \in \R^+\setminus \{0\}$, one has the lower and upper bounds 
\beq
\varrho_{\e{bd}}\big(x \big) -\varrho_{\e{bd}}\big(0\big) \, > \,  0 \; \quad  and  \quad 
  \msc{J}_{tot}(x ) \, = \, \msc{J}_{ext} \big( 0 \big) \cdot \ex{x} \, - \,   \msc{J}_{ext} \big( x \big) \, < \, 0 \;. 
\label{ecriture positivite difference densite bord asymptotique}
\enq

\end{lemme}

\Proof 

Recall the integral representation \eqref{definition varrho bd} for $\varrho_{\e{bd}}$ and observe that one has the factorisation
\beq
 \mf{r}(-\i \la) \, = \,  \mf{r}_h(-\i \la)\cdot  \mf{r}_d(-\i \la) \quad \e{and} \quad 
R(\la) \, = \, H(\la) D(\la)
\enq
 with 
\beq
 \mf{r}_h(-\i \la) \, = \, \f{  \la  }{2 (\la - \i) } \cdot \Ga^2\bigg( \ba{c}  \tfrac{1-\i\la}{2} \\  1-\tfrac{\i\la}{2} \ea \bigg)
\quad ,  \quad 
 \mf{r}_d(-\i \la) \, = \, \f{ 3 \pi \mf{b} \hat{\mf{b}} }{  2^{\i\frac{\la}{2}} \,  \mf{b}^{\i \la \mf{b} } \,  \hat{\mf{b}}^{\i \la \hat{\mf{b}} } }
\cdot \Ga\bigg( \ba{c}   1-\tfrac{\i\la}{2}  \\ 1- \i \mf{b} \la, 1- \i \hat{\mf{b}} \la \ea \bigg) 
\enq
and 
\beq
H(\la) \; = \; 2 \f{ \sinh(\pi \mf{b} \la) \, \sinh(\pi \hat{\mf{b}} \la) }{  \sinh( \tfrac{\pi}{2} \la)  } \; , \qquad 
D(\la) \; = \; \tanh^2 ( \tfrac{\pi}{2} \la) \;. 
\label{definition des fonctions H et D}
\enq
Thus, by the convolution property, it holds $\varrho_{\e{bd}}\big(x \big) \, = \, \Int{ \R }{} \dd y \mf{a}(x-y) \mf{d}(y) $ where, for $x \not= 0$,  
\beq
\mf{a}(x) \,= \, -\i \Int{ \R + \i \eps^{\prime}  }{} \f{ \dd \la }{2 \i \pi }  \f{ \mf{r}_h(-\i \la)  }{ H(\la)  } \ex{\i\la x}
\qquad \e{and} \qquad 
\mf{d}(x) \,= \, \pi  \Int{ \R + \i \eps^{\prime}  }{} \f{ \dd \la }{2 \i \pi }  \f{ \mf{r}_d(-\i \la)  }{ D(\la)  } \ex{\i\la x}\;. 
\enq
The integrals defining $\mf{a}$ and $\mf{d}$ may be taken by the residues either in $\mathbb{H}^+$, if $x>0$,  or $\mathbb{H}^{-}$ if $x<0$. 
One gets 
\beq
\mf{d}(x)\, = \, - \f{ 3 \pi }{ 4 }  \bs{1}_{\R^{-*}}(x) \; + \; \bs{1}_{ \R^{+*} }(x)  \sul{n \geq 1 }{}\bigg\{  \f{ \ex{- \frac{ n  x}{ \mf{b}  } }  }{ 2 \mf{b} }\mf{r}_d\big(  \tfrac{  n  }{ \mf{b}  } \big) 
\; + \; \f{ \ex{- \frac{ n  x}{ \hat{\mf{b}}  } }  }{ 2 \hat{\mf{b}} }\mf{r}_d\big(  \tfrac{  n  }{ \hat{\mf{b}}  } \big) \bigg\}  \; , 
\enq
where $\R^{\pm *} = \R^{\pm}\setminus \{0\}$, and 
\beq
\mf{a}(x) \, = \, - \f{ 2 }{  \pi }  \bs{1}_{ \R^{-*} }(x) \; + \; \bs{1}_{ \R^{+*} }(x)  \sul{n \geq 1 }{}  \tfrac{ 4 \mf{r}_h(2n) }{ \pi^2 }  \ex{-2n x}  \cdot 
\bigg\{ x \, +\, \f{1}{2n-1} \, - \, \f{1}{2n} \, + \, \psi\big( 1 + n \big)  \, - \, \psi\big(\tfrac{1}{2}+n \big)  \bigg\}  \; . 
\enq
Since, 
\beq
\mf{r}_d(\a) \, = \,    2^{\frac{\a}{2}} \,  \mf{b}^{\a \mf{b} } \,  \hat{\mf{b}}^{ \a \hat{\mf{b}} }  
\cdot \f{  3 \pi \mf{b} \hat{\mf{b}}  \, \Ga\big( 1+\tfrac{\a}{2}\big)  }{ \Ga\big( 1+ \mf{b} \a ,  1+ \hat{\mf{b}} \a \big) } \, >\, 0 \qquad \e{and} \qquad 
\mf{r}_h(\a) \, = \, \f{  \a  }{2 (\a - 1) } \cdot \Ga^2\bigg( \ba{c} \tfrac{1+\a}{2} \\ 1+\tfrac{\a}{2} \ea \bigg)   \, > \,0 \;, 
\enq
it is direct to see that $\mf{d}(x)>0$ on $\R^{+*}$.  One infers that $\mf{a}(x)>0$ on $\R^{+*}$ by also using that $\ln \Ga$ is strictly convex on $\R^{+*}$, 
\textit{viz}. $x \mapsto \psi(x)$ is strictly increasing on $\R^{+*}$. Moreover, differentiating term-wise the series defining $\mf{a}$ for $x>0$ yields that 
\beq
\mf{a}^{\prime}(x)\, = \, - \sul{n \geq 1 }{}  \tfrac{ 8 n   }{\pi^2 } \mf{r}_h(2n)  \cdot 
\bigg\{ x \, +\, \f{1}{2n-1} \, - \, \f{1}{n} \, + \, \psi\big( 1 + n \big)  \, - \, \psi\big(\tfrac{1}{2}+n \big)  \bigg\}  \cdot \ex{-2n x}  \; . 
\enq
One may control the sign of each summand occurring in this series by using the below identity 
\beq
\f{1-s}{x+s} \, < \, \psi(x+1)\, - \, \psi(x+s) \qquad \e{with} \qquad x>0 \;\; \e{and} \;\; s \in \intoo{0}{1}
\label{ecriture lower bound difference fct psi}
\enq
established by Alzer in \cite{AlzerInequalitiesforGammaAndPsiFct} as a direct consequence of the the strict convexity of $x \mapsto x \psi(x)$ on $\R^+$. 
When applied for $s=\tf{1}{2}$, it yields 
\beq
\f{1}{2n-1} \, - \, \f{1}{n} \, + \, \psi\big( 1 + n \big)  \, - \, \psi\big(\tfrac{1}{2}+n \big)  \, > \, \f{1}{2n+1}\, + \, \f{1}{2n-1} \, - \, \f{1}{n} \, > \, 
\f{ 1 }{ n (4n^2-1) } \, > \, 0 \;. 
\enq
Hence $\mf{a}$ is strictly decreasing on $\R^+$. 
All the above handling yield that, for $x\geq 0$,  
\beq
\varrho_{\e{bd}}\big(x \big) \, = \, \Int{ 0 }{ x } \hspace{-1mm} \dd y \, \mf{a}(x-y) \mf{d}(y) \,  - \, \f{2}{\pi} \Int{x}{+\infty} \hspace{-1mm} \dd y \, \mf{d}(y) \, - \, \f{3\pi}{4} \Int{-\infty}{0} \hspace{-1mm} \dd y \, \mf{a}(x-y) \;. 
\enq
In particular, one gets 
\beq
\varrho_{\e{bd}}\big(x \big) \, - \,  \varrho_{\e{bd}}\big(0 \big)  \, = \, \f{3\pi}{4} \Int{0}{+\infty} \hspace{-1mm} \dd y \, \Big( \mf{a}(y)\, - \,  \mf{a}(x+y) \Big) 
\, + \, \Int{ 0 }{ x } \hspace{-1mm} \dd y \, \Big( \tfrac{2}{\pi} + \mf{a}(x-y) \Big) \,  \mf{d}(y) \;.   
\enq
The expression is manifestly strictly positive for $x>0$ owing to the strict decay  of $ \mf{a}$ on $\R^+$
as well as the strict positivity of $ \mf{a}$ and $\mf{d}$ on $\R^+$\;.

It remains to deal with $\msc{J}_{\e{ext}}$.
The value of $ \msc{J}_{\e{ext}}(0)$ follows upon taking the integral defining $\msc{J}_{\e{ext}}$ \eqref{definition J ext}
by the residues in the lower-half plane, the only pole being located at $\la=-\i$.
Further, one has the factorisation $ \mf{l}(-\i \la) =  \mf{l}_h(-\i \la)\,  \mf{l}_d(-\i \la)$ where 
\beq
 \mf{l}_d(-\i \la) \, = \, \f{ \i }{ 2 \la }  2^{\i\frac{\la}{2}} \,  \mf{b}^{\i \la \mf{b} } \,  \hat{\mf{b}}^{\i \la \hat{\mf{b}} }  
\cdot  \Ga\bigg( \ba{c}  1- \i \mf{b} \la , 1- \i \hat{\mf{b}} \la  \\  1-\tfrac{\i\la}{2} \ea \bigg)  \qquad \e{and} \qquad 
 \mf{l}_h(-\i \la) \, = \, - \f{12   }{   \la (\la + \i) } \cdot  \Ga^2\bigg( \ba{c}  1-\tfrac{\i\la}{2} \vspace{1mm} \\ \tfrac{1-\i\la}{2} \ea \bigg)  \;.  
\enq
 Hence, analogously to the previous reasonings, starting from \eqref{definition J ext}  one gets that $ \msc{J}_{\e{ext}}(x) \, = \, \int_{ \R } \dd y \, \wt{\mf{a}}(x-y) \cdot  \wt{\mf{d}}(y) $ where, for $x \not= 0$,
\beq
\wt{\mf{a}}(x) \,= \, -\i \, \f{\pi^2}{4} \hspace{-1mm} \Int{ \R + \i \eps^{\prime}  }{} \hspace{-2mm} \f{\dd \la }{ 2\i\pi } \,  \mf{l}_h(-\i \la)   H(\la)  \ex{ \i \la x}
\qquad \e{and} \qquad 
\wt{\mf{d}}(x) \,= \,   \Int{ \R + \i \eps^{\prime}  }{}\hspace{-2mm} \f{\dd \la }{ 2\i\pi } \,    \mf{l}_d(-\i \la)  D(\la)  \ex{ \i \la x} \;. 
\enq
 The functions $H$ and $D$ have already been introduced in \eqref{definition des fonctions H et D}. Taking the integrals analogously to the previous case, 
one gets that 
\beq
\wt{\mf{d}}(x) \,= \, \bs{1}_{\R^{+*}}(x) \cdot  \f{4}{\pi} \sul{ n \geq 0}{} \sin^2\big[ 2\pi n \mf{b} \big]\, \mf{l}_d(2n) \cdot \ex{- 2 n x} \, > \, 0 \;, 
%
%
%
\enq
while 
\beq
\wt{\mf{a}}(x) \,= \, - 3 \pi \ex{x}  \bs{1}_{ \R^{-*} }(x) \; + \; \bs{1}_{ \R^{+*} }(x)  \sul{n \geq 1 }{}   \mf{l}_h(2n+1)   \cdot 
\bigg\{ x \, +\, \f{1}{2(n+1)} \, - \, \f{1}{2n+1} \, + \, \psi\big( 1 + n \big)  \, - \, \psi\big( \tfrac{1}{2} + n \big)  \bigg\} \cdot \ex{-(2n+1) x}   \; . 
\enq
By using the Alzer lower bound \eqref{ecriture lower bound difference fct psi}, one readily infers that  $\wt{\mf{a}}(x) \,> \,0$ on $\R^{+*}$. 
 
The above representation thus yields that 
\beq
 \msc{J}_{\e{ext}}(x) \, = \,  \Int{ 0 }{ x }\hspace{-1mm} \dd y \,  \wt{\mf{a}}(x-y) \wt{\mf{d}}(y) \,  - \, 3 \pi \Int{x}{+\infty} \hspace{-1mm} \dd y \, \ex{x-y} \, \wt{\mf{d}}(y) 
\enq
 so that $\msc{J}_{\e{tot}}$ as defined in \eqref{ecriture positivite difference densite bord asymptotique} takes the form
\beq
 \msc{J}_{\e{tot}}(x) \, = \, - \Int{0}{x} \hspace{-1mm} \dd y \, \Big\{ 3\pi  \ex{x-y} \, + \,  \wt{\mf{a}}(x-y) \Big\}\cdot  \wt{\mf{d}}(y)  \, < \, 0
\enq
for $x>0$, leading to the claim. \qed

\section{The large-$N$ behaviour of $ \mc{E}_N^{(+)}\big[ \sg_{\e{eq}}^{(N)}\big]$} 
\label{Section comportement a grand N de la fnelle energi en la mesure eq}

To state the main result of this section it is convenient to introduce the rescaled sequence $w_k$ which has constant $N\tend +\infty$
asymptotics:
\beq
w_1\, =\, 2 \ov{b}_N \tilde{w}_1\; , \quad w_2 \, = \, 2 (\ov{b}_N)^2\,  \tilde{w}_2\; , \quad \e{with} \quad 
\tilde{w}_k \,=\, 1+ \e{O}\Big( \tfrac{ 1 }{ \ov{b}_N } \Big) \; \quad \e{as} \; \; N \tend + \infty \;. 
\enq

\begin{theorem}
 
 One has the large-$N$ asymptotic behaviour 
\beq
\mc{E}_N^{(+)}\big[ \sg_{\e{eq}}^{(N)}\big] \; = \;  \f{ 3 \pi^{4} \, \mf{b} \,  \hat{\mf{b}}  \, \tilde{w}_1 }{ 4 ( \ov{b}_N)^3\, \tilde{w}_2   \, \mf{t}\Big(2\ov{b}_N\Big)  } \, + \, 
\f{ 9 \, \pi^4 \, \mf{b} \,  \hat{\mf{b}} }{ 8 ( \ov{b}_N)^4 \, \mf{t}^2\Big( 2\ov{b}_N \Big)  } 
\bigg\{    1  - \tfrac{ 2 \tilde{w}_1 }{ \ov{b}_N  \tilde{w}_2 }    \bigg\}  
\, + \,  
\e{O}\Big( \ex{-2\ov{b}_N(1-\tilde{\a} )}\Big) \;, 
\enq
where $\mc{E}_N^{(+)}$ is as defined in \eqref{definition fonctionnelle energie +}  and $\mf{t}$ is as introduced in \eqref{definition fonction frak t}.

\end{theorem}

\Proof 

By carrying out integration by parts, one may transform the original expression for $\mc{E}_N^{(+)}\big[ \sg_{\e{eq}}^{(N)}\big] $
given in \eqref{definition fonctionnelle energie +} into one which is simpler to 
evaluate in the large-$N$ limit. 
Indeed, by setting $\mc{U}_N(\xi)=\Int{a_N}{\xi}  \mc{W}_N\big[ H_N \big](\eta) \dd \eta $  and integrating by parts over $s$ gives
\bem
\; - \;  \Int{}{} \mf{w}^{(+)}\Big(\tau_N(s-t)\Big) \dd  \sg_{\e{eq}}^{(N)}(s) \dd  \sg_{\e{eq}}^{(N)}(t) \; = \; 
\, - \, \mc{U}_N(b_N)    \hspace{-1mm} \Int{a_N}{ b_N} \hspace{-1mm} \dd \eta \, \mf{w}^{(+)}\Big(\tau_N(b_N-\eta)\Big) \cdot \mc{W}_N\big[ H_N \big](\eta) \\
\; + \; 
\tau_N \Int{a_N}{b_N} \hspace{-1mm} \dd \xi \, \mc{U}_N(\xi)
\Fint{a_N}{ b_N} \hspace{-1mm} \dd \eta  \, \Big(\mf{w}^{(+)}\Big)^{\prime}\Big(\tau_N(\xi-\eta)\Big) \cdot \mc{W}_N\big[ H_N\big](\eta)  \\
\; = \;   -     \Int{a_N}{ b_N} \hspace{-1mm} \dd \eta  \, \mf{w}^{(+)}\Big(\tau_N(b_N-\eta)\Big) \cdot \mc{W}_N\big[ H_N \big](\eta)
\; + \; \f{1}{N} \Int{a_N}{b_N} \hspace{-1mm} \dd \xi \, \mc{U}_N(\xi)  V^{\prime}_N(\xi) \;.
\end{multline}
Above, I used the normalisation condition for the equilibrium measure $\mc{U}_N(b_N)=1$ along with 
the linear integral equation satisfied by its density \eqref{ecriture eqn int sing lin pour densite mesure eq}. Thus, upon integrating by parts in the last integral one gets
\bem
\; - \;  \Int{}{} \mf{w}^{(+)}\Big(\tau_N(s-t)\Big) \dd  \sg_{\e{eq}}^{(N)}(s) \dd  \sg_{\e{eq}}^{(N)}(t) \; = \; \f{\varkappa }{ N } \cosh(\ov{b}_N) \; - \; 
 \f{1}{N} \Int{a_N}{b_N}  \hspace{-1mm} \dd \xi \,  \mc{W}_N\big[ H_N \big](\xi)  V_N(\xi) \\
 \, -  \, \Int{a_N}{ b_N} \hspace{-1mm} \dd \eta  \, \mf{w}^{(+)}\Big(\tau_N(b_N-\eta)\Big) \cdot \mc{W}_N\big[ H_N \big](\eta) 
\end{multline}
Thus, all-in-all, 
\beq
\mc{E}_N^{(+)}\big[ \sg_{\e{eq}}^{(N)}\big] \; = \; \f{\varkappa }{ 2 N } \cosh(\ov{b}_N) \; + \; 
 \f{1}{ 2 N} \Int{a_N}{b_N} \hspace{-1mm} \dd \xi \,  \mc{W}_N\big[ H_N \big](\xi)  V_N(\xi)  \, -  \, \f{1}{2}\Int{a_N}{ b_N} \hspace{-1mm} \dd \eta  \, \mf{w}^{(+)}\Big(\tau_N(b_N-\eta)\Big) \cdot \mc{W}_N\big[ H_N \big](\eta)  \;. 
\enq
The rest is a consequence of Lemmas \ref{Lemme evaluation integrale densite mes eq contre VN} and \ref{Lemme evaluation integrales varpi contre mesure eq en bn} given below. \qed

\subsection{Auxiliary Lemmas}

 \begin{lemme}
 \label{Lemme evaluation integrale densite mes eq contre VN}
  
 Let $\intff{a_N}{b_N}$ correspond to the support of the equilibrium measure $\sg_{\e{eq}}^{(N)}$.  It holds,  
\beq
 \f{1}{ 2  N} \Int{a_N}{b_N} \mc{W}_N\big[ H_N \big](\xi)  V_N(\xi) \cdot \dd \xi  \; = \;   \f{ \varkappa^2 \ex{2\ov{b}_N }  }{ 8 \pi  N^2 } \cdot 
 \bigg\{  \chi_{12}^2(\i)\, +  \, 2  \Big[  \chi_{12}( \i) \chi_{11}^{\prime}(\i)  \, - \,  \chi_{11}( \i) \chi_{12}^{\prime}(\i)  \Big]  \bigg\} \;. 
\label{evaluation explicite integrale densite eq vs VN}
\enq
Moreover, for $\a^{\prime} \in \intoo{0}{1}$ and fixed, \textit{c.f.} \eqref{definition alpha prime}, the integral has the large-$N$ behaviour
\beq
 \f{1}{ 2  N} \Int{a_N}{b_N} \mc{W}_N\big[ H_N \big](\xi)  V_N(\xi) \cdot \dd \xi  \; = \;   \f{  \varkappa^2 \ex{2\ov{b}_N }  \mf{b} \, \hat{\mf{b}} }{ 16 \pi^2  N^2   } \cdot
 \f{ \Ga^2\big( \mf{b}, \hat{\mf{b}} \big) }{ \mf{b}^{ 2 \mf{b}} \cdot {\hat{\mf{b}}}^{ 2 \hat{\mf{b}} }  }  \cdot
 \bigg\{    1 - 2 \tfrac{w_1}{w_2}   \,  +\, \e{O}\Big( (\ov{b}_N )^4 \cdot \ex{ - 2 \ov{b}_N (1 - \tilde{\a}) } \Big) \bigg\} \;.
\label{ecriture DA integrale densite eq vs VN}
\enq

 \end{lemme}

 \Proof

A direct calculation shows that  
\beq
\Int{a_N}{b_N} \ex{ - \i\tau_N \la (\xi-a_N) } V_N(\xi) \dd \xi \; = \;
\f{   \i \varkappa }{ 2  \tau_N }  \sul{\sg=\pm}{}  \Bigg\{   \ex{ - \i \la \ov{x}_N}   \f{ \ex{ \sg \ov{b}_N }}{\la + \sg  \i  } \, - \, \f{ \ex{ \sg \ov{a}_N }}{\la +  \sg \i  }   \Bigg\}\; .
\enq
Thus, recalling the expression for the density of the equilibrium measure \eqref{expression explicite pour densite mesure equilibre}, one gets that  
\beq
 \Int{a_N}{b_N} \mc{W}_N\big[ H_N \big](\xi)  V_N(\xi) \cdot \dd \xi \; = \;   \f{ \varkappa^2 \ex{\ov{b}_N} }{4 \pi N } \Int{ \R + 2 \i \eps^{\prime} }{} \f{ \dd \la  }{2\i\pi}
\sul{ \sg = \pm }{} \Bigg\{   \ex{ - \i \la \ov{x}_N}   \f{ G(\la) \ex{ \sg \ov{b}_N }}{\la +   \sg \i  } \, - \,  \f{ G(\la) \ex{ \sg \ov{a}_N }}{\la +   \sg \i }   \Bigg\}
\enq
where 
\beq
G(\la) \, = \, \sul{ \sg = \pm }{}    \f{ \sg }{ \sg \i - \la  } \bigg\{ \f{ \i \sg }{ \la } \chi_{11}(\la) \chi_{12}(\sg \i) - \chi_{12}(\la) \chi_{11}(\sg \i ) \bigg\} \;.
\enq
The function $G$ has no poles at $\la=\sg \i$ by construction and decays as $\e{O}\Big( |\la|^{-\f{3}{2}} \Big)$ at infinity. 
Moreover the boundary values $G_{\pm}$ are smooth. Since, for $\la\in \R$, it holds $\ex{ - \i \la \ov{x}_N}  G_{+}(\la) \, = \,  G_{-}(\la)$ , the boundary values
admit holomorphic extensions to a neighbourhood of $\R$ in $\mathbb{H}^{\mp}$. Finally, it holds that 
\beq
\e{Res}\Big( G_{-}(\la) \dd \la, \la=0 \Big) \, = \, \chi_{11;-}(0) \sul{ \sg = \pm }{}  \sg  \chi_{12}(\sg \i)  \;= \; 0
\enq
by virtue of $\chi_{12}(\i) \, = \, \chi_{12}(-\i)$, \textit{c.f.} Lemma \ref{Lemme Ecriture diverses proprietes solution RHP chi}, , so that $G_{-}$ has no pole at $\la=0$.  Therefore,
using $a_N=-b_N$,
\beq
 \Int{a_N}{b_N} \mc{W}_N\big[ H_N \big](\xi)  V_N(\xi) \cdot \dd \xi \; = \;   \f{ \varkappa^2 \ex{\ov{b}_N} }{ 4 \pi N } \Big\{ - \ex{\ov{b}_N} G(-\i)  - \ex{-\ov{a}_N} G(\i) \Big\}
\; = \;   - \f{ \varkappa^2 \ex{2\ov{b}_N} }{ 4 \pi N } \Big\{   G(-\i) + G(\i) \Big\}\;. 
\enq
A direct calculation utilising $\chi_{12}(-\la)=\chi_{12}( \la)$ and $\chi_{11}(-\la) = \chi_{11}( \la) + \la \chi_{12}( \la)$ leads to 
\bem
G(-\la) + G(\la)  \, = \, \f{2}{\i - \la} \bigg\{ \f{ \i }{ \la } \chi_{11}( \la) \chi_{12}(\i) \, - \,   \chi_{12}( \la) \chi_{11}(\i)   \bigg\} \\  
\; + \;  \f{2}{\i + \la} \bigg\{ \f{ - \i }{ \la } \chi_{11}( \la) \chi_{12}(\i) \, - \,   \chi_{12}( \la) \chi_{11}(\i)   \bigg\} 
\, - \, \f{ 2\i }{ \i + \la }  \chi_{12}( \la) \chi_{12}(\i) \;. 
\end{multline}
This yields
\beq
  G(-\i) + G(\i) \; = \; -\chi_{12}^2(\i)\, +  \, 2  \Big[   \chi_{11}( \i) \chi_{12}^{\prime}(\i) \, - \,   \chi_{12}( \i) \chi_{11}^{\prime}(\i)   \Big]  \;. 
\enq
All of this already yields \eqref{evaluation explicite integrale densite eq vs VN}. It then remains to insert in this expression 
the large-$N$ expansion of $\chi$ given in \eqref{forme DA de chi au dessus de gamma up} and then use \eqref{ecriture relations entre les coeffs ck}
so as to get
\beq
 \f{1}{ 2  N} \Int{a_N}{b_N} \mc{W}_N\big[ H_N \big](\xi)  V_N(\xi) \cdot \dd \xi  \; = \; -  \f{  \varkappa^2 \ex{2\ov{b}_N }  }{ 8 \pi  N^2 R_{\ua}^2(\i) } \cdot
 \bigg\{    1 - 2 \tfrac{w_1}{w_2}   \,  +\, \e{O}\Big( (\ov{b}_N )^4 \cdot \ex{ - 2 \ov{b}_N (1 - \tilde{\a}) } \Big) \bigg\} \;.
\enq
Upon substituting the value of $ R_{\ua}^2(\i)$, \eqref{ecriture DA integrale densite eq vs VN} follows.  \qed

 \begin{lemme}
 \label{Lemme evaluation integrales varpi contre mesure eq en bn} 
  
  Let $\intff{a_N}{b_N}$ correspond to the support of the equilibrium measure $\sg_{\e{eq}}^{(N)}$. Then, one has the exact evaluation
\bem
\, -  \f{1}{2}  \Int{a_N}{ b_N}  \mf{w}^{(+)}\Big(\tau_N(b_N-\eta)\Big) \cdot \mc{W}_N\big[ H_N \big](\eta) \cdot \dd \eta  \\
\; = \; - \f{ \varkappa \ex{\ov{b}_N }  }{ 4   N }
\bigg\{ 1 \, + \,  \ex{-\ov{x}_N } \, + \, \chi_{22;-}(0) \Big[ 2 \chi_{11}(\i)+\i \chi_{12}(\i) \Big] \, - \, 2 \chi_{21;-}(0)  \chi_{12}(\i) \bigg\} \;. 
\label{ecriture explicite integrale densite contre varpi en bn}
\end{multline}
Moreover, for $\a^{\prime} \in \intoo{0}{1}$ and fixed, \textit{c.f.} \eqref{definition alpha prime},  one has the large-$N$ expansion
\bem
\, -  \f{1}{2}  \Int{a_N}{ b_N}  \mf{w}^{(+)}\Big(\tau_N(b_N-\eta)\Big) \cdot \mc{W}_N\big[ H_N \big](\eta) \cdot \dd \eta  \; = \;
- \f{ \varkappa    }{ 2   N } \cosh(\ov{b}_N ) \\
+\f{ \varkappa \ex{\ov{b}_N }  }{ 4   N }
\bigg\{  \Big( \f{\pi}{2} \Big)^{\f{3}{2}} \cdot \f{ w_1}{ w_2 } \cdot \mf{b}\, \hat{\mf{b}} \cdot
\f{ \Ga\big(\mf{b}, \hat{\mf{b}} \big) }{ \mf{b}^{ \mf{b} } \cdot  \hat{\mf{b}}^{ \hat{\mf{b}} }   }  \, +  \, \e{O}\Big( (\,\ov{x}_N)^4 \ex{ - \ov{x}_N (1- \tilde{\a}) }  \Big) \bigg\} \;.
\label{ecriture DA integrale densite contre varpi en bn}
\end{multline}

 \end{lemme}

\Proof

 To start with, observe that it holds
\beq
   \Int{a_N}{ b_N}  \mf{w}^{(+)}\Big(\tau_N(b_N-\xi)\Big) \cdot \ex{- \i \tau_N \la (\xi-a_N)} \cdot \dd \xi  \; = \; 
\f{\i}{2 \tau_N}\Int{\R}{} \dd  \mu \f{ R(\mu) }{ \mu (\mu-\la) } \cdot  \Big\{ \ex{-\i \ov{x}_N \la } \,- \, \ex{-\i \ov{x}_N \mu }\Big\} \;. 
\enq
Note that the integrand has no singularity at $\mu=0$ owing to the triple zero of $R$ at the origin. 

Thus, the integral of interest may be recast as 
\bem
\msc{I}_N\, = \, \, -  \f{1}{2}  \Int{a_N}{ b_N}  \mf{w}^{(+)}\Big(\tau_N(b_N-\xi)\Big) \cdot \mc{W}_N\big[ H_N \big](\xi) \cdot \dd \eta  \; = \; 
- \f{ \i \varkappa \ex{\ov{b}_N} }{4 N } \Int{\R}{} \f{ \dd \mu  }{2\i\pi} \Int{ \R + 2 \i \eps^{\prime} }{} \f{ \dd \la  }{2\i\pi}\\
\times \sul{ \sg = \pm }{}   \f{ \sg }{ \sg \i - \la  } \bigg\{ \f{ \i \sg }{ \la } \chi_{11}(\la) \chi_{12}(\sg \i) - \chi_{12}(\la) \chi_{11}(\sg \i ) \bigg\}
\f{ R(\mu) }{ \mu (\mu-\la) } \cdot  \Big\{ \ex{-\i \ov{x}_N \la } \,- \, \ex{-\i \ov{x}_N \mu }\Big\} \;. 
\end{multline}
One may take the $\la$-integral involving the integrand having the factor  $\ex{-\i \ov{x}_N \mu }$ by deforming the integration contours to $+\i\infty$. 
Since the associated  $\la$ integrand has no poles in $\mathbb{H}^+$ and decays, pointwise in $\mu$, as $\e{O}\big( |\la|^{-\f{5}{2} } \big)$ there, this 
piece of the full integrand does not contribute to $\msc{I}_N$. 
The contribution of the integrand containing the factor  $\ex{-\i \ov{x}_N \la }$ may be obtained by using the jump conditions 
$\ex{-\i \ov{x}_N \la } \cdot \chi_{1a;+}(\la) = \chi_{1a;-}(\la)$, deforming the contours to $-\i\infty$ and taking the residue at $\la=\mu$,
which is the only pole of the associated integrand in $\mathbb{H}^- + 2\i\eps^{\prime}$. Indeed, the apparent singularities at $\la= \sg \i$ and $\la=0$ are easily seen to be removable. This yields
\beq
\msc{I}_N\, = \, 
- \f{ \i \varkappa \ex{\ov{b}_N} }{4 N } \Int{\R}{} \f{ \dd \mu  }{2\i\pi} 
 \sul{ \sg = \pm }{}   \f{ \sg \, R(\mu) }{\mu(\sg \i  - \mu)  } \bigg\{ \f{ \sg \i  }{ \mu } \chi_{11;-}(\mu) \chi_{12}(\sg \i) - \chi_{12;-}(\mu) \chi_{11}(\sg \i ) \bigg\}   \;.
\label{ecriture evaluation lambda integral dans integrale contre potentiel varpi}
\enq

The integral may then be estimated further by using the jump conditions 
\beq
R(\la) \chi_{1a;-}(\la) \, = \,  \chi_{2a;+}(\la) \,  +\,  \ex{-\i \ov{x}_N \la } \cdot \chi_{2a;-}(\la)
\label{ecriture saut chi avec fct R}
\enq
Since the left hand side of \eqref{ecriture saut chi avec fct R} above has a triple zero at the origin  while, taken individually, the two factors on the right hand side are non-vanishing, 
it is convenient to slightly deform the $\mu$ integration in  \eqref{ecriture evaluation lambda integral dans integrale contre potentiel varpi} to $\R+\i\eps^{\prime}$
while understanding the symbols $\chi_{2a;-}(\mu)$ as the analytic continuation of the $-$ boundary  value $\chi_{2a;-}$ on $\R$ to $\R+\i\eps^{\prime}$. 
This yields $\msc{I}_N\, = \, \msc{I}_N^{(\ua)} \,  + \, \msc{I}_N^{(\da)}$, where
\beqa
\msc{I}_N^{(\ua)} & = & - \f{ \i \varkappa \ex{\ov{b}_N} }{4 N } \Int{\R + \i \eps^{\prime} }{} \f{ \dd \mu  }{2\i\pi}
 \sul{ \sg = \pm }{}   \f{ \sg \, }{\mu(\sg \i  - \mu)  } \bigg\{ \f{ \sg \i  }{ \mu } \chi_{21}(\mu) \chi_{12}(\sg \i) - \chi_{22}(\mu) \chi_{11}(\sg \i ) \bigg\} \\
\msc{I}_N^{(\ua)} & = & - \f{ \i \varkappa \ex{\ov{b}_N} }{4 N } \Int{\R + \i \eps^{\prime} }{} \f{ \dd \mu  }{2\i\pi}
 \sul{ \sg = \pm }{}   \f{ \sg \, \ex{-\i \ov{x}_N \la }  }{\mu(\sg \i  - \mu)  } \bigg\{ \f{ \sg \i  }{ \mu } \chi_{21;-}(\mu) \chi_{12}(\sg \i) - \chi_{22;-}(\mu) \chi_{11}(\sg \i ) \bigg\}
\eeqa
Note that it is licit to split the integral in $2$ pieces since both integrands appearing above behave as $\e{O}\big( |\mu|^{-\tf{5}{2}}\big)$.
The above integrals may then be computed by taking the residues in appropriate half-planes. In the case of  $\msc{I}_N^{(\ua)}$, the integrand
admits a single pole, which is moreover simple, at $\mu=\i$, and has a quick decay at $\infty$. Hence, applying the residue theorem  leads to
\beq
\msc{I}_N^{(\ua)} \; = \;   \f{   \varkappa \ex{\ov{b}_N} }{4 N } \Big( \chi_{21}(\i) \chi_{12}(\i) - \chi_{22}(\i) \chi_{11}( \i ) \Big)  \; = \;   - \f{   \varkappa \ex{\ov{b}_N} }{4 N }\;.
\enq
Note that the last equality follows from $\det\big[ \chi(\la) \big] \, = \, \e{sgn}\big[ \Im(\la) \big] $.

In what concerns $\msc{I}_N^{(\da)}$  the integrand has good decay properties in $\mathbb{H}^{-}$ and admits 2 poles in the strip $\Im(\la)\leq \eps^{\prime}$:
one simple at $\mu=-\i$ and one double at $\mu=0$. A direct calculation then gives:
\bem
\msc{I}_N^{(\da)}\, = \,
- \f{ \i \varkappa \ex{\ov{b}_N} }{4 N }  \Bigg\{
 \f{ \ex{-\ov{x}_N} }{ \i }  \Big[ \chi_{21}( -\i) \chi_{12}( -\i) -  \chi_{22}( -\i) \chi_{11}( -\i)  \Big]
\, - \, \f{ \Dp{} }{ \Dp{}\mu} \bigg[  \sul{ \sg = \pm }{}  \f{ \i  \ex{-\i \ov{x}_N  \mu }  }{\sg \i - \mu  }   \cdot \chi_{21;-}( \mu)  \chi_{12}(\sg \i)   \bigg]_{\mid\mu=0}  \\
\, + \, \f{1}{\i} \chi_{22;-}(0) \sul{\sg=\pm }{} \chi_{11}(\sg \i ) \bigg\}   \;. 
\label{ecriture calcul explicite IN}
\end{multline}
At this stage, one invokes Lemma \ref{Lemme Ecriture diverses proprietes solution RHP chi} and the identity
\beq
\sul{ \sg = \pm }{}  \f{ \i     }{\sg \i  - \mu  }  \; = \; \f{ - 2\i\mu}{  1+\mu^2  }
\enq
so as to ensure that there is no contribution to \eqref{ecriture calcul explicite IN} stemming from $\Dp{\mu}\Big[ \ex{-\i \ov{x}_N  \mu } \chi_{21;-}( \mu) \big]_{\mid \mu=0}$,
one eventually gets \eqref{ecriture explicite integrale densite contre varpi en bn}. The large-$N$ expansion is obtained by inserting the 
ones of the matrix $\chi$ given in \eqref{forme DA de chi au dessus de gamma up}-\eqref{forme DA de chi entre R et gamma down}. One gets 
\bem
\msc{I}_N\, = \,
- \f{   \varkappa \ex{\ov{b}_N} }{4 N }  \Bigg\{ 1+ \ex{-2\ov{b}_N} \, - \, \f{c_1}{c_2}R_{\da}(0) \Big[ \f{ 2 }{ c_2 R_{\ua}(\i) } \, + \,  \f{ \i }{ R_{\ua}(\i) } \big(\i - \tfrac{ c_{1} }{ c_{2} } \big)  \Big]
\, - \,  \f{ 2 R_{\da}(0) }{c_2 R_{\ua}(\i) } \big(\i - \tfrac{ c_{1} }{ c_{2} } \big) \; + \; \e{O}\Big(  (\, \ov{x}_N )^4  \ex{ -\ov{x}_N(1-\tilde{\a}) } \Big) \Bigg\} \\
\, = \,
- \f{   \varkappa \ex{\ov{b}_N} }{4 N }  \Bigg\{ 1+ \ex{-2\ov{b}_N} \, + \,  \f{ \i R_{\da}(0) }{w_2 R_{\ua}(\i) }  \Big[ 2\f{w_1}{w_2}  + w_1\Big( 1-\tfrac{ w_{1} }{ w_{2} } \Big) + 2 \Big( 1-\tfrac{ w_{1} }{ w_{2} } \Big)  \Big]
 + \; \e{O}\Big(  (\, \ov{x}_N )^4  \ex{ -\ov{x}_N(1-\tilde{\a}) } \Big) \Bigg\} \;. 
\end{multline}
It then remains to use that $w_1^2=w_2$ so as to get
\beq
\, -  \f{1}{2}  \Int{a_N}{ b_N}  \mf{w}^{(+)}\Big(\tau_N(b_N-\eta)\Big) \cdot \mc{W}_N\big[ H_N \big](\eta) \cdot \dd \eta  \; = \; - \f{ \varkappa \ex{\ov{b}_N }  }{ 4   N }
\bigg\{ 1 \, + \,  \ex{-\ov{x}_N }  \, + \, \i \f{ w_1 R_{\da}(0) }{ w_2 R_{\ua}(\i) } \, +  \, \e{O}\Big( (\,\ov{x}_N)^4 \ex{ - \ov{x}_N (1- \tilde{\a}) }  \Big) \bigg\} \;.
\enq
Then, direct substitutions for the constants yield \eqref{ecriture DA integrale densite contre varpi en bn}.   \qed

\section{Conclusion}
\label{Section conclusion}

This work developed a method allowing to to prove the convergence of the form factor series representation for the  vacuum two-point function of space-like separated exponentials of the field 
in the quantum Sinh-Gordon integrable field theory in $1+1$ dimensions. While the paper only discussed this specific situation, the method is general enough so as to 
allow dealing with the convergence of series arising in the description of two-point functions in the model involving other operators, be it for space of time-like separations between them. 
The method is also generalisable to the case of form factor series associated with multi-point function. Finally, since it only relies on very general properties of the form factors, the method 
should also be applicable so as to discuss convergence problems arising in more complex massive quantum integrable field theories such as the Sine-Gordon model.

\section*{Acknowledgment}

K.K.K. acknowledges support from  CNRS and ENS de Lyon. The author is indebted to A. Guionnet and F. Smirnov for stimulating discussions
on various aspects of the project.

\appendix

\section{Auxiliary results}
\label{Appendix Section resultats auxiliaires}

\subsection{A boundedness result for Sobolev spaces}

\begin{lemme}
\label{Lemme bornage produit dans norme Hs}
 Let $f$ be smooth on $\R$ and $h \in \mc{H}_s(K)$ with $K$ a compact subset of $\R$. Then, for $s>0$ it holds
\beq
\norm{ f h  }_{ \mc{H}_s(\R) }  \; \leq \; C \cdot \norm{ h  }_{ \mc{H}_s(\R) }
\enq
for some constant $C>0$ which depends on $f$.

\end{lemme}

 \Proof 
 
 Let $\varrho \in \mc{C}^{\infty}_{\e{c}}(\R)$ be such that $\e{supp}(\varrho) \subset K_1$ with $K_{\eps}=\{x \in \R \, : \,  d(x,K)\leq \eps \}$ and $\varrho_{\mid K}=1$. 
 Set $\wt{f}=f \varrho$ so that $\mc{F}[\wt{f}]$ belongs to the Schwartz class.  
Then one has that 
\beq
\mc{F}[\wt{f} h](\la) \; = \; \Int{}{} \f{ \dd \nu }{2 \pi } \mc{F}[\wt{f}](\nu) \mc{F}[h](\la-\nu) \;. 
\enq
Therefore, one has the upper bound 
\beq
\norm{ f h  }_{ \mc{H}_s(\R) }^2  \, = \, \norm{ \wt{f}  h  }_{ \mc{H}_s(\R) }^2  \; \leq \; \Int{ \R^3 }{}  \f{ \dd \la \dd \nu \dd \nu^{\prime} }{ (2\pi)^2 } \big(1+|\la| \big)^{2s}
\big| \mc{F}[\wt{f}](\nu) \big| \cdot \big| \mc{F}[\wt{f}](\nu^{\prime}) \big| \cdot  \big|  \mc{F}[h](\la-\nu) \big|  \cdot  \big|  \mc{F}[h](\la-\nu^{\prime}) \big|  \;. 
\enq
Since  
\beq
\big|  \mc{F}[h](\la-\nu) \big|  \cdot  \big|  \mc{F}[h](\la-\nu^{\prime}) \big|
\, \leq \,  \big|  \mc{F}[h](\la-\nu) \big|^2  +    \big|  \mc{F}[h](\la-\nu^{\prime}) \big|^2  
\enq
one has that 
\bem
\norm{ f h  }_{ \mc{H}_s(\R) }^2 \; \leq \;  \f{2}{ (2\pi)^2 } \Bigg\{  \Int{ \R }{}  \dd \nu  \big| \mc{F}[\wt{f}](\nu ) \big|  \Bigg\} \cdot
\Int{\R^2 }{}   \dd \la \dd \nu \,  \big(1+|\la| \big)^{2s}
\big| \mc{F}[\wt{f}](\nu) \big|   \cdot  \big|  \mc{F}[h](\la-\nu) \big|^2  \\
\, \leq \,  C_{\a}  \Int{\R^2 }{} \dd \la \dd \nu  \f{ \big(1+|\la| \big)^{2s} }{ \big(1+|\nu| \big)^{\a} }
  \cdot  \big|  \mc{F}[h](\la-\nu) \big|^2  \; = \; C_{\a} \Int{\R }{} \dd \nu  \big|  \mc{F}[h](\la) \big|^2  \mc{I}_{s,\a}(\nu)
\end{multline}
 where I used that, for any $\a>0$, $ \big| \mc{F}[\wt{f}](\nu) \big| \leq \tf{ C_{\a} }{ (1+|\nu|)^{\a} }$ since $\mc{F}[\wt{f}]$ is in the Schwartz class. Also, I have introduced 
\beq
  \mc{I}_{s,\a}(\nu) \; = \; \Int{\R }{}\dd \la \f{ \big(1+|\la| \big)^{2s} }{ \big(1+|\la - \nu| \big)^{\a} } \;. 
\enq
Clearly, $ \mc{I}_{s,\a}$ is continuous on $\R$ provided that $\a > 2s + 1$. This choice of $\a$ is assumed in the following. To estimate the large $\nu$ behaviour of $\mc{I}_{s,\a}(\nu)$
it is enough to focus on the case $\nu>1$ since $ \mc{I}_{s,\a}$ is even. Pick $\eps>0$ and small enough. Agreeing upon $K^{\e{c}} \, = \, \R \setminus K$, one has 
\bem
  \mc{I}_{s,\a}(\nu) \; =   \hspace{-5mm}  \Int{ \intff{ \nu(1-\eps) }{ \nu(1+\eps) }^{\e{c}}  }{}  \hspace{-8mm} \dd \la \, \f{ \big(1+|\la| \big)^{2s} }{ \big(1+|\la - \nu| \big)^{\a} } 
\; + \;   \Int{\nu(1-\eps) }{ \nu(1+\eps) }  \dd \la \f{ \big(1+\la \big)^{2s} }{ \big(1+|\la - \nu| \big)^{\a} }  \\
\; =  \nu^{2s+1-\a} \hspace{-5mm}  \Int{  \intff{(1-\eps) }{ (1+\eps) }^{\e{c}}  }{}  \hspace{-8mm} \dd t \, \f{ \big( \tf{1}{\nu}  + |t| \big)^{2s} }{ \big(\tf{1}{\nu} + |t - 1 | \big)^{\a} } 
\; + \;  \nu^{2s+1-\a}  \Int{-\eps }{ \eps }  \dd t  \f{ \big( \tf{1}{\nu} +1 + t  \big)^{2s} }{ \big(\tf{1}{\nu} + |t| \big)^{\a} }  \\ 
\; \leq \; \nu^{2s+1-\a}  \hspace{-5mm}  \Int{  \intff{(1-\eps) }{ (1+\eps) }^{\e{c}}  }{}  \hspace{-8mm} \dd t \,  \big( 1  + |t| \big)^{2s} \cdot |t - 1 |^{-\a}    
\; + \;  2 \nu^{2s+1-\a} (2+\eps)^{2s} \Int{ 0 }{ \eps }  \f{ \dd t }{ \big(\tf{1}{\nu} + t \big)^{\a} }  \\
\; \leq \; C  \nu^{2s+1-\a} +  \f{\nu^{2s+1-\a}}{ \a-1 } \bigg(  \nu^{\a-1} - \f{  1 }{ \big(\tf{1}{\nu} + \eps \big)^{1-\a}  }  \bigg) \; \leq \; C^{\prime} \nu^{2s} \;.
\end{multline}
The last bound allows one to conclude. \qed

\subsection{The functions $\mf{w}^{(\pm)}$}

\begin{lemme}
\label{Lemme calcul integrales elementaires}

 It holds that
\beq
\mf{w}^{(\ups)}(x) \, = \, - \Int{\R}{} \hspace{-1mm} \dd \la   \f{ R^{(\ups)}(\la)  }{ \la }  \, \ex{- \i \la x } \; ,  \quad x \not= 0\,,
\enq
with $R^{(\pm)}$ as defined in \eqref{definition R plus et moins}. Furthermore, given $\mf{v}_{ \a , \eta}$ as in \eqref{definition potentiel varpi tot et correctif v alpha eta}, $0<\a, \eta < \pi$:
\beq
\mc{F}\big[ \mf{v}_{ \a , \eta} \big](\la) \, = \, \Int{}{} \dd x  \,  \ex{\i x \la } \mf{v}_{ \a , \eta}(x) \, = \, - 4\pi
\f{   \sinh\Big( \la \tfrac{\eta-\a}{2}\Big)  \sinh\Big( \la \tfrac{\pi - \eta-\a}{2}\Big)   }{  \la \sinh\Big( \tfrac{\pi \la }{ 2 }  \Big) }  \;.
\enq

\end{lemme}

\Proof

Observe that $x \mapsto \mf{v}_{ \a , \eta}(x)$ is continous on $\R$ and behaves as $\mf{v}_{ \a , \eta}(x)\, = \, \e{O}\big( \ex{-2|x|} \big)$ as $ x \tend \pm \infty$.
Thus, upon carrying out an integration by parts one gets
\beq
\mc{F}\big[ \mf{v}_{ \a , \eta} \big](\la) \, = \,  \f{\i}{\la} \Int{\R }{} \! \dd x \ex{\i x \la} \Big\{ \coth\big[\la - \i (\pi-\a)\big] \, + \,  \coth\big[\la - \i  \a \big]
\, - \, \coth\big[\la - \i (\pi-\eta)\big] \, - \,  \coth\big[\la - \i  \eta \big] \Big\} \;.
\enq
Then, by using the $\i\pi$ quasi-periodicity of the integrand, one may recast the integration over $\R$ into one over $\Dp{}\mc{B}_{\pi}$,
with $\mc{B}_{\pi}=\Big\{ \la \in \Cx \; : \; 0 < \Im(\la) < \pi \Big\}$, this up to multiplying with a pre-factor which takes into account
the quasi-periodicity constant. Then, the integral may be simply taken by the residues associated with the poles $ \i (\pi-\a),  \i (\pi-\eta), \i \a , \i\eta$
located inside of $\mc{B}_{\pi}$. Hence, one gets

\bem
\mc{F}\big[ \mf{v}_{ \a , \eta} \big](\la)  \; = \;
\f{\i}{\la} \Int{ \Dp{} \mc{B}_{\pi} }{}    \f{ \dd x \ex{\i x \la} }{ 1 - \ex{-\pi \la} }  \Big\{ \coth\big[\la - \i (\pi-\a)\big] \, + \,  \coth\big[\la - \i  \a \big]
\, - \, \coth\big[\la - \i (\pi-\eta)\big] \, - \,  \coth\big[\la - \i  \eta \big] \Big\}  \\
\; = \;  \f{ -2 \pi \ex{ \tfrac{\pi}{2} \la } }{ 2 \la \sinh\big[ \tfrac{\pi}{2}\la  \big] } \cdot \Big\{ \ex{-(\pi-\a) }  \, + \,  \ex{-\a \la } \, - \, \ex{-(\pi-\eta)\la} - \ex{-\eta \la} \Big\}
 \; = \; - 4\pi
\f{   \sinh\Big( \la \tfrac{\eta-\a}{2}\Big)  \sinh\Big( \la \tfrac{\pi - \eta-\a}{2}\Big)   }{  \la \sinh\Big( \tfrac{\pi \la }{ 2 }  \Big) } \;.
\end{multline}

In order to discuss the formula for $\mf{w}^{(-)}$, one first observes that
\beq
\f{-1}{4\pi} \mc{F}\big[ \mf{v}_{ 2 \pi\mf{b} , \eta} \big](\la)  \; = \; - \f{\ex{-\eta |\la| } }{ 2 |\la|} \Big( 1 \, + \, \e{O}\big( \ex{-\eps |\la|} \big) \Big)
\enq
for some $\eps>0$ that is $\eta$-independent. Thus, when $x \not=0$, starting from
\bem
\mf{w}^{(-)}(x) \, = \, -\tfrac{1}{2}  \lim_{\eta \tend 0^+} \mf{v}_{ 2 \pi\mf{b} , \eta}(x) \\
\; = \; \lim_{\eta \tend 0^+} \Int{\R}{} \dd \la \ex{-\i x \la} \bigg\{     \mc{F}\big[  \tfrac{ \mf{v}_{ 2 \pi\mf{b} , \eta} }{-4\pi} \big](\la) \bs{1}_{\intff{-1}{1}}(\la)
\, + \, \Big(    \mc{F}\big[  \tfrac{ \mf{v}_{ 2 \pi\mf{b} , \eta} }{-4\pi} \big](\la) + \f{\ex{-\eta |\la| } }{ 2 |\la|}  \Big)\bs{1}_{\intff{-1}{1}^{\e{c}} }(\la)
\, - \,  \f{\ex{-\eta |\la| } }{ 2 |\la|} \bs{1}_{\intff{-1}{1}^{\e{c}} }(\la)  \bigg\} \;.
\end{multline}
The first two terms give rise to absolutely convergent integrals, uniformly in $\eta\geq 0$, and thus by a direct application of dominated convergence,
one may take the limit by simply setting $\eta=0^+$ in these terms. In what concerns the third contribution, one has
\bem
\lim_{\eta \tend 0^+} \Int{\R}{}\dd \la  \f{\ex{-\eta |\la| -\i x \la } }{ 2 |\la|} \bs{1}_{\intff{-1}{1}^{\e{c}} }(\la) \; = \; \sul{\sg=\pm}{} \lim_{\eta \tend 0^+} \Int{1}{+\infty} \dd \la \f{\ex{-\eta \la -\i x \sg \la  } }{ 2 \la}  \\
\; = \; \sul{\sg=\pm}{} \lim_{\eta \tend 0^+}  \Bigg\{  \bigg[   \f{\ex{-(\eta  +\i x \sg) \la  } }{ - 2 \la(\eta + \i\sg x) }  \bigg]_{1}^{+\infty}
\, - \,  \Int{1}{+\infty} \dd \la  \f{\ex{-(\eta  +\i x \sg) \la  } }{ 2 \la^2 (\eta + \i\sg x) } \Bigg\} \\
\; = \; \sul{\sg=\pm}{} \lim_{M \tend +\infty}  \Bigg\{  \bigg[   \f{\ex{-\i x \sg \la  } }{ - 2 \i \la \sg x }  \bigg]_{1}^{ M }
\, - \,  \Int{1}{ M } \dd \la  \f{\ex{-\i x \sg \la  } }{ 2 \i \la^2  \sg x } \Bigg\} \\
\; = \; \sul{\sg=\pm}{} \lim_{M \tend +\infty} \Int{1}{M} \dd \la \f{\ex{ -\i x \sg \la  } }{ 2 \la}
\;= \; \lim_{M \tend +\infty} \Int{-M}{M} \dd \la \f{\ex{ -\i x \la  } }{ 2 |\la| }\bs{1}_{\intff{-1}{1}^{\e{c}} }(\la) \;.
\end{multline}
Thus, by putting together all the limits, one gets the integral representation for $\mf{w}^{(-)}$. The one for $\mf{w}^{(+)}$
follows from the integral representation for $\mf{w}$ given in \eqref{definition potentiel w} and the explicit form for  $\mf{w}^{(-)}$
that was just established.  \qed

\subsection{Regularity of equilibrium measures}

In this subsection, we recall Lemma 2.5 of \cite{KozBorotGuionnetLargeNBehMulIntMeanFieldTh}, this in the context closest to our applications.
We thus consider a lebesgue-continuous probability measure on $\R^M$:
\beq
\dd \mf{p}(\bs{\ga}_M) \, = \, \f{1}{\mc{Z}_{\R^M}[T] } \pl{a<b}{M} |\ga_a-\ga_b|^2 \cdot \exp{ \f{ 1 }{ 2 } \sul{a,b=1}{M} T(\ga_a,\ga_b)  }
\enq

\begin{lemme}\cite{KozBorotGuionnetLargeNBehMulIntMeanFieldTh}
\label{Lemme rappel article BGK15}

 Assume that $T$ is holomorphic in some strip $\big\{ \ga \in \Cx\, : \, |\Im(\ga)|<\eta \big\}^2$ around $\R^2 $
 and that there exists a function $f$ such that, for $(x,y)$ large enough,
\beq
T(x,y) \leq -\big[ f(x) \, + \, f(y) \big] \qquad  and  \qquad \liminf_{x \tend \pm \infty } \Big\{ \f{ f(x) }{ \ln |x| }  \Big\} \; = \; + \infty
\enq
and that
\beq
\mc{E}_{T}[\mu]\;= \; - \Int{\R^2 }{} \Big\{ \f{1}{2} T(x,y)   \, +  \,  \ln |x-y| \Big\} \dd \mu(x) \dd \mu(y)
\enq
 admits a unique minimiser $\mu_{\e{eq}}$  on $\mc{M}^{1}(\R)$.

 Then, $\mu_{\e{eq}}$ is Lebesgue continuous and supported on a union of segments $\op{S}$, none of which is reduced to a point,
and takes the form
\beq
\f{ \dd \mu_{\e{eq}}(x) }{ \dd x } \; = \; \f{ \bs{1}_{\op{S}}(x) }{2\pi} M(x) \sg_0(x) \pl{\a \in \Dp{}\op{S} }{} |x-\a|^{\f{1}{2}} \;.
\enq
There
\begin{itemize}

 \item $x\mapsto M(x)$ is holomorphic in some open neighbourhood of $\op{S}$ in $\Cx$ and such that $M(x)>0$ for $x\in \op{S}$;

 \item $\sg_0$ is a polynomial taking non-negative values on $\op{S}$

\end{itemize}

\end{lemme}

 \section{Notations}

 \begin{itemize}

  \item $\sg^{\pm}$, $\sg^z$ refer to Pauli matrices
\beq
\sg^z \, = \,  \left( \ba{cc}  1 & 0 \\
                                0  & -1 \ea  \right)  \; ,   \quad
\sg^+ \, = \,  \left( \ba{cc}  0 & 1 \\
                                0  & 0 \ea  \right)    \quad \e{and}   \quad
\sg^- \, = \,  \left( \ba{cc}  0 & 0 \\
                                1  & 0 \ea  \right)  \; .
\enq

  \item $\mathbb{H}^{\pm}$ stands for the upper/lower half-plane. 
  
  \item Given $U$ open in $\Cx$, $\mc{O}(U)$ stands for the ring of holomorphic functions on $U$.

 \item Given an oriented curve $\Ga$, its $+$, resp. $-$, side is located to the left, resp. right, when following its orientation. $-\Ga$ refers to the curve $\Ga$ endowed with the 
 opposite orientation.

  \item Given an oriented curve $\ga \subset \Cx$, and $f \in \mc{O}(\Cx\setminus \ga)$, $f_{\pm}$ refer to the $\pm$ boundary values of $f$ on the $\pm$ side of $\ga$, whenever these exist in some suitable sense.  
  
  \item Given an open set $\mc{C}^k(U)$ stands for the space of $k$-times differentiable functions on $U$ while $\mc{C}^k_{\e{c}}(U)$ is the subspace of functions having compact support in $U$. 
  
  \item The Fourier transform is defined on $L^1(\R)$ by 
\beq
\mc{F}[f](\la) \; = \; \Int{\R}{} \dd x f(x) \ex{\i \la x} \;.
\enq

  \item $\mc{H}_s(\R)$ stands for the $s^{\e{th}}$ Sobolev space on $\R$ endowed with the norm
\beq
\norm{f}_{\mc{H}_s(\R)}^2 \, = \; \Int{\R }{} \dd \la (1+|\la|)^{2s} \big| \mc{F}[f](\la) \big|^2  \;.
\enq
  Given $K$ a closed subset of $\R$, $\mc{H}_s(K)$ correspond to the subset of $\mc{H}_s(\R)$ consisting of functions supported on $K$.

   \item One has that for two functions $f(\la)=\e{O}\big( g(\la) \big)$ when $\la \tend \la_0$ means that there exists an open neighbourhood $U$ of $\la$ and $C>0$ such that 
 $f(\la) \, \leq \, C \cdot \big| g(\la) \big|$. In case of matrix functions, the relation $M(\la)=\e{O}\big( N(\la) \big)$ is to be understood entrywise, \textit{viz}. 
 $M_{ab}(\la)=\e{O}\big( N_{ab}(\la) \big)$ for any $a,b$.

 \item $\mc{M}^{1}(\R)$ is the space of probability measures on $\R$. $\mc{M}^{(\a)}_{\mf{s}}(\R)$ is the space of signed measures on $\R$ of total mass $\a$.

 \item Symbols in bold with an index refer to vectors where the index stresses its dimensionality, \textit{viz}. $\bs{x}_{\ell}$ refers to a vector in $\R^{\ell}$. 
  
  \item Given some labelled variables $x_a$, $x_{ab}=x_a-x_b$. 
  
\item Given a set $A\subset \R$, $A^{\e{c}}$ stands for its complement, \textit{viz}. $A^{\e{c}}=\R \setminus A$ and $\bs{1}_A$ stands for the indicator function of $A$.
 
 \item Ratios of products of $\Ga$-functions are expressed by means of hypergeometric like notations:
\beq
\Ga\left(\ba{c} b_1,\dots, b_k \\ a_1,\dots, a_m \ea \right) \; = \; \f{ \pl{s=1}{k} \Ga(b_s) }{ \pl{s=1}{m} \Ga(a_s) }\;. 
\enq

 \end{itemize}

\end{document}